\documentclass[11pt]{article}
\usepackage[dvipdfmx]{graphicx}
\usepackage{amsfonts,amsmath,amssymb,epsf}
\usepackage{amsmath,braket}
\usepackage{graphicx,hyperref,color,xcolor}

\definecolor{awesome}{rgb}{1.0, 0.13, 0.32}
\definecolor{electricblue}{rgb}{0.03, 0.57, 0.82}
\definecolor{guppiegreen}{rgb}{0.0, 1.0, 0.5}
\definecolor{blue-violet}{rgb}{0.54, 0.17, 0.89}

\makeatletter
\def\p@figure{\color{awesome}}
\def\p@equation{\color{electricblue}}
\makeatother

\hypersetup{
    bookmarks=true,         
    unicode=false,          
    pdftoolbar=true,        
    pdfmenubar=true,        
    pdffitwindow=false,     
    pdfstartview={FitH},    
    pdfnewwindow=true,      
    colorlinks=true,       
    linkcolor=blue-violet,          
    citecolor=guppiegreen,        
    filecolor=blue,      
    urlcolor=guppiegreen           
}
\newcommand{\fr}{\frac}

\newcommand{\de}{\partial}
\newcommand{\be}{\begin{equation}}
\newcommand{\ba}{\begin{eqnarray}}
\newcommand{\ea}{\end{eqnarray}}
\newcommand{\ee}{\end{equation}}

\newcommand{\s}{\sqrt}

\newcommand{\ti}{\tilde}
\newcommand{\ap}{\alpha}

\newcommand{\ddd}{\cdot\cdot\cdot}
\newcommand{\no}{\nonumber \\}

\newcommand{\bea}{\begin{eqnarray}}
\newcommand{\eea}{\end{eqnarray}}
\newcommand{\bes}{\begin{equation*}}
\newcommand{\beas}{\begin{eqnarray*}}
\newcommand{\eeas}{\end{eqnarray*}}
\newcommand{\bas}{\begin{array*}}
\newcommand{\eas}{\end{array*}}
\newcommand{\ees}{\end{equation*}}

\newcommand{\ep}{\epsilon}

\newcommand{\pa}[1]{\left(#1 \right)}
\def\dd{{\mathrm{d}}}

\newcommand{\ex}[1]{\mathrm{e}^{#1}}

\usepackage{cite}
\usepackage[top=30truemm,bottom=30truemm,left=25truemm,right=25truemm]{geometry}
\usepackage{comment}
\usepackage[]{hyperref}
\usepackage{amsmath}
\usepackage{amsthm}
  \theoremstyle{definition}

\usepackage{amssymb}
\usepackage[all]{xy}
\usepackage{here}
\usepackage{mathrsfs}
\usepackage{braket} 
\usepackage{mwe}
\usepackage{color}

\def\CO{{\mathcal{O}}}
\def\CT{{\mathcal{T}}}

\usepackage[framemethod=tikz]{mdframed}

\begin{document}

\begin{titlepage}
\thispagestyle{empty}

\begin{flushright}
YITP-21-53\\
IPMU21-0035\\
RIKEN-iTHEMS-Report-21\\
\end{flushright}

\bigskip

\begin{center}
\noindent{{\LARGE \textbf{Holographic moving mirrors}}}
\vspace{1cm}

Ibrahim Akal$^{\star}$,
Yuya Kusuki$^{\star, \dagger}$,
Noburo Shiba$^\star$,
Tadashi Takayanagi$^{\star,\circ,\diamond}$,
Zixia Wei$^\star$
\vspace{1cm}\\

{\it $^\star$Center for Gravitational Physics,\\
Yukawa Institute for Theoretical Physics,
Kyoto University, \\
Kitashirakawa Oiwakecho, Sakyo-ku, Kyoto 606-8502, Japan}\\

{\it $^\dagger$ RIKEN Interdisciplinary Theoretical and Mathematical Sciences (iTHEMS),
Wako, Saitama 351-0198, Japan}

{\it $^\circ$Inamori Research Institute for Science,\\
620 Suiginya-cho, Shimogyo-ku,
Kyoto 600-8411, Japan}\\

{\it $^\diamond$Kavli Institute for the Physics and Mathematics
 of the Universe (WPI),\\
University of Tokyo, Kashiwa, Chiba 277-8582, Japan}

\end{center}

\begin{abstract}
Moving mirrors have been known as tractable setups modeling Hawking radiation from black holes. In this paper, motivated by recent developments regarding the black hole information problem, we present extensive studies of moving mirrors in conformal field theories by employing both field theoretic as well as holographic methods. Reviewing first the usual field theoretic formulation of moving mirrors, we construct their gravity dual by resorting to the AdS/BCFT construction. Based on our holographic formulation, we then calculate the time evolution of entanglement entropy in various moving mirror models. In doing so, we mainly focus on three different setups: escaping mirror, which models constant Hawking radiation emanating from an eternal black hole; kink mirror, which models an evaporating black hole formed from collapse; and the double escaping mirror, which models two constantly radiating eternal black holes. In particular, by computing the holographic entanglement entropy, we show that the kink mirror gives rise to an ideal Page curve. We also find that an interesting phase transition arises in the case of the double escaping mirror. Furthermore, we argue and provide evidence for an interpretation of moving mirrors in terms of two dimensional Liouville gravity. We also discuss the connection between quantum energy conditions and the time evolution of holographic entanglement entropy in moving mirror models.
\end{abstract}

\end{titlepage}

\newpage

\tableofcontents

\section{Introduction}

Considerations of black hole evaporation due to Hawking radiation \cite{Hawking:1974rv,Hawking:1974sw} 
have led to the important question whether unitarity is maintained in gravitational physics. 
A necessary, but not sufficient property, which may indicate that black hole evaporation is a unitary process, is the Page curve for the entanglement entropy of the Hawking radiation \cite{Page:1993df,Page:1993wv}. 
Owing to the holographic fine grained entropy formula \cite{RT,HRT,Faulkner:2013ana,Engelhardt:2014gca}, the Page curve has been obtained for conformal field theories (CFTs) coupled to gravity \cite{Penington:2019npb,Almheiri:2019psf,Almheiri:2019hni}. Similarly, this has also been realized by direct Euclidean gravity computations using quantum corrected, boundary-unitarity-imposing saddles in approximating the gravitational path integral \cite{Penington:2019kki,Almheiri:2019qdq}. 
For further progress along this direction as well as closely related work refer, for instance, to \cite{Akers:2019nfi,Rozali:2019day,Bousso:2019ykv,Nomura:2019dlz,Chen:2019iro,Kusuki:2019hcg,Chen:2019uhq,Laddha:2020kvp,Pollack:2020gfa,Liu:2020gnp,Balasubramanian:2020hfs,Marolf:2020xie,Kim:2020cds,Verlinde:2020upt,Chen:2020wiq,Gautason:2020tmk,Anegawa:2020ezn,Giddings:2020yes,Hashimoto:2020cas,Sully:2020pza,Hartman:2020swn,Hollowood:2020cou,Krishnan:2020oun,Alishahiha:2020qza,Chen:2020uac,Geng:2020qvw,Chandrasekaran:2020qtn,Krishnan:2020fer,Karlsson:2020uga,Chen:2020jvn,Chen:2020tes,Hartman:2020khs,Langhoff:2020jqa,Balasubramanian:2020coy,Balasubramanian:2020xqf,Chen:2020hmv,Nomura:2020ska,Goto:2020wnk,Matsuo:2020ypv,Geng:2020fxl,Miyaji:2021ktr,Geng:2021wcq,Anderson:2021vof,Bhattacharya:2021jrn,Miyata:2021ncm,Geng:2021iyq,Balasubramanian:2021wgd,Kawabata:2021hac,Kawabata:2021vyo}. 

Even though these studies are interesting and might provide further insights into how the black hole information paradox should be resolved, they are lacking a proper description of the Page curve from a fundamental theory of quantum gravity. For example, it has not been addressed how the structure of the Hilbert space in quantum gravity does actually look like. This would indeed be needed to explain the unitary behavior of the fine grained entropy during the evaporation process. 

In fact, the mentioned entropy computations do not use the hypothetical quantum state of the radiation in order to obtain an entropy curve which is consistent with unitarity, see e.g. \cite{Bousso:2020kmy,Akal:2020ujg}. Instead, it is the minimization-extremization procedure \cite{RT,HRT,Faulkner:2013ana,Engelhardt:2014gca} of the generalized entropy \cite{Bekenstein:1972tm,Bekenstein:1973ur,Bekenstein:1974ax} over an extended semiclassical bulk region terminating at some codimension-two surface, which is responsible for the results mentioned above. 

Recall that the generalized entropy is a finite quantity, whereas, both the gravitational entropy as well as the von Neumann entropy defined in the exterior region of the surface determining the former are separately cutoff dependent. However, combining the two pieces would result in a cancellation of such divergences and render the generalized entropy to be cutoff independent and therefore finite \cite{Susskind:1994sm,Jacobson:1994iw,Fursaev:1994ea,Demers:1995dq,Kabat:1995eq,Larsen:1995ax}. This strongly suggests that some of the information content present in the full quantum gravity description is captured by the generalized entropy. 

One may consider the latter as being measuring the degrees of freedom in the exterior region of the dividing black hole horizon surface, where the gravitational part may account for Planckian degrees of freedom regulated by quantum gravitational effects \cite{Frolov:1993ym,Susskind:1994sm,Jacobson:1994iw}. However, it is important reminding that the generalized entropy, as it has originally been introduced by Bekenstein \cite{Bekenstein:1972tm,Bekenstein:1973ur,Bekenstein:1974ax}, is a purely semiclassical concept. Similarly, this is the case for the notion of entanglement wedges \cite{Czech:2012bh,Wall:2012uf,Faulkner:2013ana,Headrick:2014cta,Jafferis:2015del,Dong:2016eik}, where the latter turn out to be crucial for the recent arguments regarding the reproduction of the Page curve. 

Nevertheless, in order to explain the purification of the radiation and its unitary entropy evolution, there has to exist a mechanism going beyond semiclassical bulk physics. This should indeed be responsible for the expected behavior around the Page time and, in particular, for making the generalized entropy finite and continuously decreasing in the presence of an old radiating black hole. In fact, just assigning some coarse grained entropy proportional to the black hole horizon surface area, namely, without addressing whether such degrees of freedom are quantum correlated among each other, or, lead to extrinsic structures, cannot be sufficient in addressing the black hole unitarity problem.

Based on algebraic arguments and related reflections on the bulk Hilbert space structure, it has recently been argued that the purification of the radiation happens via hidden quantum correlations extending across the black hole atmosphere \cite{Akal:2020ujg}. Such correlations are proposed to be existent due to the presence of certain, necessarily extended horizon operators in the emergent bulk spacetime, that are highly quantum correlated with each other and do not belong to the algebra describing the semiclassical world. During the evaporation process, these atmosphere correlations, being transferred to the outgoing low energy Hawking radiation, will give rise to a unitary entropy curve. In fact, using well established theorems and consistency assumptions one can further argue for the necessity of such near horizon quantum gravitational effects that start to be revealed around the halfway point \cite{Akal:2021csd}. 

Accordingly, in contrast to the usual thermal state predicted by Hawking, the state of the radiation would start undergoing a drastic change slightly before half of the black hole has evaporated. Particularly, restoring unitarity as proposed in \cite{Akal:2020ujg} would imply a non-factorized bulk Hilbert space structure, that is dynamically evolving during the evaporation process. Around the halfway point, the entanglement structure of the black hole system, which, by then, will be maximally correlated with the outgoing Hawking radiation, would necessarily extend across the atmosphere. Indeed, this is in line with earlier stringy scenarios drawn in \cite{Susskind:1993ws,Susskind:1994sm,Horowitz:1996nw} which already strengthened the idea that black holes should be viewed as ordinary quantum systems as seen from the far distance. However, the mechanism in \cite{Akal:2020ujg} would be more generically applicable to any quantum formulation of black hole thermodynamics \cite{Bardeen:1973gs} and may therefore explain the universal nature of the Bekenstein-Hawking entropy \cite{Carlip:2008rk}.

Taking all these considerations into account, we think that it will be helpful to explore an alternative approach for studying the information content of black hole radiation. In doing so, we may presumably be able to connect to the latter aspects in a direct manner, namely, without entirely ignoring the relevant fine grained properties of the underlying spacetime. 

Moving mirrors have been considered as a class of remarkable and tractable models of Hawking radiation from black holes \cite{Hawking:1974sw}, which are based on quantum field theories defined on spacetime backgrounds with boundaries \cite{Birrell:1982ix,Davies:1976hi}. Therefore, unitarity is obvious in moving mirror models, and we expect that comparisons between the dynamics in such setups and that of evaporating black holes will be helpful to understand unitarity in quantum gravity. 
Motivated by this, the computation of entanglement entropy in two dimensional CFTs with moving mirrors has recently been performed and a Page curve has been derived for a mirror profile which mimics black hole formation and evaporation \cite{Akal:2020twv}, both field theoretically and holographically. The gravity dual analysis has been achieved by extending the AdS/CFT duality
\cite{Ma} to the case where the CFT is defined on a manifold with a boundary \cite{Karch:2000gx,AdSBCFT,AdSBCFT2}. This extension is known as the AdS/BCFT construction. In holographic CFTs, which have classical gravity duals, there are two different contributions to holographic entanglement entropy \cite{RT,RT2,HRT} in AdS/BCFT. One comes from a connected extremal surface and 
the other candidate relies on a disconnected surface which ends on the end-of-the-world brane extending into the bulk. In \cite{Akal:2020twv}, it has been found that the entanglement entropy which contributes to the Page curve is the latter (disconnected) one. Contributions from the former (connected) one were already evaluated in earlier works \cite{Bianchi:2014qua,Hotta:2015huj,Chen:2017lum,Good:2019tnf}. It is particularly worth mentioning that moving mirror examples mostly allow relatively simple 
and analytical calculations. This can be taken as an advantage over other approaches to black hole information related problems.

\begin{figure}[h!]
  \centering
  \includegraphics[width=.6\textwidth]{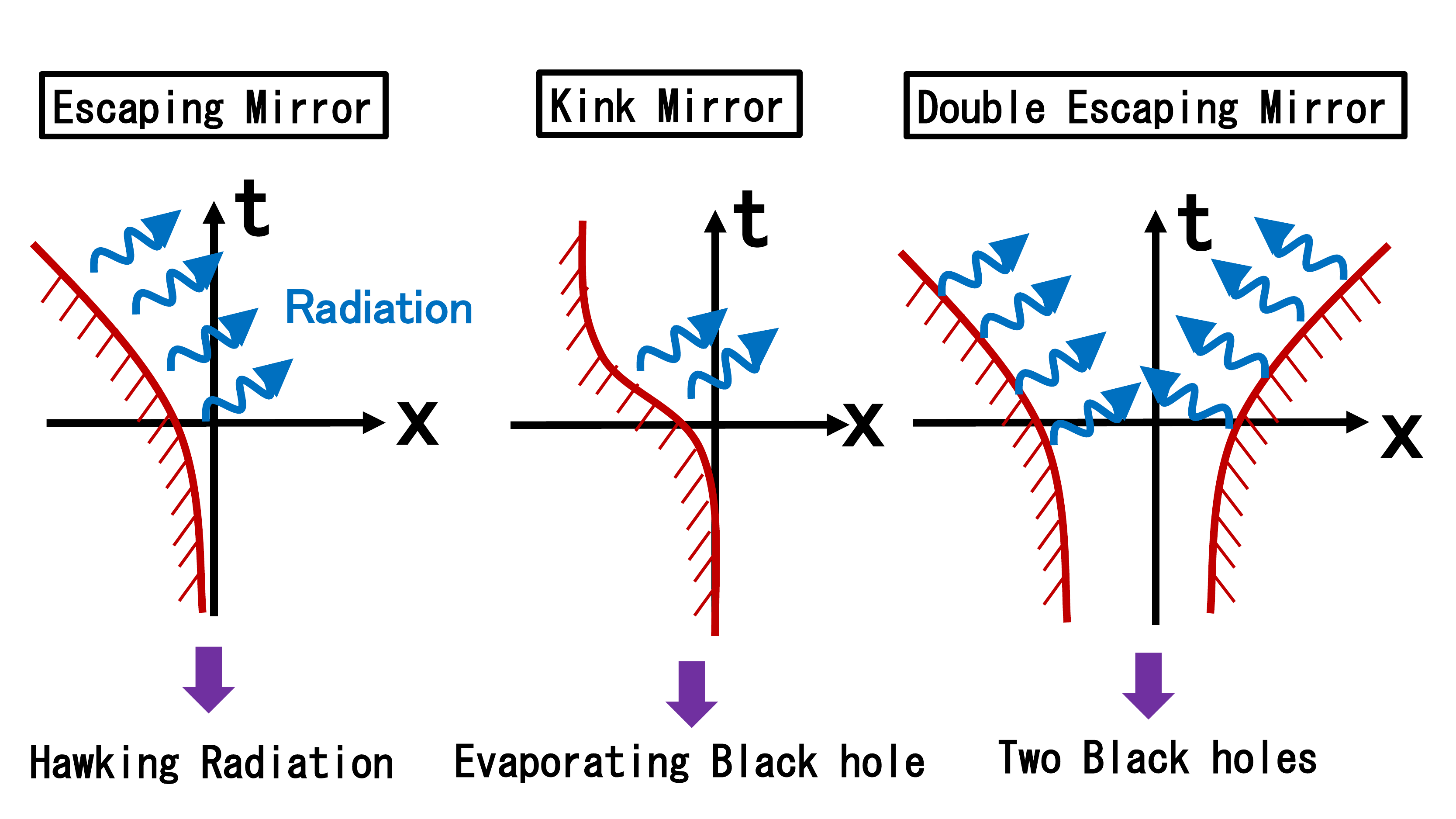}
  \caption{Sketch of the three moving mirror setups: (left) escaping mirror, (center) kink mirror, and (right) double escaping mirror. They model constant Hawking radiation, black hole formation and evaporation, and radiation from two black holes, respectively.}
\label{figmvth}
\end{figure}

In this paper, we would like to provide major extensions of our previous analysis presented in \cite{Akal:2020twv}. For doing so, we mainly focus on three different moving mirror setups: (i) escaping mirror, (ii) kink mirror, and (iii) double escaping mirror, which are depicted in Fig.~\ref{figmvth}. These can be regarded as models mimicking constant Hawking radiation from an eternal black hole, black hole formation and evaporation, and radiation emanating from two black holes, respectively.   
After presenting details of field theoretic and holographic realizations of moving mirror setups, we perform comprehensive investigations of the time evolution of entanglement entropy in all three setups. Moreover, we present a precise connection between our moving mirror models and gravity models coupled to CFTs by making use of brane-world holography and the Liouville formulation of two dimensional gravity. 
For the double escaping mirror, we observe an interesting, new phase transition behavior in our entanglement entropy calculations. We also study the properties of energy fluxes in moving mirror models. Particularly, when these give rise to unitary entropy curves, we find that the quantum null energy condition (QNEC) \cite{Bousso:2015mna,Bousso:2015wca} is saturated, while the null energy condition (NEC) turns out to be violated temporarily.
In fact, this property may be viewed as an explicit realization of more general considerations \cite{Akal:2021csd} which further advocate the mechanism discussed in \cite{Akal:2020ujg}.

The remainder of this paper is organized as follows. In Sec.~\ref{sec:mmCFT2}, we explain the description of moving mirrors in two dimensional CFTs using conformal transformations. In Sec.~\ref{sec:directCFT2}, we present a comprehensive review of direct analysis of two dimensional free scalar CFT with a moving mirror in terms of creation and annihilation operators. In Sec.~\ref{sec:AdS/BCFT-Poincare}, after briefly reviewing the AdS/BCFT construction, we work out profiles of end-of-the-world branes in the BTZ black hole background.
In Sec.~\ref{sec:hmm}, we provide extensive studies of gravity duals of a single moving mirror and calculate the time evolution of holographic entanglement entropy. We further give an interpretation of our setups in terms of two dimensional gravity coupled to a CFT on a half space by referring to brane-world holography and Liouville theory. We also argue that our moving mirror models are smoothly connected to earlier two dimensional gravity models that have led to the island picture. In Sec.~\ref{sec:dem}, we analyze the gravity dual of the double escaping mirror setup. We calculate its holographic entanglement entropy and uncover an interesting new phase transition. In Sec.~\ref{sec:u-ee}, we study properties of energy fluxes in moving mirror CFTs and discuss the connection to well known extensions of the classical NEC, such as the QNEC. In Sec.~\ref{sec:ee-cft}, we provide conformal field theoretic computations of entanglement entropy in moving mirror setups. In Sec.~\ref{sec:rbcft}, we revise our moving mirror analysis as a BCFT, and make clearer the way how the standard conformal boundary conditions are generalized into corresponding moving mirror boundary conditions. In Sec.~\ref{sec:summ}, we give a brief summary and finalize with some further comments. In appendix \ref{sec:appen}, we present details of the gravity dual associated with a conformal map. In appendix \ref{sec:moreEE}, we present more entanglement entropy results computed in CFTs with moving mirrors.

\section{Moving mirror in two dimensional CFTs}
\label{sec:mmCFT2}

In this section, we introduce the conformal map method and discuss its application in the context of various mirror profiles.

\subsection{Radiation from moving mirror}

\begin{figure}[h!]
  \centering
  \includegraphics[width=.3\textwidth]{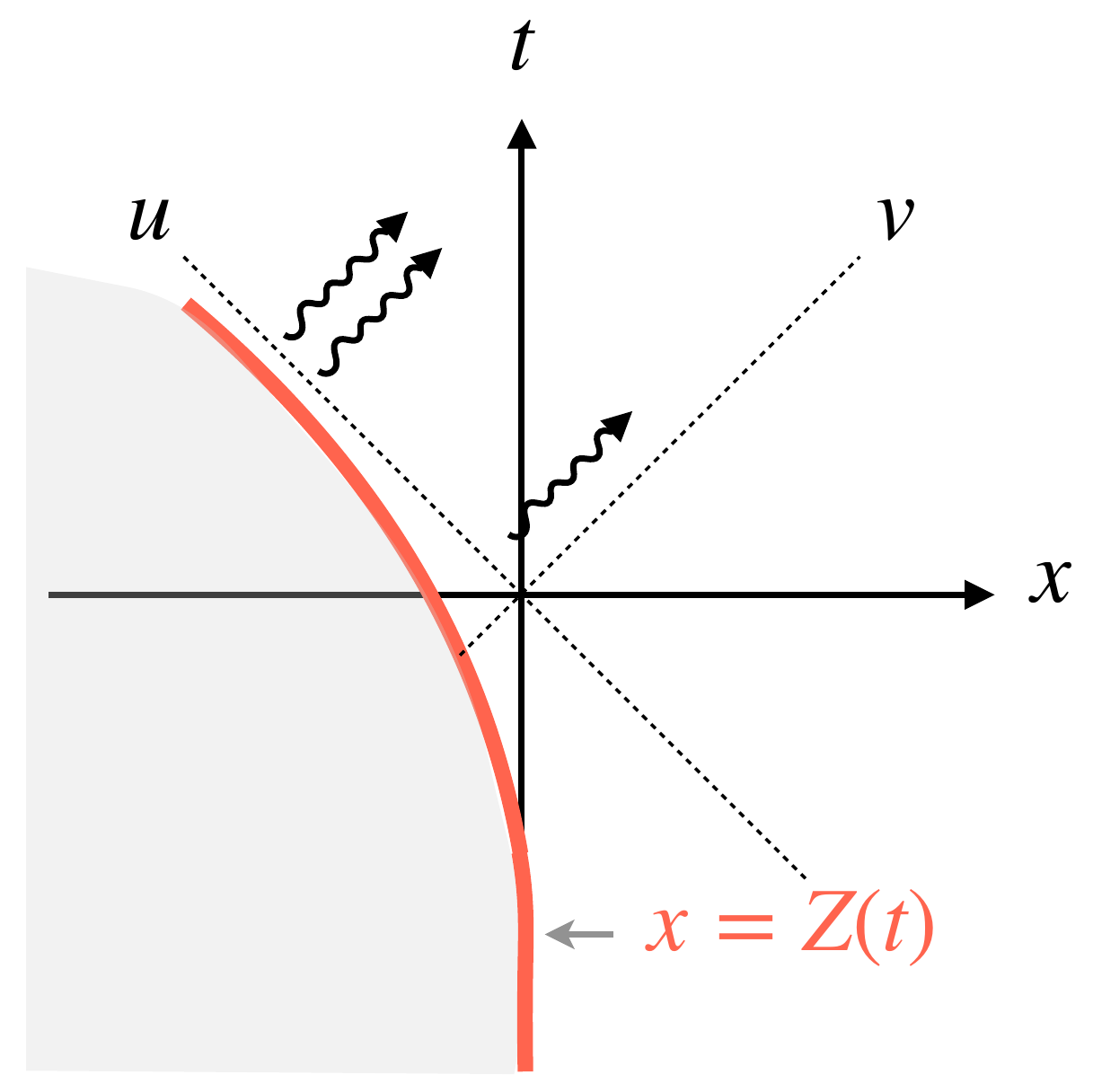}\qquad
 \includegraphics[width=.3\textwidth]{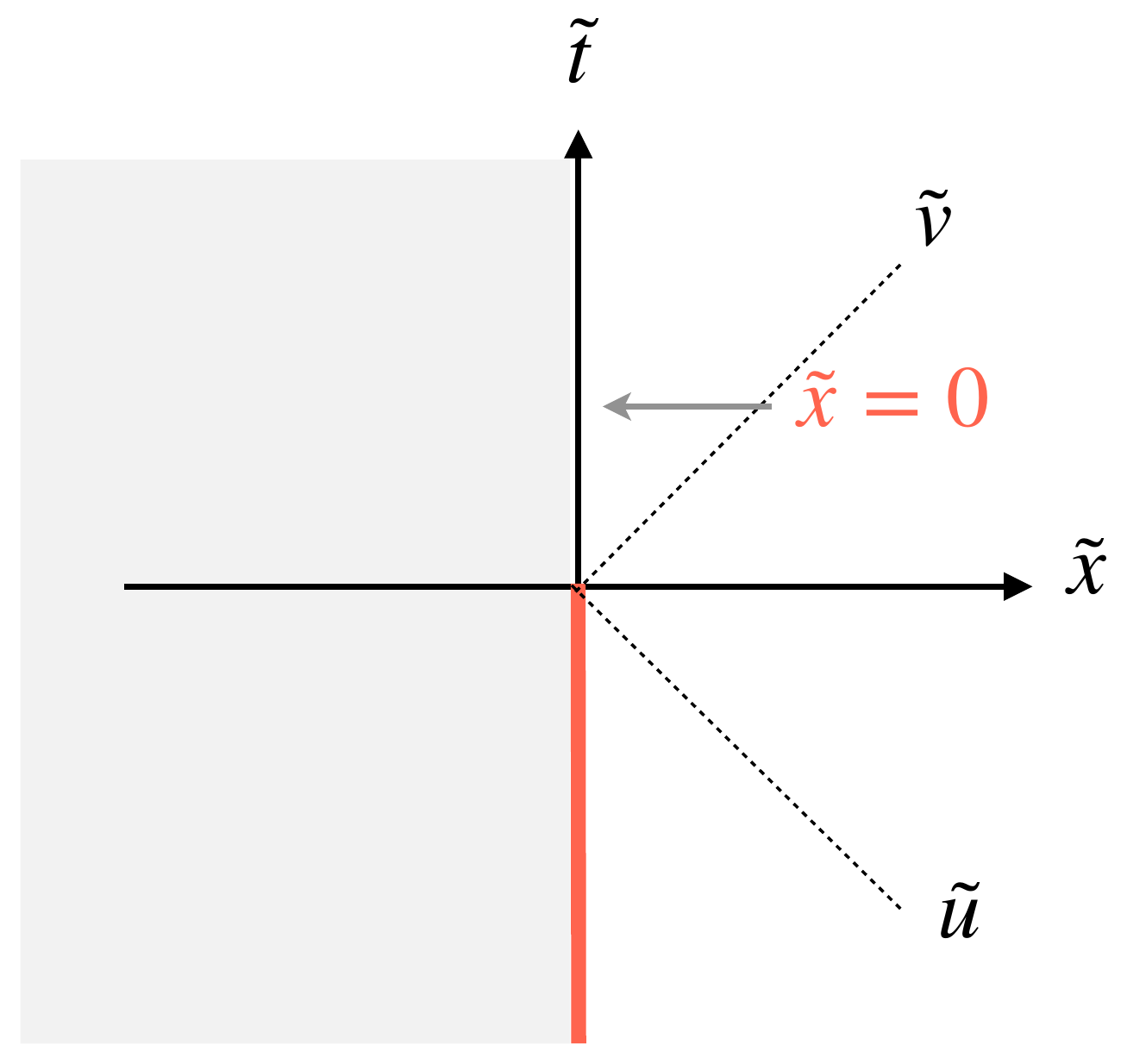}
  \caption{Sketch of conformal map from the moving mirror setup (left) to a standard setup of BCFT with a straight line boundary at $\tilde x = 0$ (right). The dotted lines correspond to the null lines in the coordinates $(t,x)$ and $(\tilde t, \tilde x)$.}
\label{MVfig}
\end{figure}

In a typical moving mirror setup, one initially places a mirror at $x=0$, so that the physical space is given by $x>0$. The location of the mirror moves according to $x=Z(t)$, as sketched in the left picture in Fig.~\ref{MVfig}. 
In such a setup, we expect the production of right moving radiation quanta giving rise to some energy flux $T_{uu}$ \cite{Birrell:1982ix}.

\subsection{Conformal map method}
\label{subsec:cmm}

A systematic method to analyze moving mirrors in two dimensional CFTs can be achieved by employing conformal transformations. We would like to consider a conformal map from the original moving mirror setup, see left picture in Fig.~\ref{MVfig}, here described by the following null coordinates
\be
\begin{split}
u =t-x,\qquad 
v =t+x,
\end{split}
\label{eq:null}
\ee
into the standard simple setup with a static mirror, see right picture in Fig.~\ref{MVfig}. The latter is described by the new coordinates 
\be
(\ti{u},\ti{v})=(\ti{t}-\ti{x},\ti{t}+\ti{x}),
\ee
so that the static boundary will be situated at $\ti{x}=\frac{1}{2}(\ti{v}-\ti{u})=0$. 

Since we are mainly interested in the radiation emitted by the moving mirror, we would like to consider the most simple case, in which only the right-movers are excited. In this case, these coordinates are related by a chiral conformal transformation of the following form
\ba
\ti{u}=p(u),\qquad \ti{v}=v,  
\label{cmapf}
\ea
where the introduced function $p$ in \eqref{cmapf} will be determined by the profile of the mirror $x=Z(t)$ as follows
\ba
t+Z(t)=p\left(t-Z(t)\right).
\label{eq:pZ-rel}
\ea
If we rewrite the relation \eqref{eq:pZ-rel} in terms of the functions $f(u)\equiv p^{-1}(u)$, we end up with
\ba
\frac{\tilde u-f(\tilde u)}{2}=Z\left(\frac{\tilde u+f(\tilde u)}{2}\right),\qquad
\frac{p(u) - u}{2}=Z\left(\frac{u+p(u)}{2}\right).
\label{eq:pZrel}
\ea
Since the state in $(\ti{u},\ti{v})$ corresponds to the CFT vacuum, we would have 
$T_{\ti{u}\ti{u}}=T_{\ti{v}\ti{v}}=0$. 

Therefore, after performing the conformal transformation \eqref{cmapf}, we can evaluate the energy flux expressed in terms of Schwarzian derivatives\footnote{
Here, we use the standard formula for the conformal transformation of the energy stress tensor,
\begin{align*}
T_{uu}=\left(\frac{d\ti{u}}{du}\right)^2 T_{\ti{u}\ti{u}}+\frac{c}{12}\{\ti{u},u\},  \label{emtens}
\end{align*}
where 
\begin{align*}
\{\ti{u},u\}=\frac{ (\de^3_u\ti{u}) ( \de_u\ti{u})  -\frac{3}{2}(\de^2_u\tilde u)^2 } { (\de_u\ti{u})^2 }.
\end{align*}
}, which takes the form
\ba
T_{uu} =
\frac{c}{24 \pi} \left( \frac{3}{2} \left(  \frac{p''(u)}{p'(u)} \right)^2 - \frac{p'''(u)}{p'(u)}  \right),
\label{eq:Tuu-gen}
\ea
with $c$ being the central charge of the underlying CFT.

Note that, since $T_{vv}$ and $T_{uv}$ are vanishing, we have $T_{tt}=T_{uu}+T_{vv}-2T_{uv}=T_{uu}$.

\subsection{Mirror trajectories}
\label{subsec:MM-profiles}

As mentioned, in the following, we would like to discuss some explicit examples for the mirror profiles in the light of the conformal map method.

\subsubsection{Rindler observer}
\label{subsec:ex1}

Consider the following function
\ba
Z(t)= \s{t^2 + \alpha^2},
\label{eq:ex1-Zprof}
\ea
which is analogous to the boundary experienced by some Rindler observer.
In such a case, we find
\ba
p(u)=- \frac{\alpha^2}{u}.
\ea
Since this is a global conformal transformation, we find the trivial result
\ba
T_{uu}=0.
\ea

\subsubsection{Perturbed Rindler observer} 
\label{subsec:ex2}

Consider now the following function
\ba
p(u) = - \frac{\alpha^2}{u} - \gamma u.
\label{eq:ex2-pfunc}
\ea
We get the profile
\ba
Z(t) = \frac{t \gamma - \sqrt{ t^2 + \alpha^2 - \gamma \alpha^2  }}{ \gamma - 1} 
\approx
 \sqrt{t^2 + \alpha^2} + \beta \left[ -t + \frac{2 t^2 + \alpha^2}{2 \sqrt{t^2 + \alpha^2}}  \right] + \mathcal{O}(\gamma^2).
\label{eq:ex2-Zprof}
\ea
The energy stress tensor reads as follows
\ba
T_{uu} = \frac{c}{12 \pi} \cdot \frac{3 \gamma \alpha^2 }{(\gamma u^2 - \alpha^2)^2}.
\ea

\subsubsection{Escaping mirror} 
\label{subsec:ex4}

\begin{figure}[h!]
  \centering
 \includegraphics[width=.08\textwidth]{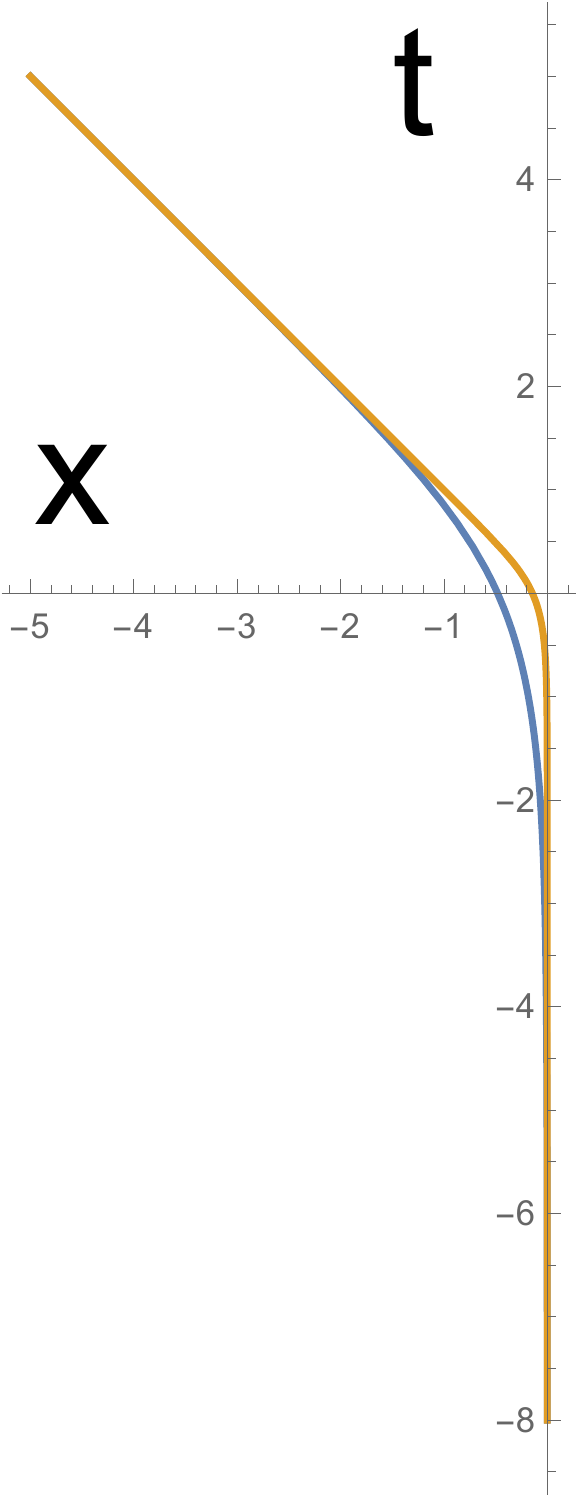}\qquad\qquad
 \includegraphics[width=.35\textwidth]{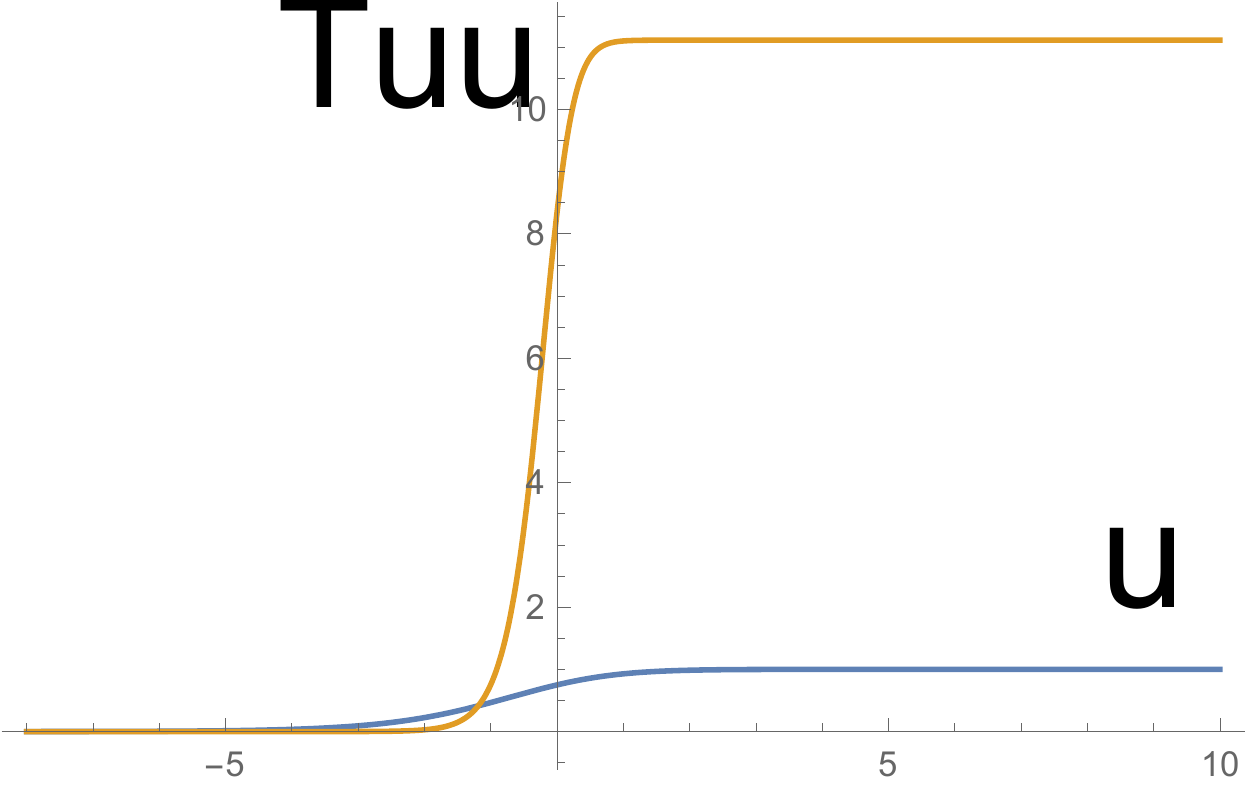}
  \caption{Depicted is the profile of the escaping mirror trajectory (left) and that of the energy stress tensor $T_{uu}(u)$ multiplied by $48\pi$ (right).
The blue and orange curves correspond to $\beta=1$ and $\beta=0.3$, respectively. }
\label{puqfig}
\end{figure}

Let us now consider 
\ba
p(u)=-\beta\log(1+e^{-u/\beta}),  
\label{exfr}
\ea
where we have introduced a positive parameter $\beta$, which will play the role of an effective temperature of the radiation.
Note that in the limit $u\to\infty$, we have
\ba
p(u)\simeq -\beta e^{-u/\beta},
\label{eq:skmm-approx}
\ea
while in the past time limit $u\to -\infty$, we have
\ba
p(u)\simeq u.
\ea
The mirror trajectory is described by 
\ba
t(u) = \frac{p(u) + u}{2},\qquad Z(u) = \frac{p(u) - u}{2},
\label{eq:t(u)Z(u)}
\ea 
which is plotted in the left panel of Fig.~\ref{puqfig}.
In the past time limit $u\to-\infty$, we have the static mirror, i.e.
\ba
Z\simeq 0,
\ea
while in the late time limit $u\to \infty$, we find
\ba
Z\simeq -t-\beta e^{-2t/\beta}.
\ea
The energy stress tensor reads
\ba
T_{uu}=\frac{c}{48\pi\beta^2}\left(1-\frac{1}{(1+e^{u/\beta})^2}\right).
\ea
It is particularly worth mentioning that the late time limit looks thermal $T_{uu}\simeq \frac{c}{48\pi\beta^2}$, similar to as one would get from a thermal density matrix with temperature $\frac{1}{2\pi\beta}$ .

Here, it is useful to consider the (infinite temperature) limit $\beta\to 0$. In this case, the mirror trajectory develops a kink of the form
\ba
Z(t)\simeq -t\theta(t),
\ea
as we can see in Fig.~\ref{puqfig}, and the energy flux is written in terms of the step function $\theta$, i.e.
\ba
T_{uu}=\frac{c}{48\pi\beta^2}\theta(t).
\ea 
\begin{figure}[h!]
  \centering
 \includegraphics[width=.08\textwidth]{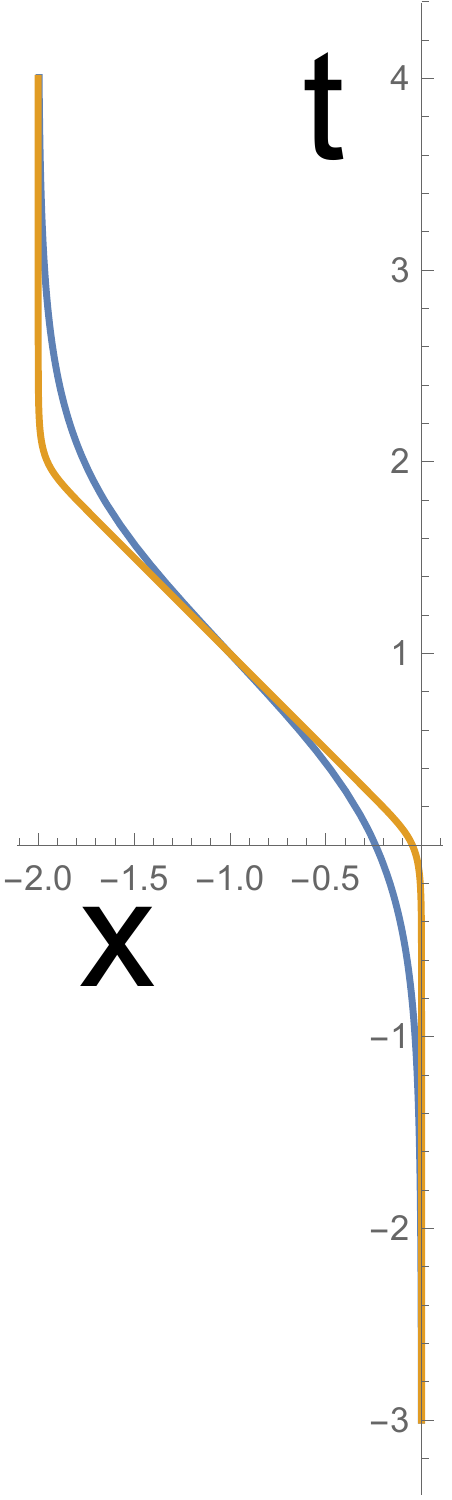}\qquad\qquad
 \includegraphics[width=.35\textwidth]{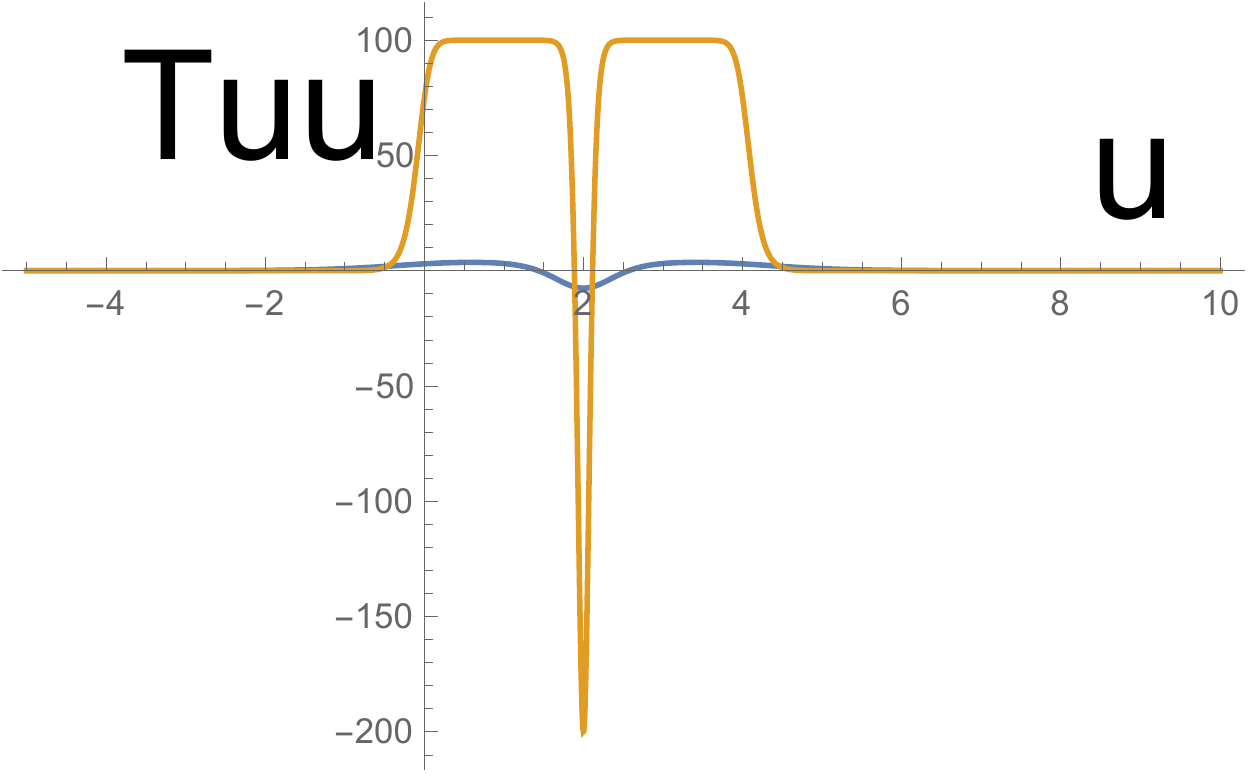}
  \caption{Shown is the profile of the kink mirror trajectory (left) and that of the energy stress tensor $T_{uu}(u)$ multiplied by $48\pi$ (right). We have set $u_0=4$. 
  The blue and orange curves correspond to $\beta=0.5$ and $\beta=0.1$, respectively.}
\label{kinkfig}
\end{figure}

\subsubsection{Kink mirror} 
\label{subsubsec:double-kink-MM}

We now consider the following extended function
\ba
p(u)=-\beta\log(1+e^{-u/\beta})+\beta\log(1+e^{(u-u_0)/\beta}), \label{kexfr}
\ea
where we assume $\beta>0$ and $u_0>0$.

This corresponds to a moving mirror trajectory $x=Z(t)$, which develops a kink-like shape as depicted in the left panel of Fig.~\ref{kinkfig}. In the $\beta\to 0$ limit, the mirror profile is described by 
\ba
&& Z(t)\simeq 
\begin{cases} 
     0  & (t<0) \\
      -t & (0\geq t \leq \frac{u_0}{2}) \\
      -\frac{u_0}{2} & (t>\frac{u_0}{2}) 
   \end{cases}.
\ea
The corresponding energy flux obtained by plugging \eqref{kexfr} into \eqref{eq:Tuu-gen} is plotted in the right panel of Fig.~\ref{kinkfig}.

\subsection{Perturbation}

In general, by using \eqref{eq:pZrel} we can write the following relation
\ba
\dot Z =   \frac{p' - 1}{p' + 1}\quad \Rightarrow\quad Z = p(t-Z) - t
\label{eq:Z(t)-from-p(u)}
\ea
in order to obtain $Z(t)$ from a given function $p(u)$,
where the dot and prime denote $t$ and $u$ derivatives, respectively.

Similarly, we may find $p(u)$ from $Z(t)$ by solving the following equation
\ba
p' = \frac{1 + \dot Z}{1 - \dot Z}\quad \Rightarrow \quad p = 2 Z \left(\frac{u + p}{2}  \right) + u.
\ea

For instance, let us assume that $p$ in \eqref{cmapf} is of the form
\ba
p(u) = s(u) + \lambda g(u),
\label{eq:pert-ans}
\ea
where $s$ and $g$ are some arbitrary functions and $\lambda$ is some real parameter.

Starting from the general expression in \eqref{eq:pert-ans} and assuming $\lambda \ll 1$, we may  expand \eqref{eq:Z(t)-from-p(u)}, so that
\ba
\dot Z \simeq \frac{s' - 1}{s' + 1} + \lambda \frac{2 g'}{[s' + 1]^2} + \mathcal{O}(\lambda^2).
\label{eq:Zdot-approx}
\ea

\subsubsection{Example 1} 

Let us consider the first term in \eqref{eq:pert-ans} to be given by
\ba 
s(u) = \alpha u.
\ea
Then, according to \eqref{eq:Zdot-approx}, we get
\ba
\dot Z  \simeq   \frac{\alpha - 1}{\alpha + 1} + \lambda \frac{2 g'}{[\alpha + 1]^2}.
\label{eq:Zdot-ex1}
\ea
Using the relation
\ba
du \simeq \frac{2}{1 + \alpha} dt,
\ea
which follows from the relation in \eqref{eq:t(u)Z(u)}, then \eqref{eq:Zdot-ex1} can be integrated as follows
\ba
Z(t)\simeq \left[  \frac{\ap-1}{\ap+1} \right] t+\frac{\lambda}{\ap+1} g\left(\frac{2 t}{\ap+1}\right).
\ea

\subsubsection{Example 2} 

As another example, let us now assume that 
\ba 
s(u) = - \frac{\alpha^2}{u}
\label{eq:s(u)-ex2}
\ea
determines the first term in \eqref{eq:pert-ans}.
Then, we get
\ba
\dot Z  \simeq   \frac{\alpha^2 - u^2}{\alpha^2 + u^2} + \lambda \frac{2 g' u^4}{[\alpha^2 + u^2]^2}.
\label{eq:ex2-Zdot}
\ea
We may write the following relation
\ba
d u \simeq \frac{2 u^2}{\alpha^2 + u^2} dt
\quad \Rightarrow\quad u - \frac{\alpha^2}{u} = 2 t.
\label{eq:ex2-dudt}
\ea
Re-expressing the LHS of \eqref{eq:ex2-Zdot} by making use of \eqref{eq:ex2-dudt}, we get
\ba
Z'  \simeq   \frac{\alpha^2 - u^2}{2 u^2} + \lambda \frac{g' u^2}{\alpha^2 + u^2}.
\ea
After integrating the latter, we find the general solution
\ba
Z(u) \simeq - \frac{1}{2} \left[ u + \frac{\alpha^2}{u} \right]
+ \lambda \int^u dy\ \frac{y^2 g'(y)}{\alpha^2 + y^2}.
\ea
By using the relation
\ba
u(t) = t - \sqrt{\alpha^2 + t^2},
\label{eq:u(t)-ex2}
\ea
which is deduced from the right equation in \eqref{eq:ex2-dudt}, we finally find
\ba
Z(t) \simeq \sqrt{t^2 + \alpha^2} 
+ \lambda \int^{u(t)} dy\ \frac{y^2 g'(y)}{\alpha^2 + y^2},
\label{eq:Z(t)-ex2}
\ea
where $u(t)$ is given in \eqref{eq:u(t)-ex2}.

For instance, if we have $\lambda \rightarrow 0$, the expression \eqref{eq:Z(t)-ex2} for $Z(t)$ becomes identical to \eqref{eq:ex1-Zprof}. If we set $g(u) = - u$ as in \eqref{eq:ex2-pfunc}, we find
\ba
\int^{u(t)} dy\ \frac{-y^2}{\alpha^2 + y^2} 
=
\alpha \mathrm{arctan}\left( \frac{ u(t) }{\alpha} \right) - u(t),
\ea
which approaches the RHS of \eqref{eq:ex2-Zprof} if one identifies the small expansion parameter $\lambda$ with $\beta$.

For instance, we may choose the function
\ba
g(y) = \alpha^2 y + \frac{y^3}{3}
\ea
appearing in \eqref{eq:pert-ans} and further assume \eqref{eq:s(u)-ex2}. Then, by using the solution in \eqref{eq:Z(t)-ex2}, we find
\ba
Z(t) \simeq \sqrt{t^2 + \alpha^2} + \lambda \frac{[t - \sqrt{\alpha^2 + t^2} ]^3}{3},
\ea
where 
\ba
p(u) \simeq - \frac{\alpha^2}{u} + \lambda \left[ u  \alpha^2 + \frac{u^3}{3} \right],
\ea
which leads to the following energy stress tensor
\ba
T_{uu} = - \frac{\lambda c}{12 \pi} \cdot \frac{10 u^2 \alpha^2 + 3 \alpha^4 - 2 u^6 \lambda + u^4 \alpha^2 \lambda}{[\alpha^2 + u^2 [u^2 + \alpha^2]]^2}.
\ea
As it is expected from the discussion in Sec.~\ref{subsec:ex1}, the latter vanishes for $\lambda \rightarrow 0$.

\section{Direct analysis of massless free scalar CFT}
\label{sec:directCFT2}

In the present section, we discuss the explicit quantization of a free massless scalar field in two dimensional flat spacetime with a moving mirror. We first calculate the expectation value of the energy momentum tensor. Afterwards, we work out the relevant Bogoliubov coefficients for the escaping mirror profile modeling a one sided eternal black hole, and compute the particle production rate. Lastly, we comment on connections to gravitational setups.

\subsection{Energy momentum tensor}
\label{subsec:emt}

In two dimensional free massless scalar field theory, the energy momentum tensor is given by \cite{Davies:1976hi}
\begin{equation}
\begin{split}
&T_{tt}=T_{xx}=\frac{1}{2}\left[\left(\partial_t \phi \right)^2 +\left(\partial_x \phi \right)^2\right], \\
&T_{tx}=T_{xt}=\frac{1}{2}\left[\left(\partial_t \phi \right)\left(\partial_x \phi \right) +\left(\partial_x \phi \right)\left(\partial_t \phi \right) \right].
\end{split}  \label{energy momentum tensor}
\end{equation}
The field equation reads\footnote{Here, $\Phi$ is used to denote an on-shell configuration of the scalar field $\phi$.}
\begin{equation}
\begin{split}
\frac{\partial^2 \Phi}{\partial u \partial v}=0.
\end{split}  \label{wave equation}
\end{equation}
The scalar field $\Phi$ shall satisfy the reflection boundary condition, i.e.
\begin{equation}
\begin{split}
\Phi(t, Z(t))=0,
\end{split}  \label{boundary condition}
\end{equation}
where $Z(t)$ denotes the moving boundary of the spatial direction, cf. Sec.~\ref{subsec:cmm}.

A complete set of positive frequency modes, i.e. solutions to \eqref{wave equation} and \eqref{boundary condition}, is given by
\ba
\phi_\omega = H(\omega v) + G(\omega u),
\label{complete set gen}
\ea
where $\omega=|k|$, with $H$ and $G$ being arbitrary functions. 
More specifically, we may write the incoming mode as
\ba
\phi^\text{in}_{\omega'} (t, x) = i(4\pi \omega')^{-1/2}\left(e^{-i\omega' v}-e^{-i\omega' (2 \tau_u - u)}\right).
\label{complete set}
\ea
For simplicity, we consider a single moving boundary, where the mirror location is determined by $2 \tau_u - u = p(u)$, see Sec.~\ref{subsec:gen-mode} for a more general setup. The variable $\tau_u$ is defined by the mirror trajectory through $\tau_u - Z(\tau_u) = u$, cf. \cite{Birrell:1982ix}. An incoming mode function describing Minkowski spacetime with a static boundary would correspond to $\tau_u = u$, hence
\ba
\phi_{\omega'}^\text{in}(t, x)=(\pi \omega')^{-1/2}\sin (\omega' x) e^{-i\omega' t}.
\label{in mode}
\ea

The field $\Phi$ can be expanded in terms of the incoming mode
\begin{equation}
\begin{split}
\Phi(t, x)=\int_0^{\infty} d\omega' \left[a_{\omega'}\phi^{\text{in}}_{\omega'}+a_{\omega'}^{\dagger}\phi_{\omega'}^{\text{in} *} \right].
\end{split}  
\label{mode expansion}
\end{equation}
Here, $a_{\omega'}$ and $a_{\omega'}^{\dagger}$ are the standard annihilation and creation operators, such that $a | 0 \rangle_\text{in} = 0$, where $| 0 \rangle_\text{in}$ is the incoming vacuum state at $\mathcal{I}^-$.

Alternatively, $\Phi$ can also be expanded in terms of the outgoing mode
\begin{equation}
\begin{split}
\Phi(t, x)=\int_0^{\infty} d\omega \left[b_{\omega}\phi^{\text{out}}_{\omega}+b_{\omega}^{\dagger}\phi_{\omega}^{\text{out} *} \right],
\end{split}  
\label{mode expansion out}
\end{equation}
where $b_{\omega}$ and $b_{\omega}^{\dagger}$ correspond to the annihilation and creation operators defined with respect to the outgoing vacuum state at $\mathcal{I}^+$. 

The outgoing mode can be expressed as
\ba
\phi^\text{out}_{\omega} (t, x) = i(4\pi \omega)^{-1/2}\left(e^{-i\omega f(v)}-e^{-i\omega u}\right),
\label{complete set out}
\ea
where $f(u) = p^{-1}(u)$.

So, forming a complete set, one can expand the positive frequency modes $\phi^\text{out}_\omega$ at $\mathcal{I}^+$ in terms of the positive frequency modes $\phi^\text{in}_{\omega'}$ at $\mathcal{I}^-$, means
\ba
\phi^\text{out}_\omega = \int d\omega'\ \left[  \alpha^*_{\omega \omega'} \phi^{\text{in}}_{\omega'} - \beta_{\omega \omega'} \phi_{\omega'}^{\text{in} *} \right],
\ea
where 
\ba
\alpha_{\omega \omega'} &= \left( \phi^\text{out}_\omega, \phi^\text{in}_{\omega'} \right),\qquad
\beta_{\omega \omega'} &= - \left( \phi^\text{out}_\omega, \phi^{\text{in} *}_{\omega'} \right)
\label{Bogo}
\ea 
are the corresponding Bogoliubov coefficients, which will be discussed in more detail below.

Similarly, one may construct
\ba
\phi^\text{in}_{\omega'} = \int d\omega\ \left[  \alpha_{\omega \omega'} \phi^{\text{out}}_{\omega} + \beta_{\omega \omega'} \phi_{\omega}^{\text{out} *} \right].
\ea

The scalar product introduced in \eqref{Bogo} normalizing the mode functions is defined as 
\ba
\begin{split}
(\phi_{1},\phi_{2})= -i \int_{\Sigma} \left[ \phi_{1} \partial_{\mu} \phi_{2}^{*} -( \partial_{\mu} \phi_{1}) \phi_{2}^{*} \right] d \Sigma^{\mu},
\end{split}  
\label{scalar product}
\ea
where $\Sigma$ is a Cauchy surface. The mode functions are orthonormal with respect to the scalar product in \eqref{scalar product}, i.e.
\ba
(\phi_{\omega},\phi_{\omega'}) &= \delta(\omega - \omega'),\qquad
(\phi_{\omega},\phi^*_{\omega'}) & = 0.
\ea
In general, we have
\ba
\begin{split}
\langle T_{\mu \nu} \rangle = \int_0^{\infty} d\omega\ T_{\mu \nu} \left[ \phi_\omega, \phi^{*}_\omega  \right].
\end{split}  
\label{energy momentum tensor 2}
\ea
According to the integrand above, the expressions in \eqref{energy momentum tensor} should be evaluated as bilinear forms on the mode functions $\phi_{\omega}$ and their complex conjugates.

Since \eqref{energy momentum tensor 2} is divergent, one may use the point splitting regularization method. Namely, instead of evaluating both modes, $\phi_\omega$ and $\phi^{*}_\omega$, in \eqref{energy momentum tensor 2} at the same point $(t, x)$, one may evaluate $\phi_\omega^{*}$ at $(t + \epsilon, x)$, where $\epsilon$ is taken to be an infinitesimally small shift parameter. 

Therefore, focusing on the expectation value for the energy momentum tensor at $\mathcal{I}^+$, the factors that need to be inserted into \eqref{energy momentum tensor 2}, where we use the mode function in \eqref{complete set}, take the form 
\ba
\begin{split}
\frac{\partial \phi^\text{in}_\omega}{\partial t}(t, x) &=\left(\frac{\omega}{4\pi} \right)^{1/2}\left(e^{-i\omega v}- p'(u)e^{-i\omega p(u)}\right), \\
\frac{\partial \phi^\text{in}_\omega}{\partial x}(t, x) &=\left(\frac{\omega}{4\pi} \right)^{1/2}\left(e^{-i\omega v}+ p'(u)e^{-i\omega p(u)}\right), \\
\frac{\partial \phi^{\text{in} *}_\omega }{\partial t}(t+\epsilon, x) &=\left(\frac{\omega}{4\pi} \right)^{1/2}\left(e^{i\omega (v+\epsilon)}- p'(u+\epsilon)e^{i\omega p(u+\epsilon)}\right), \\
\frac{\partial \phi^{\text{in} *}_\omega }{\partial x}(t+\epsilon, x) &=\left(\frac{\omega}{4\pi} \right)^{1/2}\left(e^{i\omega (v+\epsilon)}+ p'(u+\epsilon)e^{i\omega p(u+\epsilon)}\right).
\end{split}  
\label{derivatives of field}
\ea

By using \eqref{energy momentum tensor 2}, we get the following expectation value for the energy momentum tensor 
\ba
\begin{split}
\langle T_{t t} \rangle &=\langle T_{x x} \rangle = \frac{1}{4\pi} \int_0^{\infty} d\omega \left[e^{i\omega \epsilon}+p'(u)p'(u+\epsilon) e^{i\omega (p(u+\epsilon)-p(u))} \right], \\
\langle T_{t x} \rangle &=\langle T_{x t} \rangle = \frac{1}{4\pi} \int_0^{\infty} d\omega \left[e^{i\omega \epsilon}-p'(u)p'(u+\epsilon) e^{i\omega (p(u+\epsilon)-p(u))} \right].
\end{split}  \label{energy momentum tensor 3}
\ea
Finally, evaluating the integrals in \eqref{energy momentum tensor 3} gives rise to \cite{Davies:1976hi}
\ba
\begin{split}
\langle T_{t t} \rangle &=  -\frac{1}{2\pi \epsilon^2}  -\langle T_{t x} \rangle, \\
\langle T_{t x} \rangle 
&=\frac{1}{24\pi}\left[\frac{p'''}{p'}-\frac{3}{2} \left(\frac{p''}{p'}\right)^2 \right]+\mathcal{O}(\epsilon).
\end{split}  
\label{energy momentum tensor 4}
\ea

\subsection{Renormalized energy flux}
\label{subsec:ren-emt}

After having obtained the expectation values \eqref{energy momentum tensor 4}, what remains to be done is taking the limit $\epsilon \to 0$. 
This leads to the following renormalized expectation value for the energy stress tensor, cf. e.g. \cite{Birrell:1982ix},\footnote{We would like to note, that other than in the present Sec.~\ref{sec:AdS/BCFT-Poincare}, we denote the renormalized vacuum expectation value $\langle T_{\mu \nu} \rangle_\text{ren}$ simply as $T_{\mu \nu}$ throughout this paper.}
\ba
\langle T_{tt} \rangle_\text{ren}
= \lim_{\epsilon \rightarrow 0} \left[ \langle  T_{tt} \rangle - \langle T_{tt} \rangle_\text{no MM} \right]
= \frac{1}{24\pi}\left[\frac{3}{2} \left(\frac{p''}{p'}\right)^2 - \frac{p'''}{p'} \right].
\label{eq:renT}
\ea
Here, it has been used that
\ba
\langle T_{tt} \rangle_\text{no MM} = - \frac{1}{2 \pi \epsilon^2},
\ea
which corresponds to the value obtained in flat spacetime without a moving boundary, i.e. $p(u) = u$, see Sec.~\ref{subsec:emt}.

The renormalized energy flux at $\mathcal{I}^+$ from \eqref{eq:renT} should be understood as
\ba
\langle T_{tt} \rangle_\text{ren} \equiv \langle \Psi | T_{tt} | \Psi \rangle - \langle 0 | T_{tt} | 0 \rangle,
\label{ren-substract}
\ea
with
\ba
|\Psi \rangle = U | 0 \rangle
\ea
being the final state, where $| 0 \rangle = | 0 \rangle_\text{in}$ corresponds to the vacuum state at $\mathcal{I}^+$. Here, $U$ is a unitary operator.

\subsection{General mode function}
\label{subsec:gen-mode}

Consider a general positive frequency solution to \eqref{wave equation} of the form
\ba
\begin{split}
\phi^\text{in}_\omega(t, x)=i(4\pi \omega)^{-1/2}\left(e^{-i\omega q(v)}-e^{-i\omega p(u)}\right).
\end{split}  
\label{complete set2}
\ea
We calculate $\langle T_{\mu \nu} \rangle$ in the same way as in \eqref{energy momentum tensor 4}. This leads to 
\ba
\begin{split}
&\langle T_{t t} \rangle =\langle T_{x x} \rangle =  -\frac{1}{4\pi } \left[\frac{p'(u)p'(u+\epsilon)}{(p(u+\epsilon)-p(u))^2} +\frac{q'(v)q'(v+\epsilon)}{(q(v+\epsilon)-q(v))^2} \right]  \\
&=-\frac{1}{4\pi } \left[ \frac{2}{\epsilon^2}+\frac{1}{6} \left[\frac{p'''}{p'}-\frac{3}{2} \left(\frac{p''}{p'}\right)^2 \right] +\frac{1}{6} \left[\frac{q'''}{q'}-\frac{3}{2} \left(\frac{q''}{q'}\right)^2 \right] + \mathcal{O}(\epsilon) \right] , \\
&\langle T_{t x} \rangle =\langle T_{x t} \rangle = \frac{1}{4\pi } \left[\frac{p'(u)p'(u+\epsilon)}{(p(u+\epsilon)-p(u))^2} -\frac{q'(v)q'(v+\epsilon)}{(q(v+\epsilon)-q(v))^2} \right] \\
&=\frac{1}{4\pi } \left[ \frac{1}{6} \left[\frac{p'''}{p'}-\frac{3}{2} \left(\frac{p''}{p'}\right)^2 \right] -\frac{1}{6} \left[\frac{q'''}{q'}-\frac{3}{2} \left(\frac{q''}{q'}\right)^2 \right] + \mathcal{O}(\epsilon) \right],
\end{split}  
\label{energy momentum tensor 6}
\ea
where we have used
\ba
\frac{p'(u)p'(u+\epsilon)}{(p(u+\epsilon)-p(u))^2} = \frac{1}{\epsilon^2}+\frac{1}{6} \left[\frac{p'''}{p'}-\frac{3}{2} \left(\frac{p''}{p'}\right)^2 \right]+\mathcal{O}(\epsilon) .
\label{expansion}
\ea
As before, we take the limit $\epsilon \to 0$, which results in
\ba
\langle T_{t x} \rangle =\langle T_{x t} \rangle =\frac{1}{4\pi } \left[ \frac{1}{6} \left[\frac{p'''}{p'}-\frac{3}{2} \left(\frac{p''}{p'}\right)^2 \right] -\frac{1}{6} \left[\frac{q'''}{q'}-\frac{3}{2} \left(\frac{q''}{q'}\right)^2 \right]  \right].
\label{energy momentum tensor 7}
\ea
For the remaining components, we again distract the $\langle 0 | T_{tt} | 0 \rangle$ and $\langle 0 | T_{xx} | 0 \rangle$ contributions as in \eqref{ren-substract}, and find the following renormalized energy flux
\ba
\langle T_{t t} \rangle_\text{ren} =\langle T_{x x} \rangle_\text{ren} =-\frac{1}{4\pi } \left[ \frac{1}{6} \left[\frac{p'''}{p'}-\frac{3}{2} \left(\frac{p''}{p'}\right)^2 \right] +\frac{1}{6} \left[\frac{q'''}{q'}-\frac{3}{2} \left(\frac{q''}{q'}\right)^2 \right]  \right]
\label{energy momentum tensor 8}
\ea
after taking the limit $\epsilon \rightarrow 0$.

\subsection{Bogoliubov coefficients}

Let us write down the Bogoliubov coefficients more explicitly. We consider an incoming mode of the general form
\ba
\phi^\text{in}_{\omega'} (t, x) = i(4\pi \omega')^{-1/2}\left(e^{-i\omega' v}-e^{-i\omega' p(u)}\right),
\label{complete set incoming}
\ea
and an outgoing mode
\ba
\begin{split}
\phi^\text{out}_\omega(t, x)=i(4\pi \omega)^{-1/2}\left(e^{-i\omega f(v)}-e^{-i\omega u}\right),
\end{split}  
\label{complete set2 out}
\ea
where $f = p^{-1}$.

Using the definitions in \eqref{Bogo}, we obtain the following integral representations for the coefficients,
\ba
\begin{split}
\alpha_{\omega \omega'}
&=-i(2\pi)^{-1} (\omega'\omega)^{-1/2} \int_{0}^{\infty}dx \sin \omega'x \left[(\omega f'+\omega')e^{i\omega f} - (\omega p'+\omega')e^{i\omega p} \right]  \\
&= (2\pi)^{-1} \left( \frac{\omega'}{\omega}\right)^{1/2} \int_{0}^{\infty} dx \left[e^{-i\omega'x+i\omega f(x)} + e^{i\omega'x+i\omega p(-x)} \right]
\end{split}  
\label{Bogolubov coefficients alpha}
\ea
and
\ba
\begin{split}
\beta_{\omega \omega'}
&= i(2\pi)^{-1} (\omega'\omega)^{-1/2} \int_{0}^{\infty}dx \sin \omega'x \left[(\omega f'-\omega')e^{-i\omega f} - (\omega p'-\omega')e^{-i\omega p} \right] \\
&= (2\pi)^{-1} \left( \frac{\omega'}{\omega}\right)^{1/2} \int_{0}^{\infty} dx \left[e^{-i\omega'x-i\omega f(x)} + e^{i\omega'x-i\omega p(-x)} \right] .
\end{split}  
\label{Bogolubov coefficients beta}
\ea
Note that the scalar product has been calculated on the $t=0$ slice by performing integration by parts.

\subsection{Thermal spectrum}
\label{subsec:thermal-spectrum}

Now, consider the escaping mirror profile from Sec.~\ref{subsec:ex4}. We may approximate the inverse function of the asymptotic form \eqref{eq:skmm-approx} as \ba
\begin{split}
f(v) \simeq -\frac{1}{B} \ln[-Bv/A]  .
\end{split}  
\label{f ex4}
\ea
By using the expression \eqref{Bogolubov coefficients beta}, we get
\ba
\begin{split}
|\beta_{\omega \omega'}|^{2}= \left(4\pi^2 \omega \omega'\right)^{-1} e^{-\pi \omega/B} |\Gamma(1+i\omega /B)|^{2}=  \frac{1}{2\pi B \omega'} \left( \frac{1}{e^{\omega/k_B T}-1} \right) ,
\end{split}  
\label{Bogolubov coefficients beta 2}
\ea
where $B = 2 \pi k_B T$. This precisely matches with the well known black body spectrum characteristic for black hole radiation as obtained by Hawking \cite{Hawking:1974sw}. 

The average number of particles found at future null infinity $\mathcal{I}^+$ with frequency $\omega$ follows from evaluating the integral
\ba
_\text{in} \langle 0 | N_\omega^\text{out} | 0 \rangle_\text{in} = \int_0^\infty d\omega'\ |\beta_{\omega \omega' }|^2.
\ea

\subsection{Reinforcing the analogy}

The analogy between moving mirrors and black hole radiation in a gravitational setup becomes more clear if one directly compares the exact Bogoliubov coefficients on both sides. In the following, we briefly comment on two such examples. 

Indeed, as we have encountered in Sec.~\ref{subsec:thermal-spectrum}, moving mirrors commonly give rise to a thermal radiation spectrum. An appropriately chosen setup, mimicking an eternally radiating black hole, for instance, leads to a constant thermal flux of particles, see Sec.~\ref{subsec:ex4}.

A precise matching between the Bogoliubov coefficients on both sides, however, may further support our expectations put forward in Sec.~\ref{subsec:grav-int}. In there, we argue that the moving mirror setup may indeed be interpreted as a field theory coupled to gravity, thus providing a direct connection to two dimensional (quantum) gravity. 

\subsubsection{Collapsing null shell}

\begin{figure}[h!]
  \centering
 \includegraphics[width=.2\textwidth]{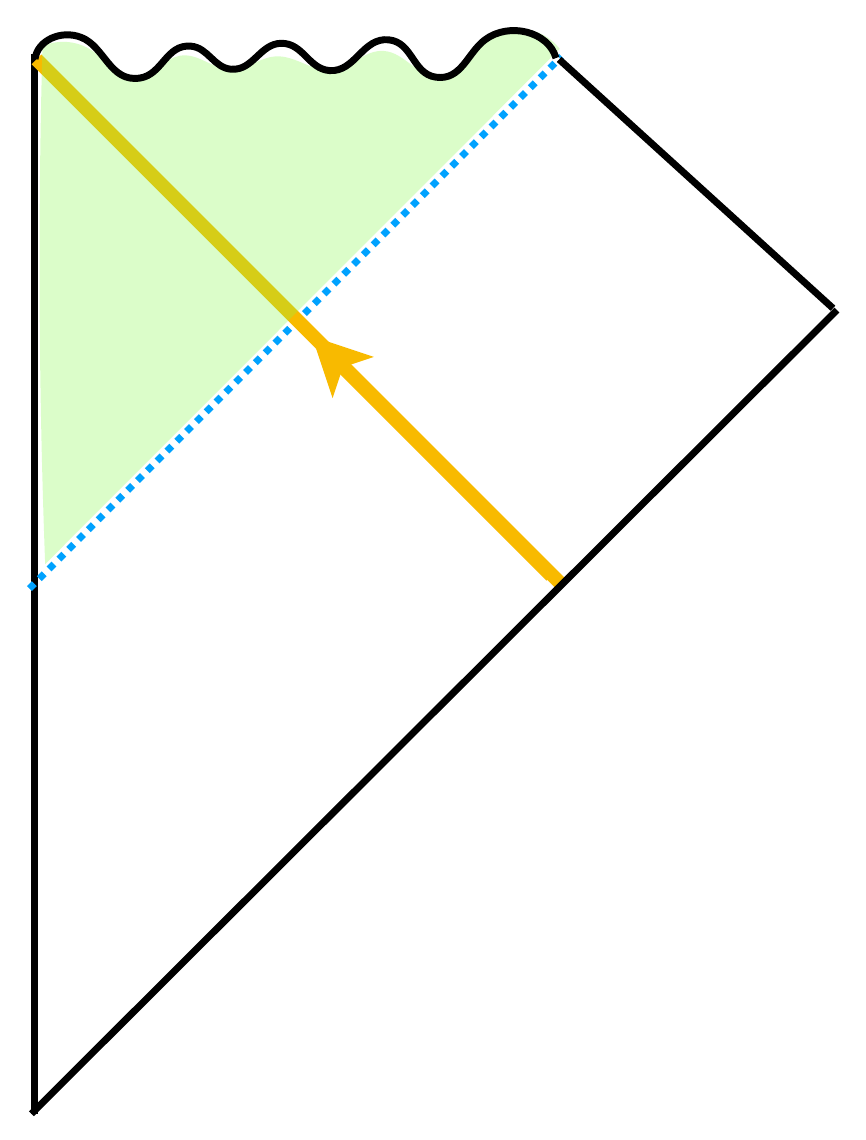}\qquad\qquad\qquad
 \includegraphics[width=.23\textwidth]{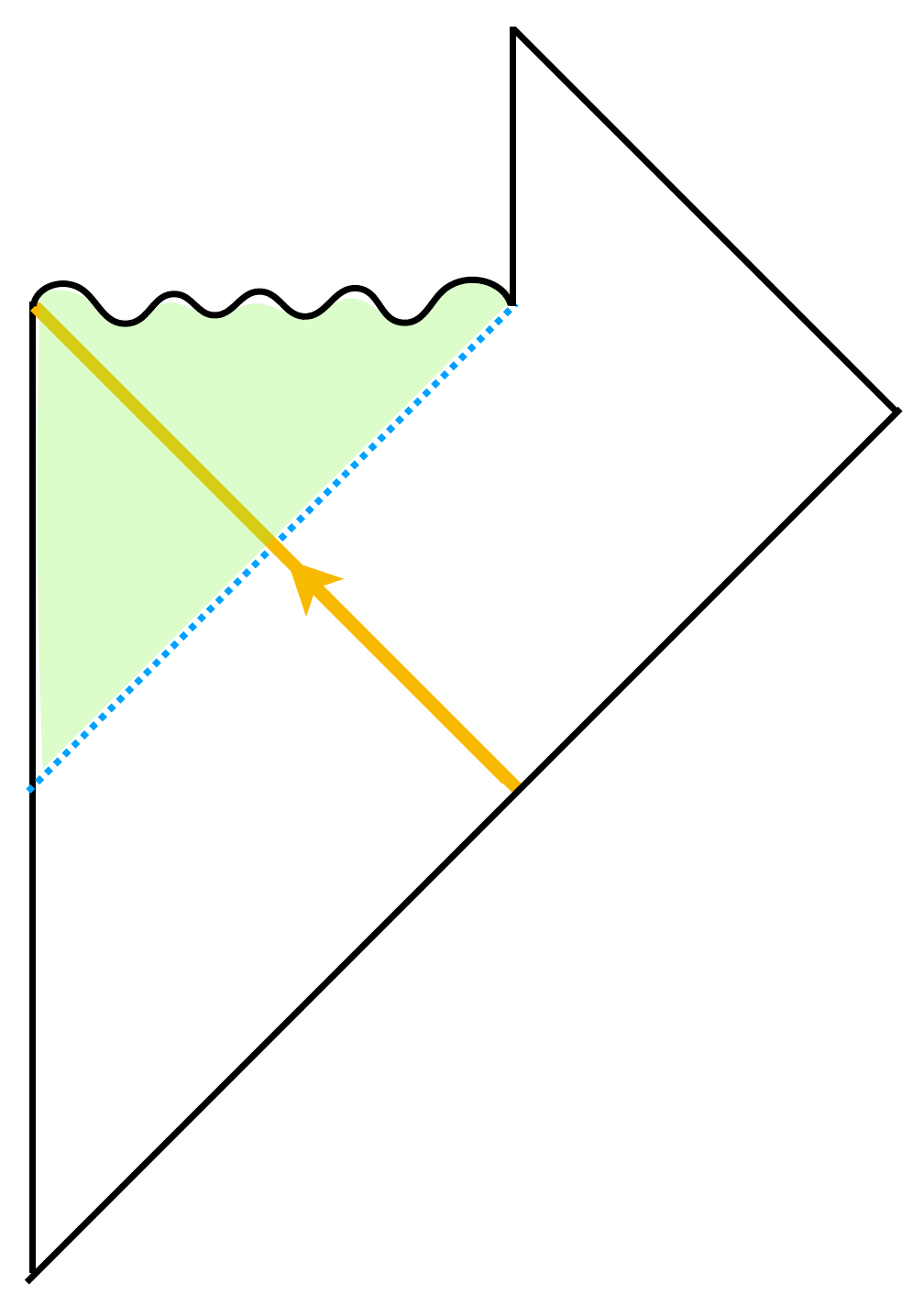}
  \caption{Left: Penrose diagram of a black hole formed from a collapsing null shell (yellow thick line). This scenario may be modeled by the escaping mirror from Sec.~\ref{subsec:ex4}. Right: Penrose diagram of the formation and evaporation of a black hole. This scenario may be modeled by the kink mirror from Sec.~\ref{subsubsec:double-kink-MM}. The region behind the event horizon is colored in green.}
\label{fig:collapseBH}
\end{figure}

A well studied case is the example of a black hole that is formed from a collapsing null shell, cf. Fig.~\ref{fig:collapseBH} left. 
Consider, for instance, the following mirror profile \cite{Carlitz:1986nh}
\ba
Z(t) = - t - \frac{W(2 e^{- 2 \kappa t})}{2 \kappa},
\label{eq:Carlitz}
\ea
where we have introduced the $W$ Lambert function\footnote{The Lambert function (also known as the product logarithm) is defined to be the inverse relation of the map
\begin{equation*}
f: z \mapsto z \exp(z),
\end{equation*}
where, in general, $z \in \mathbb{C}$. It follows
\begin{equation*}
z = W(z) \exp(W(z)).
\end{equation*}
Here, we consider the case where the argument is real. 
}. 
The profile \eqref{eq:Carlitz} is equivalent to the function
\ba
p(u) = -\frac{1}{\kappa} e^{-\kappa u}.
\ea
In this situation, the underlying integrals for the Bogoliubov coefficients are exactly solvable, and one arrives at the following two expressions 
\ba
\begin{split}
\alpha_{\omega \omega'} &= - \frac{i}{2 \pi \kappa} \sqrt{\frac{\omega}{\omega'}}
\left[ - \frac{i}{\kappa} (\omega - \omega')  \right]^{-i\omega/\kappa} \Gamma\left( \frac{i \omega}{\kappa} \right),\\
\beta_{\omega \omega'} &= - \frac{1}{2 \pi \kappa} \sqrt{\frac{\omega}{\omega'}}
\left[ - \frac{i}{\kappa} (\omega + \omega')  \right]^{-i\omega/\kappa} \Gamma\left( \frac{i \omega}{\kappa} \right).
\label{eq:cns-Bogo}
\end{split}
\ea
Here, $\Gamma$ denotes the standard Gamma function. For simplicity, we have used $\omega' \gg \omega$.

Similar to what we have already seen in Sec.~\ref{subsec:thermal-spectrum}, for the spectrum one finds
\ba
\begin{split}
|\beta_{\omega \omega'}|^{2} =  \frac{1}{2\pi \kappa \omega'} \left( \frac{1}{e^{2 \pi \omega/\kappa}-1} \right).
\end{split}  
\ea
Interestingly, the expressions \eqref{eq:cns-Bogo} indeed precisely match with the coefficients obtained for a collapsing null shell in $1+1$ dimensions, cf. e.g. \cite{Good:2016oey}. Here, the acceleration parameter $\kappa$ in \eqref{eq:Carlitz} would play the role of the black hole's surface gravity and thus its Hawking temperature.

We should note that the coefficients \eqref{eq:cns-Bogo} approximate the one obtained for the escaping mirror in Sec.~\ref{subsec:ex4}. The latter will therefore describe the same situation as well. Accordingly, the kink mirror profile introduced in Sec.~\ref{subsubsec:double-kink-MM} would then model Hawking radiation of an evaporating black hole that has initially formed from collapse. 

Quite importantly, this would even include nontrivial back-reacting effects which are expected to be necessary for unitarity. The diagram describing this scenario is sketched in the right panel of Fig.~\ref{fig:collapseBH}.

\subsubsection{CGHS black hole}

A second profile which highlights a close similarity between a moving mirror setup and black hole radiation in a two dimensional gravitational theory is described by the following function
\ba
Z(t) =  - \frac{1}{\kappa} \mathrm{arcsinh}\left( \frac{e^{\kappa t}}{2} \right),
\label{eq:mirror-CGHS}
\ea
see e.g. \cite{Good:2013lca}.\footnote{In fact, in the late time limit, this approaches the trajectory of the escaping mirror introduced before.}
As in the example before, one can explicitly compute the corresponding Bogoliubov coefficients. The one which determines the produced particle spectrum takes the form
\ba
\beta_{\omega \omega'} = \frac{1}{2 \pi \kappa} \sqrt{\frac{\omega'}{\omega}}  B\left( - \frac{i(\omega + \omega')}{\kappa} , 1 + \frac{i \omega}{\kappa}\right),
\label{eq:beta-CGHS}
\ea
such that
\ba
|\beta_{\omega \omega'}|^ 2 = \frac{1}{4 \pi^2 \kappa^2} \frac{\omega'}{\omega} \left| B\left( \frac{i(\omega + \omega')}{\kappa} , 1- \frac{i \omega}{\kappa}\right) \right|^2.
\ea
Here, we have used the notation $B(x,y) \equiv \frac{\Gamma(x)\Gamma(y)}{\Gamma(x+y)}$. Using again $\omega' \gg \omega$, the latter gives rise to the usual thermal spectrum
\ba
|\beta_{\omega \omega' }|^2 = \frac{1}{2 \pi \kappa \omega'} \left(  \frac{1}{e^{\omega/k_B T}  -1}  \right),
\ea
where $\kappa = 2 \pi k_B T$.

More interestingly, \eqref{eq:beta-CGHS} precisely coincides with the Bogoliubov coefficient which is found for the radiating black hole in the well known CGHS model \cite{Callan:1992rs},
\ba
|\beta_{\omega \omega'}|^ 2 = \frac{1}{4 \pi^2 \Lambda^2} \frac{\omega'}{\omega} \left| B\left( \frac{i(\omega + \omega')}{\Lambda} , 1- \frac{i \omega}{\Lambda}\right) \right|^2,
\ea
namely, if $\Lambda$ is identified with the acceleration parameter $\kappa$ in \eqref{eq:mirror-CGHS}, cf. e.g. \cite{fabbri2005modeling}.

\subsection{Entangled pair production}

For later analysis of entanglement entropy in the presence of a moving mirror, 
it is useful to examine the entangled pair production. 

In particular, we would like to 
locate the region where the entangled pairs are created due to the moving mirror.
This can be estimated by calculating the correlation function between the Hawking pair operator 
$\int dw\ a_{\omega}^{\mathrm{in} \dagger} a_{\omega}^{\text{in} \dagger}$ and a pair of scalar 
field operators, i.e.
\begin{equation}
\begin{split}
&_\mathrm{in} \bra{0} \phi(u_1,v_1) \phi(u_2,v_2) \int dw\ a_{\omega}^{\mathrm{in} \dagger} a_{\omega}^{\mathrm{in} \dagger} \ket{0}_\mathrm{in} \\
&=-\frac{1}{2\pi} \int dw\ \frac{1}{\omega} \left[e^{-i\omega(v_1+v_2)} - e^{-i\omega(v_1+p(u_2))} - e^{-i\omega(p(u_1)+v_2)} +e^{-i\omega(p(u_1)+p(u_2))} \right] \\
&= \frac{1}{2\pi}\ln  \left[\frac{(p(u_1)+p(u_2)-i\epsilon)(v_1+v_2-i\epsilon)}{(v_1+p(u_2)-i\epsilon)(p(u_1)+v_2-i\epsilon)} \right].
\end{split}  
\label{Bogolubov coefficients beta 21}
\end{equation}
As can be seen, the overlap gets divergent when $v_1+p(u_2)=0$ or $v_2+p(u_1)=0$.
This behavior can be explained if entangled pairs are produced at 
\ba
v+p(u)=0,  \label{entpai}
\ea
and if they are propagating in opposite directions at the speed of light.

\section{AdS/BCFT in Poincar\'e \texorpdfstring{AdS$_3$}{AdS3}}
\label{sec:AdS/BCFT-Poincare}

In what follows later, we will construct the gravity dual of CFTs in the presence of moving mirrors by employing the AdS/BCFT duality \cite{AdSBCFT,AdSBCFT2}.
As a preparation, in this section, we would like to review the AdS/BCFT construction in  Poincar\'e AdS, and present a calculation of holographic entanglement entropy 
in a simple moving mirror setup, which does not generate any energy flux. As it is mainly the case in this paper, we focus on two dimensional BCFT.

\subsection{AdS/BCFT construction}

Consider a BCFT on a two dimensional manifold $\Sigma$ with a boundary $\de \Sigma$. Its gravity dual, based on AdS/BCFT, can be constructed as follows. 

First, we consider an extension of  $\de \Sigma$
to a two dimensional surface $Q$ in a three dimensional bulk. This surface $Q$ is also called the end-of-the-world brane in recent contexts.

\begin{figure}[h!]
  \centering
 \includegraphics[width=8cm]{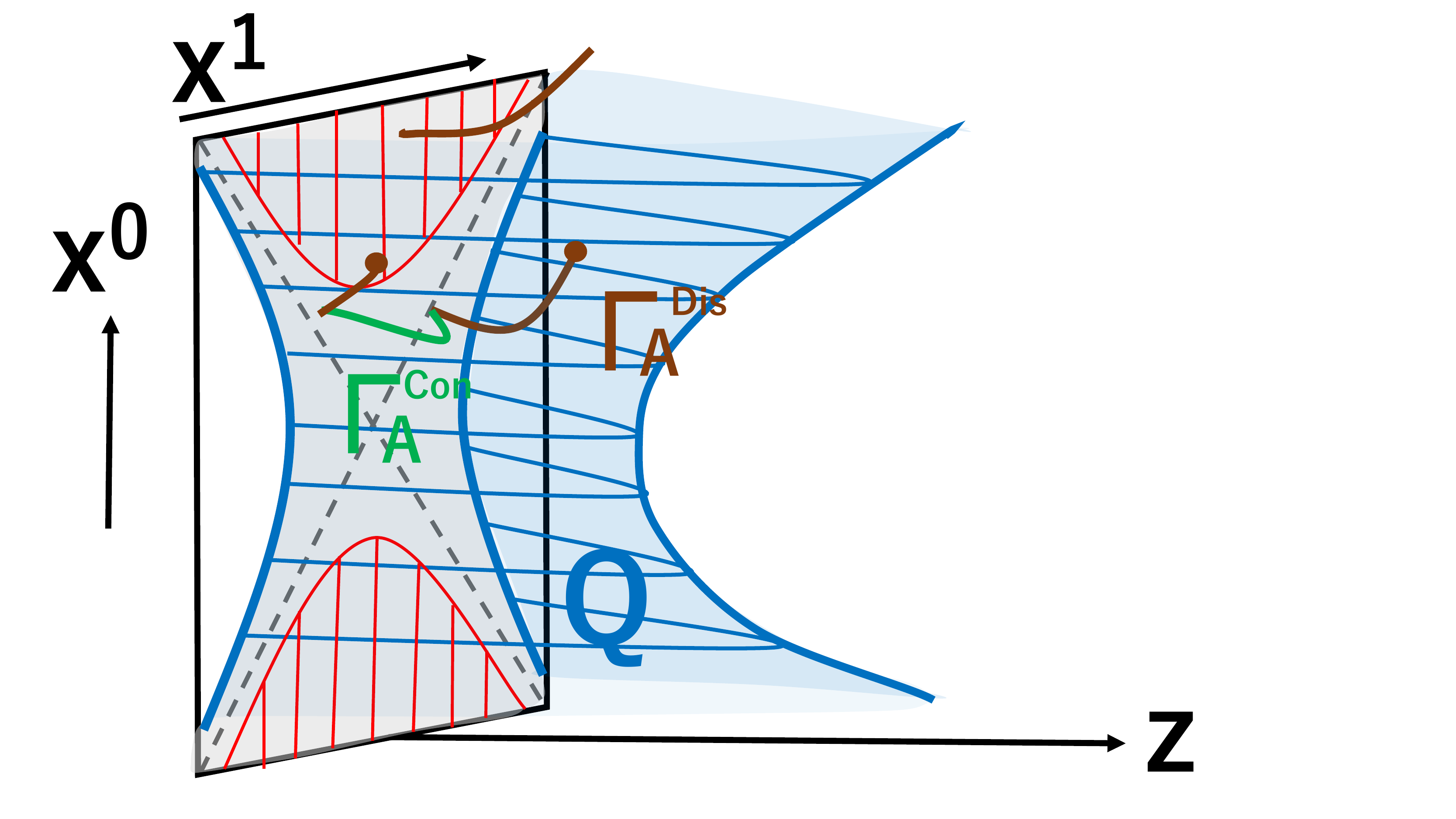}
  \caption{A simple example of AdS/BCFT dual to a two dimensional CFT defined in the region $|x_1|\leq \s{\gamma^2+x_0^2}$. The blue region in Poincar\'e 
AdS$_3$ is the gravity dual of the BCFT, enclosed by the brane $Q$.
We show the connected and disconnected geodesics. The disconnected geodesic, which departs from a point in the red shaded region, goes beyond the Poincar\'e patch.}
\label{adsbcftsfig}
\end{figure}

To preserve the boundary conformal invariance on $\de\Sigma$, 
we impose the following Neumann boundary condition
\ba
K_{ab}-h_{ab}K=-{\cal T}h_{ab}
\label{bdyein}
\ea
on the surface $Q$. The parameter ${\cal T}$ is the tension of the end-of-the-world brane $Q$ and is monotonically related to the degrees of freedom of a given BCFT which depends on the boundary condition
on $\de \Sigma$. The gravity dual of the BCFT is given by the three dimensional space called $M$, which is defined by a region enclosed by $Q$ and $\Sigma$. Refer to Fig.~\ref{adsbcftsfig} for a sketch.

The metric of $M$ is determined by solving the Einstein equation with the negative cosmological constant, which is standard in AdS/CFT, with the boundary condition \eqref{bdyein} imposed on $Q$. Note that this construction can be generalized to scenarios, where 
$\de \Sigma$ consists of multiple disconnected boundaries as well as to the higher dimensional cases. 

We can also extend the holographic entanglement entropy \cite{RT,RT2,HRT} to AdS/BCFT setups.
Consider a subsystem $A$ defined as an interval on a time-slice of BCFT. The entanglement entropy 
$S_A$ is computed by the following formula \cite{AdSBCFT,AdSBCFT2}
\ba
S_A=\mbox{min}\left\{S^{\text{con}}_A,S^{\text{dis}}_A\right\},
\label{eq:holoEE}
\ea
where 
\ba
S^{\text{con}}_A=\frac{L(\Gamma^{\text{con}}_A)}{4G_N},\qquad  S^{\text{dis}}_A
=\frac{L(\Gamma^{\text{dis}}_A)}{4G_N}.
\label{eq:holoEEarea}
\ea
Here, $S^\text{con}_A$ and $S^\text{dis}_A$ are the holographic entanglement entropies, which are proportional to the connected and disconnected geodesic lengths, denoted by $L(\Gamma^{\text{con}}_A)$ and 
$L(\Gamma^{\text{dis}}_A)$, respectively.  

The surface $\Gamma^{\text{con}}_A$ is the geodesic, which connects the two end points of the interval $A$. On the other hand, $\Gamma^{\text{dis}}_A$ is the union of the two disconnected geodesic curves which connects each of the two end points of $A$ with a point on the surface $Q$. The latter point is determined by minimizing the geodesic length. Refer to Fig.~\ref{adsbcftsfig} for a sketch of such geodesics.

\subsection{General solutions in three dimensional bulk spacetime}

In general, to determine of the metric of the gravity dual $M$, 
we need to solve a complicated back-reaction problem in general relativity.
However, in our three dimensional setups, where we consider pure gravity without any matter fields, we can easily find the shape of the surface $Q$ and the metric of $M$. This is essentially so, because any solution to the Einstein equations in three dimension is locally equivalent to Poincar\'e AdS$_3$, i.e.
\ba
ds^2=L^2 \frac{dz^2+dx_1^2-dx_0^2}{z^2}.
\ea

The general solution to \eqref{bdyein} takes the form
\ba
(z-\ap)^2+(x_1-a)^2-(x_0-b)^2=\gamma^2.
\label{solab}
\ea

When we consider the region inside this surface (\ref{solab}), this is dual to the two dimensional BCFT on
\ba
(x_1-a)^2-(x_0-b)^2\leq \gamma^2-\ap^2.
\ea
Note that this is identical to the simple moving mirror example in \eqref{eq:ex1-Zprof}, where the 
energy flux is vanishing.

In this case, the tension of the surface (\ref{solab})
 is given by 
\ba
{\cal T}=\frac{\ap}{\gamma L}.
\ea
This is depicted in Fig.~\ref{adsbcftsfig}.
If we consider the outside region of \eqref{solab}, it is dual to the BCFT on $(x_1-a)^2-(x_0-b)^2\geq \gamma^2-\ap^2$, and its tension 
is given by ${\cal T}=-\frac{\ap}{\gamma L}$. 

The standard BCFT corresponds to the values of the tension given by ${\cal T}L< 1$.
At least mathematically, we can consider also the region ${\cal T}L\geq 1$, where the surface $Q$ is floating 
in the bulk AdS. This exotic case has been discussed in \cite{WedgeHolo}, and was argued to be dual to a certain analytical continuation of the BCFT, e.g. a two dimensional CFT on a disk with an imaginary radius.

Nevertheless, from the bulk gravity viewpoint, we do not observe any obvious problem in the presence of 
such an end-of-the-world brane having tension ${\cal T}L>1$, which is still timelike, though the boundary $\de\Sigma$ of the BCFT becomes spacelike.

\subsection{Holographic entanglement entropy for a simple moving boundary}

We now want to compute the holographic entanglement entropy of a given subregion in the BCFT.
We would like to focus on the vanishing brane tension case, i.e. ${\cal T}=0$, for simplicity. By setting 
$a=0$ and $b=0$, and making use of translation invariance, 
the end-of-the-world brane $Q$ in Poincar\'e AdS is described by
\ba
z^2+x_0^2-x_1^2=\gamma^2.  
\label{bdyct}
\ea
The gravitational spacetime in the bulk ends on the brane $Q$, which is holographically dual to the CFT living on the region
$|x_1|\leq \s{\gamma^2+x_0^2}$.  This is depicted in Fig.~\ref{adsbcftsfig}.

The geodesic length $D_{12}$ between $(z,x_0,x_1)=(z_1,t_1,x_1)$ and 
$(z,x_0,x_1)=(z_2,t_2,x_2)$ is given by the standard formula
\ba
D_{12}=\cosh^{-1}\left(\frac{(x_1-x_2)^2-(t_1-t_2)^2+z_1^2+z_2^2}{2z_1z_2}\right).
\ea

We choose the subsystem $A$ to be an interval $x_1\in [l_1,l_2]$ at time $x_0=t$. 
We further consider $-\s{\gamma^2+t^2}<l_1<l_2<\s{\gamma^2+t^2}$.

The connected geodesic, which exists at any time, determines the contribution to the holographic entanglement entropy, which reads
\ba
S^\text{con}_A=\frac{c}{3}\log\frac{l_2-l_1}{\ep}, 
\label{conhee}
\ea
where $\ep$ denotes the UV cutoff.

As we explained before, in AdS/BCFT, disconnected geodesics, which end on the brane $Q$, are allowed. 
In the current setup, we can consider a geodesic $\Gamma^{\text{dis}}_A$ which connects the boundary point $(\ep,x,t)$ and a point $(\s{\gamma^2+s^2-y^2},y,s)$ on the brane $Q$. Its length takes the form
\ba
D\simeq \log\frac{\gamma^2+s^2-y^2+(x-y)^2-(t-s)^2}{\ep\s{\gamma^2+s^2-y^2}}.
\label{ww}
\ea
By extremalizing $D$ with respect to $s$ and $y$, we find
\ba
y=\frac{2\gamma^2 x}{x^2+\gamma^2-t^2},\qquad 
s=\frac{2\gamma^2 t}{x^2+\gamma^2-t^2}.  
\label{extemalp}
\ea
At first sight, to have a spacelike geodesic, which leads to the inequality $\gamma^2+s^2-y^2+(x-y)^2-(t-s)^2\geq 0$, we might require 
\ba
x^2+\gamma^2-t^2\geq 0.  
\label{condewq}
\ea
This covers only a part of the region of the space $\Sigma$ where the BCFT is defined.

However, this argument is based on the assumption that the geodesic 
ends on a point on $Q$ within the Poincar\'e patch. If we take into account the possibility that the geodesic ends on an extended surface of $Q$ in the global patch, then we can actually find that we still can have a spacelike geodesic with the desired property. The geodesic length will be given by \eqref{extemalp}, with $\s{\gamma^2+s^2-y^2}$ being replaced by $-\s{\gamma^2+s^2-y^2}$. 

\begin{figure}[h!]
  \centering
  \includegraphics[width=.35\textwidth]{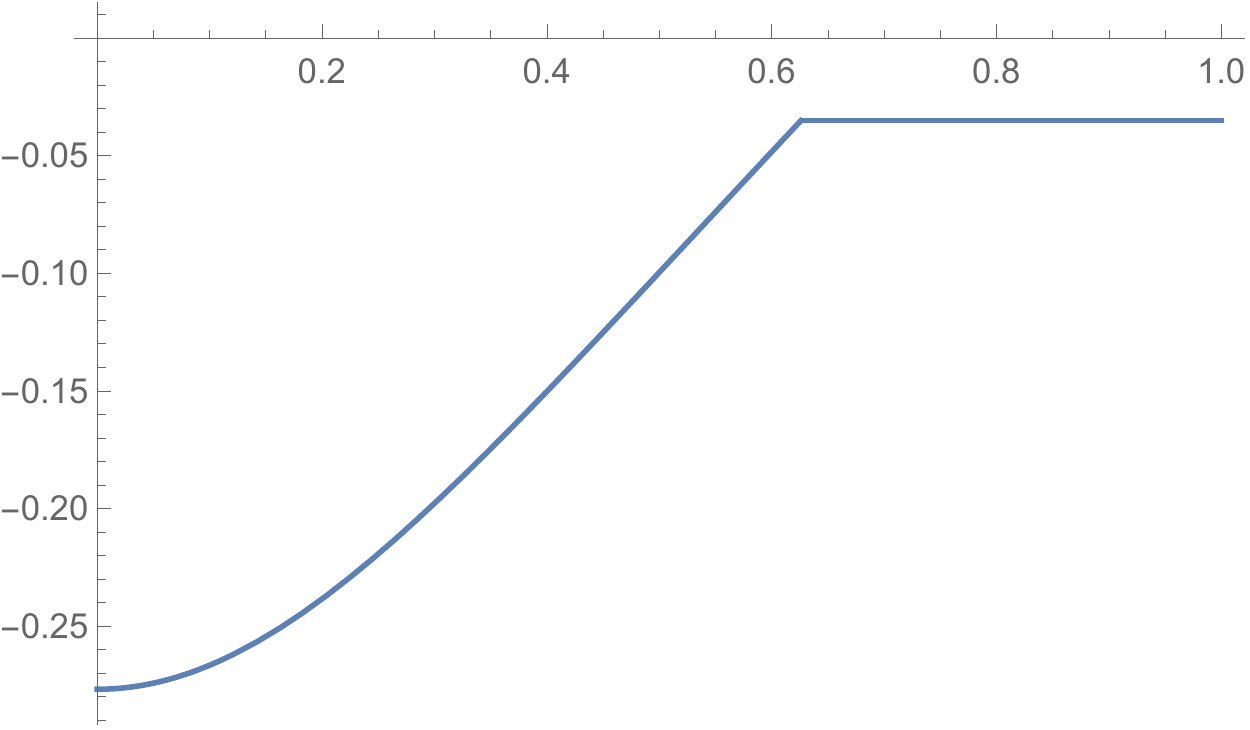}\qquad
  \includegraphics[width=.35\textwidth]{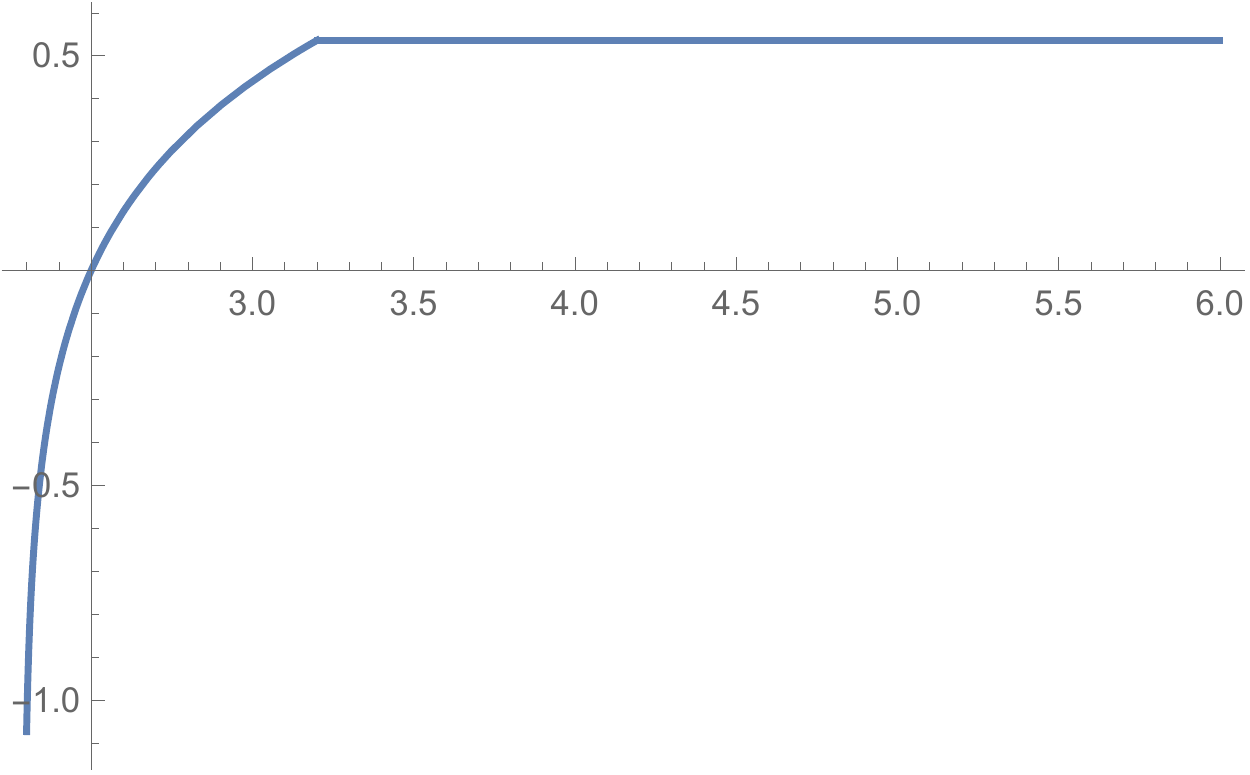}
  \caption{Time evolution of holographic entanglement entropy in the AdS/BCFT setup \eqref{bdyct}.
We plot the entropy, with the UV divergence being subtracted by setting $\ep=1$, as a function of time $t$. We set 
$\gamma=1$. 
The left and right graphs correspond to a subsystem $A$ chosen to be given by $[0,0.9]$ and $[-2.5,2.5]$, respectively.}
\label{PoHEEfig}
\end{figure}

The geodesic length for the solution in \eqref{extemalp} leads to the following holographic entanglement entropy
\ba
S^\text{dis}_A=\frac{c}{6}\log\frac{t^2-l_1^2+\gamma^2}{\ep \gamma}+\frac{c}{6}\log\frac{t^2-l_2^2+\gamma^2}{\ep\gamma},
\label{dishee}
\ea
resulting from the two disconnected geodesics.

Here, we have assumed the extension to global AdS$_3$ as we have mentioned before. Notice that the expression in \eqref{dishee} is still correct even when the inequality \eqref{condewq} gets violated.

In this way, the holographic entanglement entropy is obtained from \eqref{eq:holoEE}, where the connected and disconnected geodesic results in \eqref{eq:holoEEarea} are computed as in \eqref{conhee} and \eqref{dishee}, respectively.
It is easy to observe that the disconnected solution $S^\text{dis}$ is favored when 
\ba
t^2+\gamma^2-\frac{l_1^2+l_2^2}{2}\leq (l_2-l_1)\s{\gamma^2+\frac{(l_1+l_2)^2}{4}}.
\ea
This results in a phase transition between $S^\text{dis}_A$ and $S^\text{con}_A$ as depicted in Fig.~\ref{PoHEEfig}.

\subsection{AdS/BCFT for the BTZ black hole}
\label{sec:AdS/BCFT-BTZ}

As a next holographic moving mirror example, let us consider setups of AdS/BCFT in BTZ black hole backgrounds, i.e.
\ba
ds^2=-(r^2-r^2_0)dt^2+L^2\frac{dr^2}{r^2-r^2_0}+r^2dx^2.
\label{btzm}
\ea
In order to find the correct end-of-the-world branes $Q$, we can employ the fact that the BTZ spacetime 
can be mapped into Poincar\'e AdS$_3$ discussed in Sec.~\ref{sec:AdS/BCFT-Poincare} via the following coordinate transformations
\be
\begin{split}
x_0-x_1 &=-e^{\frac{r_0}{L}(x-t)}\s{1-\frac{r_0^2}{r^2}},\\
x_0+x_1 &=e^{\frac{r_0}{L}(x+t)}\s{1-\frac{r_0^2}{r^2}},\\
z &=\frac{r_0}{r}e^{\frac{r_0}{L}x}.
\end{split}
\label{eq:map-BTZ-Poincare}
\ee

\subsubsection{Static surfaces} 

We can map the solution in \eqref{solab} for $a=b=0$ via the transformation \eqref{eq:map-BTZ-Poincare} and obtain a static profile for the surface $Q$ in BTZ. 
\begin{figure}[t]
  \centering
  \includegraphics[width=8cm]{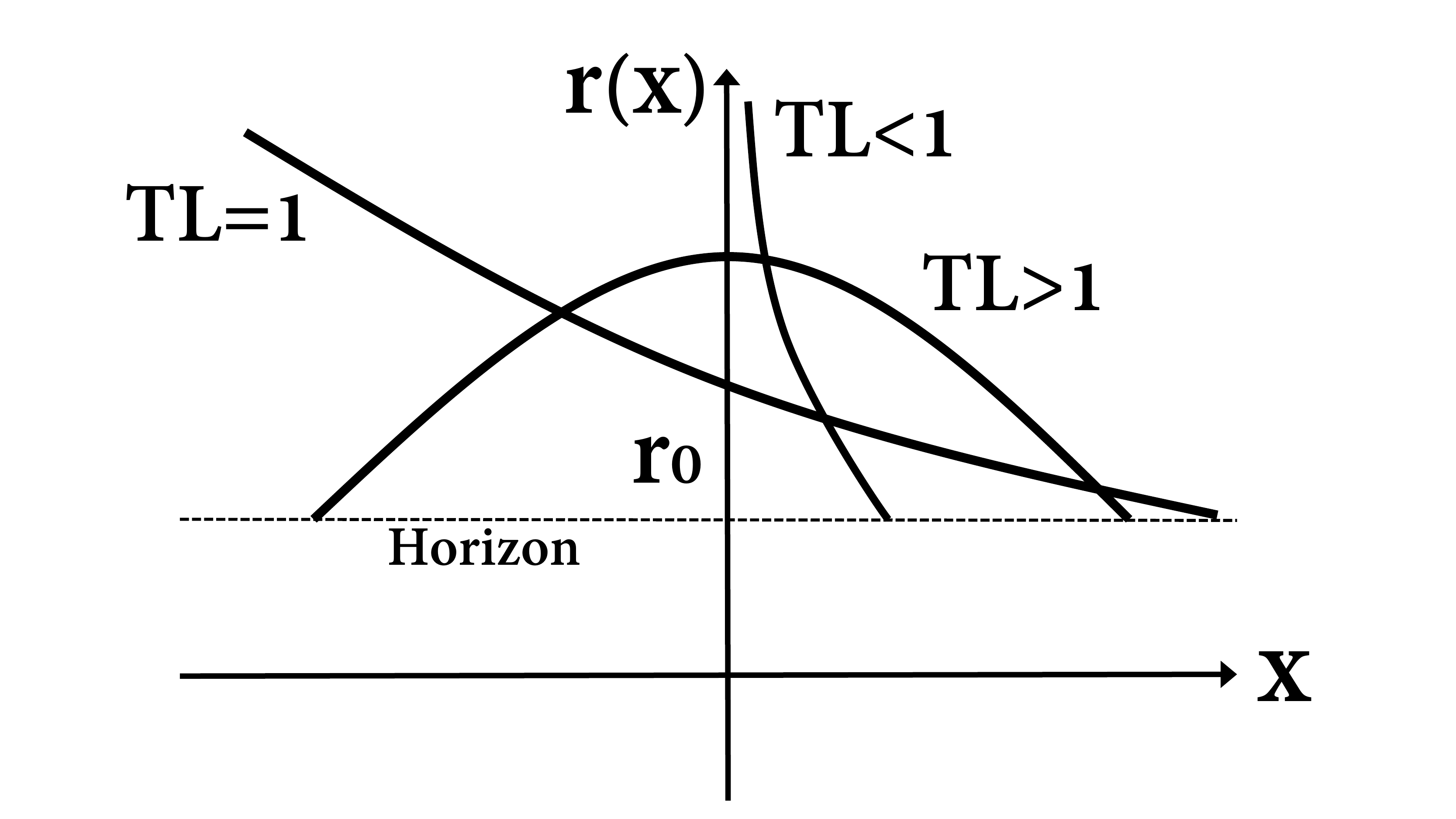}
  \caption{Shown are characteristic profiles for the end-of-the-world brane $Q$ in the BTZ background depending on the value of $L{\cal T}$.}
\label{btzbranefig}
\end{figure}
When $L{\cal T}<1$, with a translation in $x$ direction, we find 
\ba
r(x)=\frac{r_0 {\cal T}L}{\s{1-{\cal T}^2L^2}\sinh\frac{r_0 x}{L}},
\label{tenzersufg}
\ea
as depicted in Fig.~\ref{btzbranefig}.
When $L{\cal T}=1$, we obtain
\ba
r(x)=2r_0 \gamma e^{-\frac{r_0 x}{L}}.
\ea
When $L{\cal T}>1$, we get
\ba
r(x)=\frac{r_0 {\cal T}L}{\s{{\cal T}^2L^2-1}\cosh\frac{r_0 x}{L}}. 
\label{sufb}
\ea
This brane $Q$ does not extend to the AdS boundary at $r=\infty$, as opposed to 
the cases where ${\cal T}L\leq 1$.
In each case, i.e. $L{\cal T}<1$, $L{\cal T}=1$, and $L{\cal T}>1$, we realize that the world volume of $Q$ is identical to AdS$_2$, $\mathbb{R}^{1,1}$, and 
dS$_2$, respectively.

\begin{figure}[h!]
  \centering
  \includegraphics[width=0.6\textwidth]{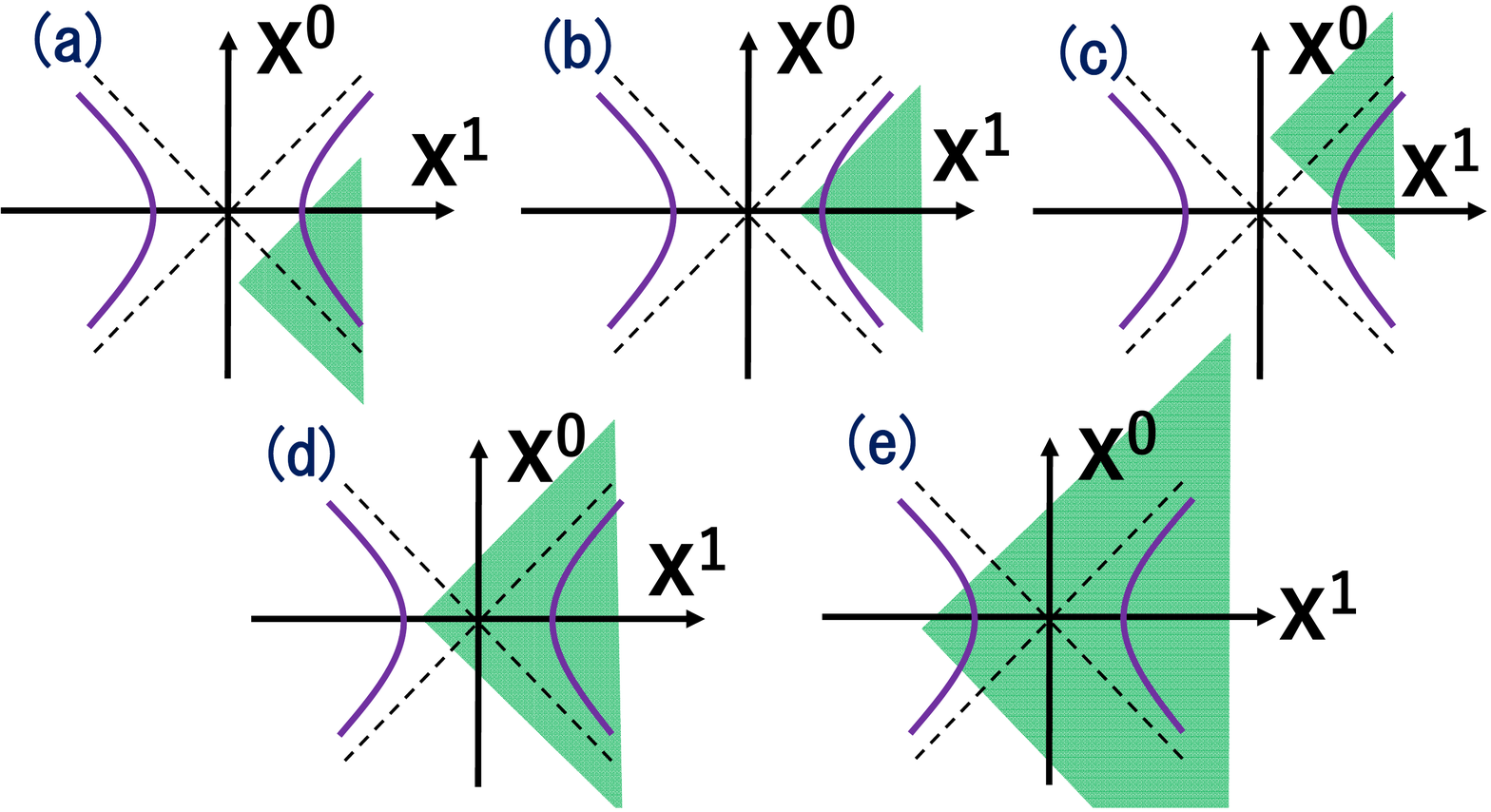}
  \caption{Shown are five different classes of the brane $Q$ with ${\cal T}=0$ at the AdS boundary $z=0$ of the BTZ background. The purple curve describes $Q$ at $z=0$. The green region is the patch covered by the outside horizon region of the BTZ black hole.}
\label{Tzerofig}
\end{figure}

\subsubsection{Time dependent surfaces} 

Next, we would like to turn to the general time dependent surface $Q$ in the BTZ black hole background.
This is parameterized by three parameters $(\ap,a,b)$, and the profile of $Q$ is explicitly given by
\ba
e^{2\frac{r_0}{L}x}-2\ap\frac{r_0}{r}e^{\frac{r_0}{L}x}-2a\cosh\frac{r_0t}{L}e^{\frac{r_0 x}{L}}\s{1-\frac{r_0^2}{r^2}}+2b\sinh\frac{r_0t}{L}e^{\frac{r_0 x}{L}}\s{1-\frac{r_0^2}{r^2}}=\gamma^2+b^2-a^2-\ap^2.\no
\ea
This is time dependent when either $a\neq 0$ or $b\neq 0$.

\begin{figure}[h!]
  \centering
  \includegraphics[width=0.6\textwidth]{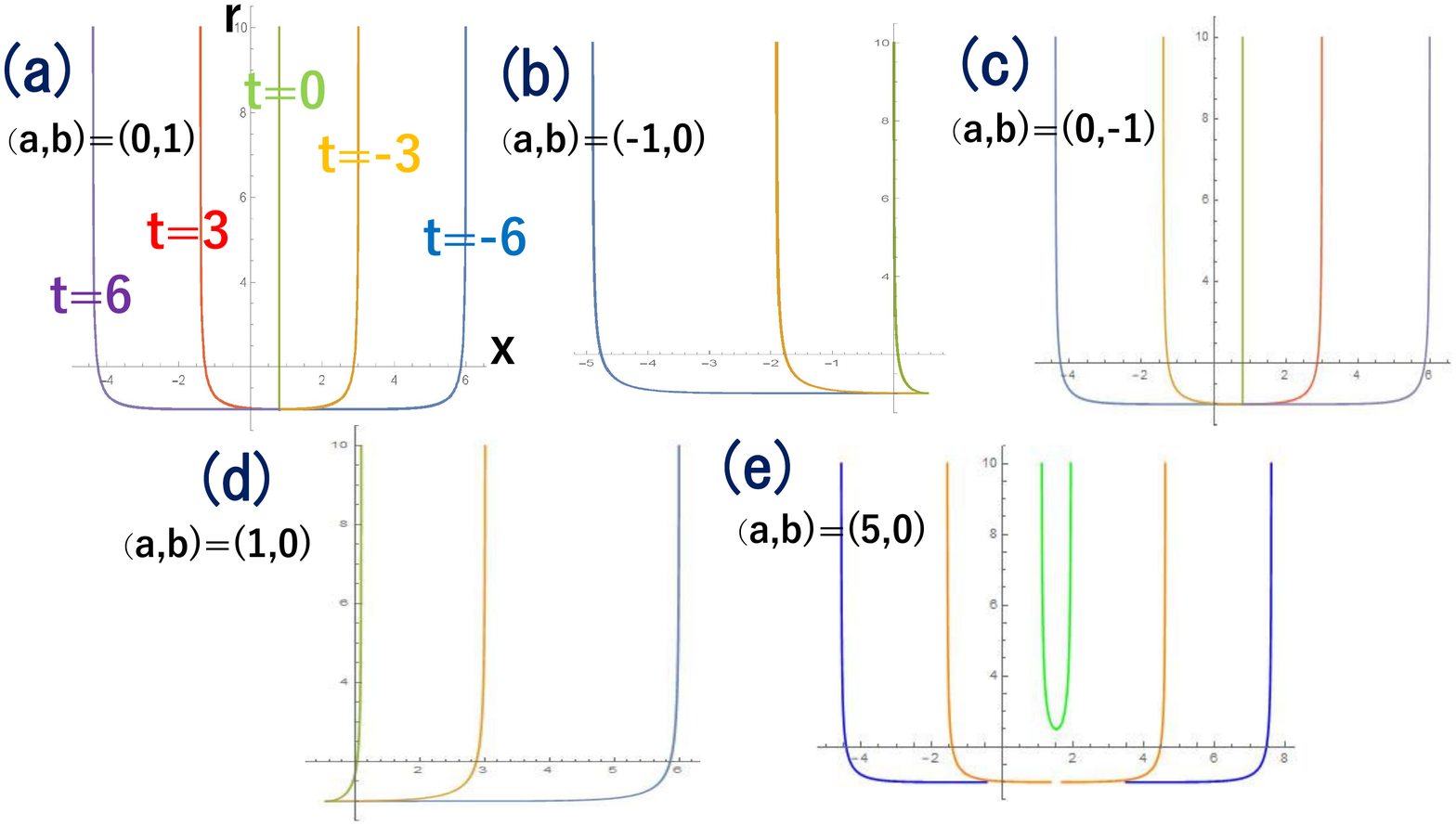}
  \caption{Plots of the brane $Q$ for five different examples corresponding to the five classes 
in Fig.~\ref{Tzerofig} with ${\cal T}=0$.  The horizontal and vertical axes correspond to the $x$ and $r$ directions, respectively. 
We have set $\ap=0$ and $\gamma=2$ with $L=r_0=1$.  The parameters $(a,b)$ we have chosen are indicated in each plot. The blue, orange, green, red and purple curves correspond to $t=-6,-3,0,3,6$. In the cases $(b),(d)$ and 
$(e)$, the curves have time reversal symmetry, i.e. $t\to -t$. Therefore, we only plot these curves for $t=-6,-3,0$.}
\label{TzeroPLfig}
\end{figure}

At the AdS boundary $z=0$, we have 
\ba
e^{2\frac{r_0}{L}x}-2a\cosh\frac{r_0t}{L}e^{\frac{r_0 x}{L}}+2b\sinh\frac{r_0t}{L}e^{\frac{r_0 x}{L}}=\gamma^2+b^2-a^2-\ap^2.
\ea

It is useful to introduce the relation between the Poincar\'e AdS$_3$ coordinates and the BTZ coordinates,
\be
\begin{split}
\frac{x_0+x_1}{x_0-x_1} &=-e^{\frac{2r_0 t}{L}},\\
z^2+x_1^2-x_0^2 &=e^{\frac{2r_0 x}{L}},\\
\frac{\s{z^2+x_1^2-x_0^2}}{z} &= \frac{r}{r_0}.
\end{split}
\ee
Remember that the outside horizon region for the BTZ black hole corresponds to 
the right wedge defined by $x_1-x_0>0$, and $x_0+x_1>0$ in Poincar\'e AdS$_3$.

We can classify the time dependent behavior of the boundary surface $Q$ according to the values of 
the four parameters $(\ap,\gamma,a,b,)$. They are qualitatively different depending on the fact whether the absolute value of the tension ${\cal T}=\frac{\ap}{\gamma L}$ is larger or smaller than $1/L$.  Therefore, below, we focus on the two different examples: (i) vanishing tension ${\cal T}=0$ case, where we choose the values $\ap=0$ and $\gamma=2$, and (ii) the case ${\cal T}=2/L$, where we choose the values $\ap=4$ and $\gamma=2$.

In the former case (i), there are five classes, depending on how the right wedge region $(x_1-x_0)(x_1+x_0)>0$ 
is situated relative to the boundary surface $Q$ at the AdS boundary $z=0$, as depicted in Fig.~\ref{Tzerofig}.
The profiles of $Q$ at various values for the time $t$ are plotted in Fig.~\ref{TzeroPLfig}.

\begin{figure}[h!]
  \centering
  \includegraphics[width=0.6\textwidth]{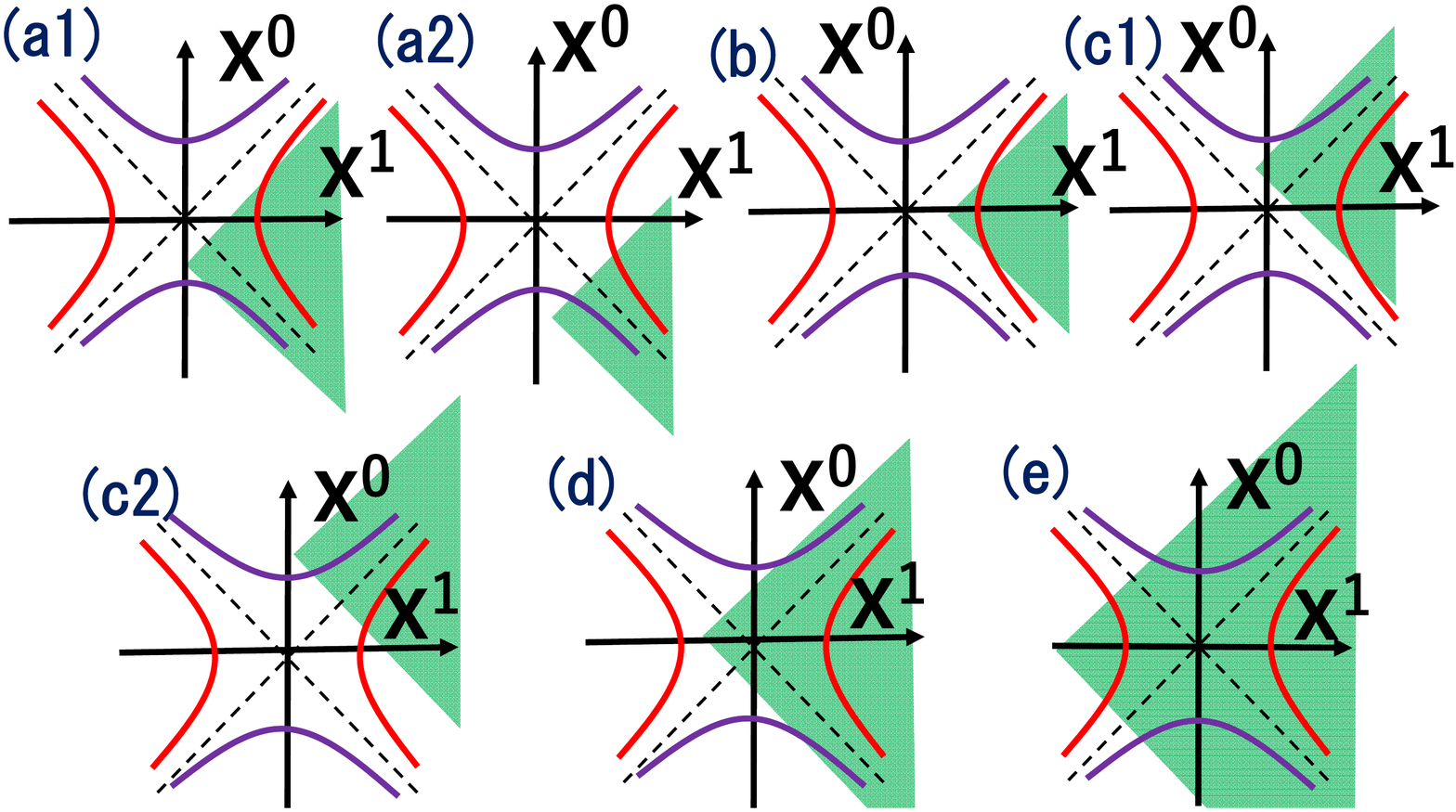}
  \caption{Shown are the seven classes of the brane $Q$ with $\mathcal{T}=2/L$ at the AdS boundary $z=0$ of the BTZ background. The purple and red curves describe $Q$ at $z=0$ and $z=\ap$, respectively.
The green region is the patch covered by the outside horizon region of the BTZ black hole.}
\label{T2fig}
\end{figure}

\begin{figure}[h!]
  \centering
  \includegraphics[width=0.6\textwidth]{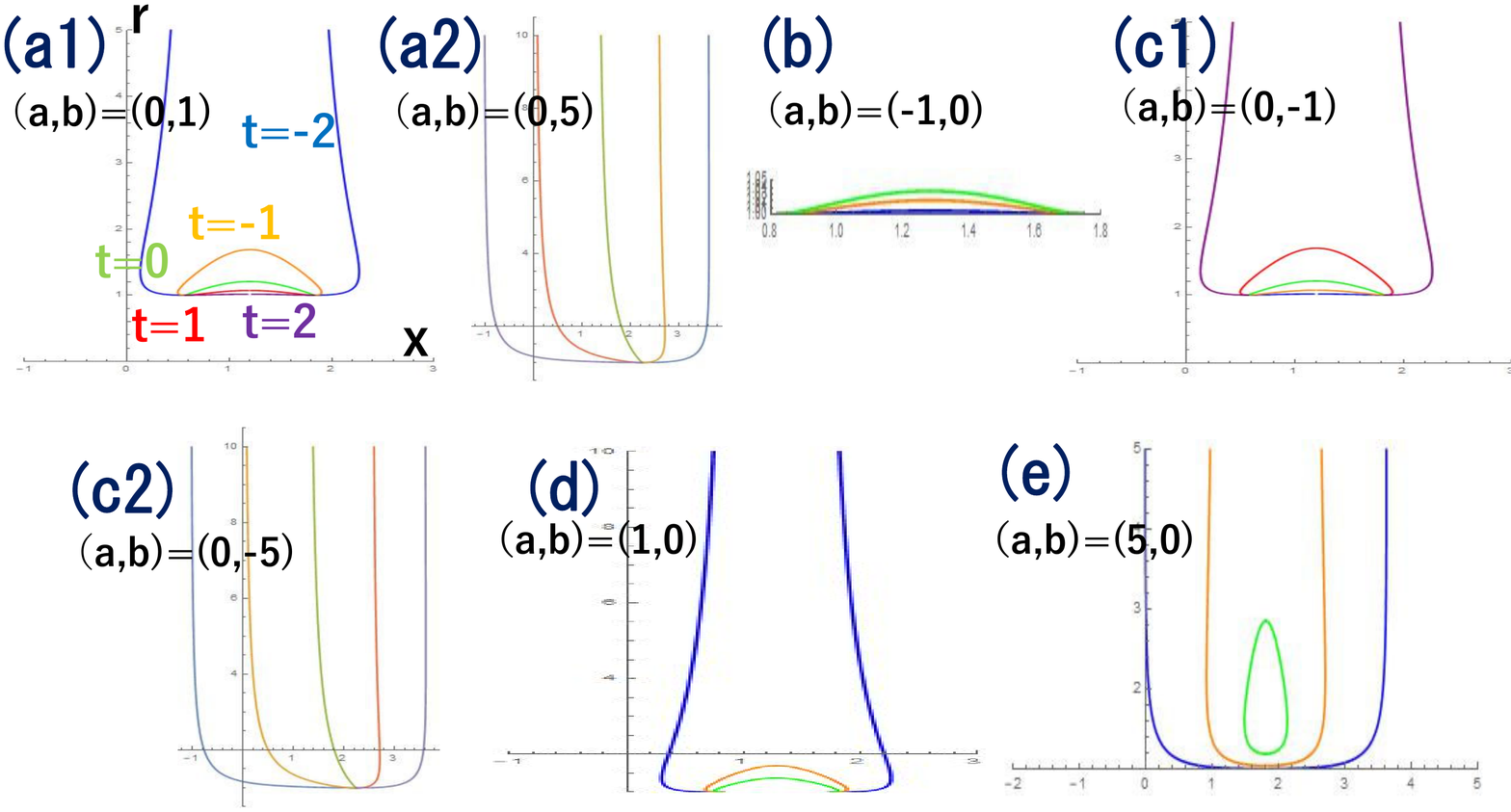}
  \caption{The plots of the brane $Q$ in seven different scenarios corresponding to the classes in Fig.~\ref{T2fig} with $\mathcal{T}=2/L$. The horizontal and vertical axes correspond to the $x$ and $r$ directions, respectively. We have set $\ap=2$ and $\gamma=4$ with $L=r_0=1$. The parameters $(a,b)$ we have chosen are indicated in each plot. The blue, orange, green, red and purple curves correspond to $t=-2,-1,0,1,2$. For the plots $(b),(d)$ and 
$(e)$, the curves have time reversal symmetry, i.e. $t\to -t$, hence, we only plot them for $t=-2,-1,0$.}
\label{T2PLfig}
\end{figure}

In the latter case (ii) with ${\cal T}=2$, depending on the location of the right wedge relative to the boundary surface $Q$ at the AdS boundary $z=0$ and the center $z=\ap$, we can classify the profiles of $Q$ into seven classes as depicted in Fig.~\ref{T2fig}. Note that the surface $Q$ looks as a spacelike boundary at $z=0$, while it becomes a timelike curve at $z=\ap$.
The profiles of $Q$ for various values of time $t$ are plotted in Fig.~\ref{T2PLfig}. A special feature of the exotic scenario with tension ${\cal T}L>1$ is that a floating brane is possible, see case (e) in Fig.~\ref{T2PLfig}. 

\section{Radiation from holographic moving mirror}
\label{sec:hmm}

In this section, starting with introducing the general procedure, we first focus on a setup with a moving mirror and construct its holographic dual. We discuss the dual bulk geometry and provide a
gravitational interpretation in the light of brane-world holography. We will discuss two different setups: (i) escaping mirror, which models Hawking radiation from a black hole and (ii) kink mirror, which mimics black hole formation and evaporation.

\subsection{Gravity dual of conformal map}
\label{sec:Banados}

Consider the Poincar\'{e} metric in AdS$_3$
\ba
ds^2=\frac{d\eta^2-dUdV}{\eta^2}, 
\label{pol}
\ea
which is dual to the vacuum of CFT$_2$. For simplicity, we have set the AdS radius to unity.
A conformal transformation of the type
\ba
\ti{u}=p(u),\ \ \ \ti{v}=q(v),
\ea
is dual to the following coordinate transformation in AdS$_3$
\ba
\begin{split}
U&=p(u)+\frac{2z^2(p')^2 q''}{4p'q'-z^2p''q''},\\
V&=q(v)+\frac{2z^2(q')^2p''}{4p'q'-z^2p''q''},\\
\eta&=\frac{4z(p'q')^{3/2}}{4p'q'-z^2p''q''},
\label{corads}
\end{split}
\ea
which is known as the so-called Ba$\tilde{\text{n}}$ados map 
\cite{Banados:1998gg,Roberts:2012aq,Shimaji:2018czt}.
The metric expressed in terms of the coordinates $(u,v,z)$ reads
\ba
ds^2=\frac{dz^2}{z^2}+T_{+}(u)(du)^2+T_{-}(v)(dv)^2-\left(\frac{1}{z^2}
+z^2T_{+}(u) T_{-}(v)\right)dudv, 
\label{metads}
\ea
where
\ba
T_{+}(u)=\frac{3(p'')^2-2p'p'''}{4p'^2},\qquad T_{-}(v)=\frac{3(q'')^2-2q'q'''}{4q'^2}
\label{emgt}
\ea
are the chiral and anti-chiral energy stress tensor, respectively.

\subsection{Moving mirror in AdS/BCFT}

As previously discussed, a single moving mirror is described by the conformal map \eqref{cmapf} from a half plane.
The gravity dual is simply obtained by employing the metric in \eqref{metads}. Since we consider the case $q(v)=v$, the metric \eqref{metads} becomes the plane 
wave metric given by
\ba
ds^2=\frac{dz^2}{z^2}+T_{+}(u)(du)^2-\frac{1}{z^2}dudv, 
\label{metkads}
\ea
where 
\ba
T_{+}(u)=\frac{3(p'')^2-2p'p'''}{4p'^2}.
\ea
Hence, according to the relations \eqref{corads}, the coordinate transformation, which maps into Poincar\'e AdS$_3$, takes the form
\ba
\begin{split}
U&=p(u),\\
V&=v+\frac{p''(u)}{2p'(u)}z^2,\\
\eta&=z \s{p'(u)}.  
\label{cordch}
\end{split}
\ea
It is also useful to introduce the coordinates $T$ and $X$ such that 
\ba
U=T-X,\ \ \ V=T+X.
\ea

Using the AdS/BCFT formulation, the mirror is dual to the brane $Q$ located at $x=Z(t)$, where $u=t-x$ and $v=t+x$ are equivalent to the equation 
\ba
X+\lambda\eta=0, \label{mirrorbc}
\ea
with $\lambda$ being related to the tension ${\cal T}$ of the brane as follows (below we set $L=1$)
\ba
\lambda=\frac{{\cal T}}{\s{1-{\cal T}^2}}.  \label{relavz}
\ea
This is equivalent to
\ba
v=p(u)-\frac{p''}{2p'}z^2-2\lambda z\s{p'}.   \label{mirrortj}
\ea
Note that the induced metric on $Q$ describes AdS$_2$ geometry,
\ba
ds^2=\frac{(1+\lambda^2)d\eta^2-dT^2}{\eta^2}.  \label{eefq}
\ea

\subsection{Holographic entanglement entropy for escaping mirror}
\label{sec:HEEes}

After having constructed the gravity dual of the moving mirror in CFT$_2$, we would first like to holographically investigate the AdS/BCFT description for the escaping moving boundary example introduced in \eqref{exfr}.

\begin{figure}[h!]
  \centering
  \includegraphics[width=.7\textwidth]{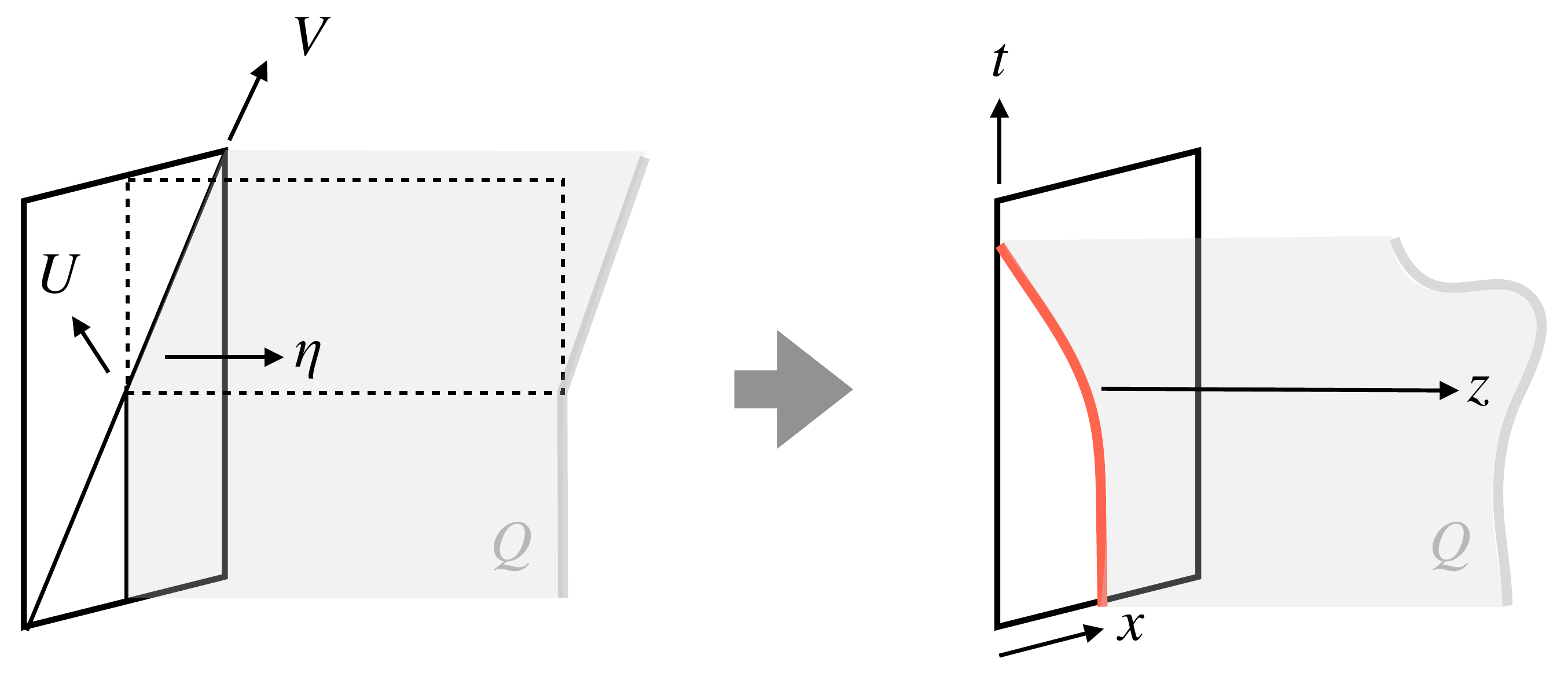}
  \caption{Sketch of the coordinate transformation in \eqref{cordch} mapping from Poincar\'e AdS$_3$ (left) to 
the gravity dual of a single moving mirror (red curve) depicted in the right panel. We can extend the geometry such that the brane $Q$ (gray, transparent) is extended to the region $U>0$ denoted by the dashed plane in the left panel.}
\label{figbmap}
\end{figure}

The gravity dual for the escaping mirror in CFT$_2$ is obtained from the following region of Poincar\'e AdS$_3$
\ba
V-U+2\lambda \eta>0, 
\label{qsura}
\ea
as depicted in the left panel of Fig.~\ref{figbmap}, where the condition \eqref{qsura} arises in AdS/BCFT due to the presence of the end-of-the-world brane $Q$.

\begin{figure}[h!]
  \centering
 \includegraphics[width=.38\textwidth]{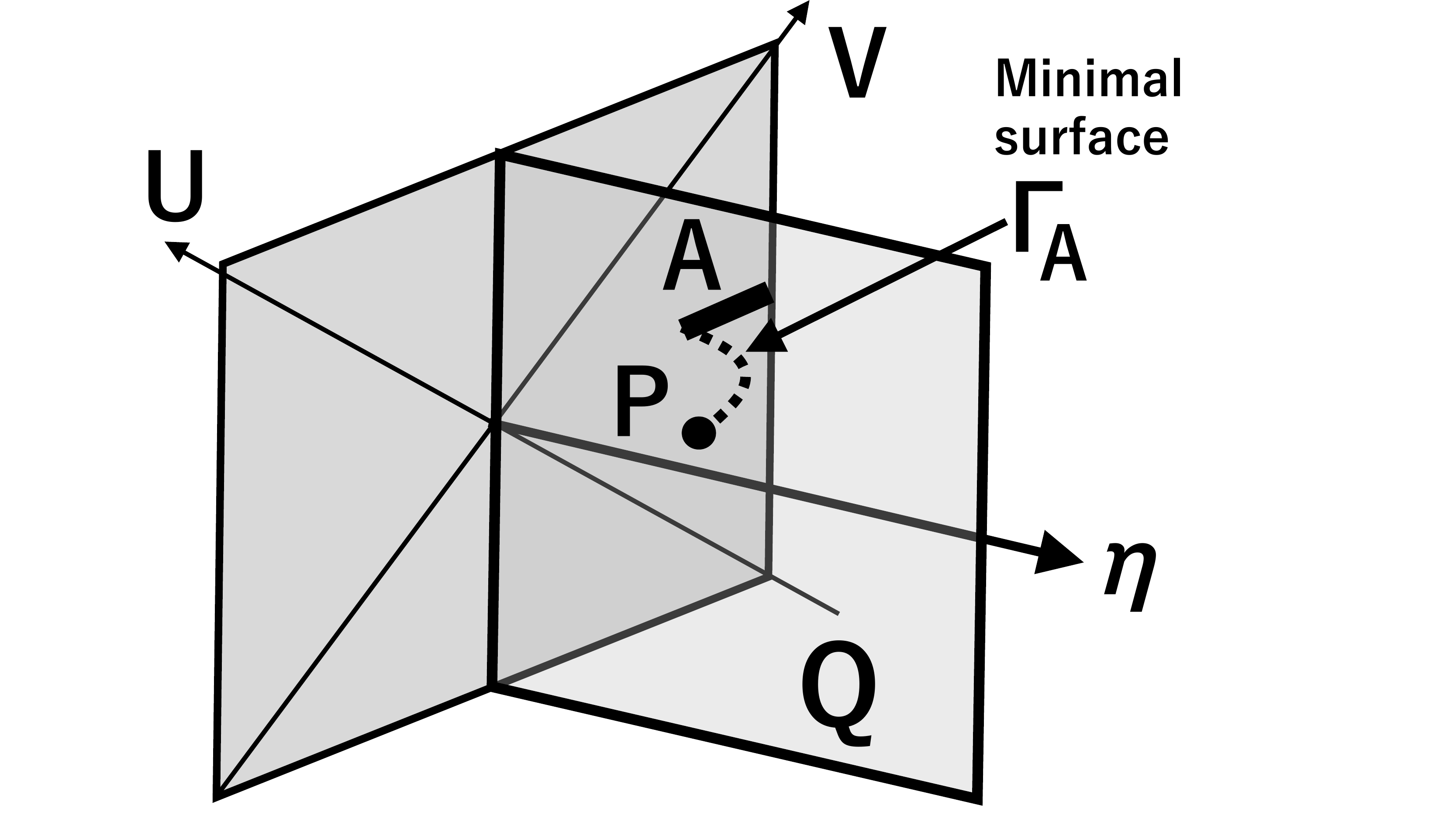}\quad
 \includegraphics[width=.25\textwidth]{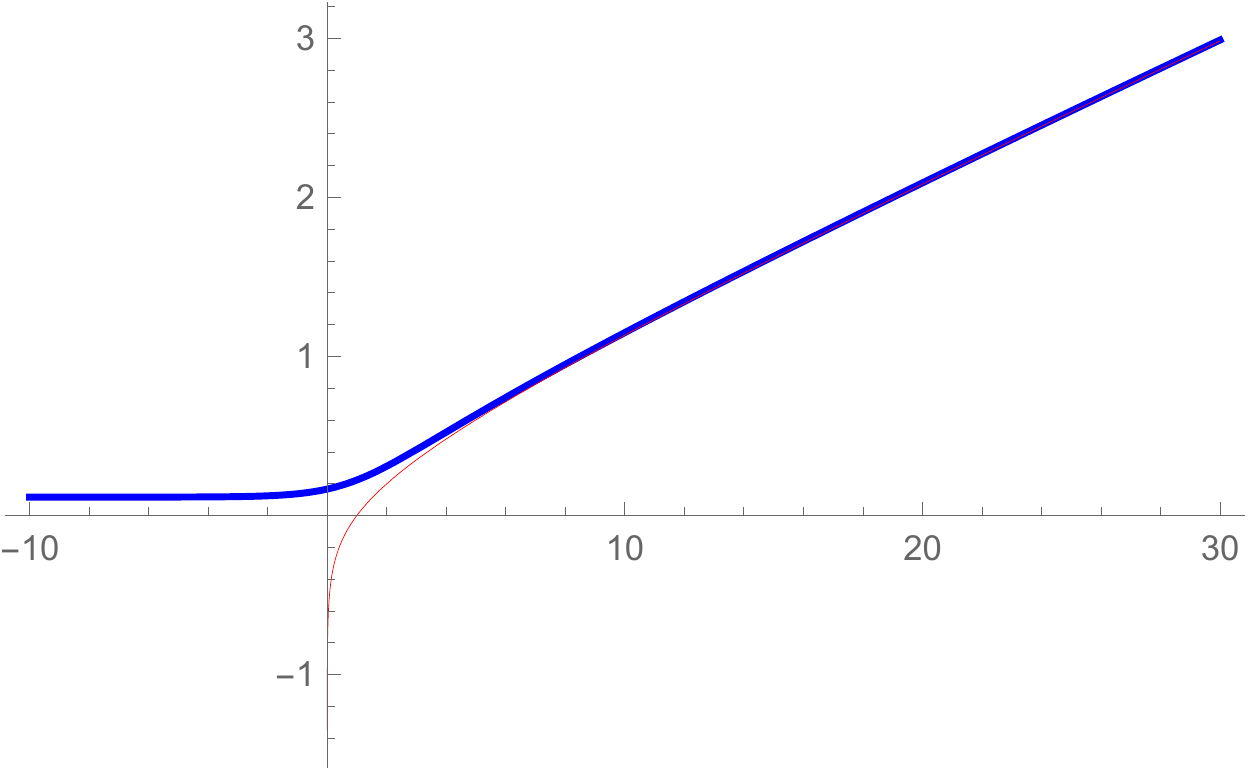}\quad
 \includegraphics[width=.32\textwidth]{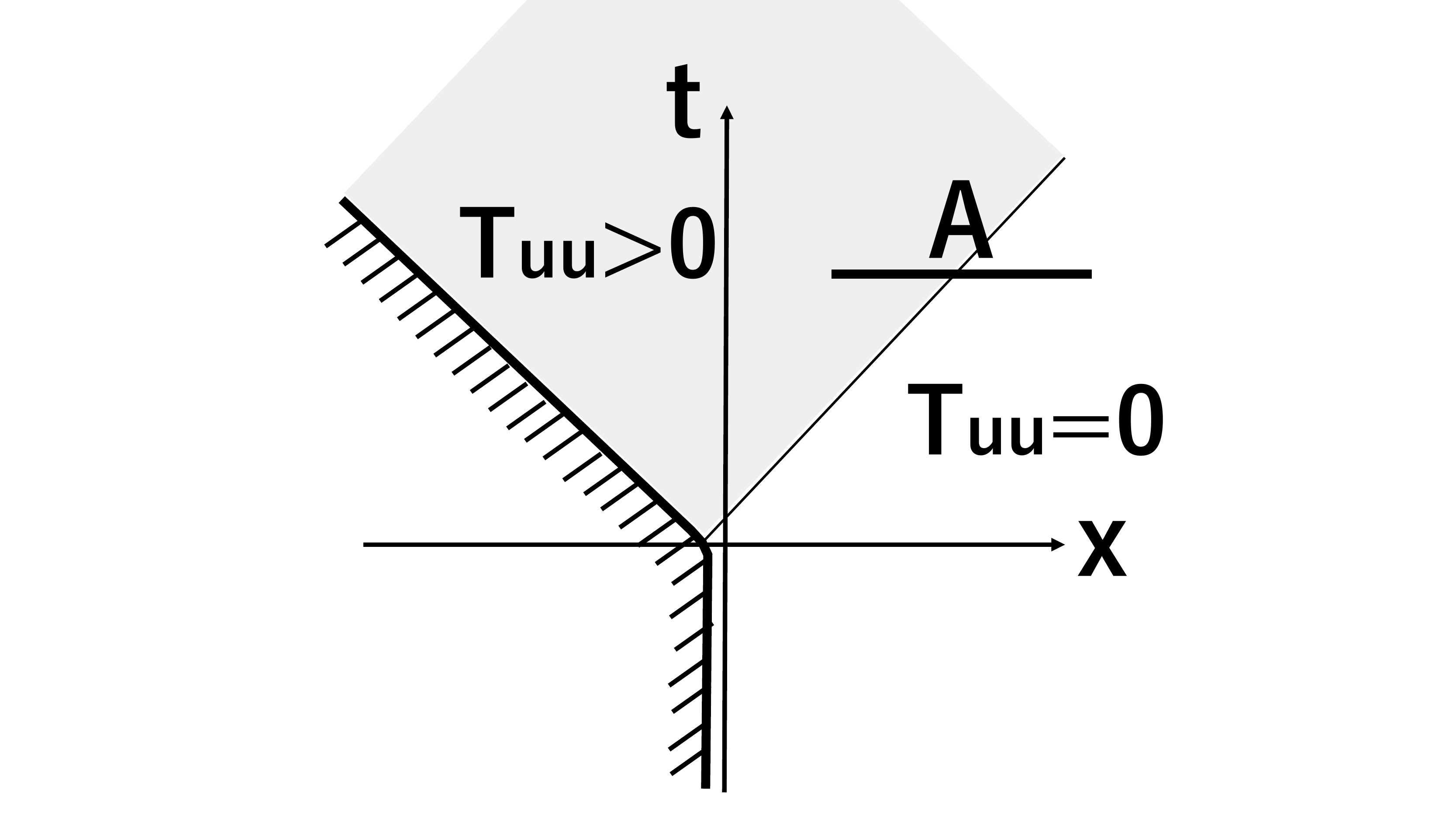}
  \caption{The left picture describes the calculation of holographic entanglement entropy in AdS/BCFT.
The blue curve in the middle shows the holographic entanglement entropy $S_A$ as a function of time for $\beta=1$, where we have set $\ep=1$. The red graph describes the approximation \eqref{apphee}. The right picture sketches the energy flux $T_{uu}$ under the time evolution of the moving mirror.}
\label{fighee}
\end{figure}

First, let us evaluate the holographic entanglement entropy. We choose the subsystem A to be a semi-infinite
line ($x>x_0$) at time $t$. By considering the disconnected geodesic $\Gamma$,\footnote{For the semi-infinite interval, the holographic entanglement entropy \eqref{eq:holoEE} is determined by the disconnected geodesic.} which starts from $x=x_0$ at time $t$ and ends on the brane $Q$ extended to 
$U>0$, we can holographically calculate the entanglement entropy by employing the Poincar\'e AdS$_3$ description as follows 
\ba
S_A =\frac{c}{6}\log\frac{V_0-U_0}{\ep\s{p'(u_0)}}+S_\text{bdy}
=\frac{c}{6}\log\frac{v_0-p(u_0)}{\ep\s{p'(u_0)}}+S_\text{bdy},
\label{sahee}
\ea
where $U_0=p(u_0)$, $V_0=v_0=x_0+t$, and $u_0=-x_0+t$ are the coordinates of the end point $P$ of the geodesic $\Gamma_A$. $S_\text{bdy}$ denotes the boundary entropy given in terms of the tension 
${\cal{T}}$ as \cite{AdSBCFT,AdSBCFT2} (we set $L=1$)
\be
S_\text{bdy}=\frac{c}{6}\log\s{\frac{1+{\cal{T}}}{1-{\cal{T}}}}.
\ee
Refer to the left panel of Fig.~\ref{fighee}, where the time evolution of this entanglement is plotted in the center. 

If we fix the end point of the subsystem $A$, i.e. $x_0$, at late time $t\to\infty$, we can approximate \eqref{sahee} as follows
\ba
S_A\simeq \frac{c}{12\beta}(t-x_0)+\frac{c}{6}\log\frac{t}{\ep}+S_\text{bdy}.  
\label{apphee}
\ea
Refer to the middle picture in Fig.~\ref{fighee}.

It is intriguing to note that this consists of a sum of a term linearly growing in $t$ and a term with logarithmic growth.
The former linear one can easily be understood by remembering that the energy flux is present 
uniformly for $u>0$. Refer to the right panel in Fig.~\ref{fighee}.

Next, it is useful to move the end point $x_0$ as a function of time. In particular, 
we take the distance 
between $x_0$ and the null trajectory to be a constant $\xi_0$,
\ba
x_0(t)=-t+\xi_0.
\label{dfferessnt}
\ea
In the late time limit $u\to \infty$,
we find the following time evolution for the entanglement entropy
\ba
S_A=\frac{c}{6}\log\frac{\left(\xi_0+\beta e^{-(t-x_0(t))/\beta}\right)
\s{1+e^{(t-x_0(t))/\beta}}}{\ep}+S_\text{bdy}.
\ea

\begin{figure}[h!]
  \centering
\includegraphics[width=.3\textwidth]{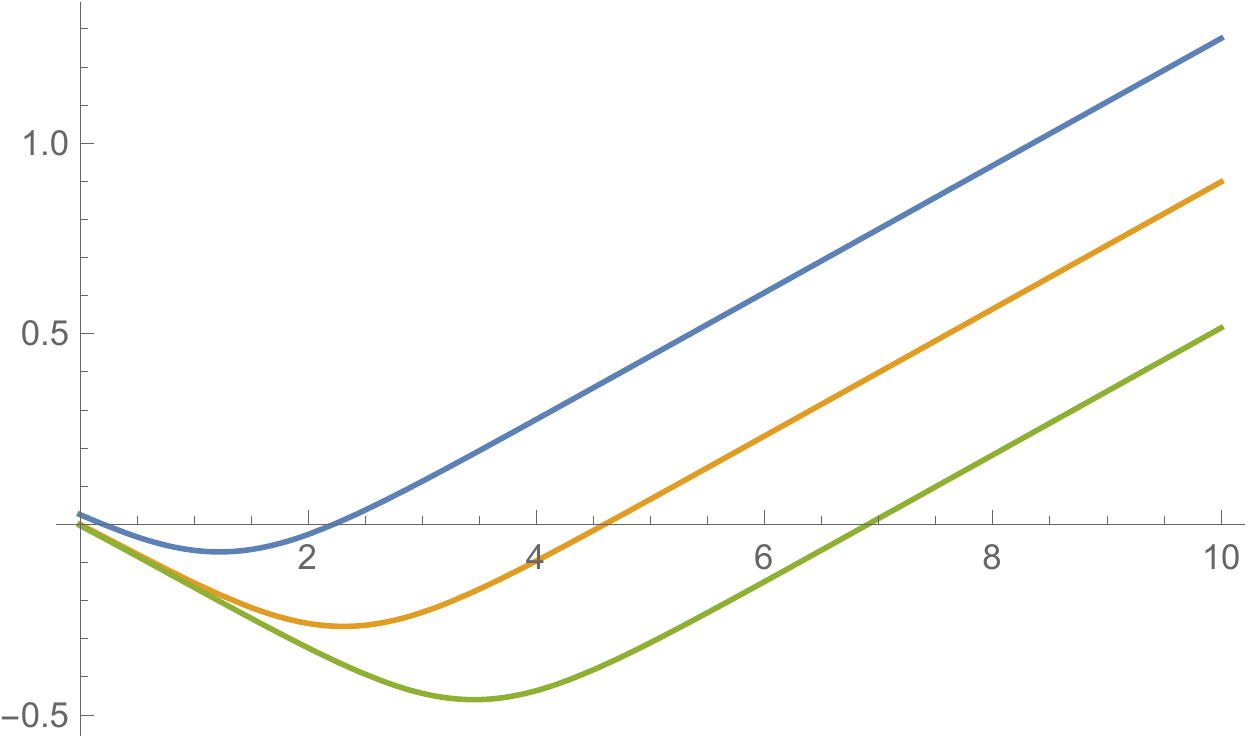}\qquad
 \includegraphics[width=.25\textwidth]{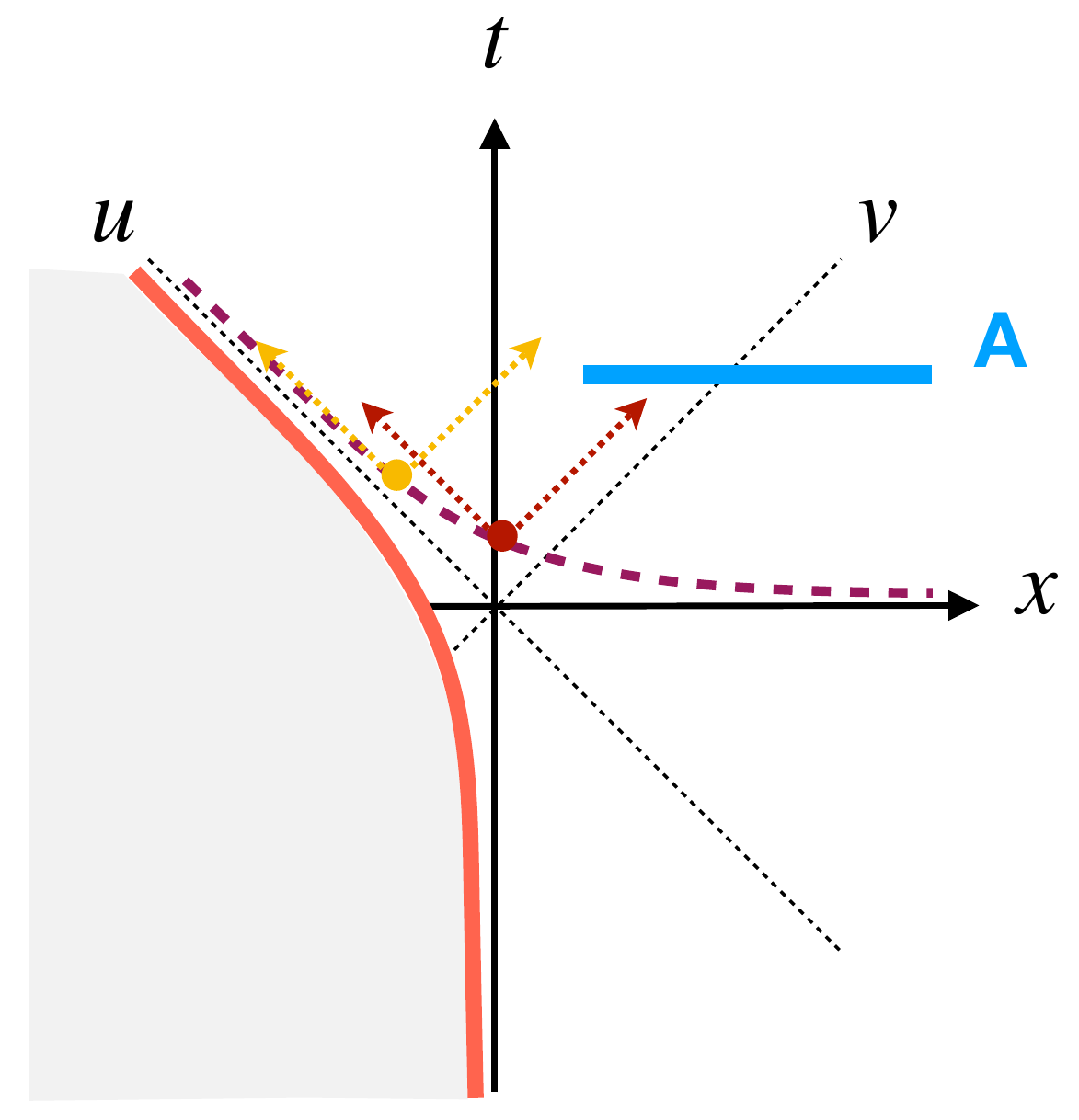}
  \caption{Left: Time evolution of holographic entanglement entropy, where the end point of 
the subsystem $A$ is chosen to be $x_0=-t+\xi_0$. The blue, yellow and green curves correspond to the values
$\xi_0=0.1$, $\xi_0=0.01$ and $\xi_0=0.001$, respectively, where we have set $\beta=1$.
Right: The quasi-particle picture of entanglement growth for the escaping mirror profile in \eqref{exfr}. 
The red thick curve is the mirror trajectory $x=Z(t)$, or equally $v=p(u)$.
The purple dashed curve corresponds to the space like curve defined by $v+p(u)=0$. The black dashed lines are the null lines described by the coordinates \eqref{eq:null}. The production of the entangled particles (red and yellow dots) occurs along the purple curve.}
\label{fighee2}
\end{figure}

In the late time regime,
\ba
\xi_0\geq \beta e^{-(2t-\xi_0)/\beta}, 
\label{condli}
\ea
we find a linear growth, i.e.
\ba
S_A\simeq \frac{c}{6\beta}t+\frac{c}{6}\log\frac{\xi_0}{\ep}+S_\text{bdy}. 
\label{apphpee}
\ea
Refer to the left graphs in Fig.~\ref{fighee2}.
Note that the condition (\ref{condli}) is equivalent to
\ba
v+p(u)>0.
\ea
Therefore, we may conclude that the entangled pair production occurs along the spacelike curve 
$v+p(u)=0$. This matches with our previous estimation of entangled pair production in the free scalar field theory given by \eqref{entpai}.
The propagation of the entangled pairs results in the linear growth of the entanglement entropy \eqref{apphpee}. This scenario is sketched in the right panel of Fig.~\ref{fighee2}.\footnote{It may seem problematic that the entangled pair production occurs along a spacelike surface far from the mirror itself. However, it can be understood in the following way. Although the excitation originates in the moving mirror, quasi-particles arise as a collective phenomenon of the excitation and hence can be distant from the mirror. In other words, it in general takes time for the excitation to behave like quasi-particles.}

\subsection{Geometry of gravity dual}

Even though, expressed in terms of the $(U,V,\eta)$ coordinates, the gravity dual is given by the region \eqref{qsura},
the gravity dual in terms of the coordinates $(u,v,z)$ only covers the region $U<0$ in Poincar\'e AdS$_3$. This is so, since $U=p(u)$ is always negative for any $u$, as we can see from \eqref{exfr}. 

\begin{figure}[h!]
  \centering
 \includegraphics[width=.55\textwidth]{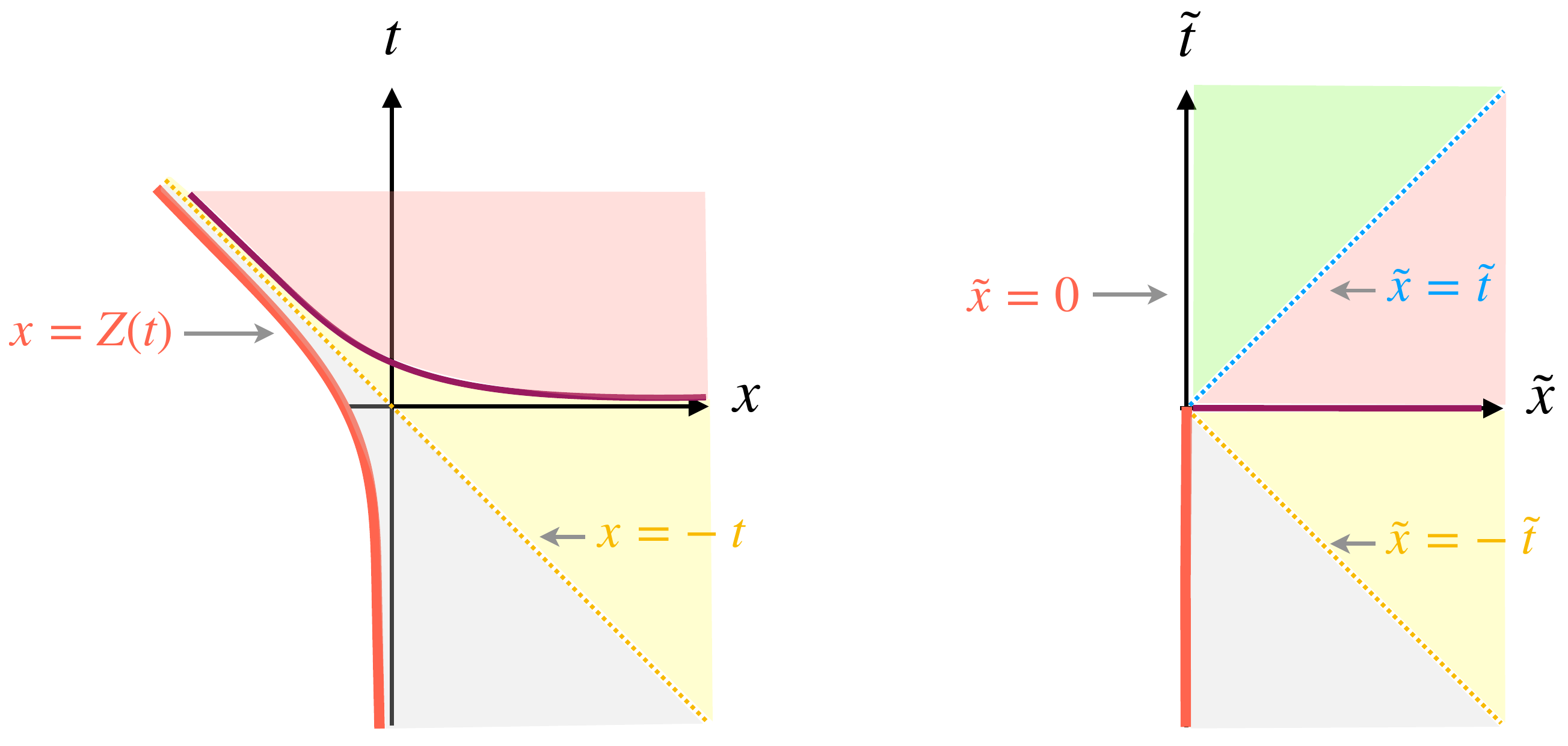}
  \caption{The global spacetime structure for the escaping mirror setup. The left and right pictures describe spacetime in terms of the $(u,v)$ and $(\ti{u},\ti{v})$ coordinates, respectively. The upper green region shown in the right panel cannot be captured in the left plot. The equivalent regions on both sides are identified with the same coloring. The right global structure, for which the corresponding Penrose diagram is depicted on the left in Fig.~\ref{figlimiqqt2}, is similar to the usual black hole diagram drawn in the right panel of the latter. Pair creation mainly occurs on the codimension one surface defined by $v+p=0$ (purple curve). After the conformal map, the purple line gets projected onto the positive horizontal axis $\tilde x \geq 0$ as shown in the right panel. 
  }
\label{figlimiqqt}
\end{figure}

To see how the global spacetime structure is affected, let us first consider the AdS boundary geometry, i.e. $z=0$, where the CFT is defined on. The coordinates $(u,v)$ are embedded into $(U=\ti{u},V=\ti{v})$ as sketched in Fig.~\ref{figlimiqqt}.

As can be seen, the region $U>0$ is missing in the  $(u,v)$ spacetime. This is analogous to the existence of an horizon in a black hole geometry. Indeed, the latter is the reason why we can mimic Hawking radiation in the moving mirror setup. This is consistent with what we have previously discussed by studying the mode functions, see Sec.~\ref{sec:directCFT2}.

Now, let us consider the bulk geometry. Note that
the brane $Q$ is also extended to $U>0$, which is missing in the $(u,v)$ coordinate. This is important in our calculation of holographic entanglement entropy.
The induced metric on $Q$ is described by
\ba
ds^2=\frac{dz^2}{z^2}+\left(\frac{p''}{zp'}+\frac{2\lambda\s{p'}}{z^2}\right)dudz
+\left(\frac{p''^2}{4p'^2}-\frac{p'}{z^2}+\frac{\lambda p''}{z\s{p'}}\right)du^2.  \label{bdyqs}
\ea
This is equivalent to the AdS$_2$ metric \eqref{eefq} via the coordinate transformation \eqref{cordch} by extending to the region $U>0$. 

\begin{figure}[h!]
\centering
\includegraphics[width=.13\textwidth]{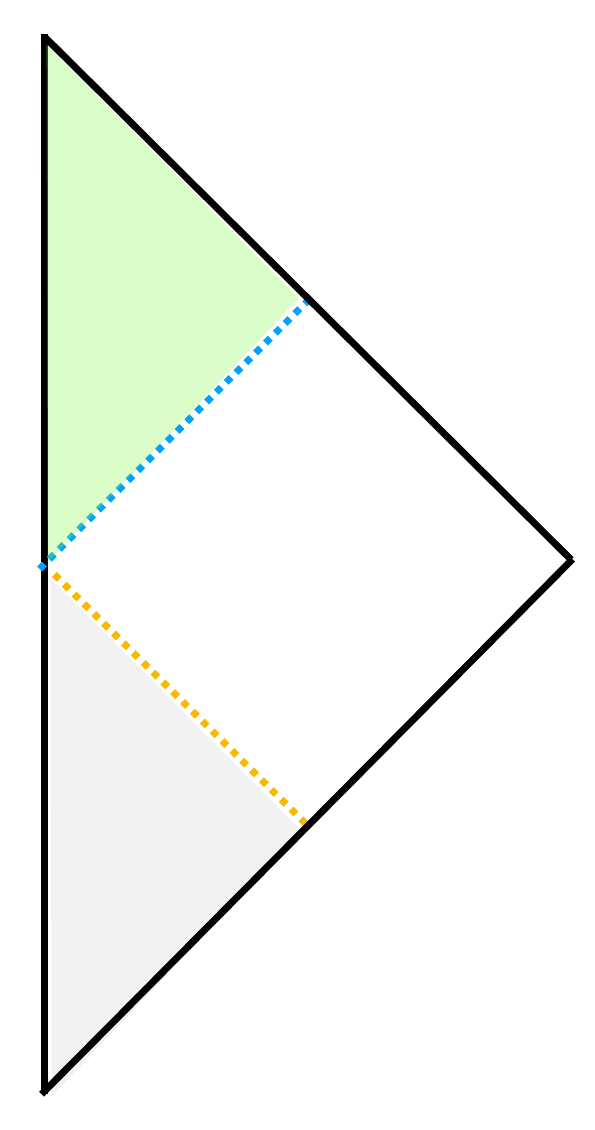}\qquad\qquad\qquad
\includegraphics[width=.18\textwidth]{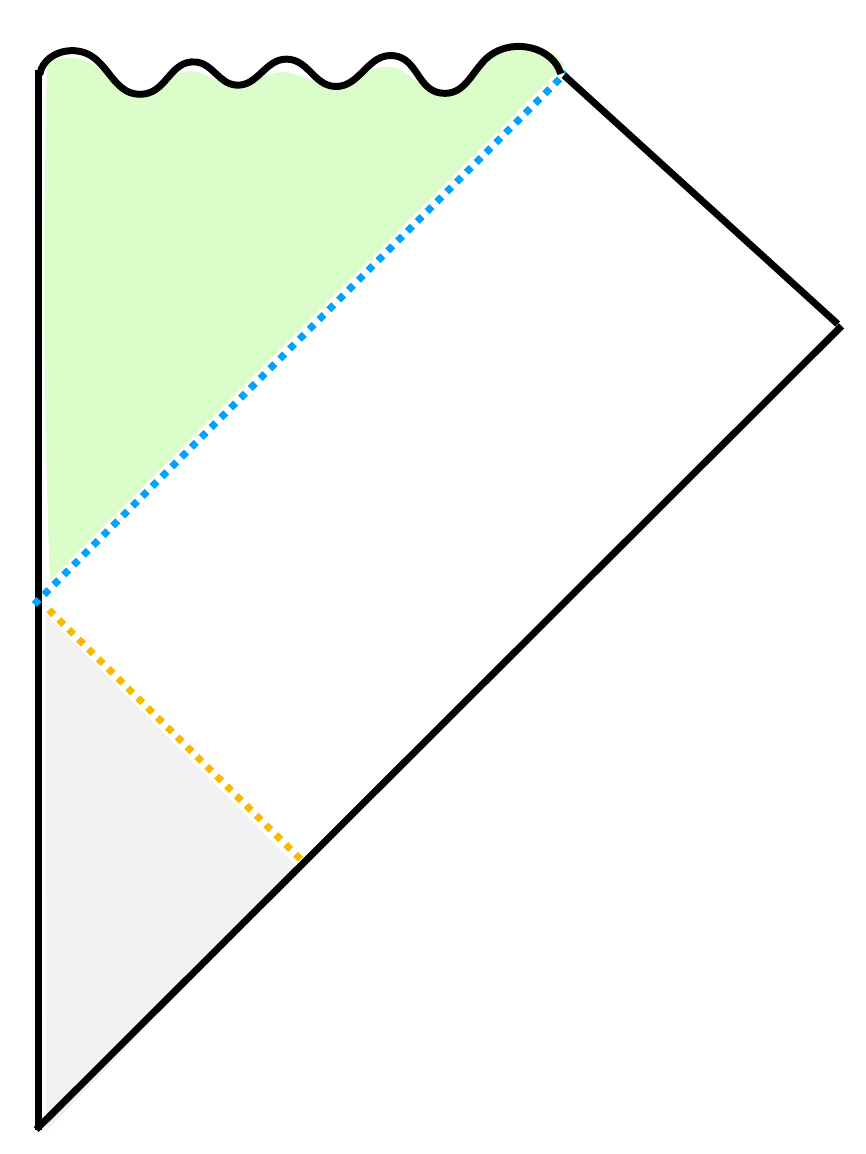}
\caption{Penrose diagram for global spacetime structure of moving mirror setup (left) and black hole geometry (right).}
\label{figlimiqqt2}
\end{figure}

\begin{figure}[h!]
\centering
\includegraphics[width=.23\textwidth]{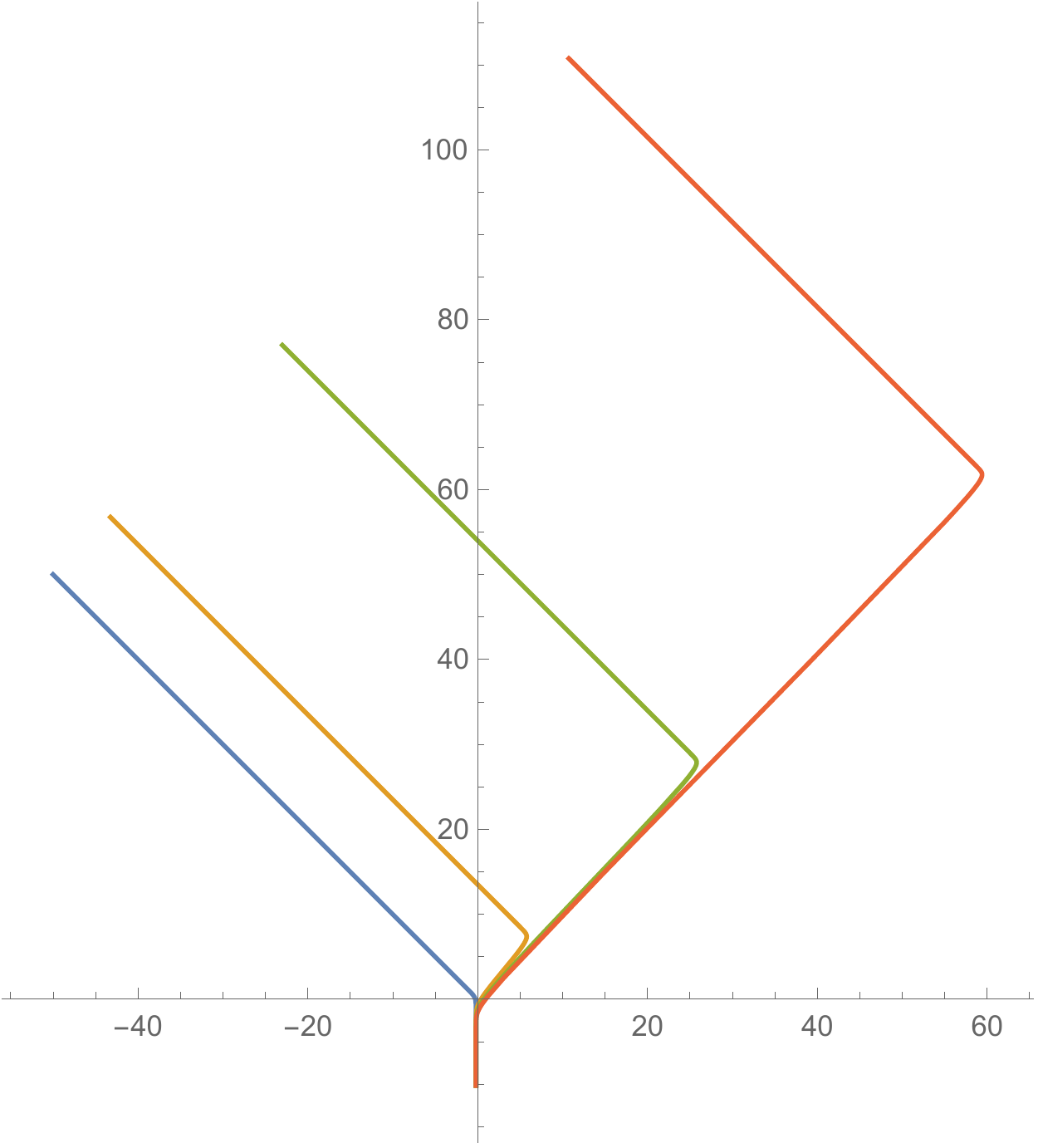}
\caption{We plot the surface $Q$ shape at fixed values $z$, where $\beta=1$. The horizontal and vertical coordinates are $x$ and $t$. We have set $\lambda=0$.
The blue, yellow, green and red curves correspond to $z=0$, $z=3$, $z=6$ and $z=9$, respectively. 
}
\label{figlimit}
\end{figure}

The brane $Q$, which for small $z$ looks as in Fig.~\ref{puqfig}, now changes its profile as $z$ increases. 
This situation is depicted in Fig.~\ref{figlimit}. In the 
$\beta\to 0$ limit, 
the brane $Q$ at a fixed value of $z$ develops a kink shape.

\begin{figure}[h!]
  \centering
 \includegraphics[width=.15\textwidth]{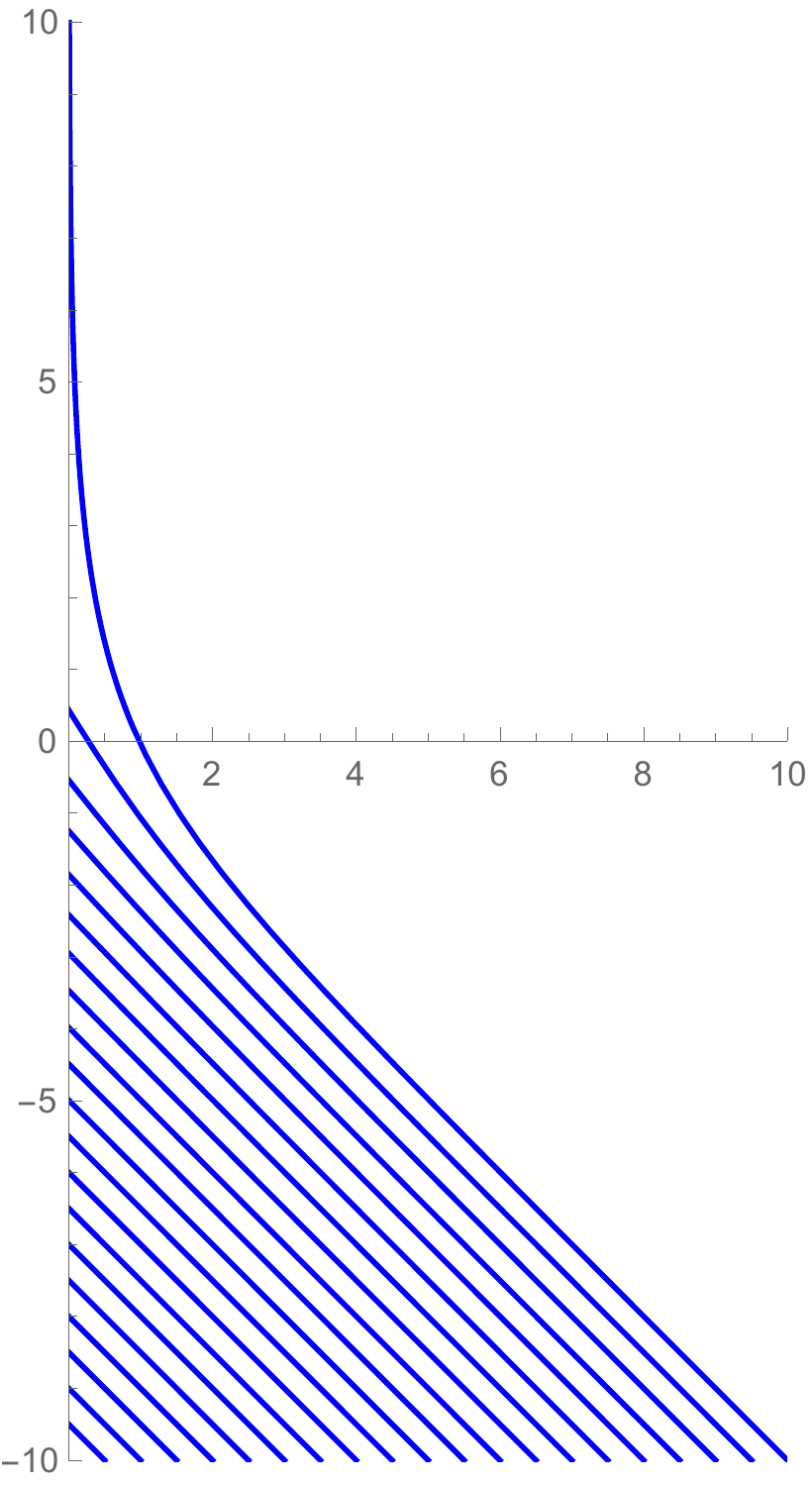}\qquad\qquad
 \includegraphics[width=.2\textwidth]{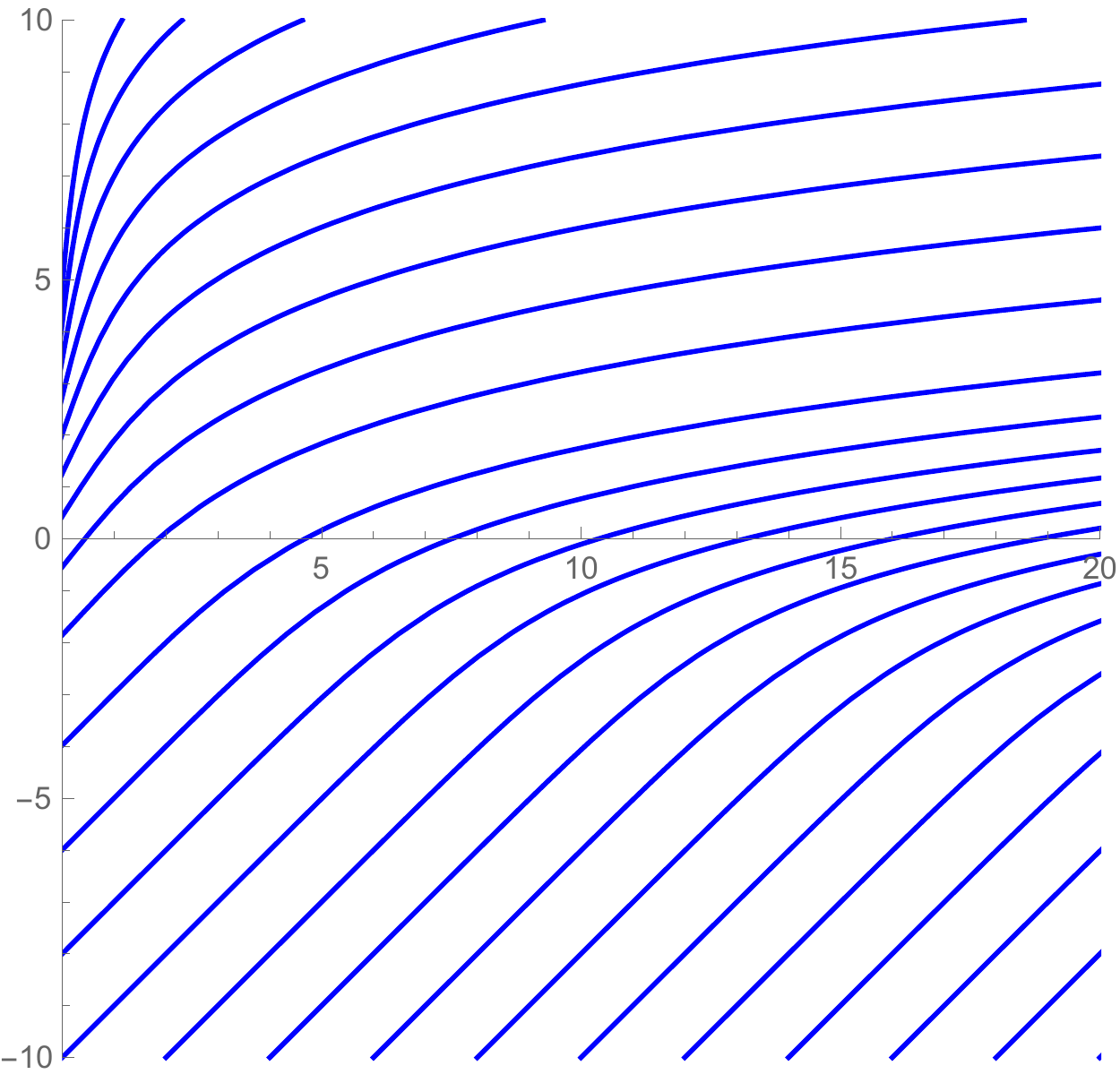}
  \caption{Depicted are null geodesics on the surface $Q$. We plot the left moving \eqref{leftm} (left) and right moving \eqref{rightm} (right) geodesics separately. The horizontal axis is the $z$ direction and the vertical one is labeled by $u$.}
\label{nullg}
\end{figure}

Although we can transform the metric on the two dimensional spacetime of $Q$, which is described by the metric \eqref{bdyqs}, into AdS$_2$, it only covers a lower half of Poincar\'e AdS$_2$. 
To understand the global embedding, let us consider the null geodesics in the metric \eqref{bdyqs}, which are given by (here, we choose $\lambda=0$ and focus on $X=0$ for simplicity)
\begin{align}
z_{L} &=\frac{-p(u)+s}{\s{p'(u)}},  
\label{leftm}\\
z_{R} &=\frac{p(u)+s}{\s{p'(u)}}, 
\label{rightm}
\end{align}
and correspond to $\eta=-T+s$ and $\eta=T+s$, respectively. Here, $s$ is an arbitrary constant. 
These null geodesics are plotted in Fig.~\ref{nullg}. 

This shows that in the $(u,z)$ metric coordinates of \eqref{bdyqs}, the region $z>\frac{-p(u)}{\s{p'(u)}}$ is not connected to the future of the boundary as depicted in 
the left panel of Fig.~\ref{nullgg}. 
Hence, we may conclude that the geometry of $Q$ has a causal horizon much like a black hole.

\subsection{Holographic entanglement entropy for kink mirror}
\label{sec:dkmm}
 
\begin{figure}[h!]
  \centering
  \includegraphics[width=0.45\textwidth]{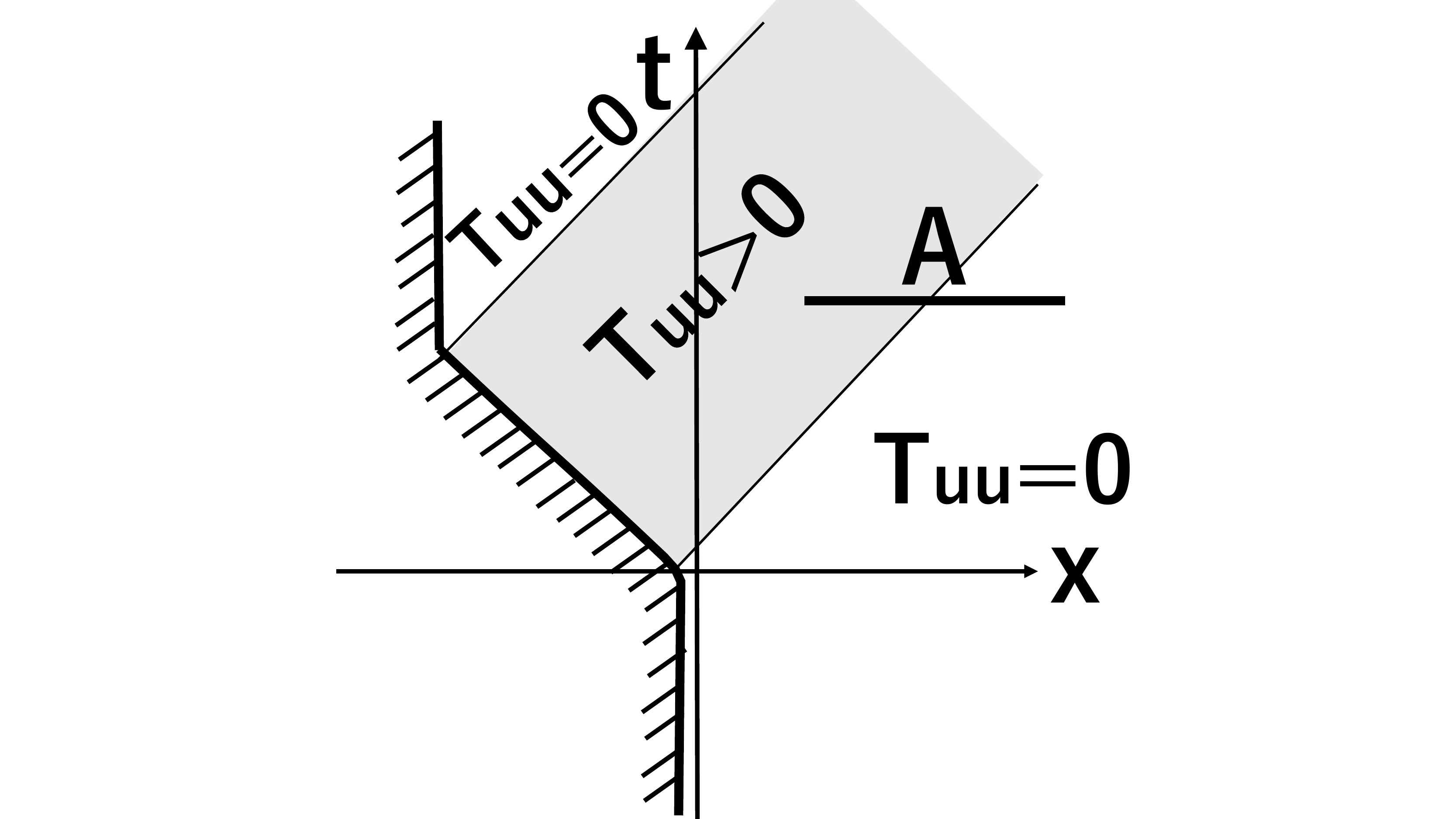}\quad
  \includegraphics[width=0.3\textwidth]{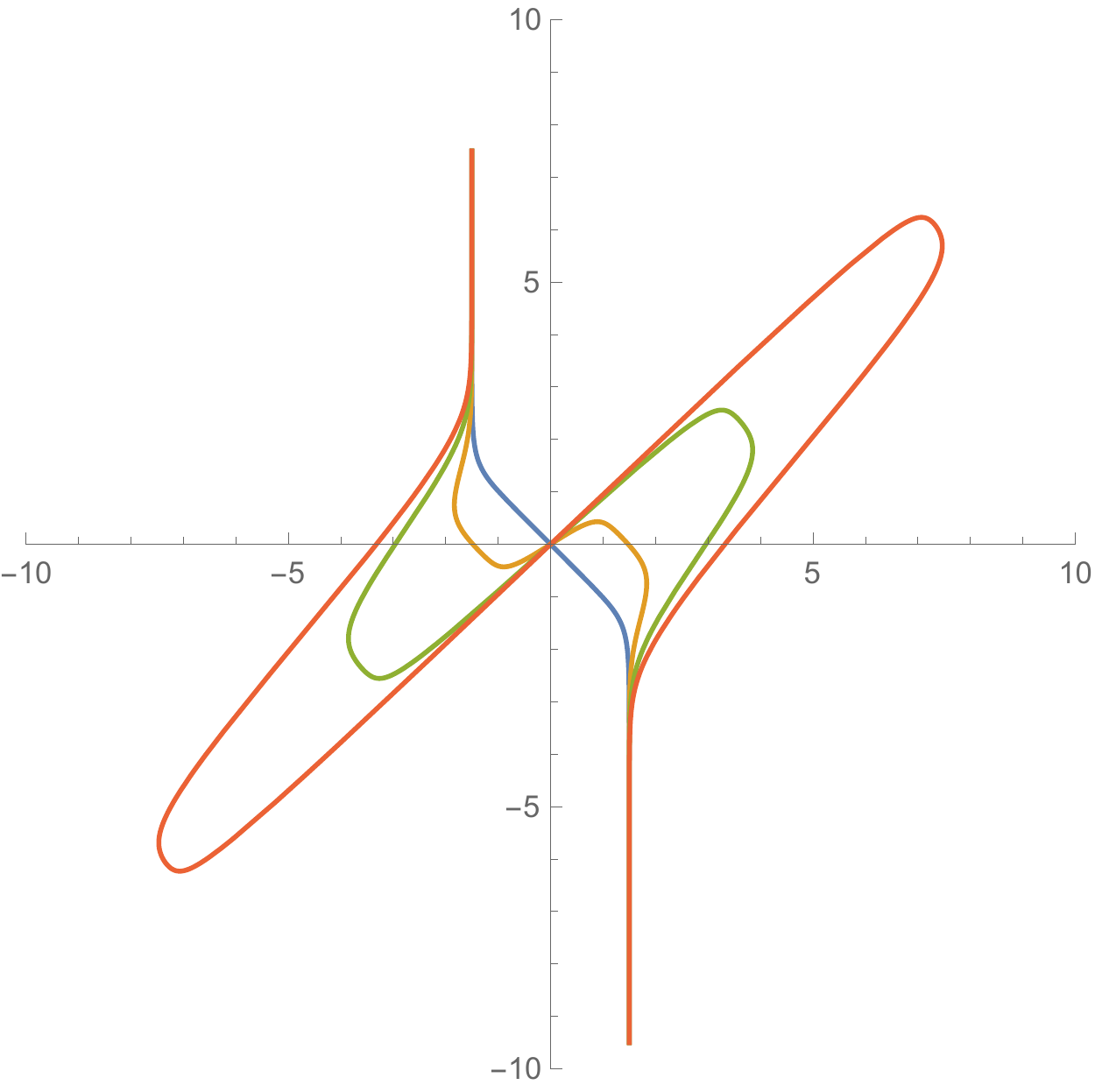}
  \caption{The left sketch shows the behavior of the energy flux $T_{uu}$.
  The profile of the brane $Q$ is depicted in the right panel. We have chosen $z=0,1,2,3$ and $\beta=1/3$. We have shifted the $x$ coordinate such that the center of the trajectory is located at the origin.}
\label{kinkheefig}
\end{figure}

Next, we would like to compute the entanglement entropy for the kink mirror \eqref{kexfr}, which models black hole formation and evaporation.
Note that in this case, we have an energy flux $T_{uu} \neq 0$ in a limited spacetime region, as depicted in the left panel of Fig.~\ref{kinkheefig}.

The gravity dual is obtained by performing the coordinate transformation \eqref{cordch}. As before, the dual geometry is simply mapped into half Poincar\'e AdS$_3$, i.e. $V-U+2\lambda \eta>0$. The profile of the end-of-the-world brane $Q$ 
in the $(x,t)$ plane at a fixed value of $z$ is depicted in the right panel of Fig.~\ref{kinkheefig}.

\begin{figure}[h!]
  \centering
\includegraphics[width=0.4\textwidth]{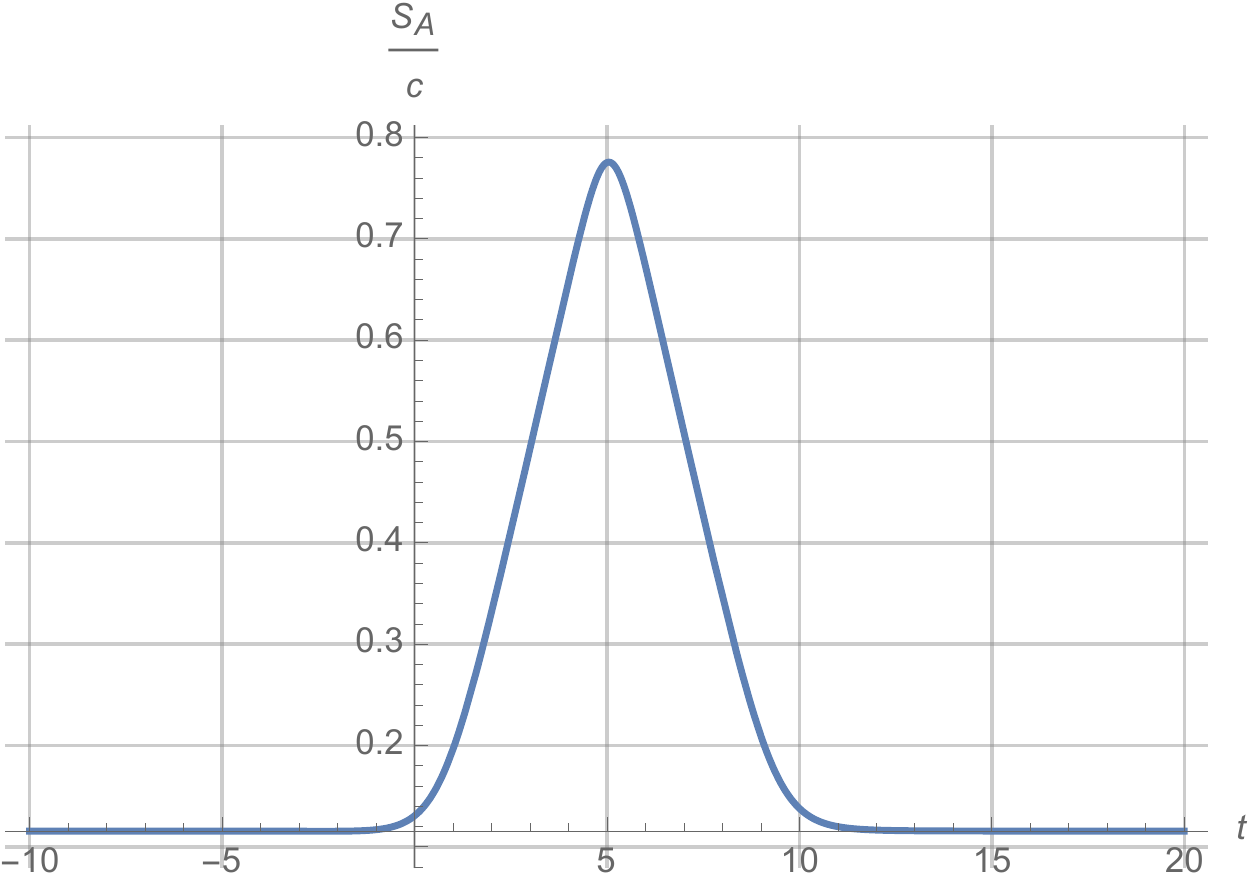}
  \caption{Time evolution of holographic entanglement entropy in the presence of a non-static, semi-infinite subregion $A=[x_0,\infty]$ with $x_0 = Z(t) + 0.1$, cf. \cite{Akal:2020twv}. We have set $\beta=1$ and $u_0=20$. Only $S_A^{\rm dis}$ is shown, since $S_A^{\rm con}$ never dominates in this case.}
\label{fig:idealPage}
\end{figure}

\begin{figure}[h!]
  \centering
  \includegraphics[width=0.65\textwidth]{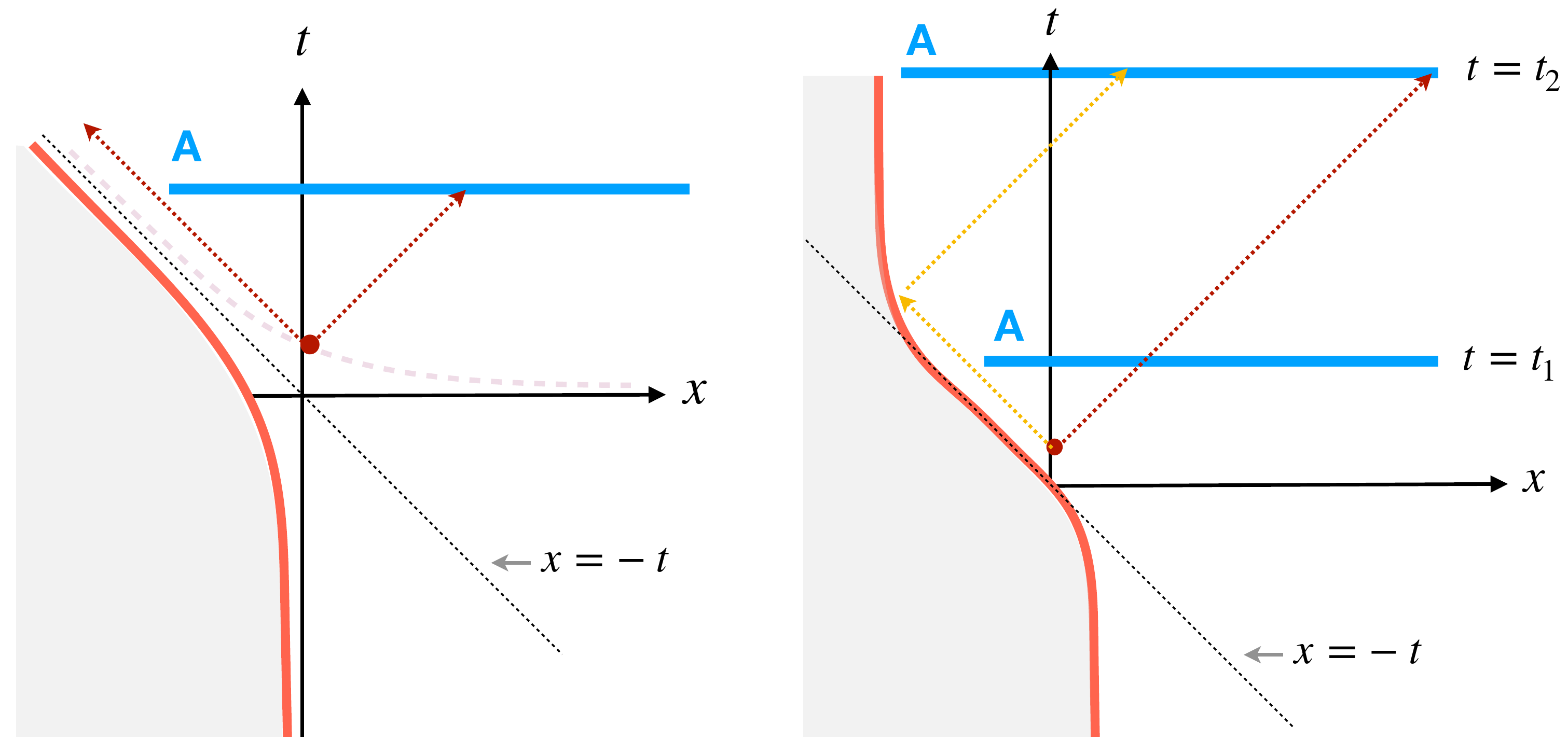}
  \caption{Sketch of propagation of entangled pairs created by the moving mirror (red thick curve). The red dot describes a point where an entangled pair is created. The particles will propagate at the speed of light in the left and right direction. 
In the left panel, the escaping mirror profile in \eqref{exfr} is depicted. In this case, one partner of the pair
propagates to the right and can be picked up in the subsystem $A$ (blue interval), while the other one goes to the left and can never reach the subsystem $A$. The right picture describes the kink mirror setup in \eqref{kexfr}. In this case, at early time $t=t_1$, the subsystem $A$ may capture only one partner of the entangled pair (red arrow). However, at later time $t=t_2$, the initially left moving partner (yellow arrow) is reflected off the mirror and can reach the subsystem $A$.}
\label{kinkeheefig}
\end{figure}

The time evolution of holographic entanglement entropy for the semi-infinite, but static subregion $A=[x_0,\infty]$,
can be computed from the expression given in \eqref{sahee}.
In particular, if we consider a non-static, semi-infinite subregion $A = [x_0,\infty]$, where $x_0 = Z(t) + 0.1$, we end up with the ideal Page curve \cite{Page:1993wv} depicted in Fig.~\ref{fig:idealPage}.\footnote{We should emphasize that if the subregion $A$ is taken to be a static, semi-infinite line as illustrated in Fig.~\ref{kinkeheefig}, the entropy curve gets altered due to remaining vacuum quantum correlations, 
i.e. the final value is increased compared to the original value.} 

\begin{figure}[h!]
  \centering
\includegraphics[width=0.4\textwidth]{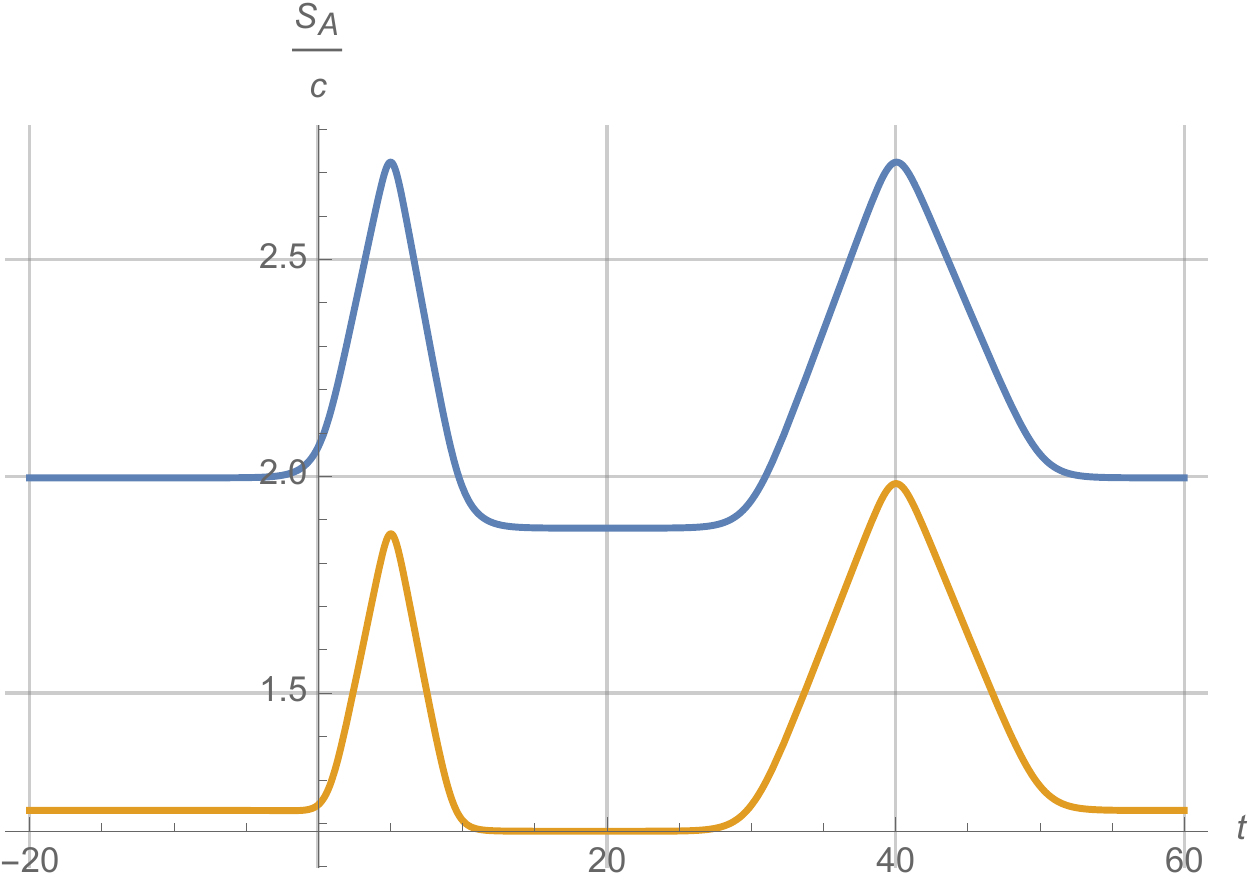}
  \caption{Time evolution of holographic entanglement entropy in the presence of a non-static, finite subregion $A=[x_0,x_1]$ with $x_0 = Z(t) + 0.1$ and $x_1 = Z(t) + 40$, cf. \cite{Akal:2020twv}. We have set $\beta=1$ and $u_0=20$. The orange curve and the blue curve show the disconnected entropy $S_A^\text{dis}$, i.e. \eqref{dishee}, and connected entropy $S_A^\text{con}$, i.e. \eqref{conhee}, respectively. For simplicity, we have set the boundary entropy as $S_{\rm bdy} = 0$. This can, however, take any value in $(-\infty, \infty)$.}
\label{fig:doublePage}
\end{figure}

On the other hand, taking the subregion $A$ to be a finite interval, we obtain two different peaks, see Fig.~\ref{fig:doublePage}. The
first one arises, if only the right moving particles are crossing the subregion $A$, see right panel of Fig.~\ref{kinkeheefig}. The second peak appears, when only the reflected particles cross the subregion $A$. As can be seen from Fig.~\ref{fig:doublePage}, the disconnected result $S_A^\text{dis}$ gives the dominant contribution according to the general formula \eqref{eq:holoEE} when the boundary entropy is smaller than a certain value (we set it to be vanishing in the plot). When the boundary entropy gets larger, the connected phase becomes dominant. For a more detailed discussion, we would like to refer to \cite{Akal:2020twv}.

In Fig.~\ref{kinkeheefig}, we provide an interpretation for the observed evolution of entanglement entropy in Fig.~\ref{kinkheefig} in terms of the propagation of entangled radiation quanta created by the kink mirror.

\subsection{Gravitational interpretation via brane-world and Liouville theory}
\label{subsec:grav-int}

\begin{figure}[h!]
  \centering
\includegraphics[width=.4\textwidth]{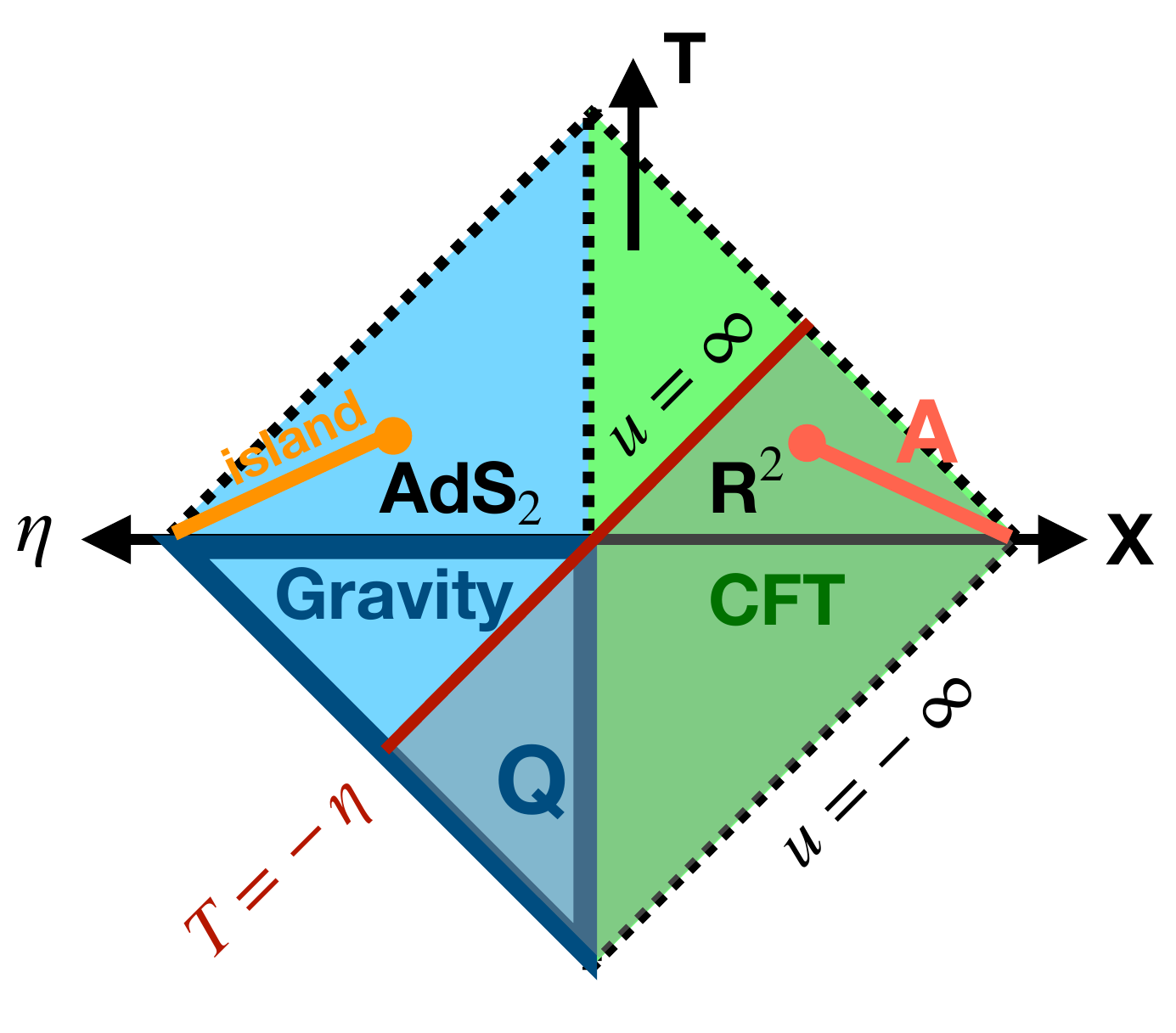}
  \caption{Global spacetime structure of gravity dual of our escaping mirror setup in form of a 
  Penrose diagram. We have chosen $\lambda=0$ for simplicity. The left and right triangle describe 
  the surface $Q$ (AdS$_2$), where the brane-world two dimensional gravity lives on, and the asymptotic AdS boundary, where the BCFT lives on, respectively. The gray shaded region below the $T=-\eta$ line is the causal past of 
  the moving mirror spacetime in the original $(u,v)$ coordinates. We also show our choice of the subsystem $A$ and the appearance of its entanglement island in the calculation of entanglement entropy.}
\label{nullgg}
\end{figure}

In the AdS/BCFT, the CFT on a manifold $M$ with a boundary $\de M$ is dual to the bulk gravity spacetime enclosed by the end-of-the-world brane $Q$ such that $\de Q=\de M$. However, there is another equivalent description, namely the CFT on $M$ may be seen as coupled to a dynamical gravity sector defined on $Q$ along the interface $\de M$. This directly follows from considerations based on brane-world holography
\cite{Randall:1999ee,Randall:1999vf,Karch:2000ct,Almheiri:2019hni}. The boundary geometry of the gravity dual of our escaping mirror setup is depicted in the right panel of Fig.~\ref{nullgg}.
In this subsection, we would like to present an interpretation of our moving mirror setup as a CFT coupled to gravity in two dimensions.

\begin{figure}[h!]
  \centering
\includegraphics[width=0.55\textwidth]{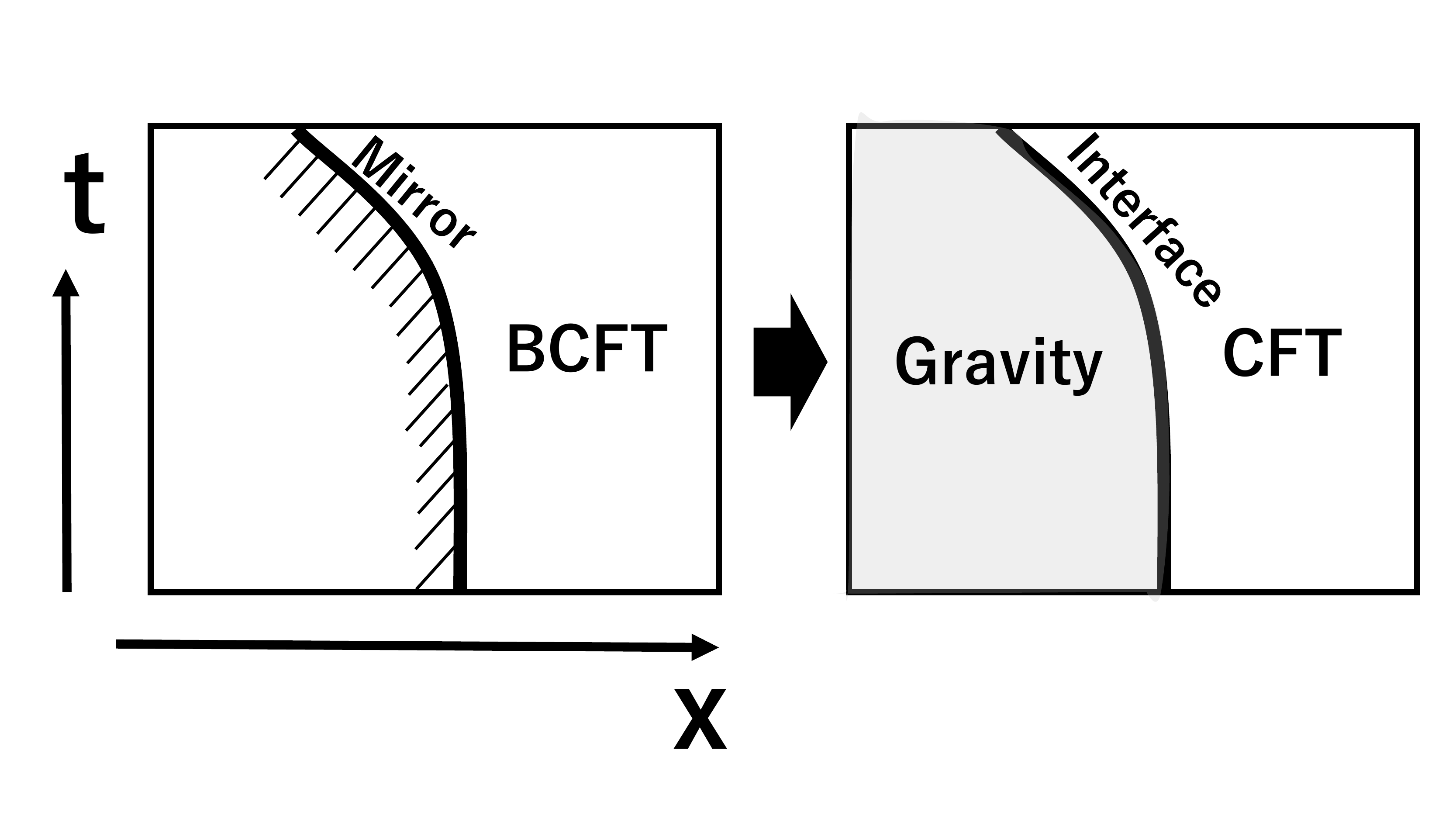}
  \caption{The brane-world interpretation of the AdS/BCFT setup implies that the the BCFT setup 
 of moving mirrors in the left picture can be regarded as a CFT coupled to a gravity along the interface, as shown in the right picture.}
\label{fig:2dgravity}
\end{figure}

Namely, for arriving at this understanding, we can regard the surface $Q$ in AdS/BCFT, where we impose the Neumann boundary condition, as the brane in a brane-world setup. This is so, simply, because their mathematical treatment is equivalent in the context of AdS/CFT.
This suggests that our two dimensional BCFT, which describes a moving mirror field theory, can be interpreted as a two dimensional CFT coupled to two dimensional gravity along an interface, which is geometrically identical to the mirror trajectory as depicted in Fig.~\ref{fig:2dgravity}.

Consider a setup of AdS$_3/$CFT$_2$, whose metric looks like 
\ba
ds^2=\frac{dz^2-e^{2\phi}dudv}{z^2}.
\ea
In the boundary limit $z\to 0$, its dual is a two dimensional CFT on a curved space
\ba
ds^2=-e^{2\phi}dudv.
\ea
In our AdS/BCFT setup, we impose the Dirichlet boundary condition in the right half of the space, while the Neumann condition is imposed in the other left part. In the latter, the gravity gets dynamical. In the presence of a matter CFT with a large central charge $c$, the two dimensional gravity dynamics is described by Liouville theory whose action is given by\footnote{Here, the Liouville mode contributes to the total partition function $Z$ of the CFT coupled to gravity as $Z\propto e^{S_L}$. Therefore, there is an overall minus sign in front of the action. This means that it can actually be identified with the timelike Liouville CFT, rather than a standard Liouville theory.}
\ba
S_L=\frac{c}{24\pi}\int dtdx\left[(\de_t\phi)^2-(\de_x\phi)^2-\mu e^{2\phi}\right]. 
\ea
The standard AdS/BCFT setup, corresponding to the static mirror, given by (\ref{mirrorbc}) and (\ref{eefq}), allows us to identify 
\ba
\mu=\frac{1}{\ep^2\lambda^2},
\ea
where $\ep$ is the UV cutoff of the radial coordinate of AdS, i.e. $\eta>\ep$ in this case. 
Indeed, the solution to the equation of motion in Liouville theory reads
\ba
e^{2\phi}=\frac{4\lambda^2\ep^2}{(V-U)^2},
\ea
which satisfies the expected boundary condition 
$e^{2\phi}=1$ along the interface $X=-\lambda \ep$, which is the intersection between the UV cutoff surface at $\eta=\ep$, and the surface $Q$ at $X+\lambda\eta=0$. 

Now, we move on to a general profile of a moving mirror, where the coordinate is written as 
$(u,v)=(t-x,t+x)$, and the mirror trajectory is given by $v=p(u)$.
In this case, we can find the solution to the Liouville equation as follows
\ba
e^{2\phi}=\frac{4\lambda^2\ep^2 p'(u)}{(v-p(u))^2}.
\label{lvmetg}
\ea
In this case, the interface between the CFT and gravity is given by 
\ba
v=p(u)-2\lambda \ep \s{p'(u)},
\ea
which we obtain by setting $z=\ep$ in (\ref{mirrortj}). In the limit $\ep\to 0$, this indeed 
coincides with the mirror trajectory.

However, note that we need to match the state in gravity with that in the CFT along the interface. The energy stress tensor of the Liouville sector is provided by the standard formula
\ba
&& T^{(LV)}_{uu}=\frac{c}{12\pi}\left(\de_u\phi^2-\de^2_u\phi\right),\no
&& T^{(LV)}_{vv}=\frac{c}{12\pi}\left(\de_v\phi^2-\de^2_v\phi\right).
\ea
For the gravity background (\ref{lvmetg}), the energy stress tensor is found as 
\ba
T^{(LV)}_{uu}=\frac{c}{24\pi}\left(\frac{3p''^2}{2p'^2}-\frac{p'''}{p'}\right),
\ \ \  T^{(LV)}_{vv}=0.
\ea
This indeed matches with the energy stress tensor of the moving mirror BCFT given by (\ref{eq:Tuu-gen}). In this way, we can determine the dynamics of the system. The latter consists of a two dimensional CFT coupled to gravity in two dimensions along the interface for which we solve the Liouville equation and impose
that the energy stress tensor is continuous along it. 

It is also intriguing to consider the holographic calculation of entanglement entropy.
Recall that in AdS/BCFT, this is given by the formula (\ref{eq:holoEE}). When the subsystem $A$ is semi-infinite,
as we have seen, the disconnected contribution is always favored. From the viewpoint of the 2d gravity interpretation, this disconnected one is equivalent to the presence of 
an island \cite{Penington:2019npb,Almheiri:2019psf,Almheiri:2019hni}, refer to Fig.~\ref{nullgg}.
In this way, in our model, where the CFT on a half plane line is coupled to 2d gravity, we always have an island region. One may think this situation is a bit different from the earlier models of evaporating black holes \cite{Penington:2019npb,Almheiri:2019psf,Almheiri:2019hni,Almheiri:2019yqk},
where a phase transition between the case without an island and that with an island 
explains the Page curve behavior. As we will clearly see in the next subsection, we actually find that our model and earlier models are naturally connected, if we view them from the viewpoint of AdS/BCFT. 

It is also useful to note that when $\lambda$ 
is very large, the 2d gravity gets weakly coupled, where the boundary entropy of the BCFT
\cite{WedgeHolo}
\ba
S_\text{bdy}=\frac{1}{4G^{(Q)}_N}=\frac{1}{4G_N}\mathrm{arcsinh}(\lambda),
\ea
gets large. Here, $G^{(Q)}_N$ is the effective Newton constant of 2d gravity on $Q$ . On the other hand, when $\lambda$ is small, the gravity gets strongly coupled and its entropy is decreased, which may be interpreted as a reflective mirror in a CFT. 

\subsection{Comparison with earlier island models via AdS/BCFT}
\label{subsec:early-isl}

In earlier models of evaporating black holes using the island prescription\footnote{As pointed out in \cite{Chen:2019uhq},
depending on the details of the model, there can exist more than 
two phases, where the turning point is not always at the phase transition point of emergence of an island.}
\cite{Penington:2019npb,Almheiri:2019psf,Almheiri:2019hni,Almheiri:2019yqk}, 
the turning point at the Page time $t=t_P$ typically 
corresponds to a
phase transition between a phase without island and that with an island. 
On the other hand, considering our model describing the moving mirror, in the 2d gravitational 
description, explained in the previous section, there is always an island region. 
Thus, one may be puzzled with these two totally different-looking situations.
However, if we consider these setups holographically based on AdS/BCFT, 
we can understand that they are actually smoothly connected as we will see below.

\begin{figure}[h!]
  \centering
\includegraphics[width=0.7\textwidth]{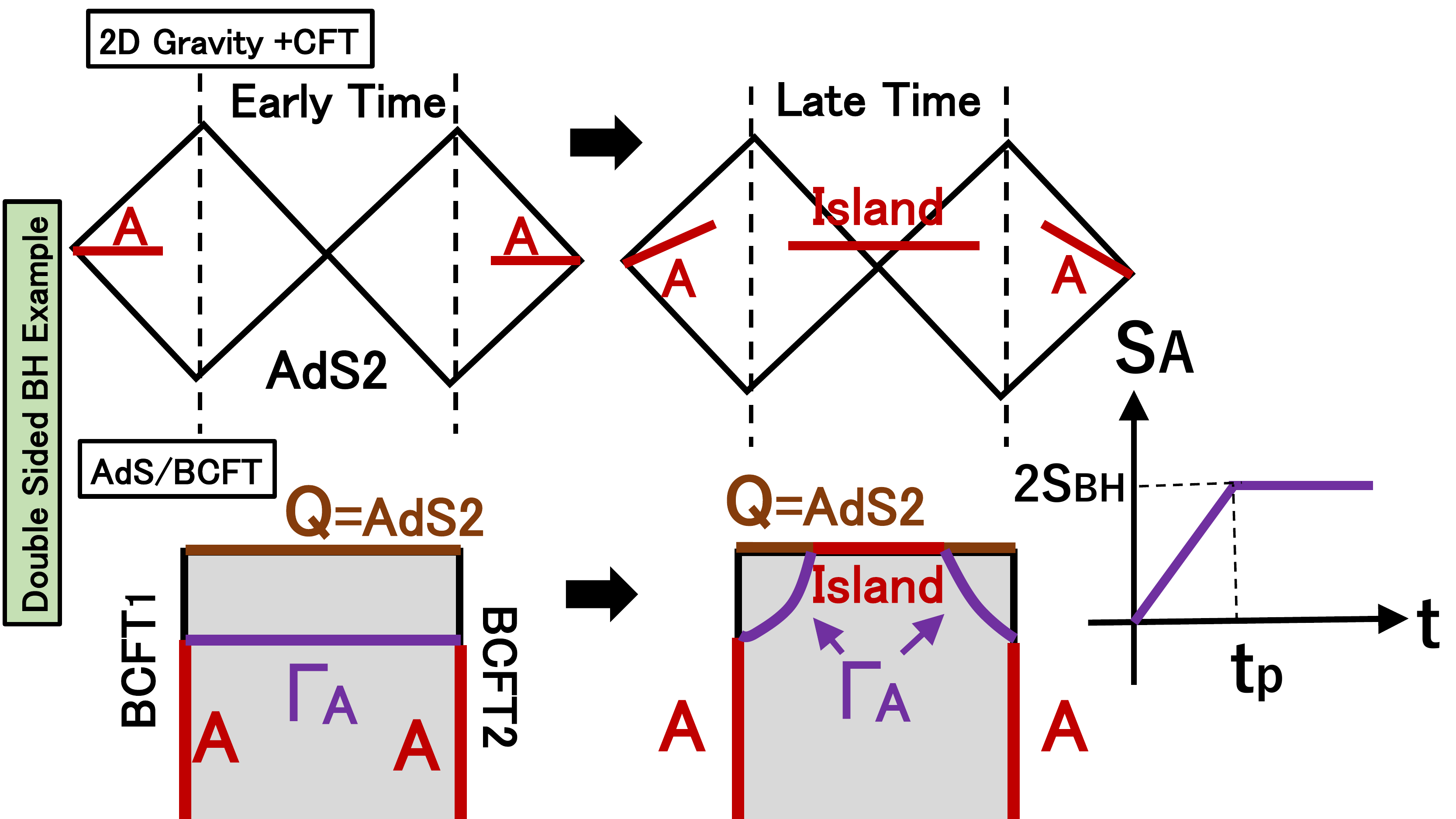}
  \caption{We sketch the gravity dual and computation of entanglement entropy in two BCFTs coupled to an eternal AdS$_2$ black hole. The two-dimensional description, summarized in the upper Penrose diagrams, is identical to the setup of \cite{Almheiri:2019yqk}. The gravity dual via the AdS/BCFT is given by the lower pictures, where $Q$ corresponds to the brane in AdS/BCFT which is dual to the boundary of the BCFT. The two BCFTs are entangled, which is dual to the Einstein-Rosen bridge in the gravity side. 
  At early time, the connected geodesic is favored and there is no island region. 
  However, at late time, the disconnected one becomes dominant and an island region appears.
  Accordingly, the entanglement entropy $S_A$ initially grows linearly, and at the Page time $t=t_P$ it saturates, as sketched in the right graph.}
\label{fig:Islandmodel}
\end{figure}

To see this, consider the setup of an eternal two dimensional black hole coupled to two CFTs on half lines, whose island prescription is given in \cite{Almheiri:2019yqk}. This is depicted in the upper pictures in Fig.~\ref{fig:Islandmodel}. We can construct its gravity dual via AdS/BCFT. It is simply given by cutting an eternal BTZ spacetime along a surface $Q$, which splits the $x$ direction into two parts as in (\ref{tenzersufg}). For vanishing 
tension ${\cal T}=0$, the surface $Q$ is simply given by $x=0$, as depicted in the lower pictures in Fig.~\ref{fig:Islandmodel}. Similar to the gravity dual of a global quantum quench \cite{Hartman:2013qma}, the holographic entanglement entropy then will grow linearly for $t<t_P$, where the connected geodesic $\Gamma^{\rm{con}}_A$ is favored. On the other hand, for later times, i.e. $t>t_P$, the disconnected geodesic $\Gamma^{\rm{dis}}_A$ will be dominating, and the entropy becomes constant. 
The disconnected geodesics end on the brane $Q$, which indeed opens up the island region.

\begin{figure}[h!]
  \centering
\includegraphics[width=0.7\textwidth]{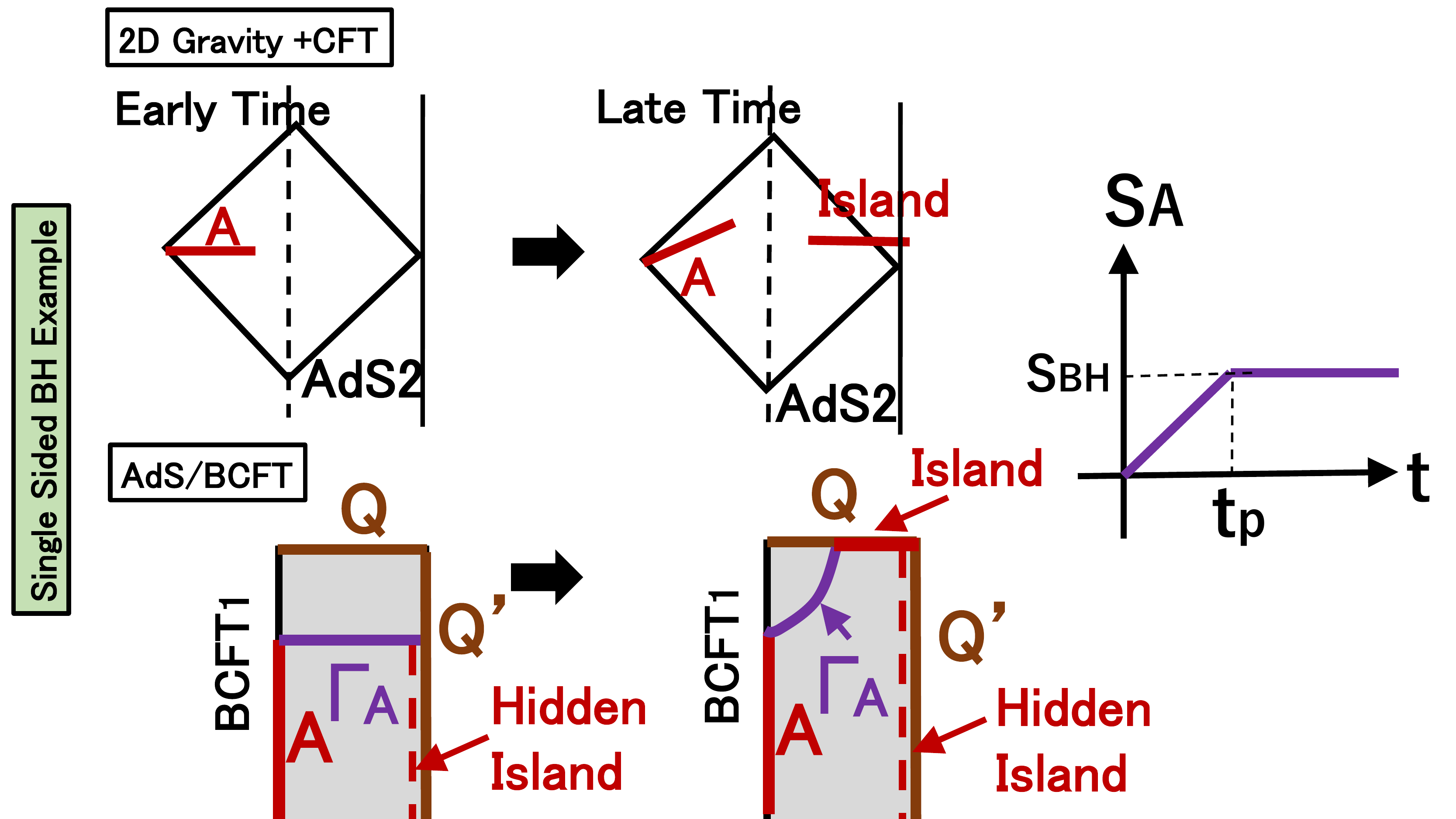}
  \caption{We sketch the gravity dual and computation of entanglement entropy in a BCFT coupled to single sided AdS$_2$ black holes. This setup is obtained from the double sided one in Fig.~\ref{fig:Islandmodel} via 
  a $Z_2$ quotient in the spacial coordinate. Its gravity dual is given by introducing an extra end-of-the-world brane $Q'$. The holographic entanglement entropy is dominated by a geodesic which ends on $Q'$ initially. At late time, the one which ends on $Q$ is preferred and an island region appears. The time evolution of entanglement entropy $S_A$ is similar to the one in the double sided example.}
  \label{fig:Islandmodel2}
\end{figure}

Now, let us take a $Z_2$ quotient of the previous setup, such that two CFTs on half spaces 
are identified. This transforms an eternal black hole into a single sided black hole, as depicted in Fig.~\ref{fig:Islandmodel2}. Its gravity dual now contains an extra brane $Q'$, which arises due to the $Z_2$ quotient, and is identical to the end-of-the-world brane in \cite{Hartman:2013qma}, in addition to the original brane $Q$. In this gravity dual setup, the holographic entanglement entropy again experiences a phase transition as depicted in the lower pictures of Fig.~\ref{fig:Islandmodel2}. At early time, the geodesic $\Gamma_A$, which ends on the extra brane $Q'$ is favored, and this gives rise to a linear growth of the entanglement entropy. At late time $t>t_P$, the geodesic $\Gamma_A$, which ends on the original brane $Q$, becomes dominating and the entanglement entropy saturates. 
Note that in this single sided setup, the geodesic $\Gamma_A$ always ends on one of the end-of-the-world branes $Q$ or $Q'$. This means that there always exists an island region, if we also count the hidden island region, see lower pictures of Fig.~\ref{fig:Islandmodel2}, which appears on the extra brane $Q'$. In this way, we can always find an island region, if we apply brane-world holography to the brane $Q'$ as well. Also note, that this construction is qualitatively similar to \cite{Almheiri:2019hni}. Thus, we can conclude that our model, describing 2d gravity coupled to a CFT on a half space, is smoothly connected to earlier island models of evaporating single sided black holes. 

\section{Double escaping mirror}
\label{sec:dem}

In this section, we generalize the previous results to the case of two oppositely placed moving mirrors in the CFT. After explaining how they can be mapped to a much more simple setup in the CFT, we construct the corresponding gravity duals of them. There are two types of gravity duals, namely, depending on how one chooses the boundary conditions on the mirrors. Finally, we will consider a holographic CFT with two oppositely escaping mirrors as a concrete example, and discuss our findings. 

\subsection{Conformal map: from double moving mirrors to strips}

\begin{figure}[h!]
  \centering
  \includegraphics[width=0.8\textwidth]{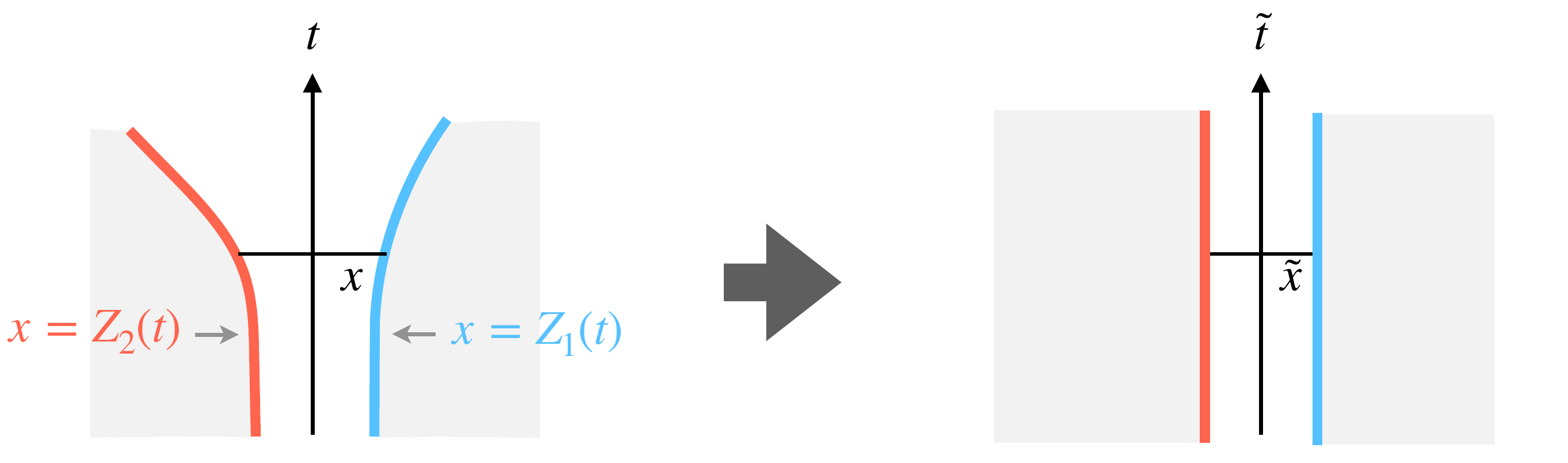}
  \caption{Sketch of conformal map from a double moving mirror setup (left) to a standard setup of BCFT with two straight line boundaries, i.e. a strip (right).}
\label{MVDfig}
\end{figure}

The trajectories of the two mirrors can be described by $x=Z_1(t)$  and $x=Z_2(t)$. We would like to conformal map these two trajectories into $\tilde{x}=L/2$ and $\tilde{x}=-L/2$, so that the resulting manifold is a strip with width $L$. See Fig.~\ref{MVDfig} for a sketch. The corresponding map is given by 
\ba
\ti{u}=p(u),\qquad \ti{v}=q(v),  
\label{cmapp}
\ea
where the functions $p$ and $q$ are determined by 
\be
\begin{split}
L &=q\left(t+Z_1(t)\right)-p\left(t-Z_1(t)\right),\\
-L &=q\left(t+Z_2(t)\right)-p\left(t-Z_2(t)\right).
\end{split}
\label{mirrorprofile}
\ee
In particular, if we assume $Z_1=-Z_2$, then we have $p=q$.

Under the conformal transformation, the energy stress tensor reads
\be
\begin{split}
T_{uu} &=\frac{c}{12\pi}\frac{3(p'')^2-2p'p'''}{4p'^2} + p'^2 T_{\ti{u}\ti{u}},\\
T_{vv} &=\frac{c}{12\pi}\frac{3(q'')^2-2q'q'''}{4q'^2} + q'^2 T_{\ti{v}\ti{v}}, 
\label{emcftl}
\end{split}
\ee
where $T_{\ti{u}\ti{u}}$ ($T_{\ti{v}\ti{v}}$) is the energy stress tensor for a strip and depends on the details of the boundary conditions. When the two boundaries of the strip have exactly the same conformal boundary conditions, the energy stress tensor is identical to that of a cylinder and $T_{\ti{u}\ti{u}}=T_{\ti{v}\ti{v}}=-\frac{\pi c}{48 L^2}$. In this case, 
\be
\begin{split}
T_{uu} &=\frac{c}{12\pi}\left(\frac{3(p'')^2-2p'p'''}{4p'^2}-\frac{\pi^2 p'^2}{4L^2}\right),\\
T_{vv} &=\frac{c}{12\pi}\left(\frac{3(q'')^2-2q'q'''}{4q'^2}-\frac{\pi^2 q'^2}{4L^2}\right). 
\label{emcftl2}
\end{split}
\ee

\subsection{Gravity duals: confined and deconfined configurations}

Now, let us consider the gravity dual of a double moving mirror setup. As we have discussed in the former subsection, since a double moving mirror setup can be mapped to a strip, it is sufficient to consider the gravity dual of a strip and perform a proper Ba$\tilde{\text{n}}$ados map to it to obtain the gravity dual of a double moving mirror. 

\begin{figure}[H]
  \centering
  \includegraphics[width=5cm]{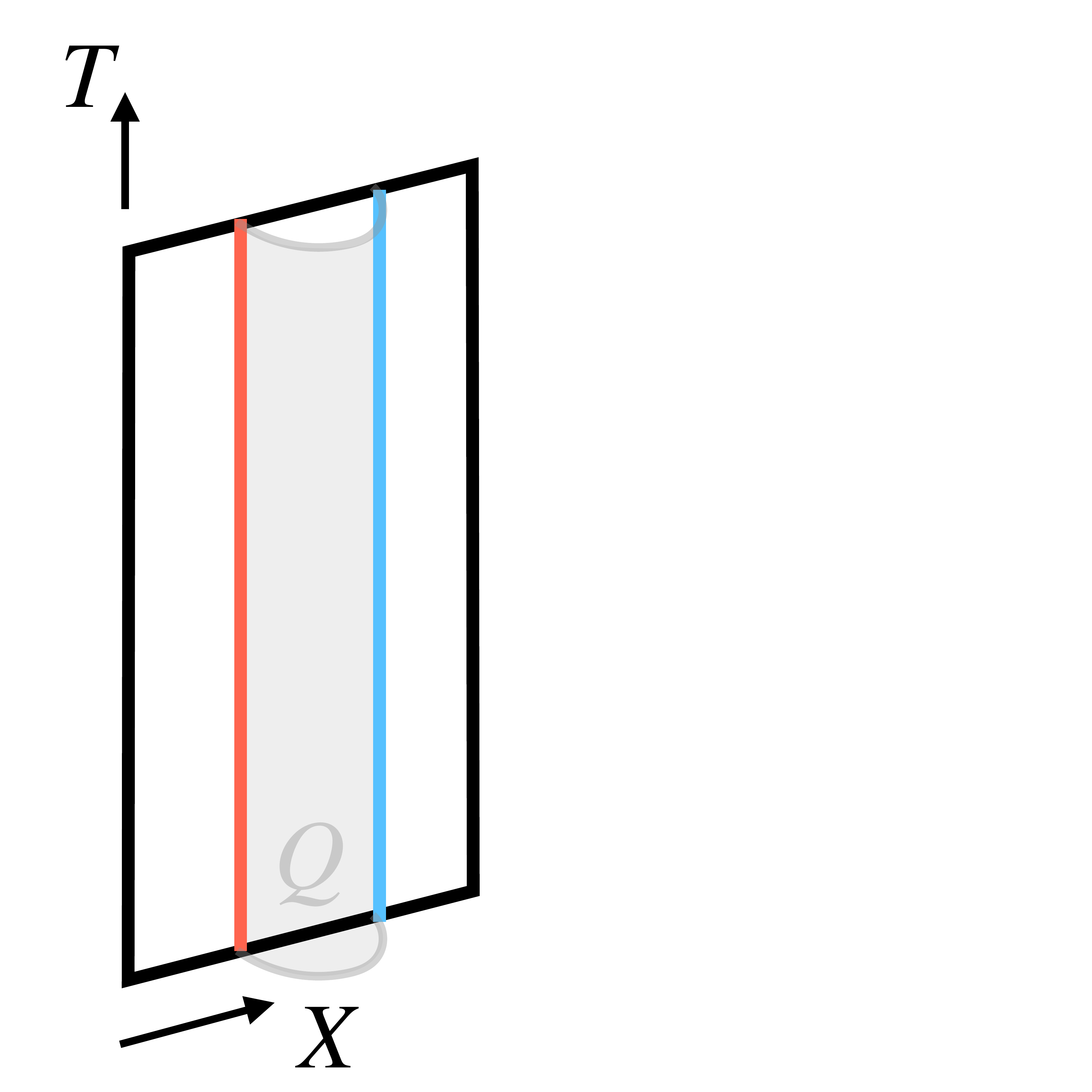}
  \includegraphics[width=5cm]{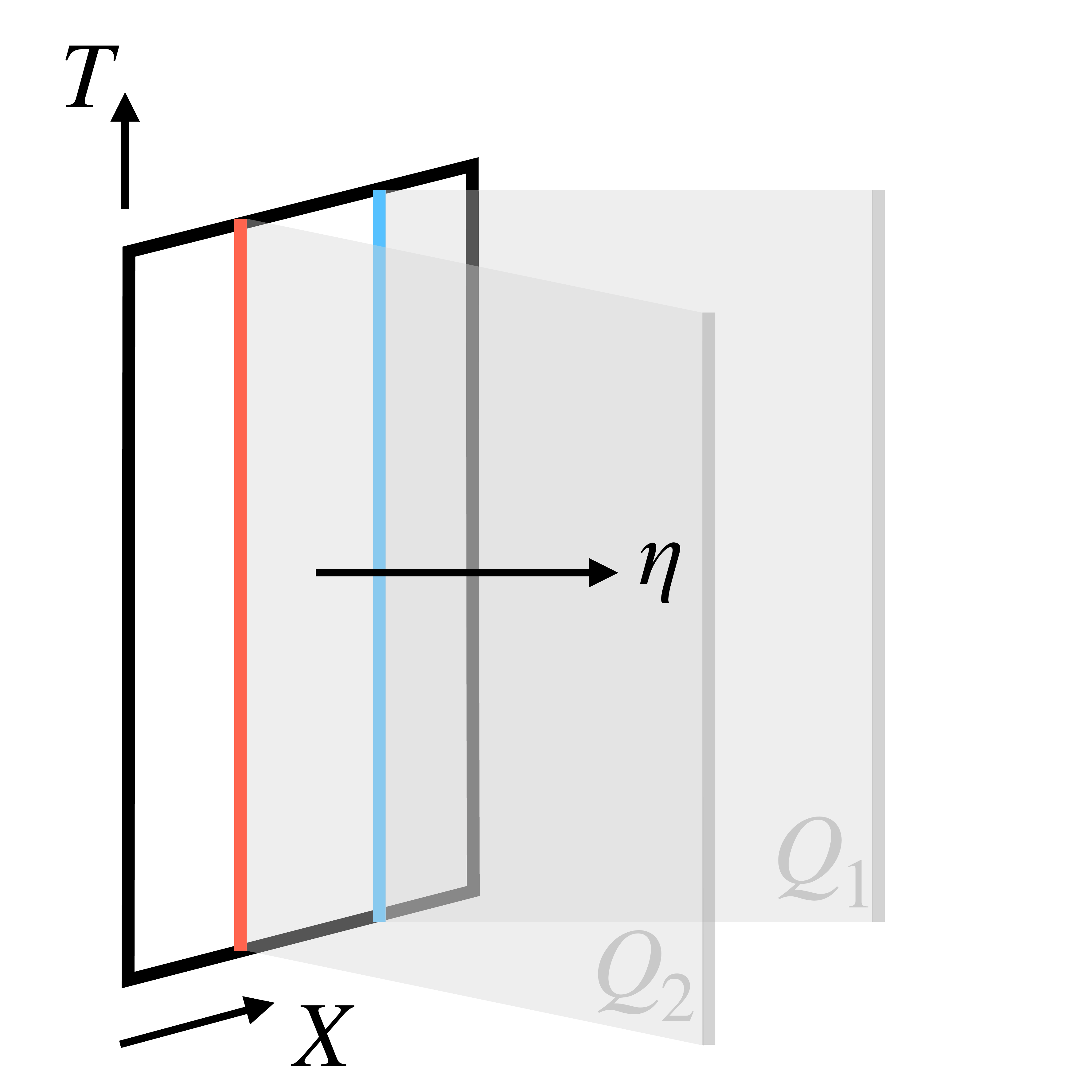}
  \caption{Two types of gravity duals for a strip. The left panel shows the confined configuration, and the right panel shows the deconfined configuration.}
\label{fig:condecon}
\end{figure}

According to AdS/BCFT, there are two types of gravity duals for a strip \cite{AdSBCFT,AdSBCFT2}. One is the confined configuration given by a portion of global AdS$_3$ with one end-of-the-world brane $Q$. The other one is the deconfined configuration given by a portion of Poincare AdS$_3$ with two end-of-the-world branes $Q_1$ and $Q_2$. See Fig.~\ref{fig:condecon} for a sketch. If the two boundaries of the strip have the same conformal boundary condition imposed on, then both configurations are allowed as on-shell configurations. In this case, the confined configuration has a smaller action and is hence physical. On the other hand, if the two boundaries of the strip have different boundary conditions, then only the deconfined configuration is allowed. Note that the confined configuration breaks supersymmetry in the bulk \cite{AdSBCFT2}. 

\subsubsection{Confined configuration: geometry and holographic entanglement entropy}

The confined bulk configuration for a strip is given by a portion of global AdS$_3$, whose metric can be expressed as
\begin{align}
    ds^2 = - \left(1+\left(\frac{\pi \eta}{2L}\right)^2\right)^2\frac{dT^2}{\eta^2} + \frac{d\eta^2}{\eta^2} + \left(1-\left(\frac{\pi \eta}{2L}\right)^2\right)^2\frac{dX^2}{\eta^2}, 
\end{align}
This can be related to the standard global AdS$_3$ metric 
\begin{align}\label{eq:corglobalads}
    ds^2 = -\left(R^2 + \frac{\pi^2}{L^2}\right) dT^2 + \frac{dR^2}{R^2 + \frac{\pi^2}{L^2}} + R^2 dX^2,
\end{align}
with 
\begin{align}
    R = \frac{1}{\eta} - \left(\frac{\pi}{2L}\right)^2 \eta . 
\end{align}
The end-of-the-world brane $Q$ is located at 
\begin{align}
    R \cos\left(\frac{\pi X}{L}\right) = \frac{\pi}{L} \frac{\CT}{\sqrt{1-\CT^2}}. 
\end{align}
Performing a proper bulk coordinate transformation, the Ba$\tilde{\text{n}}$ados formalism \cite{Banados:1998gg} guarantees that the bulk metric of a holographic double moving mirror setup can be expressed as 
\begin{align}
ds^2=\frac{dz^2}{z^2}+T_{+}(u)(du)^2+T_{-}({v})(dv)^2-\left(\frac{1}{z^2}
+z^2T_{+}(u) T_{-}(v)\right)dudv, 
\end{align}
where 
\begin{align}
T_{+}(u)=\frac{3(p'')^2-2p'p'''}{4p'^2}-\frac{\pi^2 p'^2}{4L^2},\qquad T_{-}(v)=\frac{3(q'')^2-2q'q'''}{4q'^2}-\frac{\pi^2 q'^2}{4L^2}.
\end{align}
The exact expression of this coordinate transformation is complicated and discussed in appendix \ref{sec:appen}. However, it is essential that 
\begin{align}
    &U = p(u) + \CO(z^2), \nonumber \\
    &V = q(v) + \CO(z^2), \\
    &\eta = z\sqrt{p' q'}  + \CO(z^2). \nonumber
\end{align}
Therefore, the holographic entanglement entropy can be computed in the global coordinates $(T,X,R)$ with the UV cutoff deformed by a Weyl factor $\sqrt{p' q'}$. For an interval $A$, whose two edges are given by $(u,v) = (u_1,v_1)$ and $(u,v) = (u_2,v_2)$, respectively, the holographic entanglement entropy is given by \eqref{eq:holoEE}, i.e.
\begin{align}
    S_A = \min\left\{S_A^{\rm con}, S_A^{\rm dis}\right\},
    \label{eq:HEEconfine}
\end{align}
where
\begin{align}
    S_A^{\rm con} &= \frac{c}{6} \log \left[
    \frac{(2L/\pi)^2}{\epsilon^{2} \sqrt{p^{\prime}\left(u_{1}\right) p^{\prime}\left(u_{2}\right) q^{\prime}\left(v_{1}\right) q^{\prime}\left(v_{2}\right)}}  \left|\sin\left(\frac{\pi}{2L}(\tilde{u}_1-\tilde{u}_2)
    \right)
    \sin\left(\frac{\pi}{2L}(\tilde{v}_1-\tilde{v}_2)
    \right)\right|\right], \\
    S_A^{\rm dis} &= \frac{c}{6} \log \left[\frac{(2L/\pi)^2}{\epsilon^{2} \sqrt{p^{\prime}\left(u_{1}\right) p^{\prime}\left(u_{2}\right) q^{\prime}\left(v_{1}\right) q^{\prime}\left(v_{2}\right)}} \left|\cos\left(\frac{\pi}{2L}(\tilde{u}_1-\tilde{v}_1)
    \right)
    \cos\left(\frac{\pi}{2L}(\tilde{u}_2-\tilde{v}_2)
    \right)\right|  \right] \nonumber\\
    &\quad + 2S_{\rm bdy} .
    \label{eq:HEEconfine2}
\end{align}

\subsubsection{Deconfined configuration: geometry and holographic entanglement entropy}

The deconfined bulk configuration for a strip is given by a portion of Poincar\'e AdS$_3$ with metric
\begin{align}
    ds^2=\frac{d\eta^2-dUdV}{\eta^2}. 
\end{align}
Let us use $Q_1$ ($Q_2$) to denote the brane attached on $\tilde{x} = L/2$ ($\tilde{x} = -L/2$). In this case, $Q_1$ is located at 
\begin{align}
    (X-L/2)-\lambda_1\eta=0, 
\end{align}
and $Q_2$ is located at 
\begin{align}
    (X+L/2)+\lambda_2\eta=0,
\end{align}
where $\lambda_i$ is related to the tension $\CT_i$ of $Q_i$ as 
\begin{align}
    \lambda_i=\frac{{\cal T}_i}{\sqrt{1-{\cal T}_i^2}}.
\end{align}
Performing the Ba$\tilde{\text{n}}$ados map (\ref{corads}), the deconfined bulk metric for a double moving mirror is given by (\ref{metads}). In this case, the holographic entanglement entropy can be evaluated in a similar way as in the single moving mirror cases. However, since we have one more end-of-the-world brane on which a disconnected Ryu-Takayanagi (RT) surface can end, we get more candidates. For an interval $A$, whose two edges are given by $(u,v) = (u_1,v_1)$ and $(u,v) = (u_2,v_2)$, respectively, the holographic entanglement entropy is given by 
\begin{align}
    S_A = \min\left\{ S_A^{\rm con}, S_A^{\rm dis,11}, S_A^{\rm dis,12}, S_A^{\rm dis,21}, S_A^{\rm dis,22}\right\},
    \label{HEEdeconfine}
\end{align}
where
\begin{align}
    S_A^{\rm con} &= \frac{c}{12} \log \left[\frac{\left(\tilde{u}_{1}-\tilde{u}_{2}\right)^{2}\left(\tilde{v}_{1}-\tilde{v}_{2}\right)^{2}}{\epsilon^{4} p^{\prime}\left(u_{1}\right) p^{\prime}\left(u_{2}\right) q^{\prime}\left(v_{1}\right) q^{\prime}\left(v_{2}\right)}\right], \\
    S_A^{{\rm dis},ij} &= \frac{c}{12} \log \left[\frac{\left(-\tilde{u}_{1}+\tilde{v}_{1}+(-1)^i L\right)^{2}\left(-\tilde{u}_{2}+\tilde{v}_{2}+(-1)^j L\right)^{2}}{\epsilon^{4} p^{\prime}\left(u_{1}\right) p^{\prime}\left(u_{2}\right) q^{\prime}\left(v_{1}\right) q^{\prime}\left(v_{2}\right)}\right]+ S_{{\rm bdy},i} + S_{{\rm bdy},j}.
    \label{HEEdeconfine2}
\end{align}
Here, $S_A^{{\rm dis},ij}$ is the disconnected entanglement entropy evaluated with the help of two disconnected RT surfaces: one connecting $(u_1,v_1)$ and $Q_i$, and another one connecting $(u_2,v_2)$ and $Q_j$. This feature leads to interesting phase transitions in the time evolution. 

\subsection{Confined holographic double escaping mirror}
\label{sec:EEconfine}

\begin{figure}[h!]
  \centering
  \includegraphics[width=.3\textwidth]{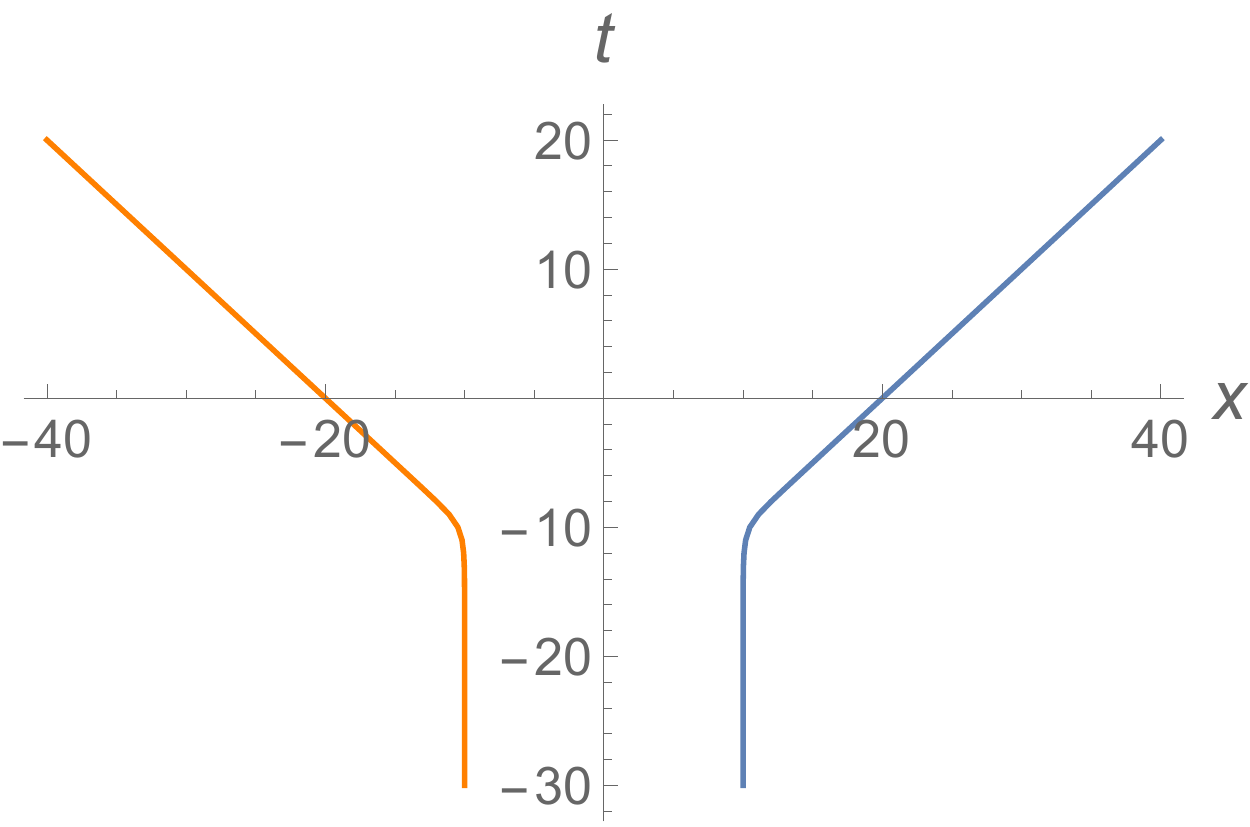}\qquad
 \includegraphics[width=.3\textwidth]{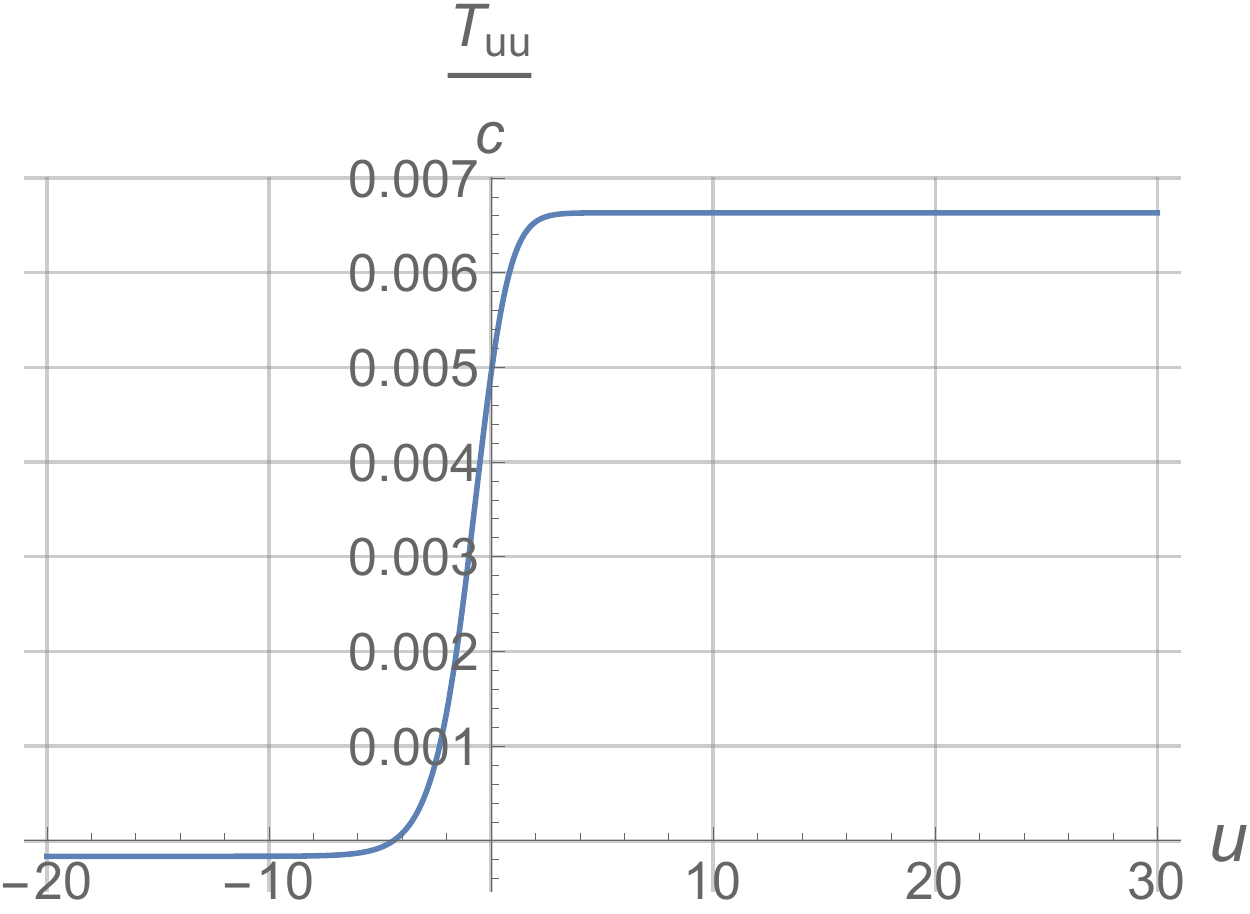}\qquad
 \includegraphics[width=.3\textwidth]{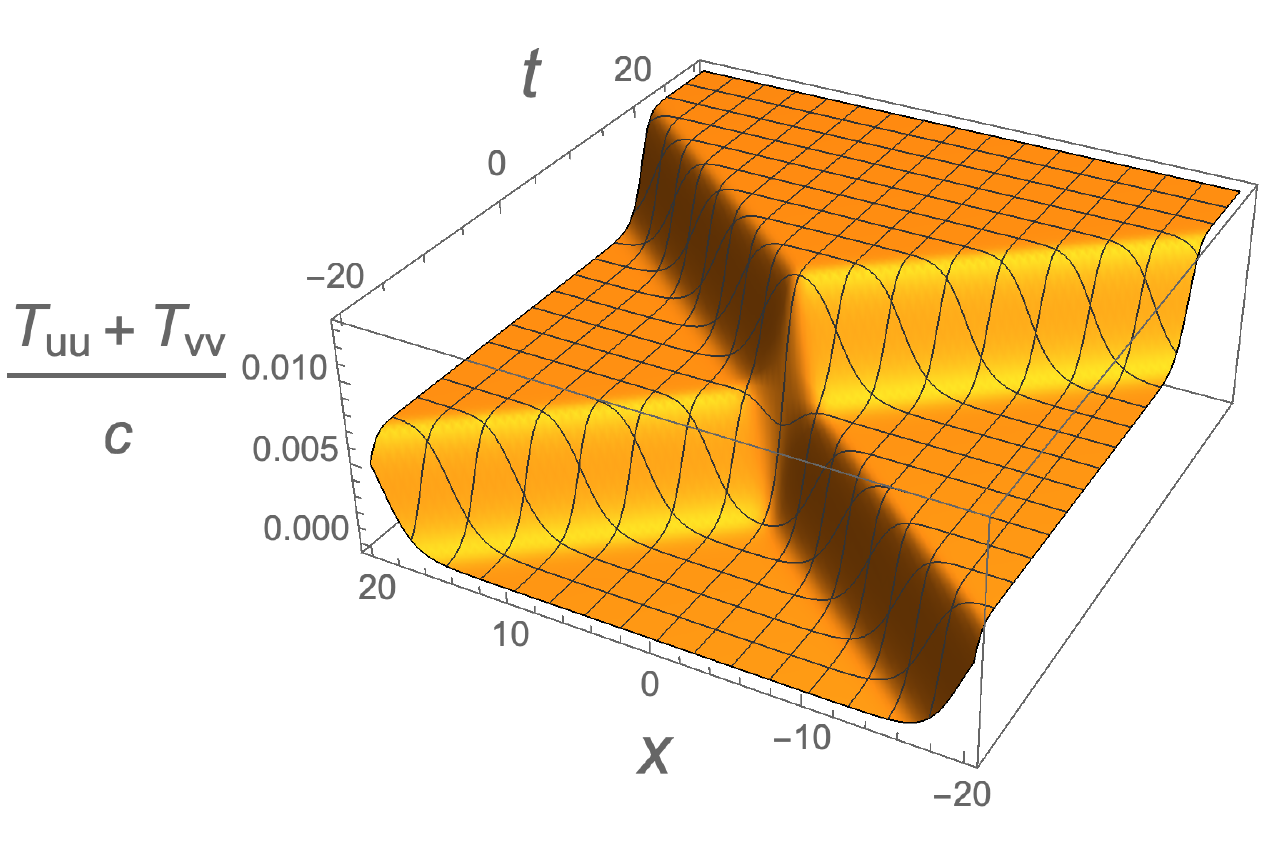}
  \caption{The left picture shows the two opposite trajectories for the double escaping mirror setup. The picture in the middle plots the energy stress tensor $T_{uu}$ as a function of retarded time $u$. The right picture shows the energy density $T_{uu}+T_{vv}$ as a function of $t$ (depth) and $x$ (horizontal). We have used $\beta=1$ and $L=20$.}
\label{dmpfig}
\end{figure}

As a concrete example, let us consider a double moving mirror, which can be mapped to a strip $-L/2 \leq \tilde{x} \leq L/2$ with
\begin{align}
    \ti{u}=p(u)=-\beta\log(1+e^{-u/\beta}), \nonumber\\
    \ti{v}=q(v)= -\beta\log(1+e^{-v/\beta}).
\end{align}
Plugging these relations into (\ref{mirrorprofile}), we get the two mirror trajectories given by 
\begin{align}
    x &= \pm Z(t),
\end{align}
where
\begin{align}
    Z(t) &= \beta \log \frac{\left(e^{L/\beta}-1\right)e^{t/\beta} + \sqrt{\left(e^{L/\beta}-1\right)^2e^{2t/\beta} +4e^{L/\beta}}}{2}.
\end{align}
We can see that in the early time limit $t\rightarrow -\infty$,
\begin{align}
    Z(t) = \frac{L}{2},
\end{align}
while in the late time limit $t\rightarrow \infty$,
\begin{align}
    Z(t) = t + \frac{L}{2} + \beta \log \left(2~\sinh \frac{L}{2\beta}\right). 
\end{align}
This setup realizes a double escaping mirror. See the left panel of Fig.~\ref{dmpfig} for a plot of the mirror trajectory. 

In this subsection, let us first discuss the holographic double escaping mirror in the confined configuration. In this case, the energy stress tensor can be obtained by plugging the conformal map into (\ref{emcftl2}):
\begin{align}
\begin{split}
    &T_{uu} = \frac{c}{12\pi}\left(\frac{e^{u/\beta}\left(2+e^{u/\beta}\right)}{4\beta^2(1+e^{u/\beta})^2}-\frac{\pi^2}{4 L^2}\left(\frac{1}{1+e^{u/\beta}}\right)^2\right), \\
    &T_{vv} = \frac{c}{12\pi}\left(\frac{e^{v/\beta}\left(2+e^{v/\beta}\right)}{4\beta^2(1+e^{v/\beta})^2}-\frac{\pi^2}{4 L^2}\left(\frac{1}{1+e^{v/\beta}}\right)^2\right).
\end{split}
\end{align}
See the middle and the right panels of Fig.~\ref{dmpfig}. 

The holographic entanglement entropy for a single interval $A$ can be also computed straightforwardly by plugging the conformal map into (\ref{eq:HEEconfine}) - (\ref{eq:HEEconfine2}). The explicit form is complicated, but it is useful to see how it behaves at late time. We will discuss this in several cases and show the corresponding numerical results. 

\subsubsection{Single interval with two edges fixed}

Let us start with the most simple case. Let $A$ be a single interval $x_0\leq x \leq x_0+l$ with two edges fixed. In the late time limit $t\rightarrow\infty$, the leading contribution reads
\begin{align}
    S_A^{\rm con} &= \frac{c}{3}\log \left(\frac{2\beta}{\epsilon}\sinh\frac{l}{2\beta}\right) + \cdots ,\\
    S_A^{\rm dis} &= \frac{c}{3}\frac{t}{\beta} + \cdots.
\end{align}
Therefore, $S^{\rm con}_A$ dominates at late time. Note that the entanglement entropy at late time is similar to that in a thermal state with temperature $1/(2\pi\beta)$. Refer to the left column in Fig.~\ref{DEsconfined} for a sketch and numerical plot of such a setup.  

\subsubsection{Single interval with two comoving edges}

Let us now consider $A$ taken to be a single interval $-Z(t)+\xi_0 \leq x \leq Z(t)-\xi_0$ with two edges comoving with the mirrors. In this setup, the whole system is divided into mirrors and ``heat bath''. In the late time limit $t\rightarrow\infty$, the leading contribution turns out to be 
\begin{align}
    S_A^{\rm con} &= \frac{c}{3}\frac{t}{\beta} + \cdots ,\\
    S_A^{\rm dis} &= \frac{c}{3}\frac{t}{\beta} + \cdots.
\end{align}
Therefore, which RT surface dominates depends on how one chooses other parameters, especially, the boundary entropy $S_{\rm bdy}$. For example, in the large tension limit $\CT\rightarrow1$, $S_{\rm bdy}$ goes to infinity and the connected RT surface dominates. In this case, $A$ does not have an island on the end-of-the-world brane $Q$. Refer to the middle column in Fig.~\ref{DEsconfined} for a sketch and numerical plot of such a setup. 

\subsubsection{Single interval with one fixed edge and one comoving edge}

\begin{figure}[H]
    \centering
    \includegraphics[width=5.2cm]{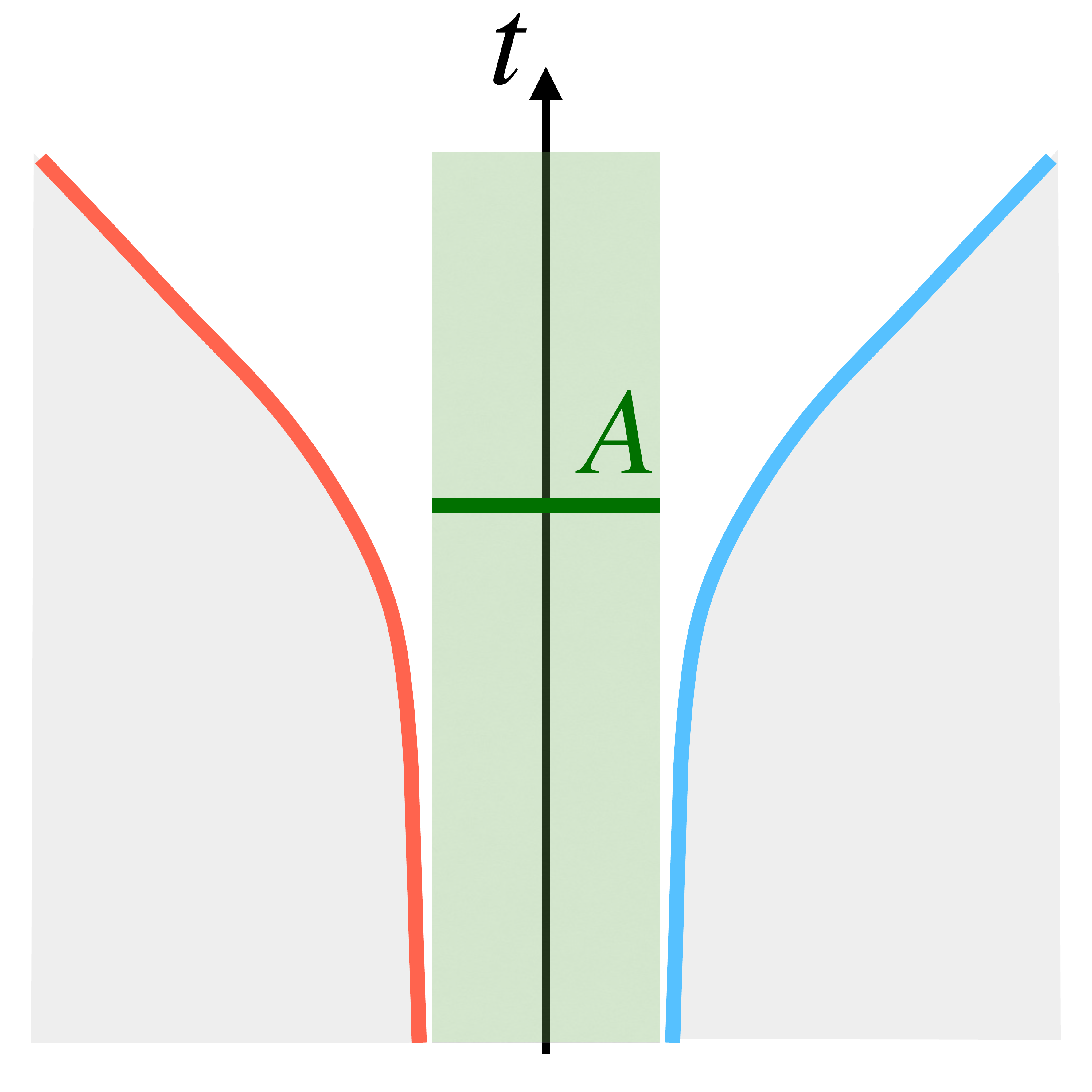}
    \includegraphics[width=5.2cm]{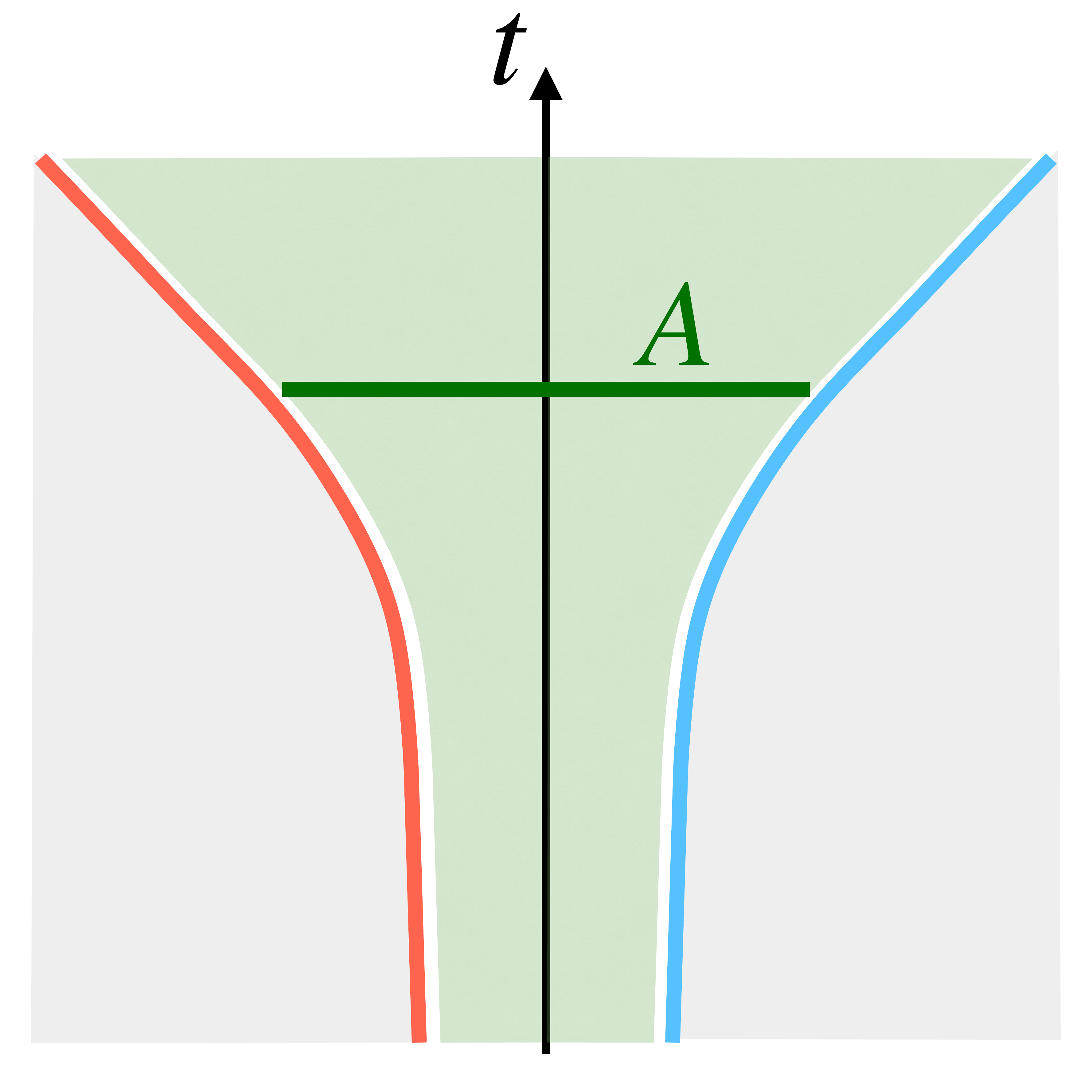}
    \includegraphics[width=5.2cm]{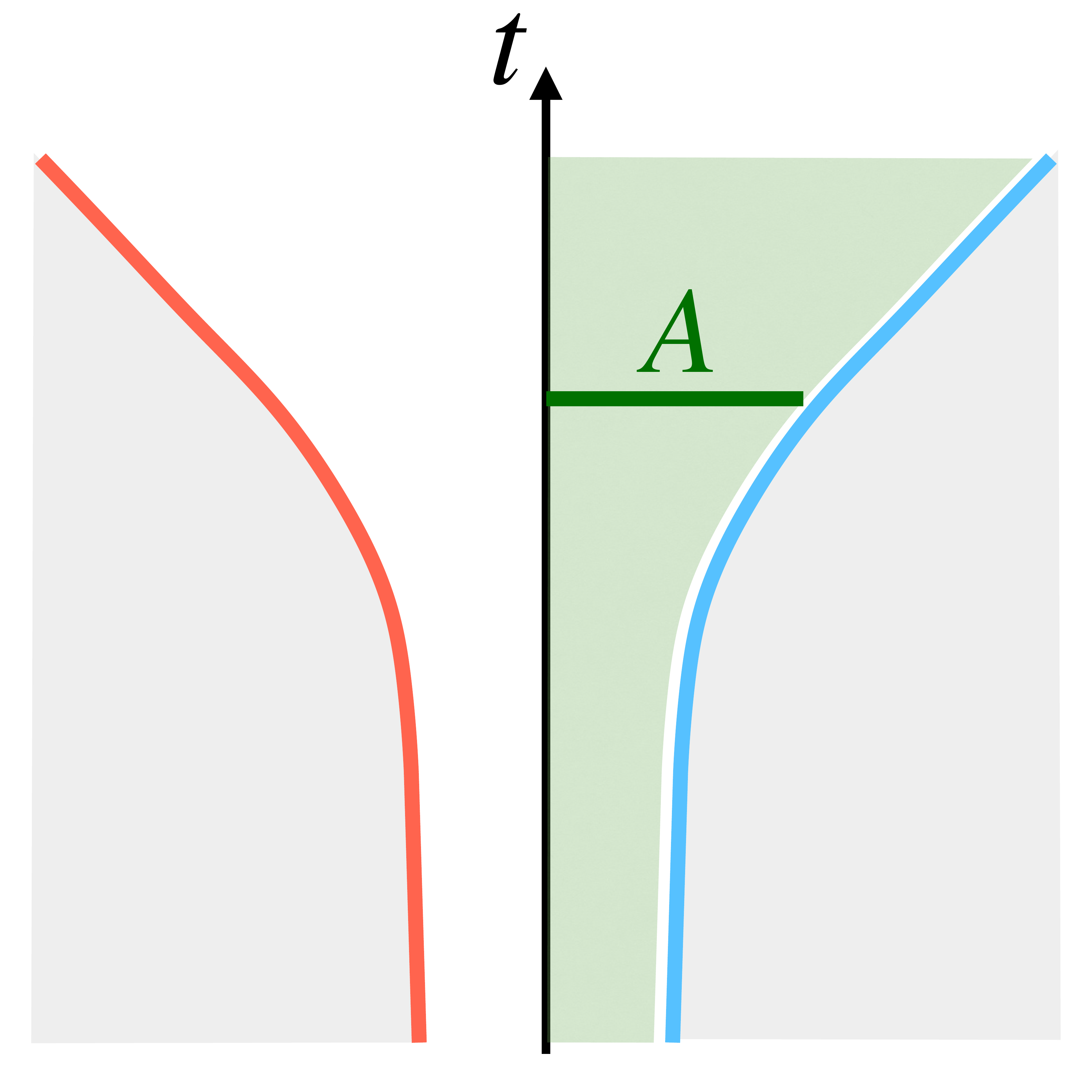}
    \includegraphics[width=5.24cm]{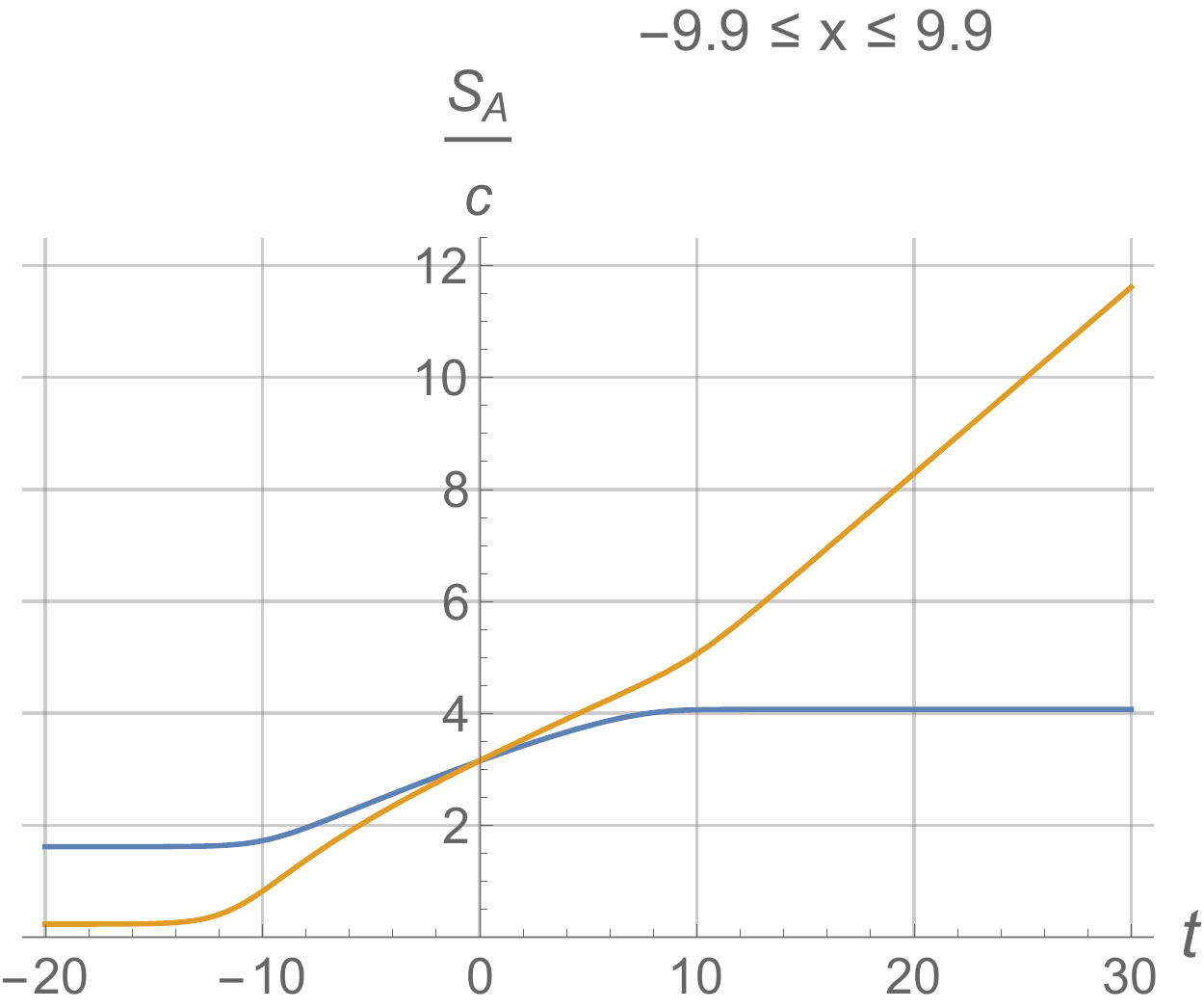}
    \includegraphics[width=5.24cm]{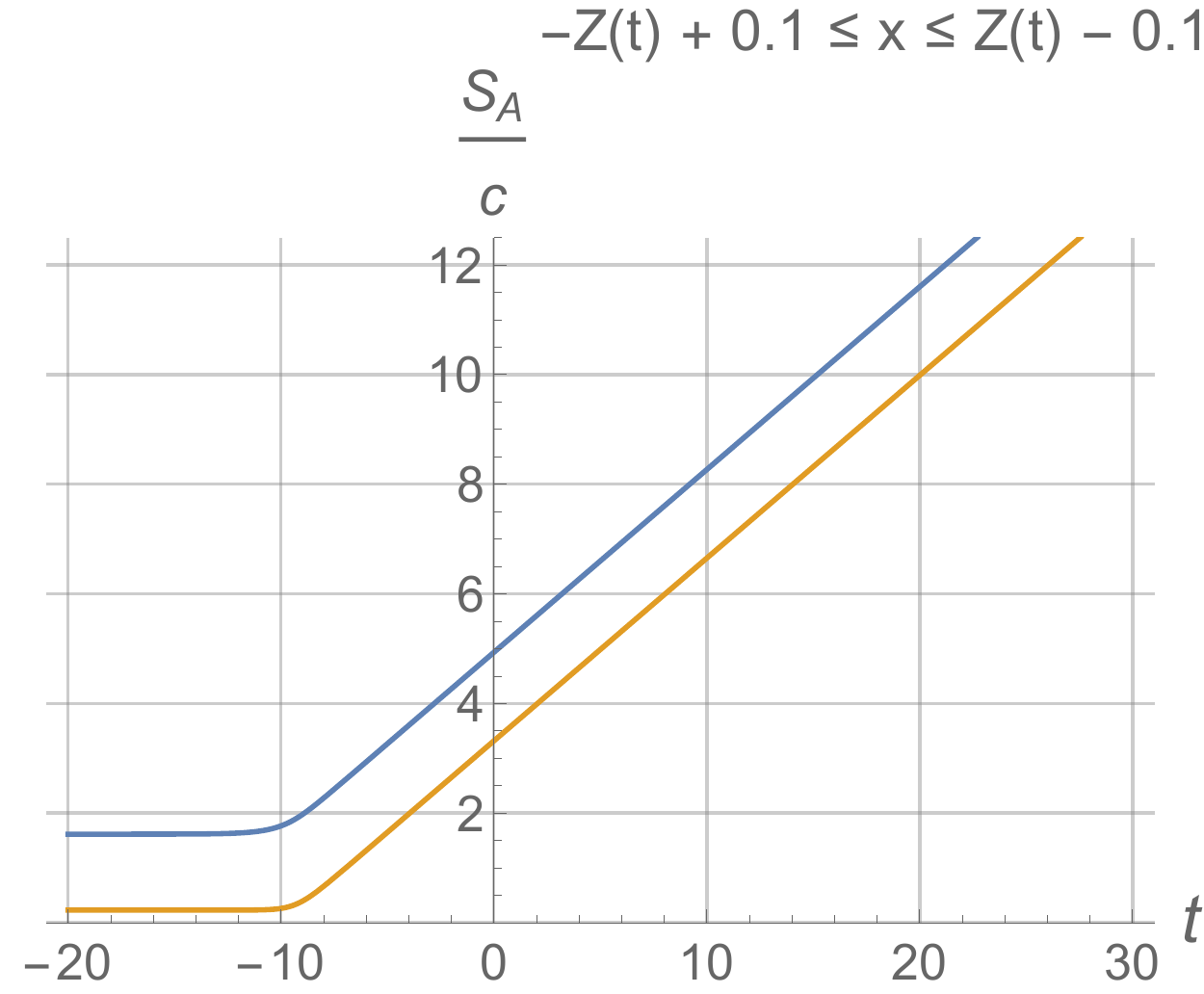}
    \includegraphics[width=5.24cm]{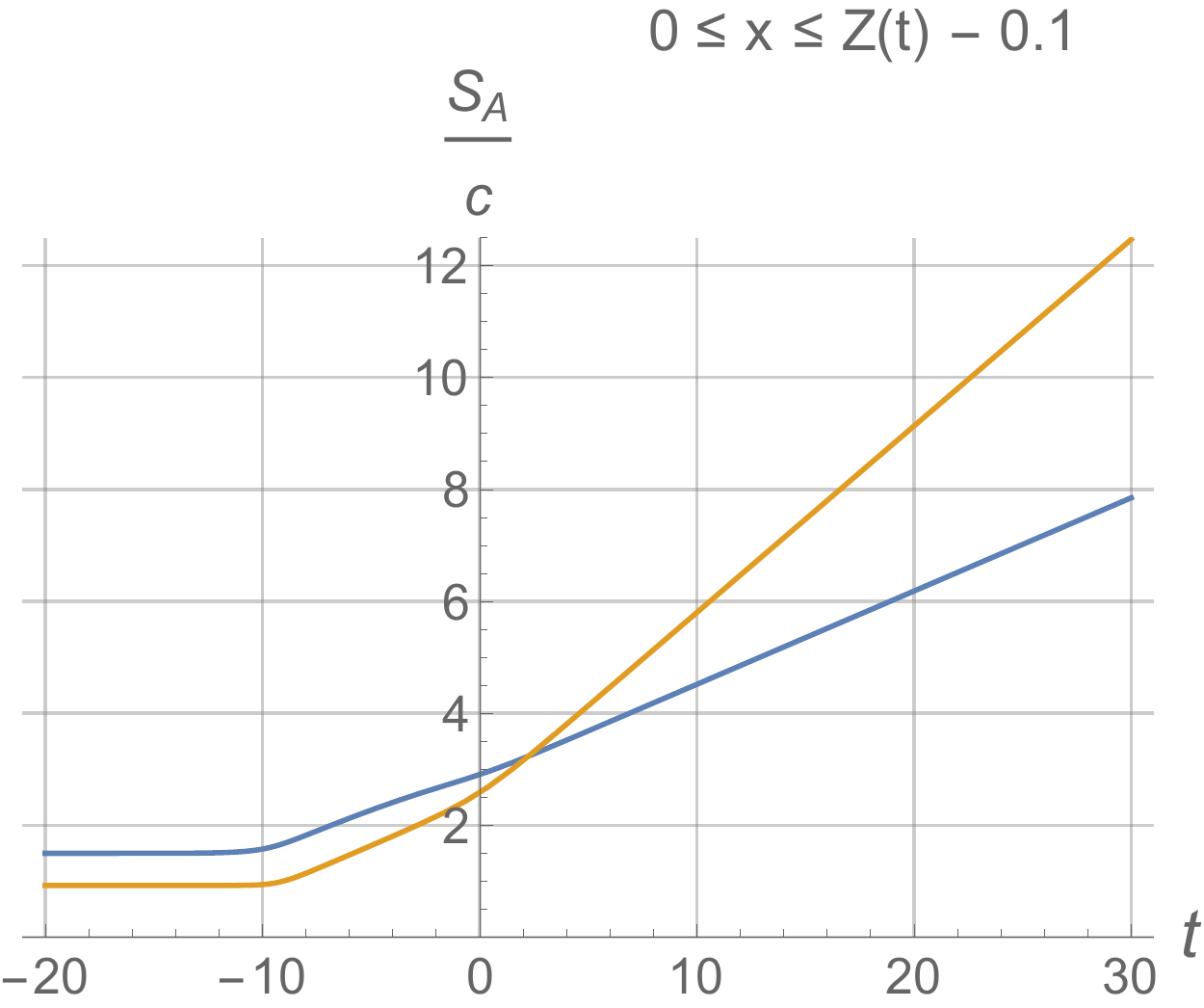}
    \caption{Sketches for different setups with double escaping mirrors and plots for holographic entanglement entropy in each case. Here, we consider the confined bulk configuration. The green solid line shows subsystem $A$ on a time slice, and the green shaded region shows the ``worldsheet" of $A$. The left column shows the case, where two edges of $A$ are fixed. The middle column shows the case, where two edges of $A$ comove with the mirror trajectory. The right column shows the case, where one edge of $A$ is fixed and the other one comoves. In all the plots, the blue (orange) curve shows the time evolution of $S_{A}^{\rm con}$ ($S_{A}^{\rm dis}$), and $\beta=1$, $L=20$, $\epsilon=0.1$. For simplicity, we have set the boundary entropy as $S_{\rm bdy} = 0$. This can, however, take any value in $(-\infty, \infty)$. How the interval is chosen is shown in each plot.}
    \label{DEsconfined}
\end{figure}

As the last setup, let $A$ be a single interval $x_0\leq x \leq Z(t)-\xi_0$. One edge of $A$ is fixed and the other one comoves with the mirror. In the late time limit, the leading contribution is 
\begin{align}
    S_A^{\rm con} &= \frac{c}{6}\frac{t}{\beta} + \cdots ,\\
    S_A^{\rm dis} &= \frac{c}{3}\frac{t}{\beta} + \cdots.
\end{align}
It is clear that $S^{\rm con}_A$ dominates at late time. Refer to the right column in Fig.~\ref{DEsconfined} for a sketch and numerical plot of such a setup. 

\subsection{Deconfined holographic double escaping mirror}
\label{sec:EEdeconfine}

\begin{figure}[h!]
  \centering
  \includegraphics[width=.4\textwidth]{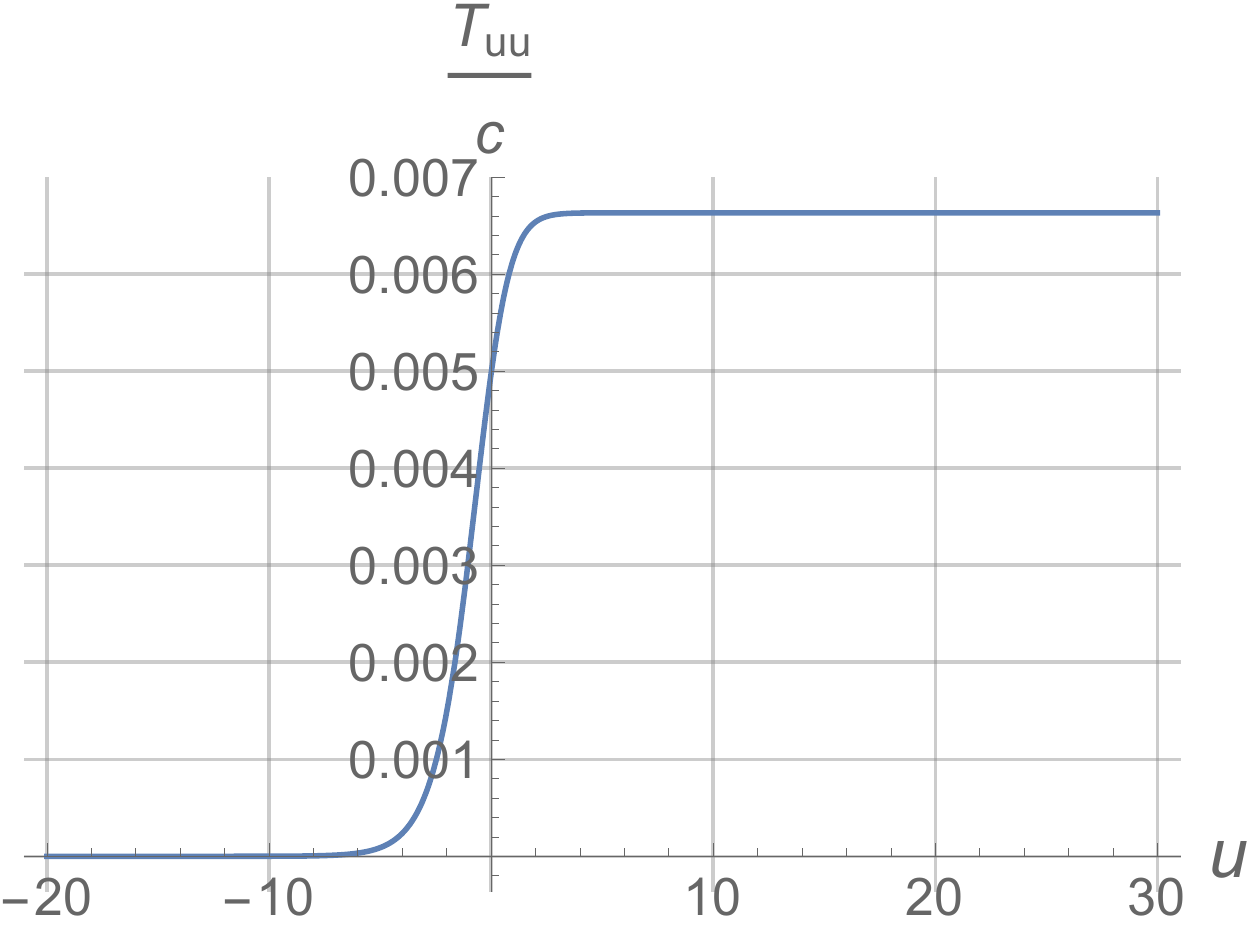}\qquad
 \includegraphics[width=.4\textwidth]{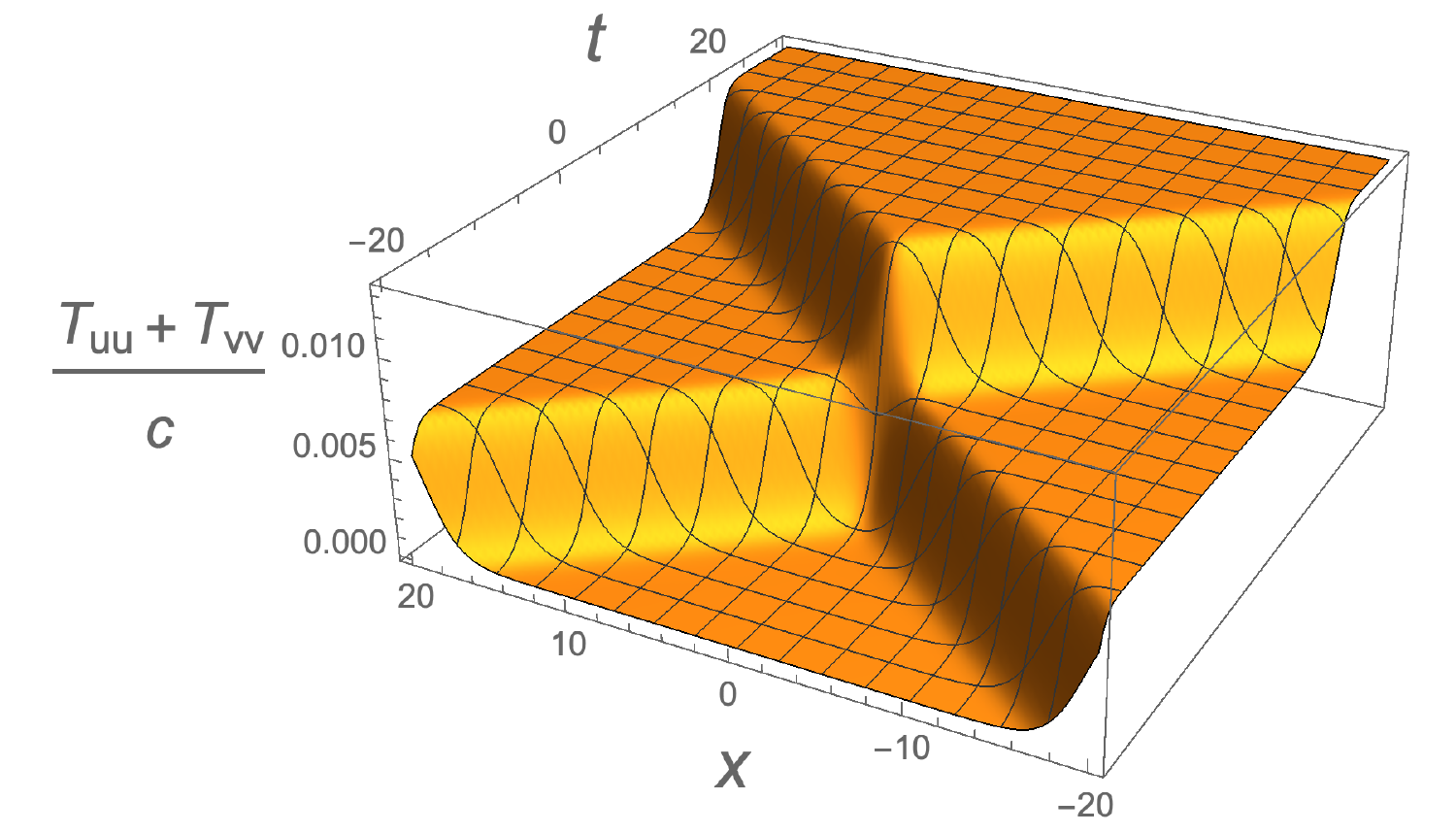}
  \caption{Energy stress tensor in the deconfined case. The picture on the left plots the energy stress tensor $T_{uu}$ as a function of $u$. The right picture shows the energy density $T_{uu}+T_{vv}$ as a function of $t$ (depth) and $x$ (horizontal). We have used $\beta=1$ and $L=20$.}
\label{dmpfig2}
\end{figure}

We would like to move on to the holographic double escaping mirror in the deconfined configuration. In this case, the energy stress tensor turns out to be 
\begin{align}
\begin{split}
    &T_{uu} = \frac{c}{12\pi}\frac{e^{u/\beta}\left(2+e^{u/\beta}\right)}{4\beta^2(1+e^{u/\beta})^2}, \\
    &T_{vv} = \frac{c}{12\pi}\frac{e^{v/\beta}\left(2+e^{v/\beta}\right)}{4\beta^2(1+e^{v/\beta})^2},
\end{split}
\end{align}
since $T_{\tilde{u}\tilde{u}} = T_{\tilde{v}\tilde{v}} = 0$.
See Fig.~\ref{dmpfig2}. The behavior is very similar to that in the confined case, see Fig.~\ref{dmpfig}.

Plugging the conformal map into (\ref{HEEdeconfine}) - (\ref{HEEdeconfine2}), we can get the holographic entanglement entropy for a single interval $A$. One can easily find that the leading contribution at late time is the same as that in the confined case. Therefore, we will not repeat it here. On the other hand, one crucial feature of the deconfined case, which differs from the confined case, is that there are two end-of-the-world branes and hence two boundary entropy values to tune. This feature generates colorful phase transitions in the process sketched in the middle. 

\begin{figure}[H]
    \centering
    \includegraphics[width=5.2cm]{DEs1.pdf}
    \includegraphics[width=5.2cm]{DEs2.pdf}
    \includegraphics[width=5.2cm]{DEs3.pdf}
    \includegraphics[width=5.24cm]{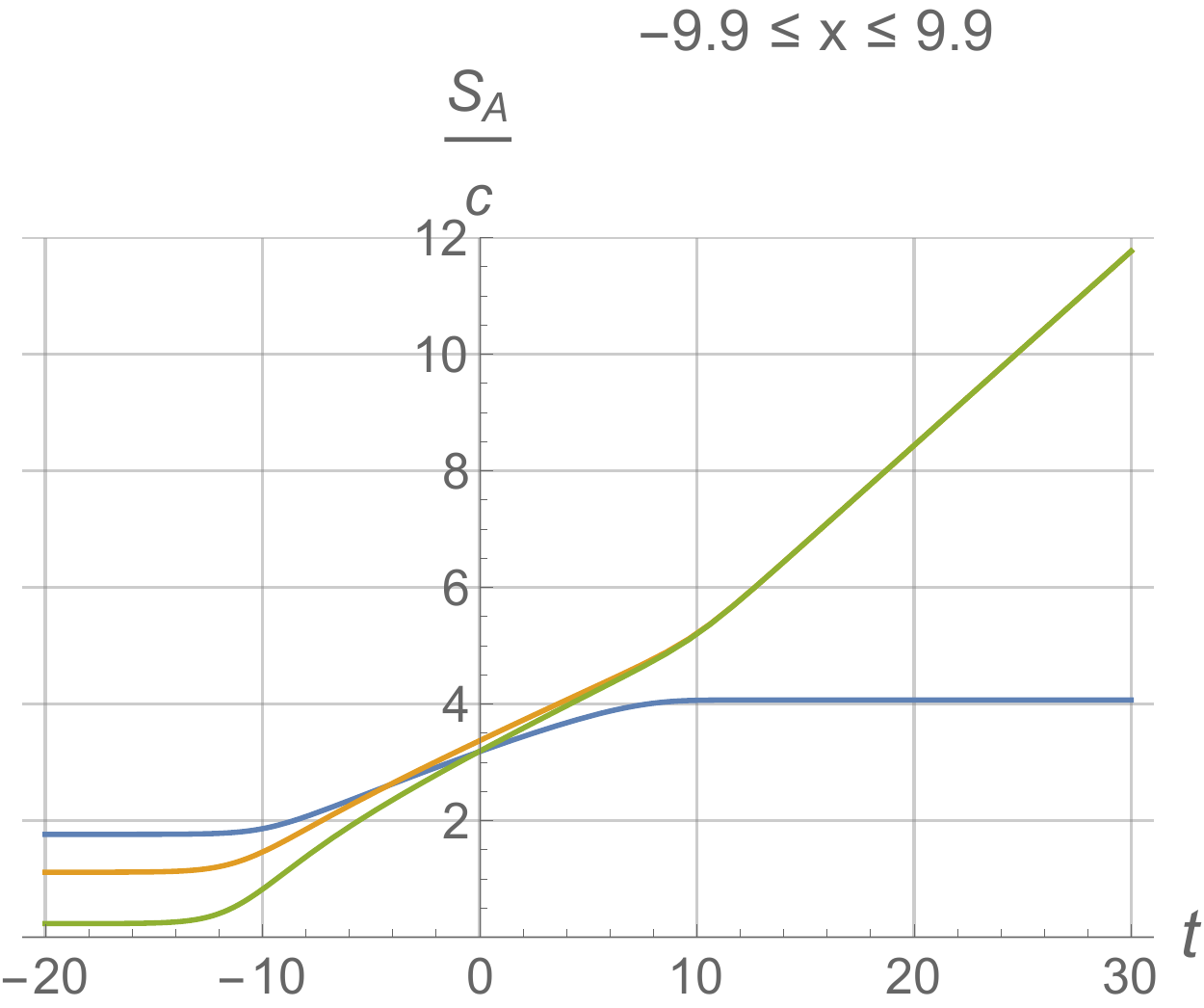}
    \includegraphics[width=5.24cm]{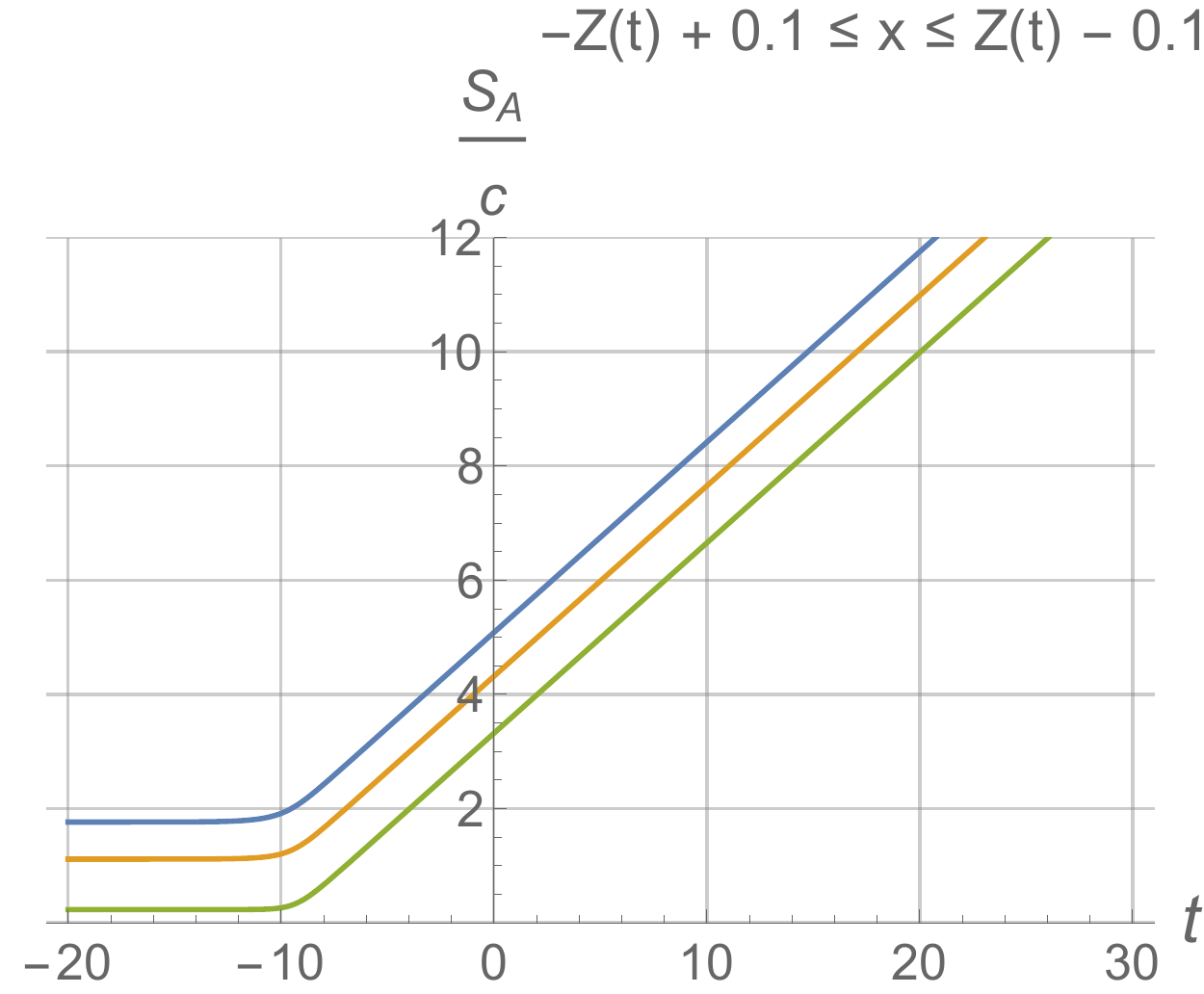}
    \includegraphics[width=5.24cm]{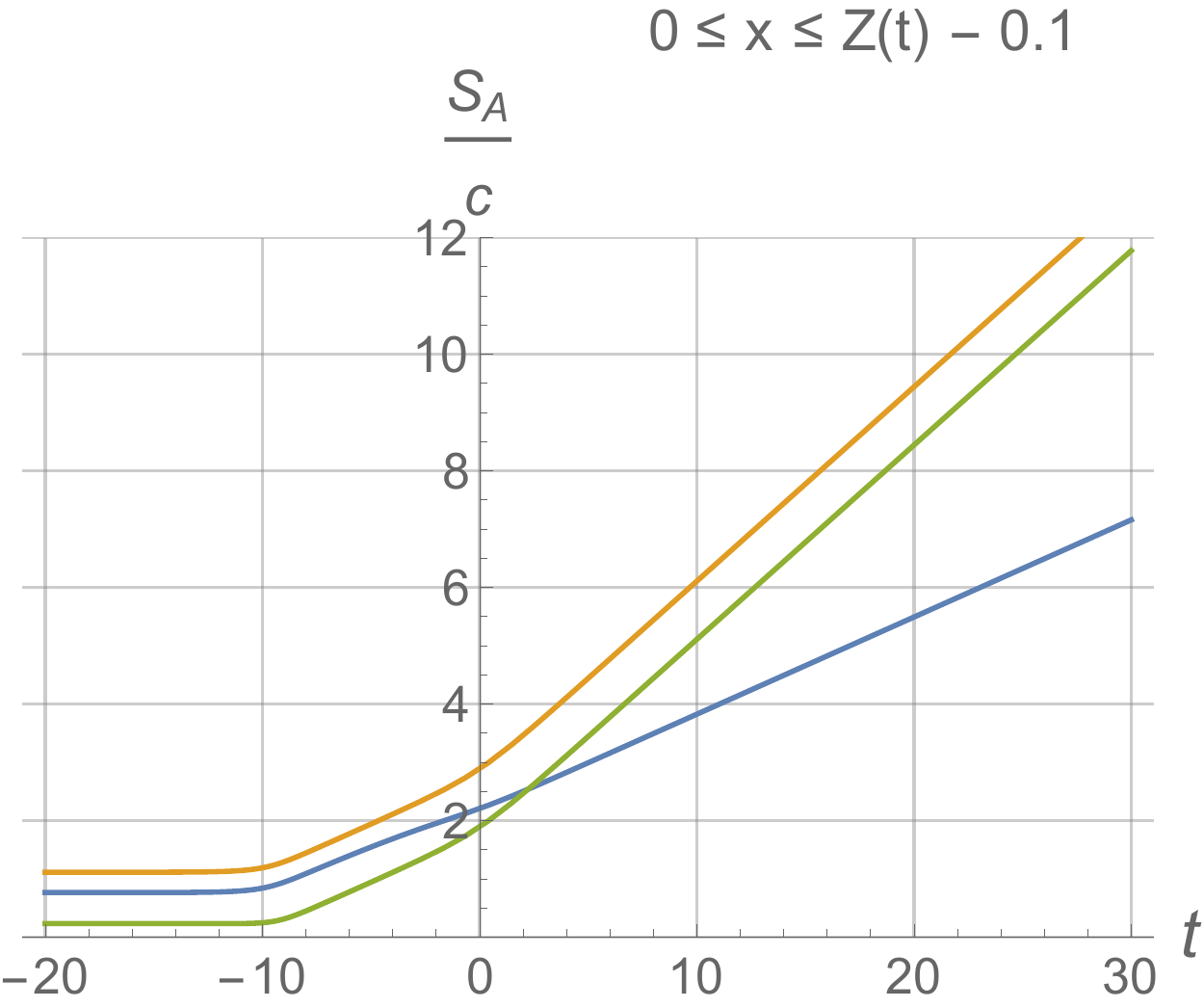}
    \caption{Holographic entanglement entropy in different setups. Here, we consider the deconfined bulk configuration. We have set $\beta=1$, $L = 20$, $\epsilon=0.1$. In all the plots, the blue, orange and green curves show $S^{\rm con}_A$, $S^{\rm dis, 11}_A$ and $S^{\rm dis, 12}_A$ respectively. For simplicity, we have set the boundary entropy as $S_{\rm bdy,1} =S_{\rm bdy,2} = 0$. These can, however, take any value in $(-\infty, \infty)$. How the interval is chosen is shown in each plot.}
    \label{DEsdeconfined}
\end{figure}

In the double escaping mirror case, we can set $S_{\rm bdy,1} \leq S_{\rm bdy, 2}$ without loss of generality thanks to the symmetry. For simplicity, let us consider a symmetric $A$ with $-x_1(t)=x_2(t) \leq x \leq x_1(t)$. In this case, although there are five candidates in (\ref{HEEdeconfine}), it is sufficient to consider $S^{\rm con}_A$, $S^{\rm dis, 11}_A$, $S^{\rm dis, 12}_A$. Refer to Fig.~\ref{DEsdeconfined} for numerical plots of different setups. Here, we have set $S_{\rm bdy,1} = S_{\rm bdy, 2} = 0$. We can see that the behavior is very similar to the confined case in Fig.~\ref{DEsconfined}, as we expect. 

\begin{figure}[H]
    \centering
    \includegraphics[width=5.2cm]{DEs2.pdf}
    \includegraphics[width=5.24cm]{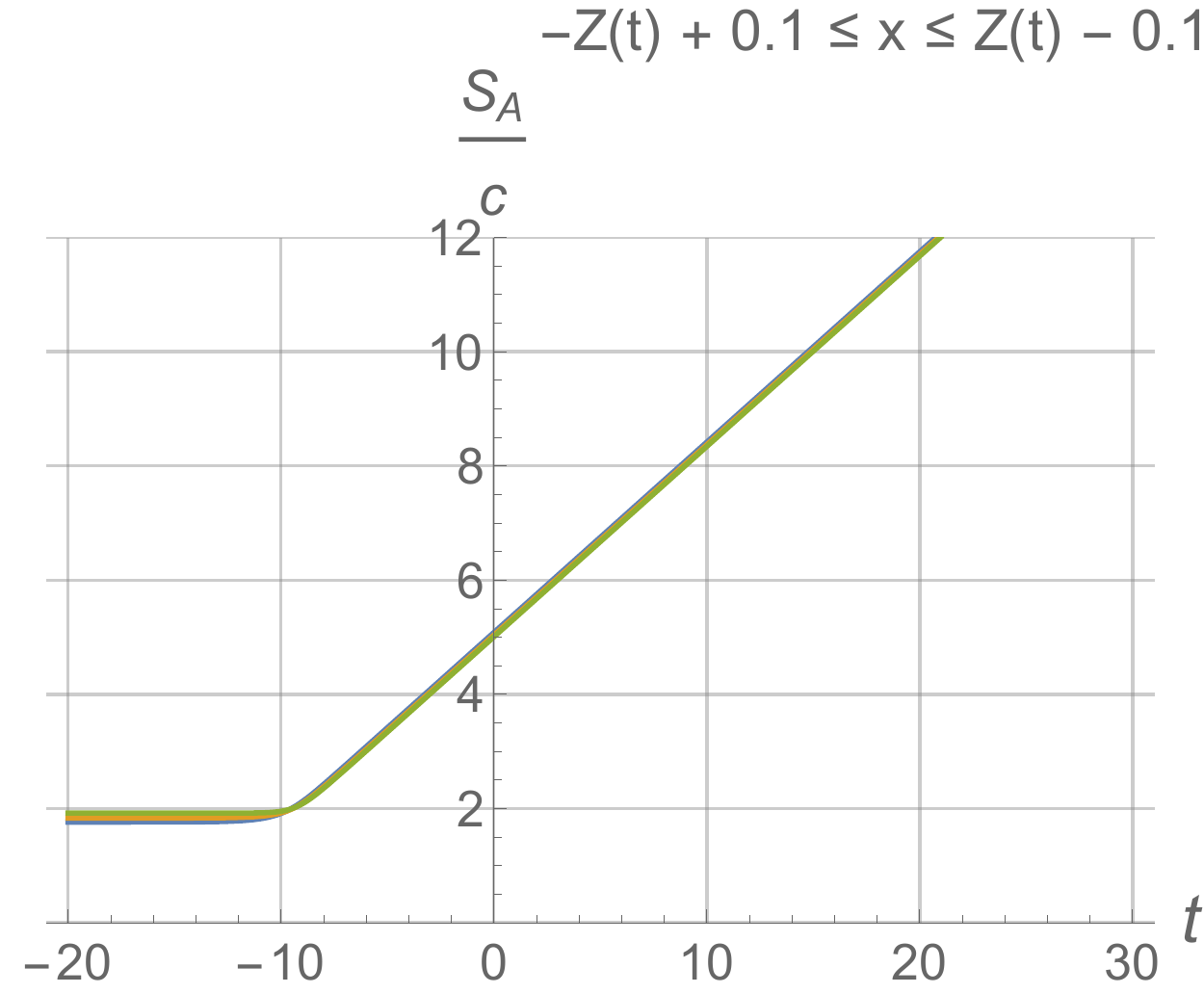}
    \includegraphics[width=5.24cm]{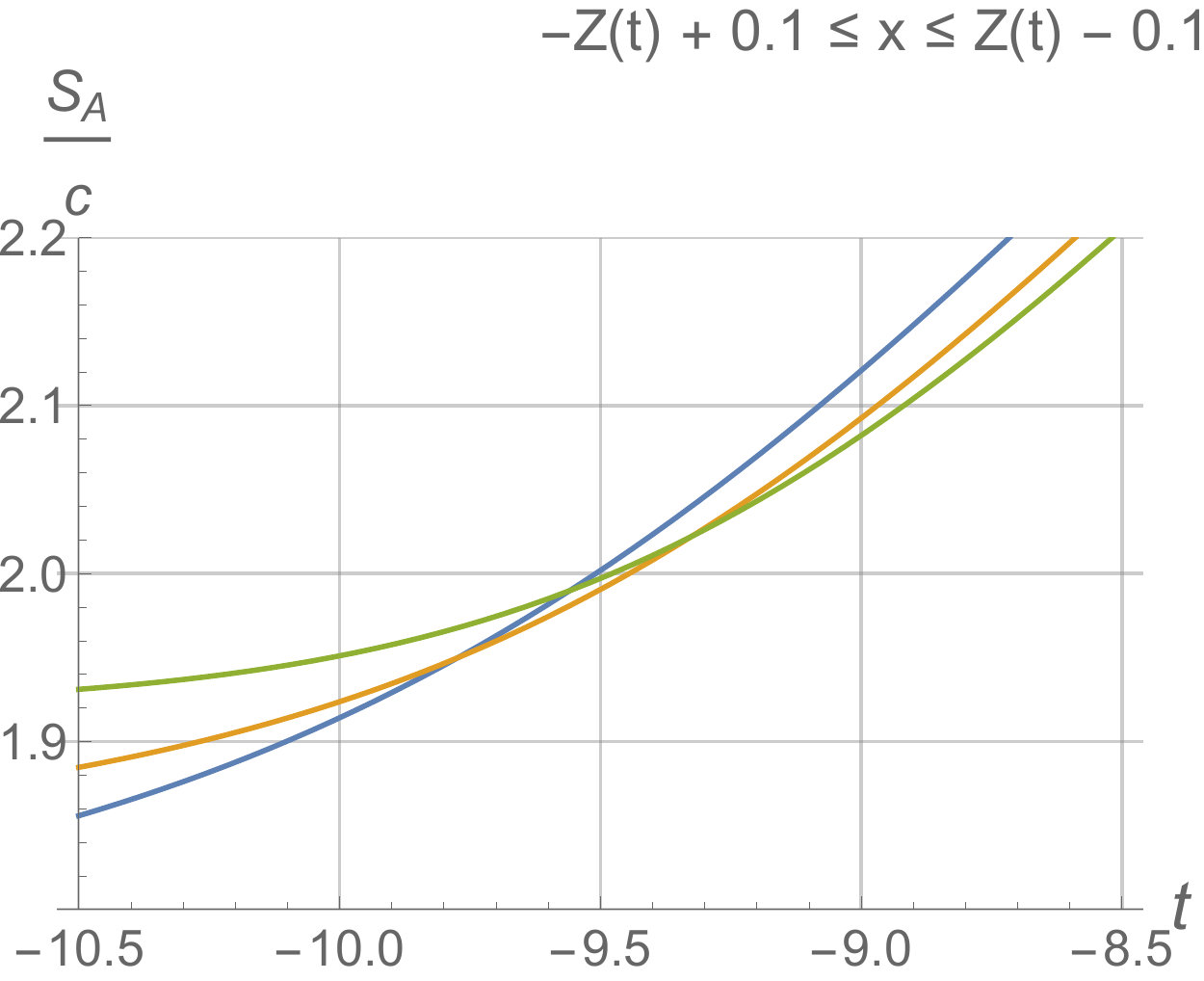}
    \caption{Holographic entanglement entropy for the heat bath. Boundary entropies are tuned, such that $S^{\rm con}_A$, $S^{\rm dis, 11}_A$ and $S^{\rm dis, 12}_A$ dominate in order. Here, $\beta=1$, $L = 20$, $\epsilon=0.1$. $S_{\rm bdy,1} = 0.36c$, $S_{\rm bdy,2} = 1.33c$.}
    \label{DEsPT}
\end{figure}

Let us then tune the boundary entropy and see what happens. Refer to Fig.~\ref{DEsPT} for a concrete example. Here, the two edges of $A$ comove with the two moving mirrors. The boundary entropies are tuned such that $S^{\rm con}_A$, $S^{\rm dis, 11}_A$ and $S^{\rm dis, 12}_A$ dominate in order.

Let us focus on the specific example shown in Fig.~\ref{DEsPT}, and consider what happens in the gravity dual. In this example, we choose $0<S_{{\rm bdy},1}<S_{{\rm bdy},2}$, and the boundary entropy is fine tuned. 

By applying brane-world holography, we can regard this setup as gravity on $Q_1$ and $Q_2$ coupled to a heat bath CFT in the middle. 
At early time, the entanglement entropy is given by the connected RT surface and hence no island exists on the brane. 
As time evolves, the disconnected RT surface, which ends on $Q_1$, dominates. In this case, the island is a finite interval on $Q_1$ sandwiched by the two end points of the disconnected RT surface. 
At late time, the disconnected RT surface, which has one end point on $Q_1$ and another on $Q_2$, dominates. In this case, the island becomes almost the whole part of the two branes. 

To understand this kind of behavior, we should note three features of the double escaping mirror setup. First of all, the left moving energy flux from the right boundary is exactly the same as the right moving energy flux from the left boundary. If we regard the radiation as information leakage from the boundary to the middle region, then this implies that the leaking speed from the two boundaries are exactly the same. 
The second point is, that the radiation from the left boundary can never reach the right boundary and vice versa in the double escaping mirror setup. As a result, the two boundaries do not interact through the released radiation. The third point is, that the only difference between the two boundaries is the boundary entropy, which quantifies degrees of freedom localized on the boundary. 

With these in mind, we can understand the phase transition behavior as follows. Initially, the middle interval of the CFT does not contain knowledge about the information localized on the two boundaries. As the mirrors move, the middle interval receives information through the radiation. The leaking rate is the same. Therefore, it simply takes more time to recover the information initially localized on boundary 2 than that on boundary 1. 

\section{Unitarity and entropy}
\label{sec:u-ee}

In this section, we further elaborate on the time evolution of holographic entanglement entropy and the outgoing energy flux in the moving mirror setup. By making use of our previous findings, we show that a unitary entropy curve implies a violation of the NEC in a finite time window, while the QNEC turns out to be saturated.

\subsection{Entropy evolution}

Consider the kink mirror introduced in Sec.~\ref{subsubsec:double-kink-MM}. 
In general, the holographic entanglement entropy \eqref{eq:holoEE} is computed from either the disconnected geodesic, leading to
\ba
S^\text{dis}_A=\frac{c}{6}\log\frac{v-p(u)}{\ep\s{p'(u)}}+\frac{c}{6}\log\frac{v'-p(u')}{\ep\s{p'(u')}},
\label{disee2}
\ea
or the connected one, which gives 
\ba
S^\text{con}_A=\frac{c}{6}\log\frac{(v'-v)(p(u)-p(u'))}{\ep^2\s{p'(u)p'(u')}},
\label{conee2}
\ea
where $(u,v)$ and $(u',v')$ are the two end points of the subsystem $A$.

As we have discussed in Sec.~\ref{sec:dkmm}, taking the latter to be a non-static, semi-infinite interval, the entropy $S_A$ gives rise to a unitary entropy curve \cite{Akal:2020twv}. This is depicted in Fig.~\ref{fig:idealPage}. Recall that in this case, the holographic entanglement entropy is always determined by the disconnected geodesic, $S_A = S_A^\text{dis}$.

\subsection{Energy inequalities}

The kink mirror can alternatively be described by
\ba
p(u)=\beta \log\frac{1+e^{\frac{u-u_0}{\beta}}}{1+e^{-\frac{u}{\beta}}},
\ea
cf. Sec.~\ref{subsubsec:double-kink-MM}.
 
In the high temperature limit, i.e. $\beta\to 0$, the (renormalized) energy flux \eqref{eq:Tuu-gen}, cf. Sec.~\ref{subsec:ren-emt},
takes the approximate form
\ba
T_{uu}\simeq \frac{c}{48\pi\beta^2}\theta(u)\theta(u_0-u)+f(u),
\label{eq:Tuu-approx}
\ea
where the function $f(u)$ gives a negative contribution and explicitly reads
\ba
f(u)\simeq -\frac{c}{16\pi\beta^2\cosh^2\left(\frac{x}{\beta}\right)}.
\ea
After integrating over $-\infty<u<\infty$, we find 
\ba
\int^\infty_{-\infty} du\ T_{uu}=\frac{c}{24\pi}\left(\frac{u_0}{2\beta^2}-\frac{3}{\beta}\right),
\ea
where the first and second term in the RHS arise from the step function and the negative function $f$ in \eqref{eq:Tuu-approx}, respectively. 

Since $\beta \ll u_0$ is assumed, we find that that the averaged NEC, i.e.
\ba
\int du\ T_{uu} \geq 0,
\ea
is satisfied. Importantly, the energy fluxes
both saturate the QNEC\footnote{
The QNEC provides a local constraint on the expectation value of the null projected energy stress tensor, $\langle T_{kk} \rangle $, which in its general form reads
\begin{align*}
2 \pi \langle T_{kk} \rangle \geq \frac{1}{\sqrt{h}} S'',
\end{align*}
where 
\begin{align*}
\langle T_{kk} \rangle \equiv \langle T_{\mu \nu} k^\mu k^\nu \rangle\qquad \forall\ k^2 = 0.
\end{align*}
Here, the primes denote ordinary derivatives with respect to an affine parameter along the corresponding generator on the null hypersurface, and $h$ corresponds to the determinant of the induced metric on the boundary of the entangling region \cite{Bousso:2015mna,Bousso:2015wca}.

Notably, in two dimensional CFT, the QNEC takes the stronger form \cite{Wall:2011kb,Bousso:2015mna}
\begin{align*}
2 \pi \langle T_{kk} \rangle \geq S'' + \frac{6}{c} (S')^2.
\end{align*}
}
\ba
\begin{split}
2\pi T_{uu} &=\de^2_u S_A+\frac{6}{c}(\de_u S_A)^2,\\
2\pi T_{vv} &=\de^2_v S_A+\frac{6}{c}(\de_v S_A)^2.
\label{eq:QNECsat}
\end{split}
\ea
Note that $T_{vv}=0$.

\subsection{Energy flux and entropy}

By rewriting the energy flux in the first line of \eqref{eq:QNECsat} in a more convenient way, and integrating over a small interval around some value $u_0/2$, we find the following expression
\ba
2\pi \int^{u_0/2+\delta}_{u_0/2-\delta} du\ T_{uu} \exp\left( \frac{6}{c}S_A \right)
=\left[e^{\frac{6}{c}S_A}\de_u S_A\right]^{u_0/2+\delta}_{u_0/2-\delta},
\label{eq:Tuu-int}
\ea
where $\beta\ll \delta\ll u_0$ has been assumed. 

We take the subregion $A$ to be a semi-infinite line for which we get a unitary entropy curve as discussed in Sec.~\ref{sec:dkmm}. Choosing $u_0$ appropriately, the RHS of \eqref{eq:Tuu-int} will become negative, so $T_{uu}<0$.

Nevertheless, to show that unitarity in the sense above implies a negative energy flux, we may more generally proceed as follows. Namely, we may first assume that the averaged NEC is satisfied and
\ba
2 \pi \int_{- \infty}^\infty du\ T_{uu} < \infty
\label{eq:energ-cons}
\ea
holds due to energy conservation.
Plugging the first line in \eqref{eq:QNECsat} into \eqref{eq:energ-cons}, we first find that
\ba
S'_A(u) = 0 \qquad (u \rightarrow \pm \infty).
\label{eq:cond-S2}
\ea
Then, using \eqref{eq:cond-S2} and \eqref{eq:energ-cons} and rewriting the energy flux which, in the present case, saturates the QNEC, leads to the following integral condition\footnote{This relation has previously been obtained in a moving mirror setup by referring to a renormalized notion of entanglement entropy \cite{Bianchi:2014qua}, see also \cite{Chen:2017lum,Good:2019tnf} for related discussions. However, we should note that \eqref{eq:E-S-rel} is a direct consequence of the fact that the energy stress tensor precisely saturates the QNEC. 

Here, we want to emphasize that our derivation of \eqref{eq:E-S-rel} is based on the holographic entanglement entropy \eqref{eq:holoEE}, which is determined by the corresponding minimal surface in the constructed bulk dual. The holographic entanglement entropy, of course, saturates the QNEC as one expects from considerations in the dual field theory. Refer, for instance, to \cite{Ecker:2017jdw,Leichenauer:2018obf} for further works on the saturation of the QNEC and, particularly, for its connection with states which are dual to Ba$\tilde {\text{n}}$ados geometries \cite{Ecker:2019ocp}.}
\ba
\int_{- \infty}^\infty du\ T_{uu} \exp\left( \frac{6}{c} S_A \right) = 0.
\label{eq:E-S-rel}
\ea
In other words, what we find is that there must exist some $u$ so that in a certain time window the energy flux violates the NEC.

\section{Entanglement entropy from CFT calculation}
\label{sec:ee-cft}

In this section, we would like to explain how to compute the entanglement entropy of a single interval $A = \{x~|~x_1\leq x \leq x_2\}$ from the CFT side in moving mirror setups. After explaining the general method, we will see how the computation on the CFT side matches with that on the bulk gravity side studied in the former sections for holographic CFT. We will also study entanglement entropy in free Dirac fermion as another explicit example. Finally, we will analytically study the entanglement entropy for a fixed interval in (single) escaping mirror setups, and compare the results with a famous class of dynamics in CFT, known as quantum quenches \cite{Calabrese:2016xau}. Results for more setups are presented in appendix \ref{sec:moreEE}.

\subsection{Calculation method in general CFT}
As usual, let us use $c$ to denote the central charge of the CFT. The $n$-th entanglement R\'{e}nyi entropy can be evaluated by following the so-called twist operator formalism\cite{Calabrese:2004eu,Calabrese:2009qy}, so 
\begin{align}
    S^{(n)}_A = \frac{1}{1-n}\log \langle\sigma_{n}(t,x_1)\bar{\sigma}_{n}(t,x_2)\rangle ,
\end{align}
where the conformal weights of the twist operators $\sigma_n$ and $\bar{\sigma}_n$ are
\begin{align}
    h_n=\bar{h}_n = \frac{c}{24} \left(n-\frac{1}{n}\right) .
\end{align}
Here, the twist operator and its correlation function are defined in a $n$-replicated theory of the original CFT. This theory is also a CFT itself, whose central charge is $nc$ and the twist operators act as primaries in it. 

As in the former sections, we introduce the following null coordinates
\begin{align}
    u = t-x,\quad v= t+x. 
\end{align}
In order to evaluate the correlation function, let us perform the following conformal transformation 
\begin{align}
    \tilde{u} = p(u), \quad \tilde{v} = q(v)
\end{align}
to map the original setup to a static mirror setup which is parameterized by 
\begin{align}
    \tilde{u} = \tilde{t}-\tilde{x},\quad \tilde{v}= \tilde{t}+\tilde{x}. 
\end{align}
Remember that for single moving mirror setups, the static mirror setup was defined on the right half plane (RHP). For double moving mirror setups, this region becomes a strip. 

A primary operator $\CO(u,v)$ transforms as 
\begin{align}
    \tilde{\CO}(\tilde{u}, \tilde{v}) = \left(\frac{d\tilde{u}}{du}\right)^{-h_{\CO}}  \left(\frac{d\tilde{v}}{dv}\right)^{-\bar{h}_{\CO}} \CO(u,v)
\end{align}
under the conformal transformation, where $h_\CO~ (\bar{h}_\CO)$ is the chiral (anti-chiral) conformal dimension of $\CO$. Accordingly, 
\begin{align}
    \langle\sigma_{n}(t,x_1)\bar{\sigma}_{n}(t,x_2)\rangle
    = \left(p'(u_1)p'(u_2)\right)^{h_n} \left(q'(v_1)q'(v_2)\right)^{\bar{h}_n} \times \langle\tilde{\sigma}_{n}\left(\tilde{t}_1,\tilde{x}_1\right)\tilde{\bar{\sigma}}_{n}\left(\tilde{t}_2,\tilde{x}_2\right)\rangle_{\rm static~mirror} ,
\end{align}
where we have introduced 
\begin{align}
    &u_i = t - x_i,\quad v_i = t + x_i, \\
    &\tilde{u}_i = p(u_i),\quad \tilde{v}_i = q(v_i), \\
    &\tilde{t}_i = \frac{\tilde{u}_i+\tilde{v}_i}{2},\quad \tilde{x}_i = \frac{-\tilde{u}_i+\tilde{v}_i}{2},
\end{align}
for $i=1,2$. 

The two point function $\langle \sigma \sigma \rangle_{\rm static~mirror}$ can be computed via a ``mirror method"\footnote{This is also called a ``doubling trick". One may doubt whether the mirror method or doubling trick can be applied to the moving mirror setups, since they are not conventional BCFTs. A rigorous formulation for CFT with moving mirrors and a justification of the mirror method will be given in Sec.~\ref{sec:rbcft}.}, i.e. fill in the region behind the mirror (boundary) with the mirror image and compute it as if the boundary does not exist. For example, $\langle \sigma \sigma \rangle_{\rm RHP} \simeq \sqrt{\langle\sigma \sigma\sigma \sigma\rangle_{\mathbb{R}^{1,1}}}$. The details will be explained in the following sections. 

\subsection{Holographic CFT}

For simplicity, we first consider single moving mirror setups. In holographic CFT, the two point function of twist operators on the RHP can be reduced to some two point functions on $\mathbb{R}^{1,1}$
\begin{align}
    &\langle\tilde{\sigma}_{n}\left(\tilde{t}_1,\tilde{x}_1\right)\tilde{\bar{\sigma}}_{n}\left(\tilde{t}_2,\tilde{x}_2\right)\rangle_{\rm RHP}  \nonumber\\
    =& \max
  \begin{cases}
    \langle\tilde{\sigma}_{n}\left(\tilde{t}_1,\tilde{x}_1\right)\tilde{\bar{\sigma}}_{n}\left(\tilde{t}_2,\tilde{x}_2\right)\rangle_{\mathbb{R}^{1,1}} \\
    g^{2(1-n)}
    \left(
    \langle\tilde{\sigma}_{n}\left(\tilde{t}_1,\tilde{x}_1\right)\tilde{\bar{\sigma}}_{n}\left(\tilde{t}_1,-\tilde{x}_1\right)\rangle_{\mathbb{R}^{1,1}}
    \langle\tilde{\sigma}_{n}\left(\tilde{t}_2,-\tilde{x}_2\right)\tilde{\bar{\sigma}}_{n}\left(\tilde{t}_2,\tilde{x}_2\right)\rangle_{\mathbb{R}^{1,1}}
    \right)^{1/2}
  \end{cases}.
\end{align}
Here, the two candidates correspond to the connected/disconnected channel contributions, respectively. $g>0$ is a constant which depends on the boundary condition on the mirror.
This allows us to evaluate the entanglement R\'{e}nyi entropy in a single moving mirror setup according to
\begin{align}
    S^{(n)}_A = \min\left\{S^{(n),{\rm con}}_A,S^{(n),{\rm dis}}_A\right\},
\end{align}
where 
\begin{align}\label{eq:EEhol}
    S^{(n),{\rm con}}_A 
    &= \frac{1}{1-n}\log\left[\left(p'(u_1)p'(u_2)\right)^{h_n} \left(q'(v_1)q'(v_2)\right)^{\bar{h}_n} \times \langle\tilde{\sigma}_{n}\left(\tilde{t}_1,\tilde{x}_1\right)\tilde{\bar{\sigma}}_{n}\left(\tilde{t}_2,\tilde{x}_2\right)\rangle_{\mathbb{R}^{1,1}} 
    \right] \nonumber\\
    &=\frac{c}{24} \left(1+\frac{1}{n}\right) 
    \log \left[ \frac{(\tilde{u}_1-\tilde{u}_2)^2(\tilde{v}_1-\tilde{v}_2)^2}{\ep^4 p'(u_1)p'(u_2) q'(v_1)q'(v_2)}  \right] \nonumber\\ 
    &\equiv\frac{c}{24} \left(1+\frac{1}{n}\right) 
    \log F^{\rm con},  \nonumber\\ 
    S^{(n),{\rm dis}}_A 
    &= \frac{1}{1-n}\log\Big[\left(p'(u_1)p'(u_2)\right)^{h_n} \left(q'(v_1)q'(v_2)\right)^{\bar{h}_n}  \nonumber\\
    &\qquad\qquad\qquad \times g^{2(1-n)}
    \left(
    \langle\tilde{\sigma}_{n}\left(\tilde{t}_1,\tilde{x}_1\right)\tilde{\bar{\sigma}}_{n}\left(\tilde{t}_1,-\tilde{x}_1\right)\rangle_{\mathbb{R}^{1,1}}
    \langle\tilde{\sigma}_{n}\left(\tilde{t}_2,-\tilde{x}_2\right)\tilde{\bar{\sigma}}_{n}\left(\tilde{t}_2,\tilde{x}_2\right)\rangle_{\mathbb{R}^{1,1}}
    \right)^{1/2} 
    \Big] \nonumber\\
    &=\frac{c}{24} \left(1+\frac{1}{n}\right) 
    \log \left[ \frac{(\tilde{u}_1-\tilde{v}_1)^2(\tilde{u}_2-\tilde{v}_2)^2}{\ep^4 p'(u_1)p'(u_2) q'(v_1)q'(v_2)}  \right] 
    + 2\log g   \nonumber\\
    &=\frac{c}{24} \left(1+\frac{1}{n}\right) 
    \log F^{\rm dis} 
    + 2 S_{\rm bdy}
\end{align}
are the candidates for the entanglement entropy computed from the connected/disconnected channels of the twist operators' correlation function. The entanglement entropy $S_A$ is given by the $n\rightarrow1$ limit in these expressions. 

The above results are justified when assuming vacuum block dominance in holographic BCFT\cite{Sully:2020pza}. Clearly, computing the contribution from the connected (disconnected) channel corresponds to evaluating the connected (disconnected) RT surface in the gravity dual.

Double moving mirror setups can be discussed in a similar way by replacing ``right half plane'' with ``strip'', and ``$\mathbb{R}^{1,1}$'' with ``cylinder'' in the discussion above. As a result, the entanglement entropy will be given by (\ref{HEEdeconfine}) - (\ref{HEEdeconfine2}).

\subsection{Free Dirac fermion CFT}

Let us consider the free Dirac fermion CFT as another concrete example.

For single moving mirror setups, by performing the described mirror method, we can compute the two point function of twist operators on the RHP as follows\footnote{See, for example, Sec.~3.1.5 of \cite{Casini:2009sr} for the correlation functions of twist operators in free Dirac CFT.}
\begin{align}
    \langle\tilde{\sigma}_{n}\left(\tilde{t}_1,\tilde{x}_1\right)\tilde{\bar{\sigma}}_{n}\left(\tilde{t}_2,\tilde{x}_2\right)\rangle_{\rm RHP} 
    &= \left( \langle
    \tilde{{\sigma}}_{n}\left(\tilde{t}_2,-\tilde{x}_2\right)
    \tilde{\bar{\sigma}}_{n}\left(\tilde{t}_1,-\tilde{x}_1\right)
    \tilde{\sigma}_{n}\left(\tilde{t}_1,\tilde{x}_1\right)  \tilde{\bar{\sigma}}_{n}\left(\tilde{t}_2,\tilde{x}_2\right) \rangle_{\mathbb{R}^{1,1}}
     \right)^{1/2} \nonumber\\
    &\propto \left( \frac{(\tilde{u}_1-\tilde{v}_2)(\tilde{v}_1-\tilde{u}_2)}{(\tilde{u}_1-\tilde{u}_2)(\tilde{v}_1-\tilde{v}_2)(\tilde{u}_1-\tilde{v}_1)(\tilde{u}_2-\tilde{v}_2)} \right)^{2h_n}.
\end{align}
Since the central charge is $c=1$ for the Dirac fermion, we have 
\begin{align}\label{eq:EEfermion}
    S^{(n)}_A 
    &= \frac{1}{24} \left(1+\frac{1}{n}\right) 
    \log \left[ \frac{1}{\ep^4 p'(u_1)p'(u_2) q'(v_1)q'(v_2)} \left( \frac{(\tilde{u}_1-\tilde{u}_2)(\tilde{v}_1-\tilde{v}_2)(\tilde{u}_1-\tilde{v}_1)(\tilde{u}_2-\tilde{v}_2)}{(\tilde{u}_1-\tilde{v}_2)(\tilde{v}_1-\tilde{u}_2)} \right)^2  \right]  \nonumber\\
    &\equiv \frac{1}{24} \left(1+\frac{1}{n}\right) 
    \log F,
\end{align}
where $\ep$ is a UV cutoff corresponding to the lattice distance. The entanglement entropy is given by the $n\rightarrow1$ limit.\footnote{See \cite{Reyes:2021npy} for a computation of entanglement entropy in single moving mirror setups for free Dirac fermion CFT via an alternative method.}

For double moving mirror setups, if we impose the same boundary condition on the two boundaries of the strip, then
\begin{align}
    &\langle\tilde{\sigma}_{n}\left(\tilde{t}_1,\tilde{x}_1\right)\tilde{\bar{\sigma}}_{n}\left(\tilde{t}_2,\tilde{x}_2\right)\rangle_{\rm strip} 
    = \left( \langle
    \tilde{{\sigma}}_{n}\left(\tilde{t}_2,-\tilde{x}_2\right)
    \tilde{\bar{\sigma}}_{n}\left(\tilde{t}_1,-\tilde{x}_1\right)
    \tilde{\sigma}_{n}\left(\tilde{t}_1,\tilde{x}_1\right)  \tilde{\bar{\sigma}}_{n}\left(\tilde{t}_2,\tilde{x}_2\right) \rangle_{\rm cylinder}
     \right)^{1/2} \nonumber\\
    \propto& \left( \left(\frac{\pi}{2L}\right)^2 \frac{\cos\left(\frac{\pi}{2L}(\tilde{u}_1-\tilde{v}_2)\right)\cos\left(\frac{\pi}{2L}(\tilde{v}_1-\tilde{u}_2)\right)}{\sin\left(\frac{\pi}{2L}(\tilde{u}_1-\tilde{u}_2)\right)\sin\left(\frac{\pi}{2L}(\tilde{v}_1-\tilde{v}_2)\right)\cos\left(\frac{\pi}{2L}(\tilde{u}_1-\tilde{v}_1)\right)\cos\left(\frac{\pi}{2L}(\tilde{u}_2-\tilde{v}_2)\right)} \right)^{2h_n}.
\end{align}
Here, we map the double mirror setup to a strip with width $-L/2\leq \tilde{x} \leq L/2$. Accordingly, the entanglement R\'enyi entropy in a double moving mirror setup is given by 
\begin{align}\label{eq:EEDMfermion}
    &S^{(n)}_A \nonumber\\
    = &\frac{1}{24} \left(1+\frac{1}{n}\right) \nonumber\\ 
    &\times\log \Bigg[ \frac{(2L/\pi)^4}{\ep^4 p'(u_1)p'(u_2) q'(v_1)q'(v_2)}  \nonumber\\
    &~~~~~~~~\times \left( \frac{\sin\left(\frac{\pi}{2L}(\tilde{u}_1-\tilde{u}_2)\right)\sin\left(\frac{\pi}{2L}(\tilde{v}_1-\tilde{v}_2)\right)\cos\left(\frac{\pi}{2L}(\tilde{u}_1-\tilde{v}_1)\right)\cos\left(\frac{\pi}{2L}(\tilde{u}_2-\tilde{v}_2)\right)}{\cos\left(\frac{\pi}{2L}(\tilde{u}_1-\tilde{v}_2)\right)\cos\left(\frac{\pi}{2L}(\tilde{v}_1-\tilde{u}_2)\right)} \right)^2  \Bigg],
\end{align}
and the entanglement entropy follows from taking the limit $n\rightarrow1$ limit. 

\subsection{Fixed interval in escaping mirror and quantum quenches}

Let us here analytically study the entanglement entropy of a fixed interval $A=[x_1,x_2]$ in a (single) escaping mirror setup both for holographic CFT and the free Dirac fermion. 

As introduced in Sec.~\ref{subsec:ex4}, an escaping mirror can be mapped to a RHP via 
\begin{align}
    &\tilde{u} = p(u) = -\beta \log(1 + e^{-u/\beta}), \\
    &\tilde{v}= q(v) = v.
\end{align}
The mirror trajectory is given by 
\begin{align}
    \tilde{u}-\tilde{v} = 0 
    ~\Longleftrightarrow~
    x = -\beta \sinh^{-1}\left(\frac{e^{t/\beta}}{2}\right) \equiv Z(t).
\end{align}
It is straightforward to find out that
\begin{align}
    &p'(u) = \frac{e^{-u/\beta}}{1+e^{-u/\beta}}, \\
    &q'(v) = 1.
\end{align}
Now, we can continue with studying the entanglement entropy $S_A$. 

\subsubsection{Holographic CFT}

For a holographic CFT, plugging the previous relations into (\ref{eq:EEhol}), we find
\begin{align}
    F^{\rm con} = &\frac{1}{\epsilon^4}
    \cdot \frac{1 + e^{-(t-x_1)/\beta}}{e^{-(t-x_1)/\beta}} 
    \cdot \frac{1 + e^{-(t-x_2)/\beta}}{e^{-(t-x_2)/\beta}} 
    \cdot \left[\left(\beta\log\frac{1+e^{-(t-x_2)/\beta}}{1+e^{-(t-x_1)/\beta}}\right)(x_2-x_1)\right]^2 , \\ 
    F^{\rm dis} = &\frac{1}{\epsilon^4}
    \cdot \frac{1 + e^{-(t-x_1)/\beta}}{e^{-(t-x_1)/\beta}} 
    \cdot \frac{1 + e^{-(t-x_2)/\beta}}{e^{-(t-x_2)/\beta}} 
    \nonumber \\ 
    &\times \left[{\left(t+x_1+\beta\log(1+e^{-(t-x_1)/\beta})\right)\left(t+x_2+\beta\log(1+e^{-(t-x_2)/\beta})\right)}\right]^2.
\end{align}
We would like to study some typical limits of this result. 

Firstly, at early times, i.e. $t<0\leq x_1<x_2 \ll -t$, we obtain
\begin{align}
    S^{\rm con}_A &\approx \frac{c}{3} \log \frac{x_2-x_1}{\epsilon}, \\
    S^{\rm dis}_A &\approx \frac{c}{6} \log \frac{2x_1}{\epsilon} + \frac{c}{6} \log \frac{2x_2}{\epsilon} + 2 S_{\rm bdy}.
\end{align}
Secondly, the intermediate behavior for times $0 < x_1\ll t \ll x_2$ and a large interval $A$ turns out to be given by
\begin{align}
    S^{\rm con}_A &\approx \frac{c}{12\beta} (t-x_1) + 
    \frac{c}{3} \log \frac{x_2}{\epsilon}, \\
    S^{\rm dis}_A &\approx \frac{c}{12\beta}(t-x_1) + \frac{c}{6}\log\frac{t}{\ep} + \frac{c}{6}\log\frac{2x_2}{\ep} + 2 S_{\rm bdy}.
\end{align}
Finally, at late times, i.e. $0<x_1<x_2\ll t$, we get
\begin{align}
    S^{\rm con}_A &\approx \frac{c}{6}\log\frac{x_2-x_1}{\epsilon } + \frac{c}{6}\log\left(\frac{2\beta}{\ep }\sinh{\frac{(x_2-x_1)}{2\beta}}\right), \\
    S^{\rm dis}_A &\approx \frac{c}{12\beta}(2t-x_1-x_2) + \frac{c}{3}\log\frac{t}{\ep} + 2 S_{\rm bdy}.
\end{align}

\subsubsection{Free Dirac fermion}

For the free Dirac fermion, plugging the conformal transformation above into (\ref{eq:EEfermion}), gives
\begin{align}
    &\tilde{u} = p(u) = -\beta \log(1 + e^{-u/\beta}), \\
    &\tilde{v}= q(v) = v,
\end{align}
and we get, 
\begin{align}
    F = &\frac{1}{\epsilon^4}
    \cdot \frac{1 +  e^{-(t-x_1)/\beta}}{ e^{-(t-x_1)/\beta}} 
    \cdot \frac{1 +  e^{-(t-x_2)/\beta}}{ e^{-(t-x_2)/\beta}} 
    \cdot \left[\left(\beta\log\frac{1+e^{-(t-x_2)/\beta}}{1+e^{-(t-x_1)/\beta}}\right)(x_2-x_1)\right]^2
    \nonumber \\ 
    &\times \left[\frac{\left(t+x_1+\beta\log(1+e^{-(t-x_1)/\beta})\right)\left(t+x_2+\beta\log(1+e^{-(t-x_2)/\beta})\right)}{\left(t+x_1+\beta\log(1+e^{-(t-x_2)/\beta})\right)\left(t+x_2+\beta^{-1}\log(1+e^{-(t-x_1)/\beta})\right)}\right]^2.
\end{align}
Some typical limits are discussed below. 

Let us begin by noting that the initial behavior of the entanglement entropy can be read out by studying the case, where $t<0\leq x_1<x_2 \ll -t$. In this limit, we get
\begin{align}
    S_A \approx \frac{1}{6} \log \frac{(x_2-x_1)^2\cdot 4x_1x_2}{\left(x_1+x_2\right)^2 \cdot \epsilon^2}. 
\end{align}
Note that, in a static mirror setup, the entanglement entropy reads
\begin{align}
    S_A^{\rm static} = \frac{1}{6} \log \frac{(x_2-x_1)^2 \cdot 4x_1x_2}{\left(x_1+x_2\right)^2 \cdot \epsilon^2}. 
\end{align}
The initial behavior of the entanglement entropy in the moving mirror setup coincides with that in the static mirror setup. 

Let us now consider the intermediate behavior for a large interval $A$ when $0 < x_1\ll t \ll x_2$. We find
\begin{align}
    F \approx &\frac{1}{\epsilon^4}\cdot e^{(t-x_1)/\beta} \cdot 1 \cdot \left(2t x_2\right)^2,
\end{align}
and hence 
\begin{align}
    S_A \approx \frac{1}{12\beta}(t-x_1) + \frac{1}{6}\log\frac{t}{\ep} + \frac{1}{6}\log\frac{2x_2}{\ep}.
\end{align}
Note that there is both a linear time evolution and a logarithmic time evolution in this intermediate limit. 

Finally, at late times, i.e. $0<x_1<x_2\ll t$, we find
\begin{align}
    F \approx &\frac{1}{\epsilon^4}
    \cdot  e^{(t-x_1)/\beta} 
    \cdot  e^{(t-x_2)/\beta}
    \cdot \left[\beta\left(e^{-(t-x_2)/\beta}-e^{-(t-x_1)/\beta}\right)(x_2-x_1)\right]^2
    \nonumber \\
    = & \frac{\beta^2}{\epsilon^4}\cdot(x_2-x_1)^2 \cdot \left(2\sinh{\frac{(x_2-x_1)}{2\beta}}\right)^2,
\end{align}
and therefore, 
\begin{align}
    S_A = \frac{1}{6}\log\frac{x_2-x_1}{\epsilon } + \frac{1}{6}\log\left(\frac{2\beta}{\ep }\sinh{\frac{(x_2-x_1)}{2\beta}}\right).
\end{align}
Here, we would like to note that the latter can be written as
\begin{align}
    S_A = \frac{1}{2} ({\rm ~EE~at~vacuum~+~EE~at~temperature~}1/(2\pi\beta)~).
\end{align}
This is because only the chiral sector is heated up by the radiation caused by the moving mirror. 

\subsubsection{Comparison with quantum quenches}

Comparing the behavior of the entanglement entropy in holographic CFT and that for the free fermion, a common feature is that the intermediate behavior for $0<x_1\ll t \ll x_2$ and a large interval behaves like
\begin{align}\label{eq:TEMM}
    S_A = \frac{c}{12\beta} t + \frac{c}{6} \log \frac{t}{\epsilon}+\cdots = \frac{\pi c}{6} \frac{1}{2\pi\beta} t + \frac{c}{6} \log \frac{t}{\epsilon}+\cdots. 
\end{align}
Here, $1/(2\pi \beta)$ is the effective temperature in the escaping mirror setup, as we have seen in the previous subsection. Let us compare this with another class of dynamics that is called quantum quench. 

First of all, a linear time evolution is a typical feature of a global quench setup \cite{Calabrese:2005in,Calabrese:2016xau}. In general, a global quench leads to some dynamics for which one first prepares an initial state, which is the ground state of a gapped Hamiltonian, and then lets it evolve under a gapless Hamiltonian. In CFT, a global quench can be simulated as 
\begin{align}
    \ket{\psi(t)} = \mathcal{N} e^{-itH}e^{-\alpha H}\ket{B},
\end{align}
where $H$ is the CFT Hamiltonian, $\ket{B}$ is a boundary state\cite{Cardy2004}, $\alpha$ is a cutoff, and $\mathcal{N}$ is a normalization factor. For a fixed interval $A=[x_1,x_2]$, and $0<x_1\ll t \ll x_2$, one has
\begin{align}\label{eq:TEGQ}
    S_A = \frac{\pi c}{6\alpha}t + \cdots = 4\cdot\frac{\pi c}{6}\frac{1}{4\alpha}t + \cdots,
\end{align}
Here, $1/(4\alpha)$ is known as the effective temperature in the global quench setup. 

Comparing (\ref{eq:TEMM}) and (\ref{eq:TEGQ}), we can see that the radiation emitted from an escaping mirror and that in a global quench setup is similar. It is also straightforward to understand the factor $4$ in the following way. After a global quench, left moving radiation and right moving radiation is emitted from all the spatial points and moves at the speed of light. Therefore, there are four origins for the entanglement generation. One is the radiation generated in $A$ with the left moving mode being outside and the right moving mode being inside. The other three are given by swapping ``left" and ``right", and/or changing ``in $A$" to ``outside of $A$" in the previous description. On the other hand, in the escaping mirror setup, there is only one origin for the entanglement generation. The radiation is always generated outside of $A$ (next to the mirror), and only the right moving mode goes into $A$. 

Second, a logarithmic time evolution proportional to $\log (t/\epsilon)$ is known as a typical feature in a local joining quench setup \cite{Calabrese:2007mtj,Asplund:2013zba}. A local joining quench is achieved by preparing first two ground states on two semi-infinite intervals, and then joining them together, i.e. letting them evolve under a CFT Hamiltonian on an infinite line. This setup can be expressed as 
\begin{align}
    \ket{\psi(t)} = \mathcal{N} e^{-itH}e^{-\alpha H}(\ket{0}_L \otimes \ket{0}_R),
\end{align}
where $\alpha$ is again a cutoff. For a fixed interval $A=[x_1,x_2]$, and $0<x_1\ll t \ll x_2$, we have 
\begin{align}\label{eq:TELQ}
    S_A = \frac{c}{6} \log \frac{t}{\alpha}+ \frac{c}{6} \log \frac{t}{\epsilon}+ \cdots.
\end{align}
In holographic CFT, these two terms can be understood as a shock wave effect and a joining effect, respectively\cite{Shimaji:2018czt,Caputa:2019avh}. Here, the shock wave effect means that the joining dynamics causes a shock wave which stretches the RT surface in the bulk gravity, and the joining effect means that the bulk gradually becomes connected and produces more entanglement between the left side and the right side. In more general language, the $\log(t/\alpha)$ comes from a shock produced by a local excitation, while the $\log(t/\epsilon)$ occurs, because $A$ can be entangled with more region after joining the two sides together.

Comparing (\ref{eq:TEMM}) and (\ref{eq:TELQ}), we can see that the second terms have the same form. Therefore, the second term of (\ref{eq:TEMM}) should be understood as a sort of ``joining effect". This is indeed the case, since the moving mirror enlarges the system and produces more region for $A$ to be entangled with. 

Having these observations in mind, an escaping mirror can be qualitatively regarded as a sequence of local joining quenches, i.e. joining infinitesimal intervals to a semi-infinite interval successively. This will push the boundary farther and farther. This sequence collectively generates a global quench-like radiation which gives the linear $t$ term, while the $\log(t/\epsilon)$ term can be understood as a joining effect. 

It is an interesting future direction to study the relation between moving mirrors and quantum quenches in more detail. 

\section{Radiative BCFT}
\label{sec:rbcft}

In this section, we formally define our CFT as a particular generalization of (Euclidean) BCFT.

\subsection{Generalization of BCFT}

A normal BCFT can be specified by the position of the boundary, which can be described by a solution of the following equation,\footnote{The solution to this equation can be expressed by
\begin{equation*}
x=Z(t).
\end{equation*}
}

\begin{equation}
\label{eq:bdy.eq}
G(z) = \bar{z},
\end{equation}
where with the reality condition $\bar{z}=z^*$, the real part of its solution is linearly dependent on the imaginary part.
In other words, the profile of the BCFT is encoded in the function $G(z)$.
For example, a CFT on a upper half plane (UHP) is specified by $G(z)=z$, whose solution leads to the boundary on a real axis.

In any CFT, we have a stress tensor field $T(z)$, therefore we also have to determine the boundary condition on $T(z)$.
We usually assume that there is no energy and momentum flux on the boundary, which leads to the boundary condition
\begin{equation}\label{eq:BT}
T(z) - \bar{T}(\bar{z})=0, \ \ \ \ \ \ \text{on }\ \ \ G(z)=\bar{z}.
\end{equation}
This condition breaks a part of conformal symmetry.
Nevertheless, we can make use of the conformal Ward identity by using the ``mirror method''.
This is a brief summary of a systematic approach to a normal BCFT \cite{Cardy2004}.

Let us focus on our case. In our setup, we have the ``radiative boundary'', which means that the right-moving radiation and left-moving radiation do not cancel with each other and consequently break the condition (\ref{eq:BT}).
Alternatively, we have non-trivial stress tensors associated to the radiation, which we will denote as
\begin{equation}
\begin{aligned}
\left\{
    \begin{array}{ll}
      &T(z)=\fr{c}{12}R(z)   ,\\
       &\bar{T}(\bar{z})=\fr{c}{12}\bar{R}(\bar{z}),\\
    \end{array}
  \right.\\
\end{aligned}
\end{equation}
in particular, which do not need to satisfy the condition $R(z)=\bar{R}(\bar{z})$.
In this case, we cannot rely on the mirror method. Therefore, we have to introduce a new approach to calculate a correlator.
One may also wonder if one can construct a consistent BCFT with any $R(z),\bar{R}(\bar{z})$.
We answer these questions in the following subsections.

\subsection{Calculation method}

In practical calculations in a CFT, we usually treat holomorphic and anti-holomorphic conformal maps independently.
Even though these generators do not map $\mathbb{R}^2$ to  $\mathbb{R}^2$, this map is still a well-defined map on $\mathbb{C}^2$.
For example, if we are interested in CFT in Minkowski space, we consider the Wick rotation of a CFT in Euclidean space.
This Wick rotation leads to independent parameters $z$ and $\bar{z}$ beyond the reality $\bar{z}=z^*$.
Even though the CFT looses the reality condition, as we have seen in many studies, this Lorentzian CFT works well.

Here, we also treat holomorphic and anti-holomorphic parts independently in a similar way, in order to accomplish calculations of radiative BCFT correlators.
Let us consider conformal maps $w=f(z)$ and $\bar{w}=g(\bar{z})$, where $g$ is not necessarily equal to $f^*$.
Under this map, the stress tensor is transformed as
\begin{equation}
T(z)=\pa{\fr{\dd f(z)}{\dd z}}^2T'(w) + \fr{c}{12}\{f,z\},
\end{equation}
where $\{f,z\}$ is the Schwarzian derivative.
If we consider particular conformal maps $f$ and $g$, which satisfy
\begin{equation}\label{eq:R}
\begin{aligned}
\{f,z\}&=R(z),\\
\{g,\bar{z}\}&=\bar{R}(\bar{z}),
\end{aligned}
\end{equation}
then the profile of the radiative BCFT can be re-expressed as
\begin{itemize}

\item position of boundary
\begin{equation}
G\circ f^{-1}(w) = g^{-1}(\bar{w}),
\end{equation}

\item boundary condition on $T$
\begin{equation}\label{eq:BT2}
T'(w)-\bar{T'}(\bar{w})=0.
\end{equation}
\end{itemize}
In particular, we have $T'(w)=T'(\bar{w})=0$.
The boundary condition (\ref{eq:BT2}) is just the same as (\ref{eq:BT}),
which means that we can utilize the mirror method for this CFT analytically continued from the radiative BCFT.
Note that the condition (\ref{eq:R}) does not uniquely determine $f$ and $g$, because we have an ambiguity attributed to M{\"o}bius transformations.

When we evaluate a BCFT correlator, we usually map our CFT to a CFT on an UHP for convenience.
Here, we also parallel this approach.
This transformation can be realized by a real map $h(w)$ and $h^*(\bar{w})$.
To avoid cumbersome expressions, we will redefine the function $f$ by a combination of $f$ and $h$ (and $g$ by $g$ and $h^*$). 
That is, we redefine $f$ and $g$ such that
\begin{equation}
G\circ f^{-1}(w) = g^{-1}(\bar{w}),
\end{equation}
which has the solution $w=\bar{w}$. If we assume that $G$ is an analytic function, then we can conclude
\begin{equation}
G\circ f^{-1} = g^{-1}.
\end{equation}

Above, we show this condition by an explicit calculation.
We can also explain it in simpler words.
What we did in the above derivation is:
\begin{itemize}
\item
consider a conformal map preserving the ``degeneracy'' of the solution of (\ref{eq:bdy.eq}),
\item
map it into the usual BCFT framework by the conformal map.
\end{itemize}

\subsection{Possible radiation}

Let us answer the following question: ``Which $R$ and $\bar{R}$ are allowed from the consistency condition of BCFT?''.
For this, we would like to introduce the chain rule of the Schwarzian derivative,
\begin{equation}
\{f\circ g, z\} = \{f, g(z)\}\pa{\fr{\dd g(z)}{\dd z}}^2 + \{g,z\}.
\end{equation}
By this chain rule, we can show
\begin{equation}\label{eq:R2}
R(z)=\bar{R}(\bar{z})\pa{\fr{\dd G(z)}{\dd z}}^2 + \{G,z\},
\end{equation}
where we have used (\ref{eq:R}) and 
\begin{equation}
\{f, z\} = \{ g, \bar{z}\}\pa{\fr{\dd G(z)}{\dd z}}^2 + \{G,z\}.
\end{equation}
That is, only $\bar{R}(\bar{z})$ can be freely chosen, and $R(z)$ is uniquely fixed by $\bar{R}(\bar{z})$.
If the condition (\ref{eq:R2}) is not satisfied, then $G(z)=\bar{z}$ (with the reality condition $\bar{z}=z^*$) has no solution, so it does not make sense.

In this article, we mainly focus on the chiral radiation $\bar{R}(\bar{z})=0$.
This assumption allows us to simplify the expression of the chiral radiation as
\begin{equation}
R(z) = \{G,z\}.
\end{equation}

\subsection{Correlation function}

Now, since we have a general calculation method for the radiative BCFT, we can evaluate correlators.
We first map a correlator in the radiative BCFT into the usual BCFT by using the prescription above,
\begin{equation}
\begin{aligned}
\braket{\prod_i O_i (z_i, \bar{z_i})}_{\text{rad}}
&=
\pa{\fr{\dd f^{-1}(w_i)}{\dd w_i}}^{h_i}
\pa{\fr{\dd g^{-1}(\bar{w}_i)}{\dd \bar{w}_i}}^{\bar{h}_i}
\braket{\prod_i O_i (w_i, \bar{w_i})}_{\text{UHP}}\\
&=
\fr{
\pa{\fr{\dd G(z)}{\dd z}}^{\bar{h}_i}
}{
\pa{\fr{\dd f(z_i)}{\dd z_i}}^{h_i}
\pa{\fr{\dd f(\bar{z}_i)}{\dd \bar{z}_i}}^{\bar{h}_i}
}
\braket{\prod_i O_i (w_i, \bar{w_i})}_{\text{UHP}}.
\end{aligned}
\end{equation}
This result means that we can decompose the contribution from the boundary into two parts, i.e.
shape of the boundary and polarization of radiation. For example, if we consider entanglement entropy in the radiative BCFT, the general expression can be written as
\begin{equation}
S_A=\lim_{n \to 1} \fr{1}{1-n}\log \pa{\braket{\prod_i \sigma_{n} (w_i, \bar{w_i})}_{\text{UHP}}}
-
\fr{c}{12}\log \pa{\fr{\dd G(z)}{\dd z}}
+
\fr{c}{12} \log \pa{\fr{\dd f(z_i)}{\dd z_i} \fr{\dd f(\bar{z}_i)}{\dd \bar{z}_i}}.
\end{equation}
The first part is just the universal contribution.
The second part represents the dependence on the shape of the boundary, and the third part represents the dependence on the polarization of the radiation.

\subsection{Boundary state}

We can interpret our state as ``descendants'' of Cardy states.
Let us consider the most simple case.
Our generalized boundary is obtained by a chiral conformal map,
which means that the corresponding boundary state can be expressed by, for example, 
\begin{equation}\label{eq:descendantCardy}
\ex{\epsilon L_{k}}\ket{B},
\end{equation}
where $\epsilon$ is an infinitesimal positive number and $\ket{B}$ is the Cardy state.
We can show that this state has non-zero energy flux in the following way.
The Cardy state is composed of the Ishibashi states,
\begin{equation}
\ket{j}\rangle \equiv \sum_{N} \ket{j;N} \otimes   U \overline{\ket{j;N}},
\end{equation}
where $\ket{j;N}$ is a state in the Verma module $j$ labeled by $N$, and $U$ is an antiunitary operator.
For a general state $\ket{j;a}\otimes U \overline{\ket{j;b}}$, we can show
\begin{equation}
\bra{j;a} \otimes U\bra{j;b}  (L_n - \bar{L}_{-n})   L_{k} \ket{j}\rangle
=
\bra{j;a} [L_n, L_k] \ket{j;b}
\neq 0.
\end{equation}
It means that states like (\ref{eq:descendantCardy}) have non-zero flux.

What we have evaluated just reads like
\begin{equation}
\bra{0} O(z,\bar{z})\cdots  \ex{\epsilon L_{k}}\ket{B}.
\end{equation}
This type of correlators can be evaluated by the Ward identity.
This is the mechanism for generalizing a boundary with vanishing flux to a radiative boundary, and the reason why we have used the Ward identity in our calculation.

It might be possible to think of attaching Virasoro generators on boundary states as the ``boundary graviton excitation'' in the 2D-gravity dual to the 1D boundary.
This interpretation would provide a new understanding of the connection to the welding setup \cite{Almheiri:2019qdq}.

\section{Summary and discussion}
\label{sec:summ}

In this paper, we have extended our previous studies presented in \cite{Akal:2020twv} in several new directions. Starting with a review of moving mirrors in two dimensional CFTs, we have extensively discussed how the problem can be tackled by using the technology of conformal transformations. This has been done for the case of various mirror profiles which capture essential features of black hole physics. Particularly, we have considered three different scenarios, which we have named escaping mirror, kink mirror, and double escaping mirror, see Fig.~\ref{figmvth}. Specifically, the double escaping mirror has appeared as a novel setup which was not discussed in our previous analysis in \cite{Akal:2020twv}. 

One of our main goals in the present work was to make the connection between the physics of moving mirrors and black hole radiation more evident. For doing so, we have first readdressed certain examples for which the Bogoliubov coefficients can be computed in an explicit way. The fact, that those do perfectly match with the one obtained in well studied gravitational backgrounds, can indeed be considered as further support for interpreting field theories in the presence of moving mirrors as gravitational setups. 

In order to strengthen these claims, we have particularly made use of the AdS/BCFT construction. By focusing first on the single moving mirror case, we have been able to compute the entanglement entropy holographically for various scenarios which mimic essential features of Hawking radiation. In particular, we have shown that our setups can indeed be understood as a CFT coupled to two dimensional gravity. This understanding has found major support by relations to brane-world holography and, especially, by an interesting reformulation in terms of Liouville theory. In addition, this novel understanding has allowed us to argue that our moving mirror models are closely related to previously studied two dimensional gravitational setups which have led to the island picture. 

By employing our holographic construction of the moving mirror CFT, we have in detail studied the double escaping mirror. From the point of view of brane-world holography, this setup precisely corresponds to the situation, where two dynamically gravitating end-of-the-world branes are individually radiating into an enclosed spacetime region. Specifically, we have calculated the holographic entanglement entropy for a finite subregion situated between the two gravitational sectors, and uncovered an interesting new phase transition between differently extending minimal surfaces determining the entropy.

In addition, we have also discussed properties of the produced energy flux in our moving mirror setups and examined the connection to energy conditions. More specifically, for setups which give rise to a unitary entropy curve, we have seen that while the QNEC is always saturated, the NEC turns out to be temporarily violated as a consequence of nontrivial quantum effects.

In order to verify our holographic constructions and the related entanglement entropy calculations, we have also given a detailed conformal field theoretic analysis in the corresponding moving mirror setups. In addition to these confirmations, we have discussed how a usual CFT with a moving mirror may be regarded as a BCFT. In particular, we have clarified how the usual conformal boundary condition would need to be generalized in form of a new moving mirror boundary condition.

We would like to note that the simplicity of our holographic moving mirror models would allow to compute many interesting quantities without much difficulty for various gravitating systems. For instance, as we have explicitly done in the present work, this simplification turned out be extremely helpful in dealing with the situation, where two gravitational systems are considered. By employing the double escaping mirror, we have been able to treat this problem analytically, without the necessity of handling complicated technicalities such as the conformal welding problem as well as the appearance of replica wormholes. The former is known to be difficult to solve in general. Recently, a setup of two entangled gravitating systems has also been studied in \cite{Balasubramanian:2021wgd}. 

One interesting aspect, that we want to particularly highlight, is the proposed connection between two dimensional (quantum) gravity and moving mirrors. Starting with the earlier work by Davies and Fulling, quantum field theories in the presence of moving mirrors have generally been accepted as being tractable toy models mimicking certain features of black hole radiation. Since then, there have been worked out a lot of interesting results in the literature along this direction. However, here, we have pushed this picture even further. Namely, we have argued and provided evidence that such field theoretic constructions may indeed be interpreted as proper gravitational setups describing the physics of evaporating black holes from the perspective of a far distant observer. Nevertheless, there is still one big issue, which has not been taken into account in moving mirror models. This is the black hole singularity. It will be a very intriguing future problem to treat the effect of black hole singularity in a moving mirror 
setup.

Let us finally note that the moving mirror constructions, as we have introduced here, may also provide a simple framework for exploring intriguing aspects of non-equilibrium processes, which are of high interest in condensed matter physics.

All in all, there are many interesting questions along these lines to which we would like to come back in future work.

\section*{Acknowledgements}
We are grateful to  Kanato Goto, Tatsuma Nishioka, Jonathan Oppenheim, Shan-Ming Ruan, Yoshiki Sato, Tetsuya Shiromizu and Tomonori Ugajin for useful discussions and comments.
IA is supported by the Japan Society for the
Promotion of Science (JSPS) and the Alexander von Humboldt (AvH) foundation. 
IA and TT are supported by Grant-in-Aid for JSPS Fellows No.~19F19813.
YK and ZW are supported by the JSPS fellowship.
YK is supported by Grant-in-Aid for JSPS Fellows No.~18J22495.
TT is supported by the Simons Foundation through the ``It from Qubit'' collaboration,
and by Inamori Research Institute for Science and 
World Premier International Research Center Initiative (WPI Initiative) 
from the Japan Ministry of Education, Culture, Sports, Science and Technology (MEXT). 
NS and TT were supported by JSPS Grant-in-Aid for Scientific Research (A) No.~16H02182. 
TT is also supported by JSPS Grant-in-Aid for Challenging Research (Exploratory) 18K18766, JSPS Grant-in-Aid for Scientific Research (A) No.~21H04469 and and Grant-in-Aid for Transformative Research Areas (A) No. 21H05187.
ZW is supported by the ANRI Fellowship and Grant-in-Aid for JSPS Fellows No.~20J23116.

\appendix

\section{More on gravity dual of conformal map}
\label{sec:appen}

In Sec.~\ref{sec:Banados}, the Ba$\tilde{\text{n}}$ados map (\ref{corads}) is introduced as a gravity dual of a conformal map. A shortage of this map is that it requires the metric in the $(U,V,\eta)$ coordinates to be Poincar\'{e} AdS$_3$. On the other hand, we sometimes need to map a setup to another one which is not necessarily Poincar\'{e} AdS$_3$. For example, in Sec.~\ref{sec:dem}, we map a double moving mirror setup to a strip, whose gravity dual can be global AdS$_3$ when the confined configuration is picked. To this end, we would like to slightly extend the Ba$\tilde{\text{n}}$ados map (\ref{corads}) in the present section.

Let us consider three CFT coordinate choices $(u_0,v_0)$, $(u_1,v_1)$ and $(u_2,v_2)$, which are connected to each other via
\begin{align}
\label{eq:bdytrans}
    u_0 = f(u_1),\ \ \ v_0 = g(v_1), \\
    u_1 = p(u_2),\ \ \ v_1 = q(v_2).
\end{align}
The gravity dual of the CFT in terms of $(u_i,v_i)$ can be parameterized by $(u_i,v_i,z_i)$, where the coordinates $(u_i,v_i)$ are identified with the bulk coordinates at $z_i=0$.

If the metric of the gravity dual of the CFT expressed in the $(u_0,v_0)$ coordinates is Poincar\'e, i.e.
\begin{align}
    ds^2=\frac{dz_0^2-du_0dv_0}{z_0^2},
\end{align}
then the gravity dual in the $(u_1,v_1,z_1)$ coordinates is described by the metric
\begin{align}
    ds^2=\frac{dz_1^2}{z_1^2}+T_{+,1}(u_1)(du_1)^2+T_{-,1}(v_1)(dv_1)^2-\left(\frac{1}{z_1^2}+z_1^2T_{+,1}(u_1) T_{-,1}(v_1)\right)du_1dv_1, 
\end{align}
where
\begin{align}
    T_{+,1}(u_1)=\frac{3(f'')^2-2f'f'''}{4f'^2},\qquad T_{-,1}(v_1)=\frac{3(g'')^2-2g'g'''}{4g'^2},
\end{align}
and the bulk coordinates are related by
\ba
\begin{split}\label{eq:bulktrans}
u_0&=f(u_1)+\frac{2z_1^2(f')^2 g''}{4f'g'-z_1^2f''g''},\\
v_0&=g(v_1)+\frac{2z_1^2(g')^2f''}{4f'g'-z_1^2f''g''},\\
z_0&=\frac{4z_1(f'g')^{3/2}}{4f'g'-z_1^2f''g''}. 
\end{split}
\ea
By definition, the bulk coordinate transformation (\ref{eq:bulktrans}) coincides with the boundary CFT's conformal map (\ref{eq:bdytrans}) at $z_1 = 0$. 

Similarly, the gravity dual in the $(u_2,v_2,z_2)$ coordinates is given by
\begin{align}
    ds^2=\frac{dz_2^2}{z_2^2}+T_{+,2}(u_2)(du_2)^2+T_{-,2}(v_2)(dv_2)^2-\left(\frac{1}{z_2^2}+z_2^2T_{+,2}(u_2) T_{-,2}(v_2)\right)du_2dv_2, 
\end{align}
where
\begin{align}
    &T_{+,2}(u_2)=\frac{3((f\circ p)'')^2-2(f\circ p)'(f\circ p)'''}{4(f\circ p)'^2}, \\ &T_{-,2}(v_2)=\frac{3((g\circ q)'')^2-2(g\circ q)'(g\circ q)'''}{4(g\circ q)'^2}.
\end{align}
and the bulk coordinates are related by
\ba
\begin{split}
u_0&=(f\circ p)(u_2)+\frac{2z_2^2((f\circ p)')^2 (g\circ q)''}{4(f\circ p)'(g\circ q)'-z_2^2(f\circ p)''(g\circ q)''},\\
v_0&=(g\circ q)(v_2)+\frac{2z_2^2((g\circ q)')^2(f\circ p)''}{4(f\circ p)'(g\circ q)'-z_2^2(f\circ p)''(g\circ q)''},\\
z_0&=\frac{4z_2((f\circ p)'(g\circ q)')^{3/2}}{4(f\circ p)'(g\circ q)'-z_2^2(f\circ p)''(g\circ q)''}. 
\end{split}
\ea
Combining these two maps, we can get a map between $(u_1,v_1,z_1)$ and $(u_2,v_2,z_2)$. Although the exact map is complicated, it is essential that 
\begin{align}
\begin{split}
    &u_1 = p(u_2) + \CO(z_2^2), \\
    &v_1 = q(v_2) + \CO(z_2^2), \\
    &z_1 = z_2\sqrt{p'q'} + \CO(z_2^2).
\end{split}
\end{align}
This asymptotic behavior plays a crucial role when computing the holographic entanglement entropy, since it tells us how the UV cutoff should be deformed under the conformal transformation. This is, however, almost trivial on the CFT side. This just tells how Weyl factors are counted under a conformal transformation.

When $T_{+,1} = T_{-,1} = -\pi^2/4L^2$, $f$ and $g$ can be chosen to be 
\begin{align}
    f(u_1) = e^{i\frac{\pi}{L}u_1},\ \ \ g(u_1) = e^{-i\frac{\pi}{L}v_1}. 
\end{align}
Accordingly, 
\begin{align}\label{eq:holEM}
    T_{+,2}(u)=\frac{3(p'')^2-2p'p'''}{4p'^2}-\frac{\pi^2 p'^2}{4L^2},\qquad T_{-,2}(v)=\frac{3(q'')^2-2q'q'''}{4q'^2}-\frac{\pi^2 q'^2}{4L^2}.
\end{align}
When the setup expressed in terms of $(u_1,v_1)$ is a (Lorentzian) strip, this choice of $f$ and $g$ means transforming it to a Euclidean strip and then mapping it to a half plane. 

\subsection{Ba$\tilde{\text{n}}$ados map at leading order}

In many situations, the asymptotic behavior at the asymptotic boundary of AdS spacetime is sufficient to evaluate physical quantities such as holographic entanglement entropy. To this end, it is sufficient to apply the original Ba$\tilde{\text{n}}$ados map (\ref{corads}) to the leading order of the bulk direction. 

Let us again take global AdS$_3$ as an example. Here, we use the metric 
\ba
ds^2=-\frac{dT^2}{\zeta^2}+\frac{d\zeta^2}{\zeta^2\left(1-\frac{\pi^2\zeta^2}{L^2}\right)}+\frac{1-\frac{\pi^2\zeta^2}{L^2}}{\zeta^2}dX^2,
\ea
which is related to the metric (\ref{eq:corglobalads}) via $\zeta = 1/R$. The asymptotic behavior at $\zeta\sim0$ is Poincar\'e AdS$_3$, so we can perform the Ba$\tilde{\text{n}}$ados map \eqref{corads} at leading order of $\zeta$. Writing $U=T-X$ and $V=T-X$, and applying the coordinate transformation \eqref{corads}, we find that the metric behaves in the boundary limit $z\to 0$ as
\be
\begin{split}
ds^2 &\simeq \frac{dz^2}{z^2}+T_{+}(u)(du)^2+T_{-}(\bar{v})(dv)^2-\left(\frac{1}{z^2}
+z^2T_{+}(u) T_{-}(v)\right)dudv\\
&+\frac{\pi^2}{L^2}p'q'dz^2-\frac{\pi^2}{4L^2}(q'dv-p'du)^2+\mathcal{O}(z^2)\\
&\simeq  \frac{dy^2}{y^2}+\left(T_+(u)-\frac{\pi^2}{4L^2}p'^2\right)du^2+\left(T_-(u)-\frac{\pi^2}{4L^2}q'^2\right)(\bar{v})(dv)^2 \nonumber\\
&~~~~~~-\left(\frac{1}{y^2}
+y^2T_{+}(u) T_{-}(v)\right)dudv +\mathcal{O}(y),
\end{split}
\ee
where we have introduced the coordinate transformation 
\ba
y=z+\frac{\pi^2}{4L^2}p'q' y^3+\ddd.
\ea
Although this coordinate transformation is not exact, it allows us to read out the asymptotic behavior including the holographic energy stress tensor. Here, the holographic energy stress tensor $\hat{T}_{\pm}$ is given by
\be
\begin{split}
\hat{T}_+(u) &=T_+(u)-\frac{\pi^2}{4L^2}p'^2,\\
\hat{T}_-(u) &=T_-(u)-\frac{\pi^2}{4L^2}q'^2,
\end{split}
\ee
where $T_{\pm}$ is given by the previous one introduced in \eqref{emgt}.  
This agrees with the CFT result \eqref{emcftl2} and the result (\ref{eq:holEM}) obtained from the exact map.

\section{More on entanglement entropy}
\label{sec:moreEE}

In this appendix, we present more results for entanglement entropy in both holographic CFT and free Dirac fermion in different setups.

\subsection{Escaping mirror}

\begin{figure}[H]
    \centering
    \includegraphics[width=8cm]{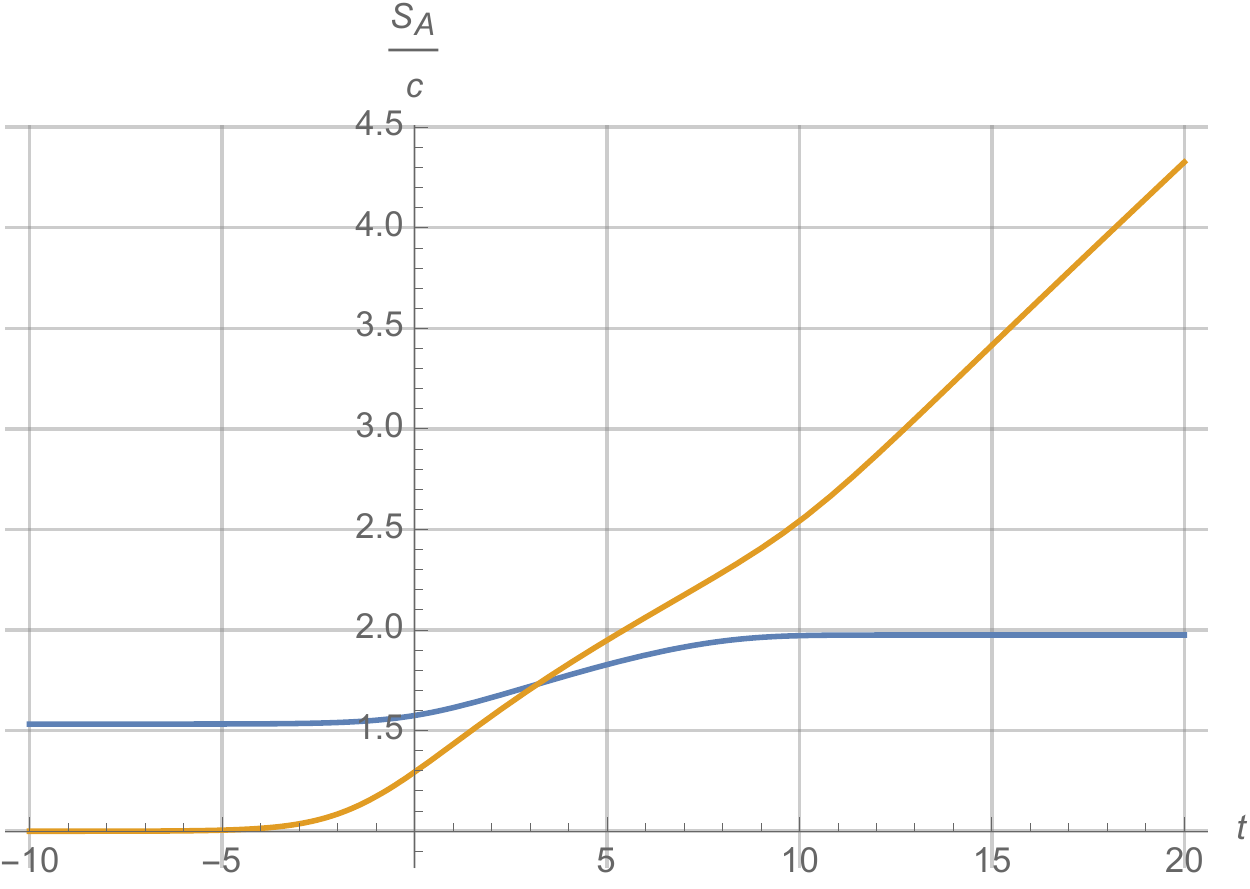}
    \includegraphics[width=8cm]{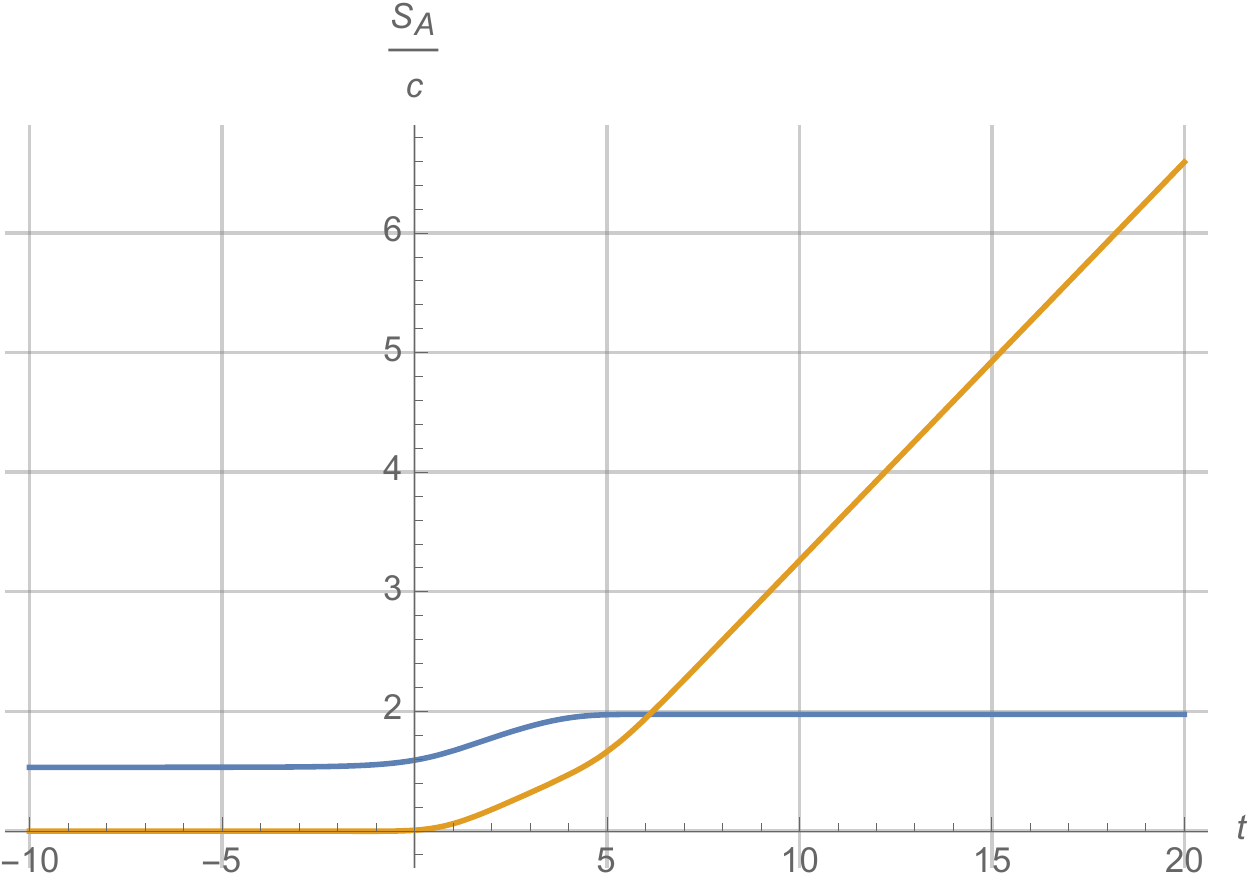}
    \includegraphics[width=8cm]{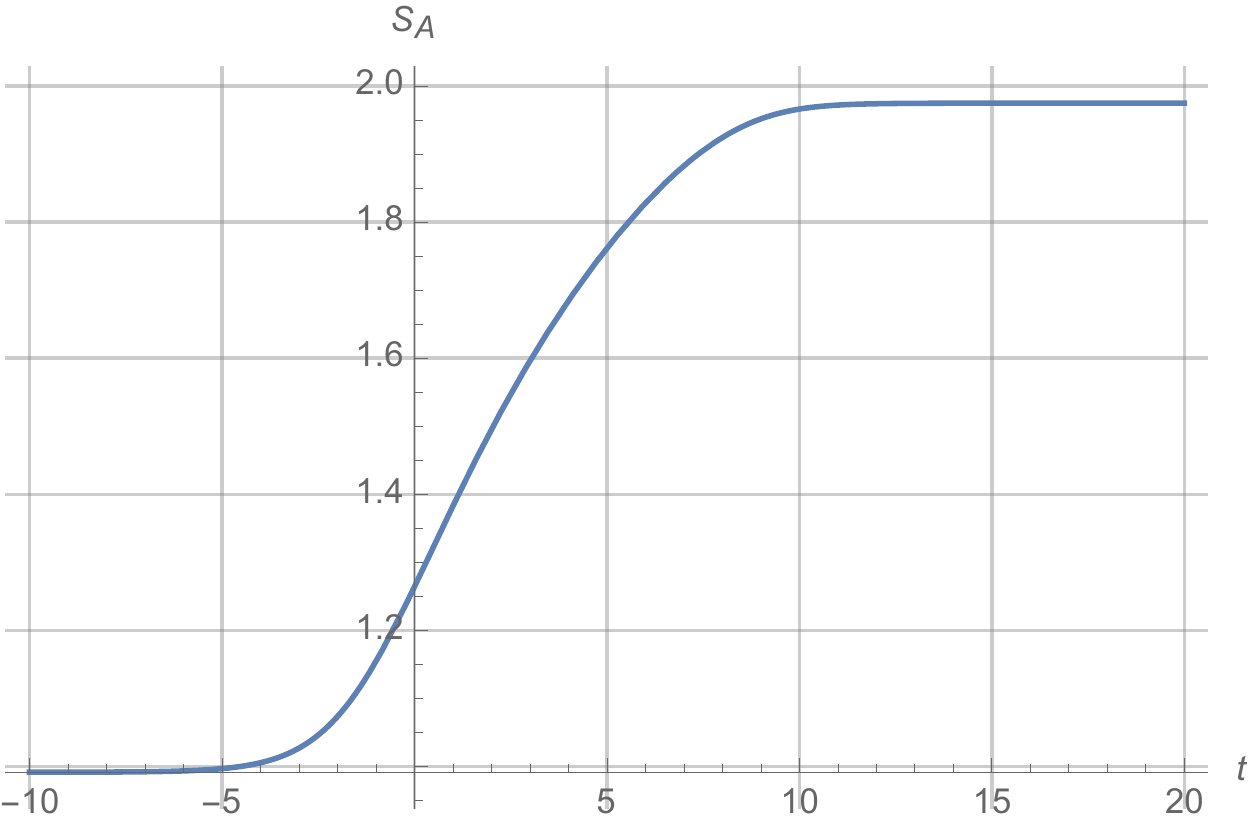}
    \includegraphics[width=8cm]{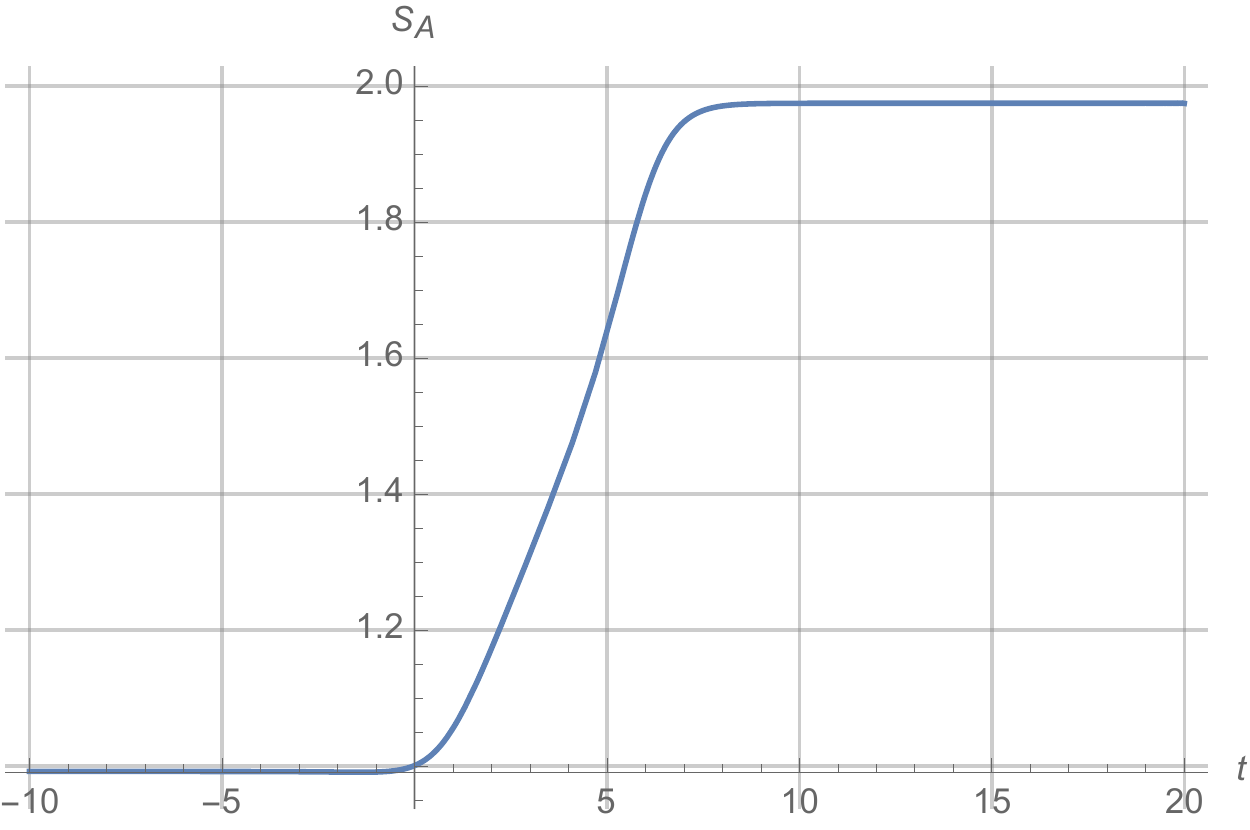}
    \caption{Entanglement entropy (regularized by $c$) under time evolution in the (single) escaping mirror setup for holographic CFT (upper half) and free fermion (lower half). The left column corresponds a fixed interval $A=[0.1,10]$. The right column corresponds a comoving interval $A=[Z(t)+0.1,Z(t)+10]$. In the plots for holographic CFT, the blue curve and the orange curve show the connected entanglement entropy and the disconnected entanglement entropy (where the boundary entropy is set to $S_{\rm bdy}=0$ for simplicity), respectively. We have set $\beta=1$, $\epsilon=0.1$.}
    \label{fig:SMEEhol}
\end{figure}

Plugging the following conformal transformation into (\ref{eq:EEhol}) and (\ref{eq:EEfermion}),
\begin{align}
    &\tilde{u} = p(u) = -\beta \log(1 + e^{-u/\beta}), \\
    &\tilde{v}= q(v) = v,
\end{align}
$S_A$ can be computed in the holographic CFT and for the free Dirac fermion, respectively, see Fig.~\ref{fig:SMEEhol}.

\subsection{Kink mirror}

\begin{figure}[H]
    \centering
    \includegraphics[width=8cm]{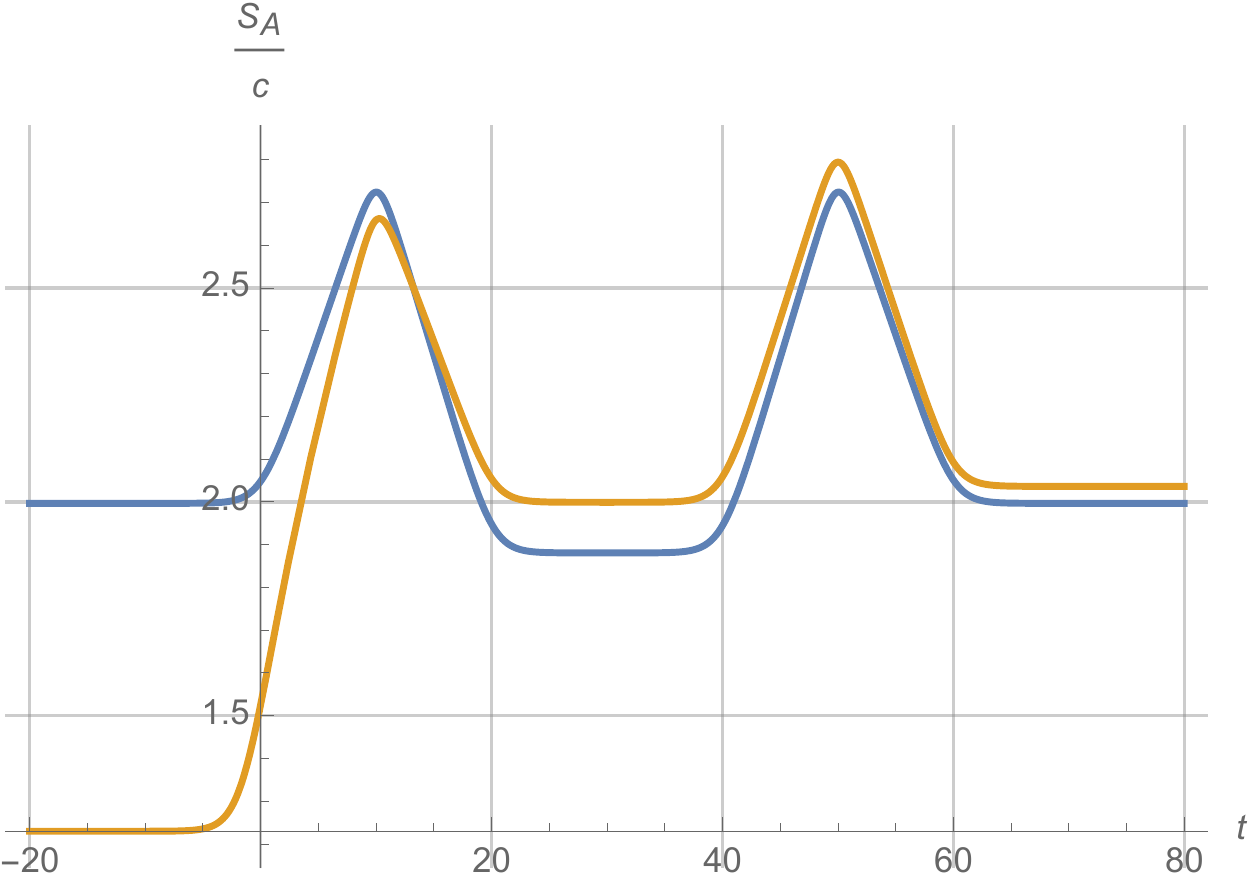}
    \includegraphics[width=8cm]{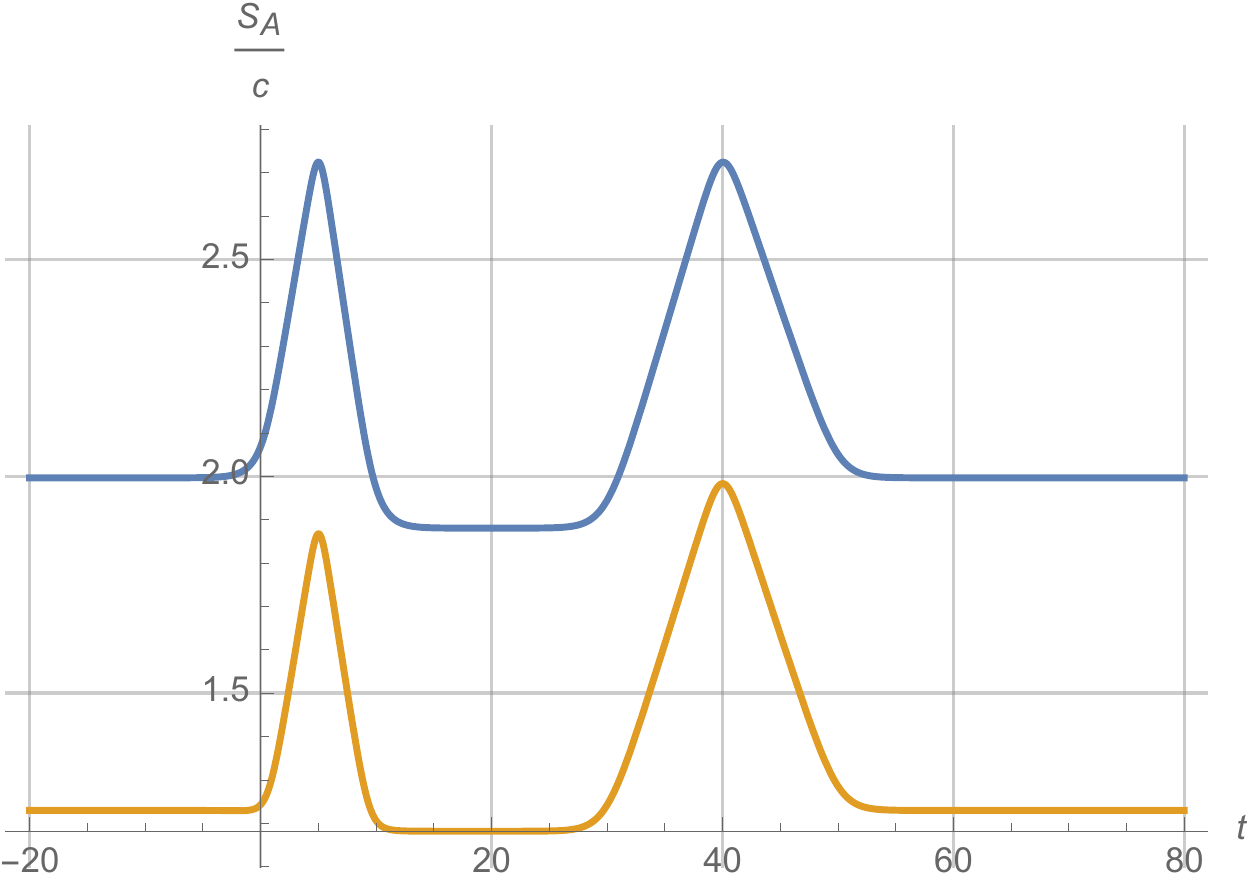}
    \includegraphics[width=8cm]{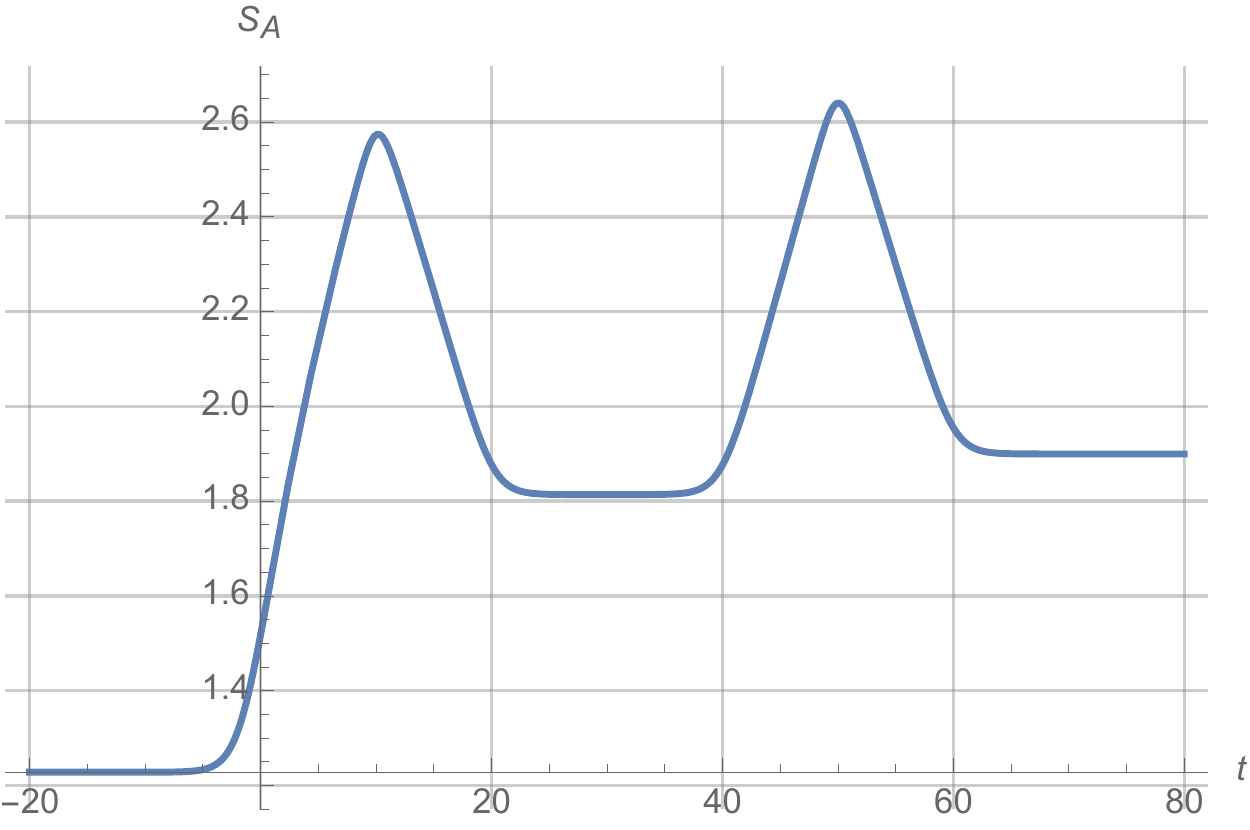}
    \includegraphics[width=8cm]{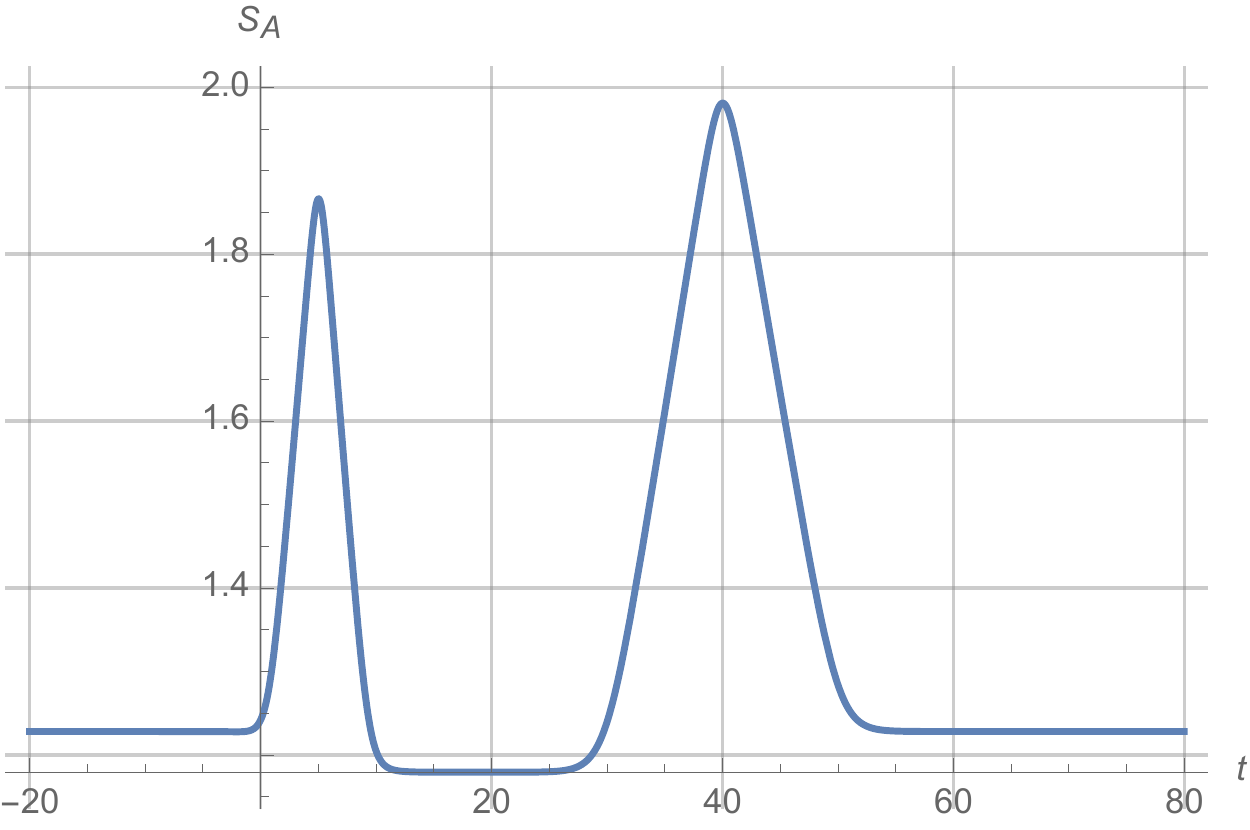}
    \caption{Entanglement entropy (regularized by $c$) under time evolution in the (single) kink mirror setup for holographic CFT (upper half) and free fermion (lower half). The left column corresponds a fixed interval $A=[0.1,40]$. The right column corresponds a comoving interval $A=[Z(t)+0.1,Z(t)+40]$. In the plots for holographic CFT, the blue curve and the orange curve show the connected entanglement entropy and the disconnected entanglement entropy (where the boundary entropy is set to $S_{\rm bdy}=0$ for simplicity), respectively. We have set $\beta=1$, $u_0 = 20$, $\epsilon=0.1$.}
    \label{fig:SkMEEhol}
\end{figure}

For the (single) kink mirror, plugging the following conformal transformation into (\ref{eq:EEhol}) and (\ref{eq:EEfermion}),
\begin{align}
    &\tilde{u} = p(u) = -\beta \log(1 + e^{-u/\beta})+\beta \log(1 + e^{(u-u_0)/\beta}), \\
    &\tilde{v}= q(v) = v,
\end{align}
$S_A$ can be computed in holographic CFT and free Dirac fermion CFT, respectively. Refer to Fig.~\ref{fig:SkMEEhol}. The behavior can be qualitatively explained by the quasi-particle picture depicted in Fig.~\ref{kinkeheefig}. Comoving intervals experience a Doppler effect and hence the two peaks have different widths. 

\subsection{Double escaping mirror}

\begin{figure}[H]
    \centering
    \includegraphics[width=5.2cm]{DEs1.pdf}
    \includegraphics[width=5.2cm]{DEs2.pdf}
    \includegraphics[width=5.2cm]{DEs3.pdf}
    \includegraphics[width=5.24cm]{DEs12.pdf}
    \includegraphics[width=5.24cm]{DEs22.pdf}
    \includegraphics[width=5.24cm]{DEs32.pdf}
    \includegraphics[width=5.24cm]{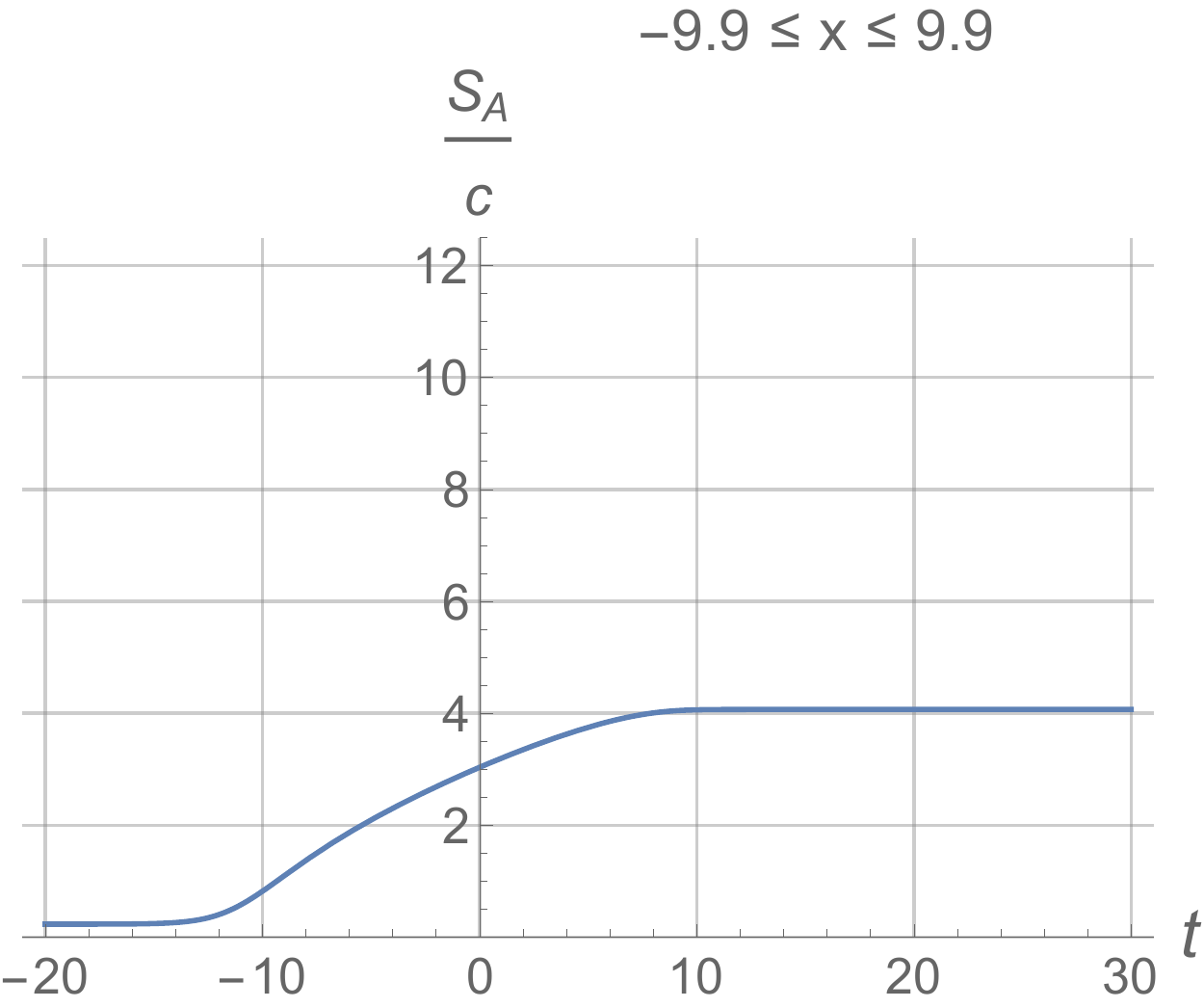}
    \includegraphics[width=5.24cm]{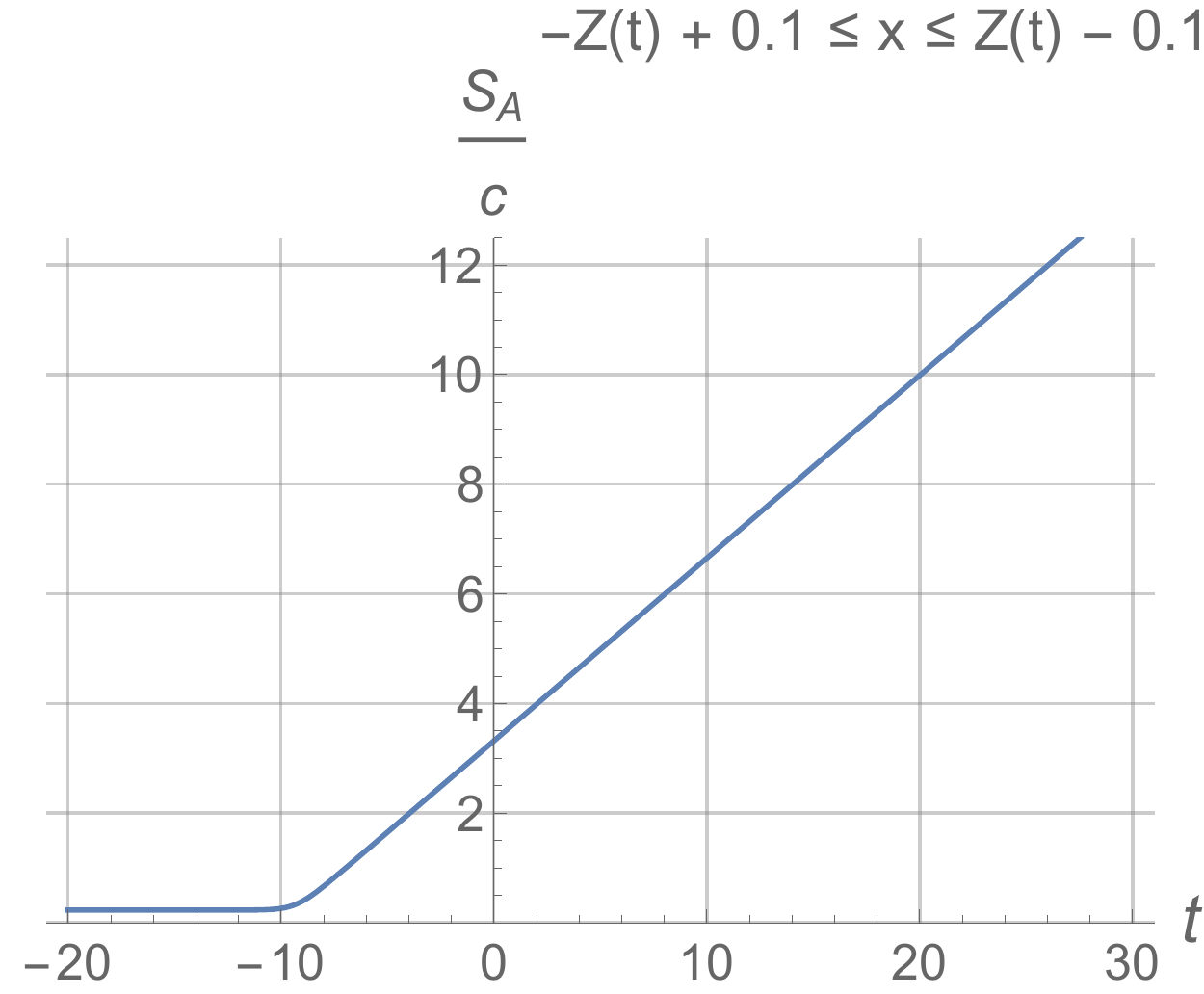}
    \includegraphics[width=5.24cm]{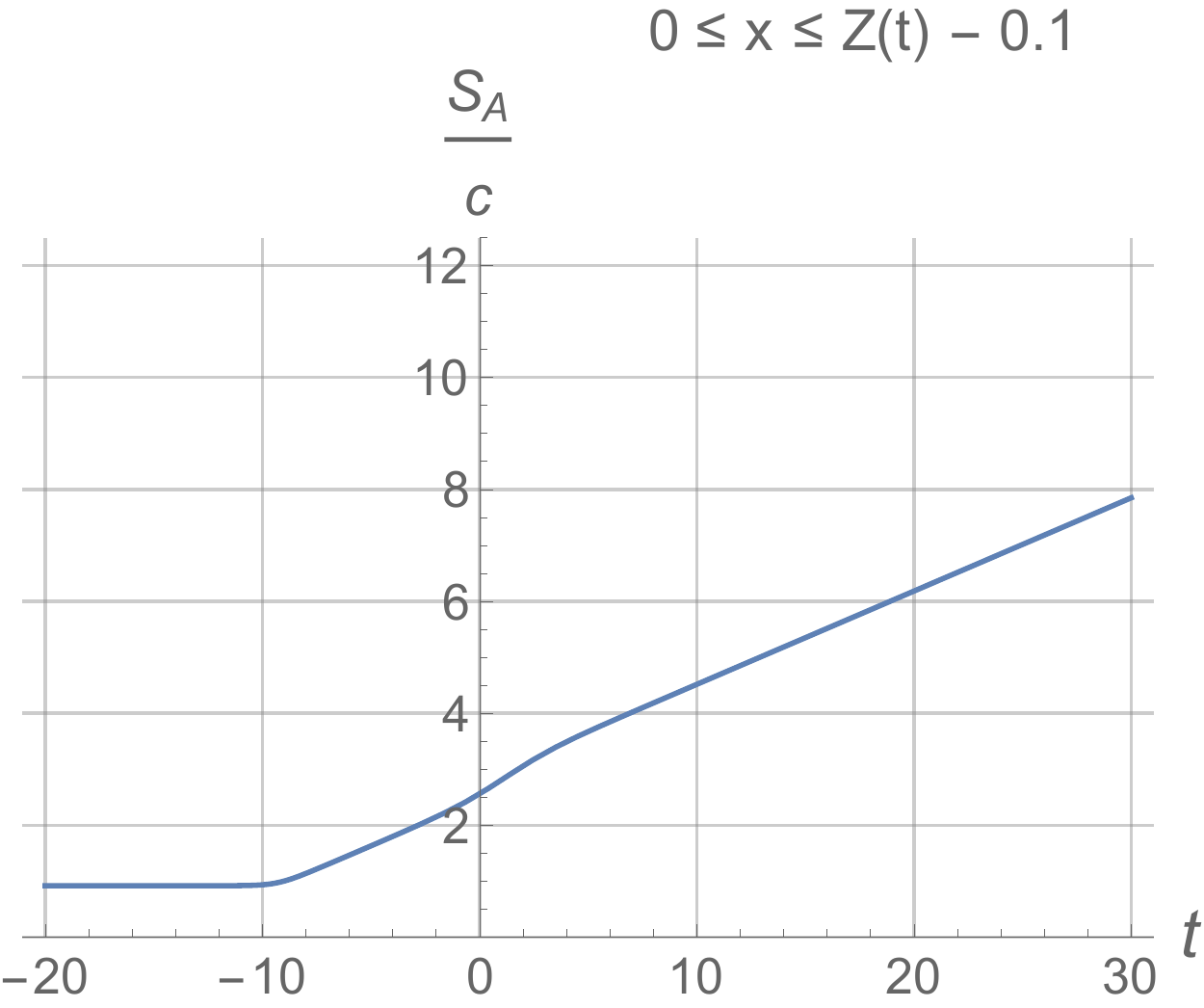}
    \caption{Entanglement entropy in a double escaping mirror setup for holographic CFT (the second row) and free fermion (the third row). The left column shows the case, where two edges of $A$ comove with the mirror trajectory. The middle column shows the case, where two edges of $A$ comove with the mirror trajectory. The right column shows the case, where one edge of $A$ is fixed and the other one comoves. We have set $\beta=1$, $L=20$, $\epsilon=0.1$.}
    \label{fig:DEsfermion}
\end{figure}

Plugging the following conformal transformation into (\ref{HEEdeconfine}) - (\ref{HEEdeconfine2}) and (\ref{eq:EEDMfermion}),
\begin{align}
    &\tilde{u} = p(u) = -\beta \log(1 + e^{-u/\beta}), \\
    &\tilde{v}= q(v) = -\beta \log(1 + e^{-v/\beta}),
\end{align}
$S_A$ can be computed in holographic CFT (for the confined configuration) and free Dirac fermion, respectively. Refer to Fig.~\ref{fig:DEsfermion}. A part of Fig.~\ref{fig:DEsfermion} has already been shown in Fig.~\ref{DEsconfined}. 

\bibliographystyle{JHEP}
\bibliography{JoinCFTs_bib}

\end{document}